\documentclass[acmsmall]{acmart}

\setcopyright{cc}
\setcctype{by}
\acmJournal{PACMHCI}
\acmYear{2025} \acmVolume{9} \acmNumber{7}  \acmArticle{CSCW487}
 \acmMonth{11} \acmPrice{}\acmDOI{10.1145/3757668}
\usepackage{multirow} 
\usepackage{microtype}
\usepackage{tikz-network}
\usepackage{tikz}
\usetikzlibrary{arrows.meta, positioning}
\usepackage{xcolor}
\definecolor{myblue}{RGB}{135,206,235}
\newcommand{\ie}{\textit{i.e.,~}}
\newcommand{\eg}{\textit{e.g.,~}}

\usepackage{booktabs}
\usepackage{graphicx}
\usepackage{hyperref}

\usepackage{bm}
\renewcommand*{\vec}[1]{\bm{#1}}
\usepackage{mathtools}
\usepackage{longtable}
\usepackage[]{lipsum}

\usepackage{subcaption}

\usepackage{url}
\usepackage{enumitem}
\setlist[description]{leftmargin=0in,labelindent=\parindent}
\usepackage{amsmath}
\usepackage{float}
\usepackage{bbm}
\usepackage{verbatim}
\usepackage{pgf}
\usepackage[]{cleveref}

\crefname{section}{\S}{\S} 
\crefname{subsection}{\S}{\S} 
\crefname{subsubsection}{\S}{\S}
\Crefname{subsubsection}{\S}{\S}

\usepackage{tabto}
\usepackage[rightcaption]{sidecap}
\usepackage[]{ragged2e}
\usepackage{tabularx}

\AtBeginDocument{%
  }

\acmISBN{978-1-4503-XXXX-X/18/06}

\newcommand{\method}{ConvTreeTrans}

\begin{document}

\title[The Chilling]{The Chilling: Identifying Strategic Antisocial Behavior Online and Examining the Impact on Journalists}

\author{Yian Wang}
\affiliation{%
  \institution{University of Illinois, Urbana-Champaign}
  \city{Urbana}
  \state{Illinois}
  \country{USA}
}
\email{yian3@illinois.edu}
\author{Mukhilshankar Umashankar}
\email{mu9@illinois.edu}
\affiliation{%
  \institution{University of Illinois, Urbana-Champaign}
  \city{Urbana}
  \state{Illinois}
  \country{USA}
}

\author{Eshwar Chandrasekharan}
\affiliation{%
  \institution{University of Illinois, Urbana-Champaign}
  \city{Urbana}
  \state{Illinois}
  \country{USA}
  }
\email{eshwar@illinois.edu}

\author{Hari Sundaram}
\affiliation{%
  \institution{University of Illinois, Urbana-Champaign}
  \city{Urbana}
  \state{Illinois}
  \country{USA}
  }
\email{hs1@illinois.edu}
\renewcommand{\shortauthors}{Yian Wang, Mukhil Shankar, Eshwar Chandrasekharan, and Hari Sundaram}

\begin{abstract}
On social platforms like Twitter, strategic targeted attacks are becoming increasingly common, especially against vulnerable groups such as female journalists. Two key challenges in identifying strategic online behavior are the complex structure of online conversations and the hidden nature of potential strategies that drive user behavior.
To address these, we develop a new tree-structured Transformer model that categorizes replies based on their hierarchical conversation structures, offering insights into the latent strategies underlying these interactions.
Extensive experiments demonstrate that our proposed classification model can effectively detect different user groups---namely attackers, supporters, and bystanders---and their latent strategies.
To demonstrate the utility of our approach, we apply this classifier to real-time Twitter data and conduct a series of quantitative analyses on the interactions between journalistswith diverse cultural backgrounds and different groups of users---attackers, supporters, and bystanders.    
Our classification approach allows us to not only explore strategic behaviors of attackers but also those of supporters and bystanders who engage in online interactions.
When examining the impact of online attacks, we find a strong correlation between the presence of attackers’ interactions and \textit{chilling effects}, where journalists tend to slow their subsequent posting behavior.
Additionally, we find that attackers tend to negatively influence the posting behavior of other users within these conversations. As conversations deepen, replies often deviate from original posts and get more toxic. 
This paper 
provides a deeper understanding of how different user groups engage in online discussions and highlights the detrimental effects of attacker presence on journalists, other users, and conversational outcomes. 
Our findings underscore the need for social platforms to develop tools that address coordinated toxicity and foster healthier conversation dynamics. By detecting patterns of coordinated attacks early, platforms could limit the visibility of toxic content to prevent escalation. Additionally, providing journalists and users with tools for real-time reporting and de-escalation could empower them to manage hostile interactions more effectively. Enhanced moderation tools targeting coordinated behaviors, particularly among attackers, could ensure a safer environment for vulnerable groups like female journalists, ultimately supporting constructive discussions and resilient online communities.
\end{abstract}


\received{October 2024}
\received[revised]{April 2025}
\received[accepted]{August 2025}
\maketitle

\section{Introduction} \label{sec:1}
In this paper, we focus on understanding the strategic behaviors of users on social media, particularly in the context of coordinated attacks against female journalists. \textcolor{black}{Prior works~\cite{Posetti2021,possetti2022chilling,Posetti2023}\footnote{https://www.aspistrategist.org.au/smart-asian-women-are-the-new-targets-of-ccp-global-online-repression/} have demonstrated} that female journalists from different countries facing a disproportionately high volume of such attacks, with women journalists attacked for their identity, including their gender, religious and sexual orientation, and receive violent, physical threats. This has a \textit{chilling}  effect on their interactions with their followers. We show the extent and nature of some these attacks from our dataset in Table~\ref{tab:attack}. \footnote{The table shows examples of misogynistic and racially motivated attacks. These may be distressing to the reader. We include them to underscore the harm to women journalists in online spaces.} An Intriguing suggestion in these works~\cite{Posetti2021,Posetti2023} is that these attacks are coordinated, possibly at the behest of the state ``Abuse against [..] comes at a very high speed, sometimes within seconds of her posting a tweet, and there can be a visible spike in abusive replies within one to two minutes of a post. This pattern is highly unusual and it could signal coordinated campaigns of abuse.''~\cite{Posetti2023}. In addition to these attacks, a considerable number of users engage in debates with attackers to support the journalists, often employing similar strategies within their respective groups. 

In this paper, we ask \textit{can we identify whether these groups of users (\textit{e.g.,} attackers and supporters) are engaging in coordinated behavior, and if so, can we characterize the strategies they utilize to manipulate social media?} Uncovering patterns of coordination among attackers can help platforms take action to mitigate harm, including improving platform design, developing novel moderation tools addressing groups of attackers and giving journalists more control of their audience.

    \begin{table}[t]
        \small
    \centering

    \begin{tabularx}{\columnwidth}{@{}>{\RaggedRight\arraybackslash}p{0.48\columnwidth}>{\RaggedRight\arraybackslash}p{0.48\columnwidth}@{}}
     \toprule
    \textsc{Journalist A} & \textsc{Journalist B}\\
    \midrule
    \textit{``your mother is so pitiful'' }\textcolor{gray}{8:09 PM · Nov 30, 2022} & \textit{``A yellow face promoting a morden yellow peril.'' }\textcolor{gray} {2:16 AM · Feb 13, 2022} \\\addlinespace[0.5em]
    \textit{``Isn't it tiring to wear a mask every day and pretend to be a good person? You are just a liar.''} \textcolor{gray} {9:13 AM · Dec 13, 2022} & \textit{``Until you two anti-china fanatics got hurt in china nobody trust your china-bad messages'' }\textcolor{gray}{9:44 PM · Feb 12, 2022}\\\addlinespace[0.5em]
    \textit{``You're doing fine being an American dog!''} \textcolor{gray}{2:03 AM · Dec 1, 2022} & \textit{``Funny banana, you have no guts to write nasty but true things about the States, badmouth China is somthing hypocritical indeed''} \textcolor{gray}{8:58 AM · Feb 12, 2022}\\
    \bottomrule
    \end{tabularx}
    \caption{Examples of attacks against female journalists on Twitter.Those replies are personal attacks against the journalists, so we consider them as attackers. Please see~\cref{sec:defnition} for a more detailed definition.}
    \label{tab:attack}
    \vspace{-0.33in}
    \end{table}

Prior work has investigated this dynamic between attackers and supporters by collecting and examining users' online behaviors~\cite{saveski2021structure,  zafarani2013connecting, cao2014uncovering, cai2023hate}. 
These studies provide valuable insights into the nature of online interactions and the responses they elicit, but they do not focus on the \textit{structure} of these responses. As various users respond to a tweet, they form reply trees, which originate from the initial tweet and branch through subsequent replies. 
The methods described above struggle to capture these complex structures and are unlikely to perform effectively in classification tasks on large social platforms like Twitter.
A different group of works concentrates on identifying coordination through user behaviors~\cite{sharma2021identifying, weber2021amplifying, keller2020political, mariconti2019you}. These works typically focus on single coordination strategies, for example, the timing of posts, to spread disinformation. However, attackers often collectively use composite strategies beyond just timing, covering more complex factors like the content of posts, user preferences, and the viewpoints expressed in earlier tweets. 

To address these challenges, we propose four research questions:
\begin{itemize}
    \item [\textbf{(RQ1)}]How can we identify distinct groups corresponding to attackers, supporters, and bystanders from empirical data?
    \item [\textbf{(RQ2)}]What strategies are used by the three groups described above? 
    \item [\textbf{(RQ3)}] What are the effects of attacker presence and intensity on journalists (e.g., their activity levels, posting frequency) within the tweet replies and across tweets?
    \item [\textbf{(RQ4)}] How does the presence of attackers (and supporters) affect the conversational outcomes? 
\end{itemize}

\begin{figure}[htbp]
    \centering
    \includegraphics[width=0.95\columnwidth]{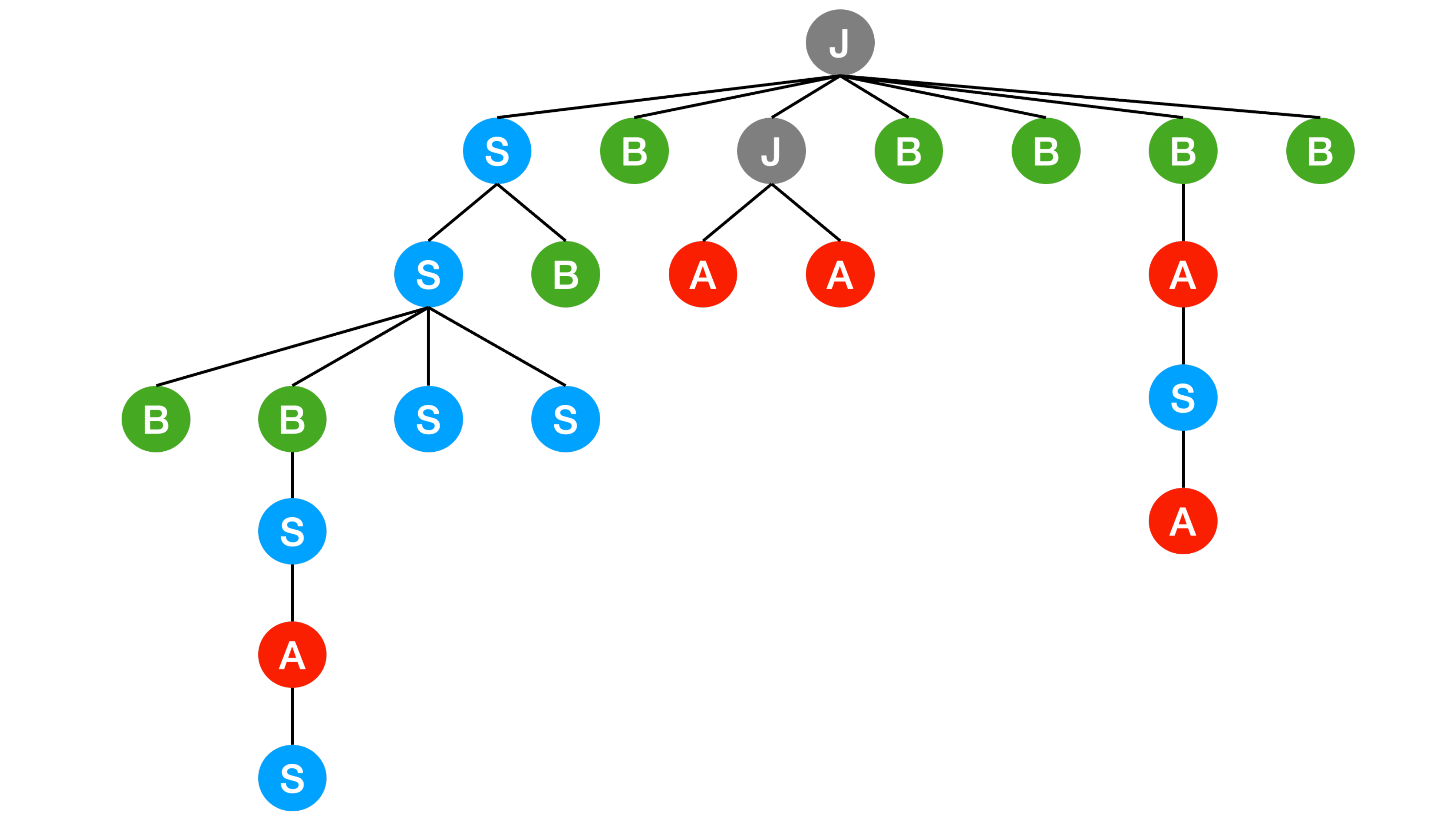}
    \caption{An example conversation tree. The root node denotes a post created by a journalist; the remaining nodes signify replies or subsequent replies to initial responses. Red nodes denote attackers towards the journalist, blue nodes denote supporters, and green nodes represent bystanders. See \cref{sec:1}}
    \label{fig:tree}
\end{figure}

We address these questions by first developing a novel transformer-based framework, \method~to \textit{jointly} identify user groups (RQ1) and group strategies (RQ2) from the structure of conversations. Our unit of analysis is the conversation tree (\Cref{fig:tree}), which captures the branching nature of online conversations. The central insight is that classifying the participant as an attacker, supporter, or bystander requires us to attend to both local and global context in the tree structure. We then use these labels to identify the strategies employed by different user groups. Notice that the behavior (\eg posts or replies) are observable, but the strategies are not observable (\eg reply to a similar user). We use the term strategy to refer to the latent behavior of the user groups. Extensive experiments show that our framework outperforms state-of-the-art models for identifying strategies and classifying user accounts (RQ1).




Our contributions are as follows:
\begin{description}
\item[Online User Classification and Strategy Discovery: ]To the best of our knowledge, we are the first to uncover \textit{hidden} and composite strategic behaviors on social networks. Previous work~\cite{sharma2021identifying} has primarily focused on identifying a single strategy (e.g. temporal coordination) that is used to spread disinformation, without accounting for the conversational structure or targeted attacks on individuals. We developed \method, a transformer-based framework designed for tree-structured data, which enables the identification of attackers, supporters, and bystanders, along with their strategies in online discussions. While we have applied our framework towards understand strategic behavior directed against journalists, the framework can be applied to other domains.

\item[Distinct strategies across three groups:] Our extensive analysis uncovers diverse strategic behavior patterns across three user groups: attackers, supporters, and bystanders. 
Our insight is to distinguish between alignment of the topic of the reply of the user with the original journalist post and the toxicity of the reply. Doing so, allows identifying bystanders, who are neutral towards the original post. This distinction is crucial for understanding the dynamics of online conversations and identifying the strategies employed by different user groups.

\item[Impact of Attackers:] We find that attackers are coordinated, targeting specific users and topics to maximize their impact and influence the conversation. In contrast, bystanders exhibit a more passive engagement pattern, while supporters do not demonstrate a clear strategic approach (RQ2). 
We observe a strong correlation between online attacks and changes in journalists' posting behavior, with a statistically significant difference in the average time they take to respond after receiving toxic replies (RQ3).
We find that as conversations increase (either over time or depth), there are fewer supporters and attackers, and more bystanders and the toxicity of the conversation increases, and the time between replies decreases. Further, approximately a third of the attackers and supporters begin to attack each other, suggesting that they are attacking each other and engaged in toxic conversations (RQ4).
\end{description}

\textit{Our impact \footnote{We plan to publish the code and data once the paper is accepted for publication.}:} By identifying these strategic antisocial behaviors, our work can detect early coordinated attacks and inform the development of platform policies aiming to regulate harmful interactions, ultimately contributing to healthier online environments. Our framework have implications for improving platform design, such as real-time toxicity detection and conversational interventions that prevent discussions from escalating. 

\section{Related Work} \label{sec:7}
In our work, we model strategic behavior on social networks via tree-structured transformers to discover strategies of coordinated attackers. We provide a brief overview of related past work.

\subsection{Online Antisocial Behavior on Twitter}
Several studies have focused on identifying toxic or anti-social behavior on Twitter through lexical analysis. For example, \citet{pavalanathan2017multidimensional} constructed a theoretically motivated lexicon of stance markers and applied multidimensional analysis to identify underlying stances. Similarly, \citet{addawood2017stance} developed a machine learning based method that leverages various features including lexical and Twitter-specific content to classify users' opinions. However, considering the complexity of online conversations, it is crucial to also consider structural information from social networks into these analyses. For instance, \citet{saveski2021structure} examined how Twitter conversation structures are related to toxicity scores at different levels, which is helpful in predicting the toxicity of future replies. \citet{cao2014uncovering} proposed a detection framework to identify large groups of malicious accounts across online social networks. Additionally, \citet{sharma2021identifying} introduced a model that jointly models account activities and latent group behaviors based on temporal information. While these studies are effective in detecting toxic or coordinated behavior, they do not explore the underlying composite strategies that users employ in social interactions.  

\subsection{Coordination Among Users}
Coordination among users on social platforms has been extensively studied in game theory. In simultaneous coordination games, agents benefit from selecting the same strategy, as shown by \citet{cooper1990selection} and \citet{gale1995dynamic}. We apply this coordination framework to model user interactions on Twitter. Beyond game theory, a substantial body of work explores coordination within networks \cite{arditti2021equilibria, tomassini2010coordination, weber2021amplifying, schoch2022coordination}. For instance, \citet{keller2020political} use a principal-agent model to identify disinformation campaigns and their influence on public opinion. Similarly, \citet{jackson2002formation} model network-based coordination by forming edges within an interaction network, and reveals multiple stochastically stable states through simulation. Unlike classic game theory approaches, our study does not assume that users are inherently inclined to coordinate; instead, we analyze real-world Twitter data to uncover coordination strategies among users.

Other studies focus on detecting "hate raids," a form of group harassment in video communities. \citet{mariconti2019you} investigate harassment against YouTube content, proposing an ensemble model to identify videos likely to attract coordinated attacks. In live-streaming communities, \citet{cai2023hate} examine harassment received by users and suggest moderation techniques to mitigate this behavior. Our work differs from these works by focusing on coordination specifically on Twitter—a platform with a much larger user base and more intricate conversation structures compared to other online communities. Moreover, we consider not only attackers and supporters but also the bystanders who are often overlooked but play a critical role in shaping interactions.

\subsection{Discovering Strategic Behavior on Social Networks}
While many works focus on modeling users' behaviors on social networks, only a few attempt to explore the strategies behind these behaviors.
~\citet{xu2012modeling} used Twitter users' posting behavior to infer their motivation to create and share content.
Similarly, \citet{dong2014inferring} inferred demographic attributes and social strategies of online users by their behaviors.
These prior works used visible user attributes and behaviors to make assumptions regarding hidden attributes, like strategies. However, these works do not consider the collaboration or coordination of these users. 
In one such application, \citet{xiao2020discovering} proposed a dynamic attention network model to discover collaborative content production strategies. In this work, collaboration is obvious---co-authorship on a piece of content is public information. In our work, we tackle the more difficult task of discovering strategic behavior among users who coordinate secretly. 

\subsection{Prior Work Using Tree-based Transformers}
The Transformer model~\cite{vaswani2017attention} on attention mechanisms has proven to be highly effective in learning language representations. 
However, its absolute positional encoding struggles to capture positional information in other structured data such as graphs and trees. 
To address this issue,~\citet{shaw2018self} introduced relative positional encoding to Transformer models; ~\citet{shiv2019novel} extended this to tree-structured data. 
~\citet{peng2021integrating} further proposed to learn structure and context together by introducing bias in the self-attention of code context with the underlying tree structure, then ~\citet{peng2022rethinking} incorporated both global and local information. In our work, we build this existing work on tree-based Transformers, incorporating both local and global position encoding. Specifically, we apply this mechanism to conversation trees.

\section{Problem Statement} \label{sec:2}
In this section, we describe the problem this paper aims to address. We begin with a problem overview---this includes the general tasks that must be achieved. Then, we give a formal problem definition. 
\subsection{Problem Overview} \label{sub:2.1}
Consider a journalist $A$ who posts a tweet $t_0$ and obtains $n$ replies $\{t_i\}_{i=1}^n$. We associate attributes with each reply including language, number of retweets, replies, and likes, number of impressions, whether it includes media content like URLs, and so on.
To describe reply behaviors of social media users in response to a post on Twitter, we must first complete two tasks: identification of user groups, and discovery of their strategies.

\textbf{Group Identification: }
In our research, we categorize replies to a journalist’s post into three distinct groups: those that support the journalist’s viewpoint (supporters), those that oppose it (attackers), and those that remain neutral (bystanders). \textcolor{black}{Throughout the remainder of the paper, we refer to posts as attackers, supporters, or bystanders unless further clarification is needed. One advantage of labeling at the post level rather than at the individual level is that we can see how attitudes shift for a person; e.g., a supporter can become an attacker/bystander over the duration of the conversation.}

Nowadays, online attacks are often observed on social networks where groups of users conduct various malicious activities including spreading misinformation, posting toxic replies, and harassing other users. Attackers may use these strategies to distribute spam, fake news, or manipulate public opinion through coordinated campaigns~\cite{cao2014uncovering}. \textcolor{black}{Prior research in online harassment has frequently identified distinct participant roles, notably attackers, supporters, and bystanders. In the context of hate speech and harassment research, attackers are defined as those who initiate or participate in abusive behavior—using hateful language, identity-based slurs, or coordinated tactics to target individuals or communities.~\citet{cai2023hate} have shown that attackers are identified actively initiating or participating in abusive behavior, frequently through coordinated actions involving mass bots or human collaboration to amplify harm, as observed in hate raids on Twitch, where attackers flood channels with hateful or abusive messages through human-bot coordinated attacks.} \textcolor{black}{At the same time, social networks serve as venues for targeted harassment, where certain groups, such as female journalists, are especially affected. These journalists face a high volume of attacks not only because of their gender but also due to their role in sharing opinions, reporting controversial issues, and engaging with the public. Their visibility makes them frequent targets of tactics such as discrediting their work, silencing their voices, or undermining their credibility.}

\textcolor{black}{In contrast, supporters are those who actively oppose these attacks through positive engagement and interventions to protect or support the targeted individuals.  Prior work~\cite{blackwell2017classification} demonstrates that supporters are those who can provide emotional support or moderation assistance, helping mitigate the impact of harassment incidents. Finally, bystanders are users who neither actively participate in the attack nor engage in supportive behavior. Although often overlooked, bystanders can significantly influence the dynamics of a conversation by their presence or passive responses.
In~\cite{blackwell2017classification}, bystanders are explicitly defined as potential allies who can actively support harassment victims through platforms like HeartMob. \citet{cai2023hate} treat bystanders—such as Twitch viewers—as largely passive observers affected by hate raids, and although~\citet{cao2014uncovering} do not explicitly mention bystanders; they treat regular users as passive recipients of malicious content, with protection handled entirely through platform-level interventions.}

However, classifying these replies is challenging, especially when there are no explicit labels indicating the stance. To address this problem, we model each conversation as a tree structure, which allows us to capture the interactions among replies more effectively. By leveraging the capabilities of the Transformer model~\cite{vaswani2017attention}, this approach provides a deeper understanding of the underlying structure of conversation trees, facilitating more accurate classification and analysis.


\textbf{Strategy Discovery: }
We assume that when deciding whether to reply to a post, a user makes a strategic decision by taking not only the post but also other factors into account.
These other factors might include the topic of the post, the individuals they are responding to, and the timing of the replies.
Within this context, we face several challenges. First, the number of attributes to consider is extensive. For example, when a user decides which post to respond to, they typically review all previous replies. This process involves not just seeing what has been said, but also understanding the context and content of those responses. While the topic and tone of replies are visible, the strategic thought process behind these choices is not. Understanding these strategies is crucial as it helps us understand why users choose to engage with certain posts and not others, which allows us to comprehend the dynamics of discussion on social networks, including how opinions are formed and spread.
Therefore, we can ask:
\begin{quote}
\textit{Is it possible to identify attackers, supporters, and bystanders within these conversations? Do the strategies employed by each group differ, and how do attackers influence the behaviors of supporters, bystanders, and journalists?}
\end{quote}
\subsection{Problem Definition} \label{sec:2.2}
Consider a post $t_a$ made by journalist $A$, and let $t_i$ represent a reply to this post. Considering the dynamics of this response along with its metadata, our objective is to determine the group to which $t_i$ is affiliated, that is, to optimize the probability that the reply $t_i$ is associated with the class $C_k$ represented as $P(C_k|t_i)$ where $k$ denotes attackers, supporters or bystanders. This allows us to construct a tree-like structure for each post by journalist $a$, which can capture both the depth and breadth of the interactions that originate from the initial post. This structured approach, along with the content information, enables us to categorize responses, as well as identify specific strategies used in each reply. To better represent a reply $t_i$, we incorporate both its metadata, denoted as $x_i$, and its structural information. The metadata includes features such as language, creation time, number of retweets, replies, likes, views, toxic scores, topics, and the presence of URLs. In addition to these features, we also capture global information, denoted as $p_i$, which includes the path from the root to node $i$, as well as the local interaction between the reply $t_i$ and its parent node $j$, represented as $r_{ij}$. This combined approach allows for a richer representation of each reply within the conversation.

We model users' observed actions as a generative process with two assumptions: (1) The actions are generated by a mixture model where a strategic decision changes the posterior distribution over the action space, i.e., $P(a_i|s_j, z_i, C_i)$ where $s_j$ denotes the $j$-th strategy, $z_i$ represents the representation of reply $i$ and $C_i$ the class that reply $i$ belongs to. (2) While the set of strategies, denoted as $\mathcal{S}$, is common to all, each individual's strategy is private and only the actions are visible.

For example, when a user decides which post to respond to, they might select a post with a high number of likes or simply reply to posts at random. In a more formal sense, the selection of any strategy $s \in \mathcal{S}$ influences the subsequent distribution of attribute values. We also hypothesize that when users coordinate their efforts, they likely engage in negotiations through private communication channels, such as direct messages or private messaging platforms. These interactions, not publicly observable, allow them to reach a consensus on their strategies, suggesting they share similar distributions of tactics.

Thus we can characterize the likelihood of an action $a_i$ with the sum of the total probability over all strategies:
\begin{equation}
p(a_i|z_i, C_i) = \sum_{s_j \in \mathcal{S}}p(a_i|s_j, z_i, C_i)p(s_j|z_i, C_i)
\end{equation}
where $a_i \in [0, 1]$ represents the replying action. 
Now the problem is to define $p(a_i|s_j, z_i, C_i)$ and $p(s_j|z_i, C_i)$ for each reply $t_i$.

Formally, our objective is to classify tweet replies and uncover strategies underlying users' behaviors. This can be articulated as follows:

\textbf{Input:} (1) Features of replies $\textbf{x}_i$ for each reply $i$; (2) global path $\mathbf{p}_i$ for each reply $i$; (3) local path $\mathbf{r}_{ij}$ for each reply $i$ and its parent node $j$; (4) a joint model that minimizes both the loss for classification and that for strategy discovery. 

\textbf{Output:} (1) classification result for each response, categorizing them as attackers, supporters, or bystanders to a journalist's posts; (2) a strategy distribution $P(S)_i$ for each reply $i$: $P(s_j|z_i, S_i)$.

\section{Datasets} \label{sec:5.1}
\subsection{Dataset Description}
Now we detail the dataset we collected for our study, before examine each of the four research questions. We collected social media posts by 13 female journalists, identified in news reports~\cite{possetti2022chilling}\footnote{Lnk: https://www.aspistrategist.org.au/smart-asian-women-are-the-new-targets-of-ccp-global-online-repression/} as targets, through Twitter's streaming API service. To assemble this dataset, we initially reviewed female journalists globally listed on Wikipedia who communicate in English. Subsequently, we set a criterion of having at least 11,000 followers and selected those active within a specific time range from 2018 to 2022. \textcolor{black}{We selected the threshold of 11,000 followers to ensure sufficient data volume for and robust classification and reliable quantitative analysis, though we acknowledge this criterion inevitably excludes journalists with smaller yet significant followings who may also experience targeted harassment.}

\textcolor{black}{Our selection focused on high-profile journalists representing diverse cultural backgrounds, including one from the US, five from China, three from India, two from the UK, one from Pakistan, and one from Lebanon. And they come from major media outlets such as \textit{The New Yorker}, \textit{The Economist}, \textit{The New York Times}, and others. These journalists were chosen because they frequently report on controversial topics and, as a result, receive a significant volume of toxic replies. Their prominence in public discourse and engagement with divisive issues make them highly relevant for studying online harassment.}
\textcolor{black}{We selected these journalists based on two key criteria: follower count and activity period. We ensured that they had a substantial online presence, making them more likely to engage in public discourse and attract both supportive and adversarial interactions. Additionally, we considered their sustained activity on the platform, allowing us to analyze long-term patterns of interactions and responses to online harassment. 
While we did not explicitly categorize them by region or specific topics, our selection prioritizes journalists who are frequently targeted in online spaces. We recognize that this selection does not capture the full demographic diversity of female journalists experiencing online harassment. However, our goal was to identify journalists with sufficient visibility and engagement to allow for meaningful analysis. Expanding the dataset to include a broader range of journalists is an important direction for future work.}

This collection of 3,080,699 tweets encompassed not only the original posts initiated by these journalists but also encompassed all replies they garnered ranging from 2018 to 2022. \textcolor{black}{We present an example of the journalist’s activity timeline in Figure~\ref{fig:timeline_alice}, with a more detailed breakdown of original posts and replies provided in Appendix~\ref{append:data}. In general, the journalist's posting frequency varies significantly over time, with distinct peaks and valleys. And some large spikes in replies suggests that a specific post or event drew massive attention or controversy during that time.} The dataset consists solely of quotes and replies towards the journalists, with quotes making up only 0.1\% of the data. Therefore, we did not make a distinction between the two. The number of replies varies among different journalists. The count of replies each journalist received ranged from a minimum of $379$ to a maximum of $747, 776$, while the number of conversation trees constructed from their posts fluctuated between 16 and 4,679. 
This serves as a significant indication that our dataset spans from well-known to lesser-known journalists. The data's input features encompass a variety of factors, such as the language used in the tweets, the settings for replies, and various measures of tweet popularity. These measures include the number of retweets, replies, likes, views, and quotes each tweet received. 
\textcolor{black}{Moreover, the dataset also includes the annotated topic of each tweet, such as \textit{Art and Culture, Politics}, and \textit{New Business}, enabling us to computationally determine alignment based on explicit topic relevance rather than implicit alignment, sarcasm, or nuanced disagreement. 
To quantify alignment, we first concatenate all assigned topics of a tweet, generate embeddings for these concatenated topic strings, and then calculate the cosine similarity between the embeddings of the original post and its reply.} 
\textcolor{black}{Besides, to better capture the alignment of users with journalists' original stances, we also include the similarity between replies and the original tweets utilizing SentenceBERT~\cite{reimers-2019-sentence-bert}.}
Furthermore, we used the toxicity score as a feature in our classification model. These scores were derived using the Google Perspective API\footnote{\url{https://perspectiveapi.com/}}, which analyzes text to detect potentially harmful or abusive content.  
\begin{figure}
    \centering
    \includegraphics[width=0.8\linewidth]{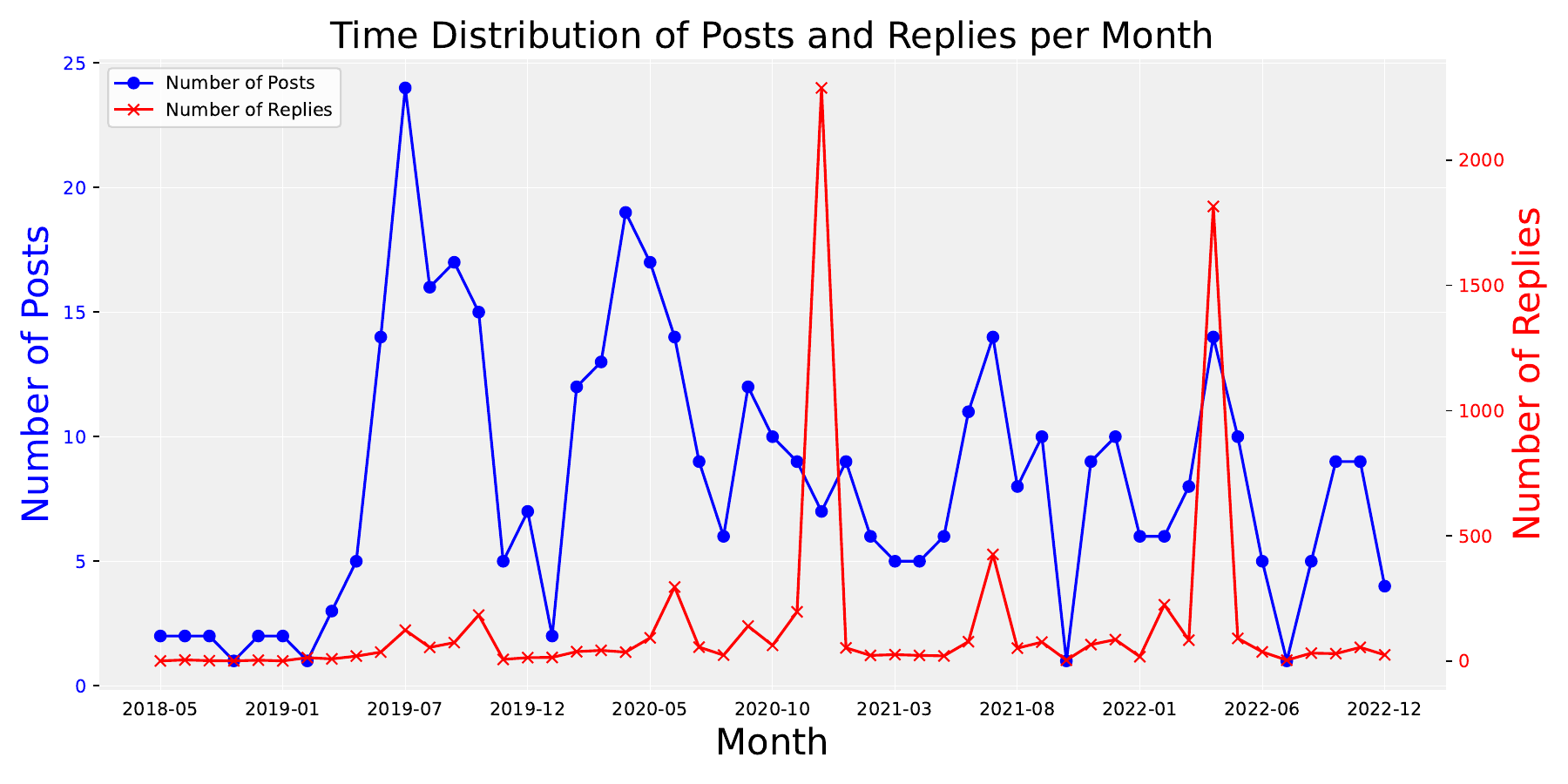}
    \caption{\textcolor{black}{An example of the journalist's activity timeline. The blue line represents the number of the journalist posting per month and the red line represents the number of replies received by the journalist per month.}}
    \label{fig:timeline_alice}
\end{figure}

\textcolor{black}{While we use the Perspective API to assess toxicity in tweet replies, we acknowledge its limitations in capturing cultural and contextual nuances. The API models are trained on millions of annotated comments from sources like Wikipedia and The New York Times, where 3–10 human raters assign toxicity labels based on predefined guidelines. The final toxicity score reflects the proportion of raters who marked a comment as toxic. While this approach ensures broad coverage, variations in annotator perception, dialect differences, and contextual ambiguity may lead to biases in toxicity classification. Future work could explore context-aware models to improve robustness across diverse online interactions.}

\textcolor{black}{We also acknowledge the potential concern regarding class imbalance in our model, with $12.6\%$ attackers, $55.0\%$ bystanders and $26.7\%$ supporters. Given that bystanders are the most prevalent class, we recognize the risk of the model being biased toward overrepresented classes.}

\textcolor{black}{To mitigate this issue, we employed undersampling during training to balance the dataset and ensure that no single class disproportionately influences the model’s predictions. Specifically, we adjusted the sample sizes to prevent the model from overfitting to the dominant class while still maintaining sufficient data for learning meaningful patterns. Additionally, we considered evaluation metrics beyond accuracy, such as F1-score per class, to assess the model’s performance across different categories and ensure fair representation.}

\subsection{Manual Annotation to Categorize Replies}\label{sec:defnition}
Our dataset did not contain pre-existing categorizations indicating whether the responses were supportive or antagonistic towards the journalists. To address this, we undertook a manual labeling process based on defined standards. We categorized replies that directly attacked the journalists as attackers, those that clearly supported the journalists as supporters, and all other replies as bystanders. 
\textcolor{black}{To ensure diversity and representativeness in our labeled dataset, for each journalist, we randomly sampled entire conversations with suitable lengths rather than individual replies. Once a conversation was selected, we labeled all tweets within that conversation, resulting in a total of approximately 5,000 labeled replies per journalist. 
To ensure labeling accuracy, we first thoroughly read through numerous posts by the journalists to understand their original stance before proceeding to label the tweets. 
However, it was still difficult for us to identify all statements that could be sarcastic or have implying meanings.}

The labeling was conducted by two annotators. Initially, both annotators independently labeled the same set of 5,000 replies to measure the reliability of the labels, achieving an initial agreement score of 70\%. After discussing discrepancies and refining our labeling criteria, we reach an agreement score of 100\% and proceeded to label the remaining replies individually. 

As shown in Table~\ref{tab:labeling-standards}, we categorized replies based on their content and tone. That is, we distinguished between replies that \textit{aligned with the topic of the original post }(\ie on the same topic), and \textit{toxicity of the replies}. \textcolor{black}{To clarify, "topic alignment" specifically refers to whether a reply is topically relevant to the journalist’s original post. During the labeling phase, annotators read each reply and evaluated whether its content was directly related to the topics of the original journalist’s tweet. Replies clearly addressing the same topic (e.g., discussing politics if the original tweet was about politics) were marked as aligned. Conversely, replies considered off-topic, unrelated, or only loosely relevant were marked as non-aligned. Annotators also took into consideration implied meaning, sarcasm, and subtle or indirect disagreement through examining the context closely. (To clarify, we made an error in our initial categorization explanation. The corrected definitions are clearly listed as follows and more detailed examples are shown in~\Cref{tab:labeling-standards})}
\begin{itemize}
    \item \textcolor{black}{Respond to the journalist:}
    \begin{itemize}
        \item \textcolor{black}{Attacker: Replies that are toxic, including personal attacks on the journalist.}
        \item \textcolor{black}{Supporter: Replies that are non-toxic and align with the journalist’s viewpoint.}
        \item \textcolor{black}{Bystander: Replies that are non-toxic and on-topic but neutral. }
    \end{itemize}
    \item \textcolor{black}{Respond to non-journalist posts}
    \begin{itemize}
        \item \textcolor{black}{Attacker: Toxic replies that are either directly relevant to the topic of the journalist’s post, or are personal attacks directed towards the journalist.}
        \item \textcolor{black}{Supporter: Replies that are non-toxic and align with the journalist’s viewpoint.}
        \item \textcolor{black}{Replies that are toxic but neither directed at the journalists nor personal attacks against the journalist, or replies that are either non-toxic and on-topic but neutral towards the journalist’s post.}
    \end{itemize}
\end{itemize}

\begin{table}[h]
    \centering
    \begin{tabular}{lp{5cm}p{5cm}}
        \toprule
        \textbf{Label} & \textbf{Example Text} & \textbf{Justification}\\
        \hline
        \textbf{Attacker} & "You are a cold-blooded animal, how can you have feelings?" & Toxic and directed personally at the journalist, making it an attack.\\
        \textbf{Attacker} & "Are you pretending to be a cultural person again? Do it all day, are you tired?" & Contains a sarcastic, mocking tone that targets the journalist personally.\\
        \textbf{Supporter} & ""What a terrific quote! Thank you for sharing with us "" & Positive and supportive of the journalist’s post, without toxicity.\\
        \textbf{Supporter} & "Nice find! I’ll put this on my reading list." & Engages positively with the content, showing alignment with the journalist’s topic.\\
        \textbf{Bystander} & "Yes, 1992.  As I state to the idiot before, dkan who probably wasn't even born when that happened." &Toxic but unrelated to the journalist or the original topic, classifying it as a bystander response.\\
        \textbf{Bystander} & "Dead on.  Now I'll never get it out of my head." & Neutral comment without clear alignment or attack, making it a bystander response.\\

        \hline
    \end{tabular}
    \caption{\textcolor{black}{Examples of replies and their labels based on our labeling criteria. Attackers and Supporter exhibit topic alignment with the original journalist post, whereas bystanders are neutral towards the topic of the original journalist post. Notice that topic alignment is different from toxicity of the individual's response to the journalist's post (See \Cref{sec:5.1} for details on how we arrived at our labeleing critera).}}
    \label{tab:labeling-standards}
\end{table}
\textcolor{black}{As we can see from the two bystander replies in Table~\ref{tab:labeling-standards}, the first one—despite being toxic—was not directed at the journalist, so we labeled it as a bystander reply. The second one, although it expresses approval, is also not directed at the journalist, so we labeled it as a bystander. This indicates that not all toxic replies come from attackers, nor do all approving replies come from supporters.}
\section{Identifying Attackers, Supporters and Bystanders (RQ1)} \label{sec:3}
\begin{figure}[htbp]
    \centering
 \includegraphics[width=0.9\textwidth]{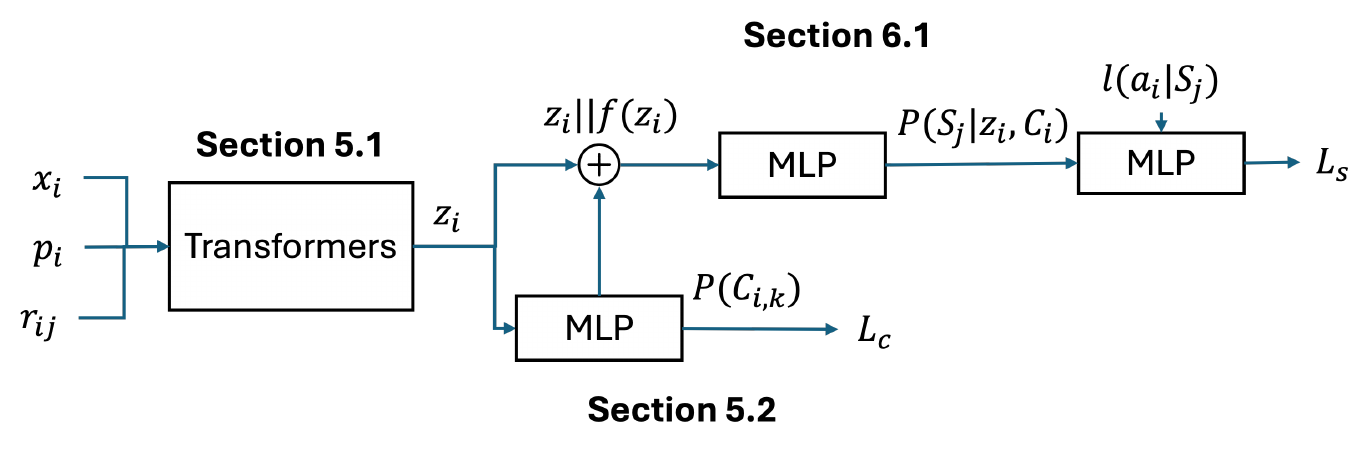
}
    \caption{Our framework. $\bigoplus$ denotes concatenation, $x_i, p_i, r_{ij}$ denote the metadata, global path and local path for node $i$. The metadata includes features such as language, creation time, number of retweets, replies, likes, views, toxic scores, topics, and the presence of URLs. Global path includes the path from the root to node $i$, and the local path refers to the path between the reply $t_i$ and its parent node $j$. $z_i$ gives the output of the Transformer model, $L_c$ and $L_s$ are loss for the classification and strategy discovery modules.}
    \label{fig:frame}
\end{figure}

To model the users' replying activities and discover their strategic behaviors, we introduce our proposed model \method. Our model consists of three primary components: a transformer built with a tree-like structure to depict conversations, a classification block, and a system designed to identify users' strategic actions on Twitter. The framework is illustrated in Figure~\ref{fig:frame}. In this section, we delve into the model's architecture~\cref{sec:Classification Model Architecture}, account classification~\cref{sec:account_classification} and the results of our experiments~\cref{sec:5.6}. The figure also shows our strategy identification model; we shall discuss that part of the model in~\cref{sub:The model}, where we discuss RQ2.
\subsection{Group Identification Model }\label{sec:Classification Model Architecture}

We depict the interactions within a journalist's post by constructing a conversation tree, where the journalist's initial post forms the root, and any direct replies to this post are considered a first-level child. Responses to these initial replies are classified as second-level children, and the pattern continues. By representing each reply as a node, we ensure that the tree structure contains no cycles, and every node has a distinct path leading to the root. We begin by introducing the concept of conversation trees, and then proceed to assign a specific position to each node. After obataining node positions, we integrate position encoding into the Transformer's self-attention mechanism, applying it on both local and global scales.

Drawing inspiration from~\cite{peng2022rethinking}, we utilize coordinates to specify the position of each node. Rather than solely relying on the order among siblings and the total number of siblings for a node, we also consider the node's level or depth within the conversation tree as crucial for the interpretation of the tree's structure and the dynamics of the conversation. Consequently, we define the coordinates for each node as follows:
\begin{equation}
    Coor(i) = \left\{ 
    \begin{aligned}
        & Coor(Pa(i)) + \{(o_i, l_i, s_i)\} & \text{if}\; i \neq root\\
        & \{(1, 1, 1)\} & \text{if}\; i = root\\
    \end{aligned}
    \right.
\end{equation}
where $Pa(i)$ denotes the parent of node $i$, while $o_i$, $l_i$, and $s_i$ denote the order, level, and number of siblings of node $i$, respectively. In our case, the order of the node is sorted by its created time.

Subsequently, we transform every coordinate into a vector. This process involves initially converting each coordinate to a scalar value, followed by obtaining the corresponding vector based on this scalar. With the given coordinates, we represent each node through a vector sequence denoted as $H_i$:
\begin{equation}
    H_i = [h_{i, 1}, h_{i, 2}, ...h_{i, n}]
\end{equation}
where $n$ is the level of node $i$ and $h_{i, j}$ is the embedding vector of the $j$th coordinate in the list.

After obtaining a formal representation of the tree structure for Transformer integration, we transform this structure into a sequence and get positional encodings for each node. Existing studies have delved into the advantages of applying relative positional encoding~\cite{shaw2018self}, employing a disentangled attention mechanism~\cite{ke2020rethinking}, as well as the combined approach suggested by~\cite{peng2022rethinking}. Given that absolute global encoding and relative local encoding each focus on distinct parts of a conversation tree, we follow the method proposed in~\cite{peng2022rethinking}, which introduced a global bias via absolute encoding and a local bias through relative encoding, subsequently merging these components into a  Transformer model.

Given the vector sequence for nodes, we can denote the global vector $p_i$ for node $i$ as follows:
\begin{equation}
    p_i = f(concat(H_i))
\end{equation}
where for each node $i$, we concatenate vectors in the list $H_i$ and subsequently apply linear layers, denoted as $f(\cdot)$, for their vectorization. The local position vector $r_{ij}$ for node $i$ is :
\begin{equation}
    r_{ij} = \left\{ 
    \begin{aligned}
        & \sum_m h_{i, m} - \sum_n h_{j, n} & \text{if} \; Pa(i) = j \\
        & \vec{0} & \text{if} \; Pa(i) \neq j\\ 
    \end{aligned}
    \right.
\end{equation}
If two nodes are connected, we initially aggregate all vectors within $\{h_i\}$ and $\{h_j\}$, then proceed with the subtraction. If there is no adjacency between two nodes, the local position vector is set to zero, indicating that only the one-hop local relationships are captured in this phase of encoding.
Finally, we integrate $p_i$ and $r_{ij}$ into Transformer by adding them to the attention score $\alpha_{ij}$:
\begin{equation}
\label{eq:att}
\begin{aligned}
    \alpha_{ij} &= \frac{1}{\sqrt{d}}[(x_iW_q)(x_iW_k)^T \\
    &+ (p_iW_q^p)(p_jW_k^p)^T \\
    &+ (x_iW_q)(r_{ij}W_k^r)^T + (r_{ji}W_q^r)(x_jW_k)^T]
\end{aligned}
\end{equation}
where $x_i$ denotes the input and $W_q, W_k, W_q^p, W_k^p, W_q^r, W_k^r$ are projection matrices, and $\frac{1}{\sqrt{d}}$ is the scaling factor applied for attention score $\alpha_{ij}$ before softmax. Here we set the scaling factor to be $\frac{1}{\sqrt{2}}$ rather than $\frac{1}{\sqrt{4}}$ given that the scale of the relative positional score is much smaller due to its sparsity, it cannot match the magnitude of the first two terms. In Eq.~\ref{eq:att}, the initial line represents the original self-attention score, the second line reflects the attention score derived from the absolute positions of nodes $i$ and $j$, and the final line illustrates the attention score based on the relative positions of nodes $i$ and $j$. And the final output of the model is:
\begin{equation}
    z_i = \sum_j \frac{\text{exp}(\alpha_{ij})}{\sum_{k=1}^n \text{exp}(\alpha_{ik})} (x_j W_v)
\end{equation}
The main difference of our positional encoding compared to the approach presented in~\cite{peng2022rethinking} is that we describe each node position by three dimensions, including not only the sibling orders and child number of its parent but also the level of each node since we believe that when analyzing users' replying behaviors, the depth of the conversation is essential.

\subsection{Account Classification}
\label{sec:account_classification}
We incorporate two linear layers followed by a softmax layer on top of the Transformer decoder to estimate which class $z_i$ belongs to (see Figure~\ref{fig:frame}). We use $f(\cdot)$ to represent the two linear layers and $\xi(\cdot)$ as the nonlinear layer. We use $L_1$ normalization to ensure that the output $\xi(f(z_i))$ is a valid distribution of node $i$ belonging to a certain class. Subsequently, in each iteration, we aim to reduce the negative log-likelihood:
\begin{equation}
    L_c = -\sum_i \text{log}\, P(C_i|z_i)
\label{eq: c_loss}
\end{equation}
where $C_i$ denotes the predicted class for node $i$.

\textcolor{black}{We acknowledge the computational challenges associated with tree-structured Transformers, especially when handling large datasets or deep conversation trees. To manage this, we set a maximum depth of 20 for conversation trees, ensuring that excessively deep threads were truncated during training. This decision was based on empirical observations of typical conversation depths while balancing computational efficiency. However, we did not apply additional pruning or sparsity techniques, such as sparse attention, which we recognize as a potential limitation. Exploring these techniques could further enhance scalability and efficiency.}

Now, we present and analyze experimental outcomes derived from the application of our methods.



\subsection{Results for Group Classification} \label{sec:5.6}
Given that only a portion of our dataset is labeled, we implement a 5-fold cross-validation strategy on this subset to ensure the robustness and generalizability of our model. Following the cross-validation process, the trained model makes predictions on the unlabeled portion of the dataset, extending our analysis to the entire dataset. For the main results, we obtain classification metrics from the labeled replies of all journalists and report the averaged Accuracy, Recall, and F1 score. The interested reader can refer to~\cref{Group Classification Baselines} for the details of the baselines used in our experiments and in ~\cref{append:each} we presented results for journalists from different cultures.

\begin{table}[h]
\begin{tabular}{lccc}
\toprule
\multicolumn{1}{l}{\textsc{Method}} & \textsc{Accuracy}  & \textsc{Recall}  & F\textsubscript{1} \\ \midrule
\multicolumn{1}{l}{THP~\cite{zuo2020transformer}} & 0.4679      &0.4815          & 0.4746        \\
\multicolumn{1}{l}{AMDN-HAGE~\cite{sharma2021identifying}}    & {0.5382}          & {0.5719}          &{0.5545}          \\
\multicolumn{1}{l}{TreeTransformer~\cite{peng2022rethinking}} & {0.7876}          & {0.8015}          &{0.7944}        \\  \addlinespace[0.25em]
\multicolumn{1}{l}{\method\ (Ours) }  & {\textbf{0.8215}} & {\textbf{0.8363}} & {\textbf{0.8288}} \\
\multicolumn{1}{l}{\method\ without positional embeddings} &{0.7370}          & {0.7395}          & {0.7417}          \\
\multicolumn{1}{l}{\method\ without global path}            & {0.7677}          & {0.7938}          & {0.7805}          \\
\multicolumn{1}{l}{\method\ without relative path}          & {0.7895}          & {0.8031}          &{0.7962}         \\ \bottomrule
\end{tabular}
\caption{Evaluation of classification results across all journalists. \method\ achieves the highest classification scores overall. Of note, is the work by~\citet{sharma2021identifying} that detected coordination of IRA accounts that spread disinformation on Twitter during the 2016 election. It's relatively poor performance may be due to its inability to model conversation tree structures effectively. \method\ without positional embeddings refers to the attention score \( \alpha_{ij} \) that includes only the first term in Eq.~\ref{eq:att}. \method\ without the global path denotes the attention score that includes the first and third terms in Eq.~\ref{eq:att}, while \method\ without the relative path includes only the first and second terms. }
\label{tab:acc_results_table}
\end{table}
Table~\ref{tab:acc_results_table} provides results of model evaluation against the baselines. We compare the Accuracy, F1, and Recall scores. We observe that \method\  outperforms other methods. For THP~\cite{zuo2020transformer} and AMDN-HAGE~\cite{sharma2021identifying} that detect temporal coordination when IRA spreads disinformation on Twitter, the likely reason for their limitations is their inability to model tree structures effectively. This suggests that these models do not capture the hierarchical and interconnected structure of tree-like data, which is crucial for understanding the complex relational patterns. As for the TreeTransformer, the results indicate that our novel node coordinates incorporating node levels are a key factor in successfully modeling tree structures. This highlights the importance of considering the depth or hierarchical level of nodes in the tree. 

\subsubsection{Ablation Study} In addition to establishing benchmarks with baseline models, we conduct a comparative analysis of \method\ against its variations to explore the significance of positional encodings in its architecture. We experiment with modifications in the self-attention mechanism of the Transformer model, where we selectively omit information about either global or local paths. The outcomes of these ablation studies are presented in Table~\ref{tab:acc_results_table}.

The full \method\ model exhibits a superior capability in accurately representing the structure of conversation trees. Notably, we discern that the variant using only global paths outperforms using only local paths by 2\%. This indicates that the inclusion of conversation length, which is captured by global paths, might be more effective in representing the conversation trees. This could be attributed to a user's attention being distributed across the whole dialogue of a thread rather than just the neighboring replies.
This finding underscores the intricate relations between global context and local interactions in shaping user responses within conversational threads.

\subsubsection{User Classification} 
\begin{figure}[htbp] 
    \centering
    \begin{subfigure}[b]{0.3\linewidth}
        \includegraphics[width=\linewidth]{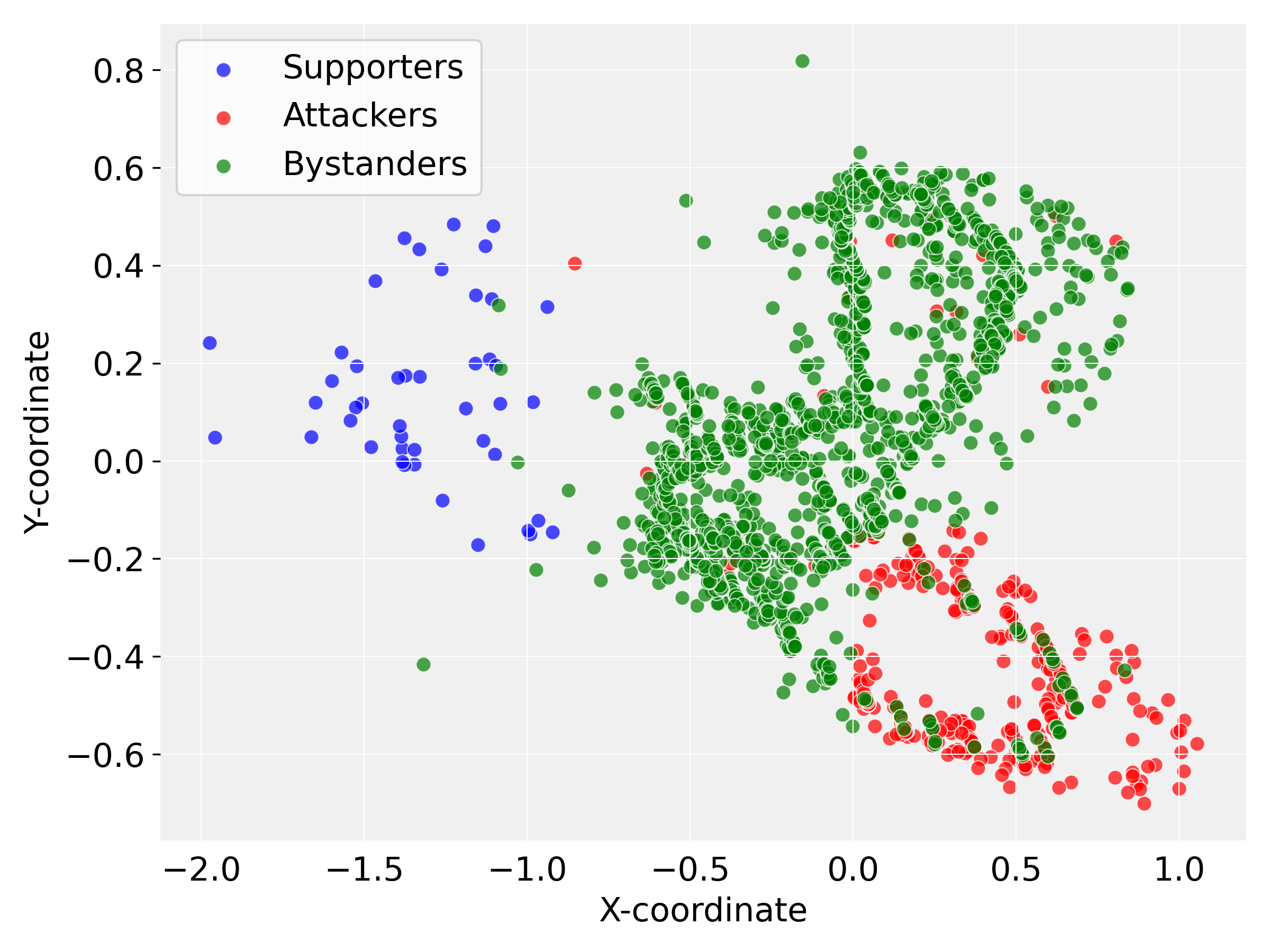}
        \caption{Clusters for Journalist A.}
        \label{fig:csub1}
    \end{subfigure}
    \hspace{0.5cm} 
    \begin{subfigure}[b]{0.3\linewidth}
        \includegraphics[width=\linewidth]{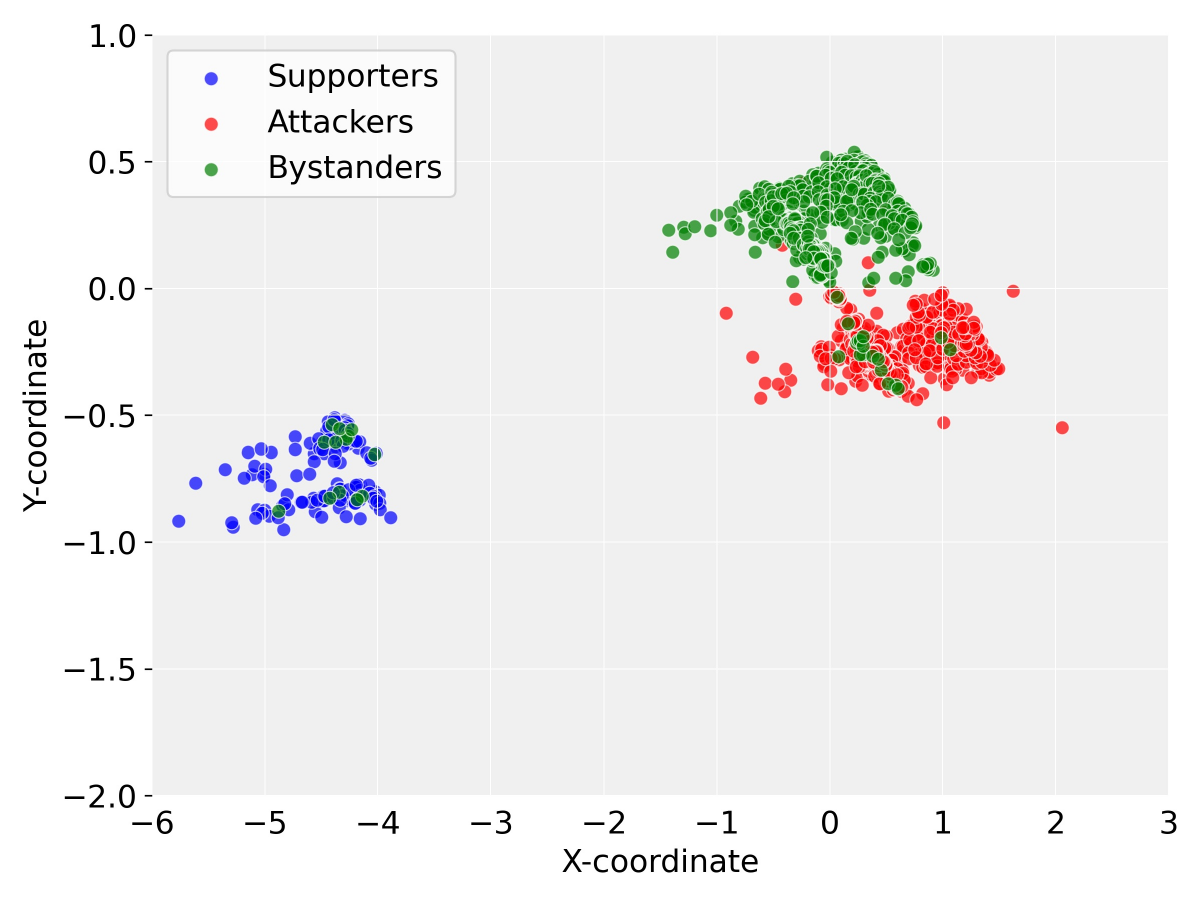}
        \caption{Clusters for Journalist B.}
        \label{fig:csub2}
    \end{subfigure}
    \hspace{0.5cm} 
    \begin{subfigure}[b]{0.3\linewidth}
        \includegraphics[width=\linewidth]{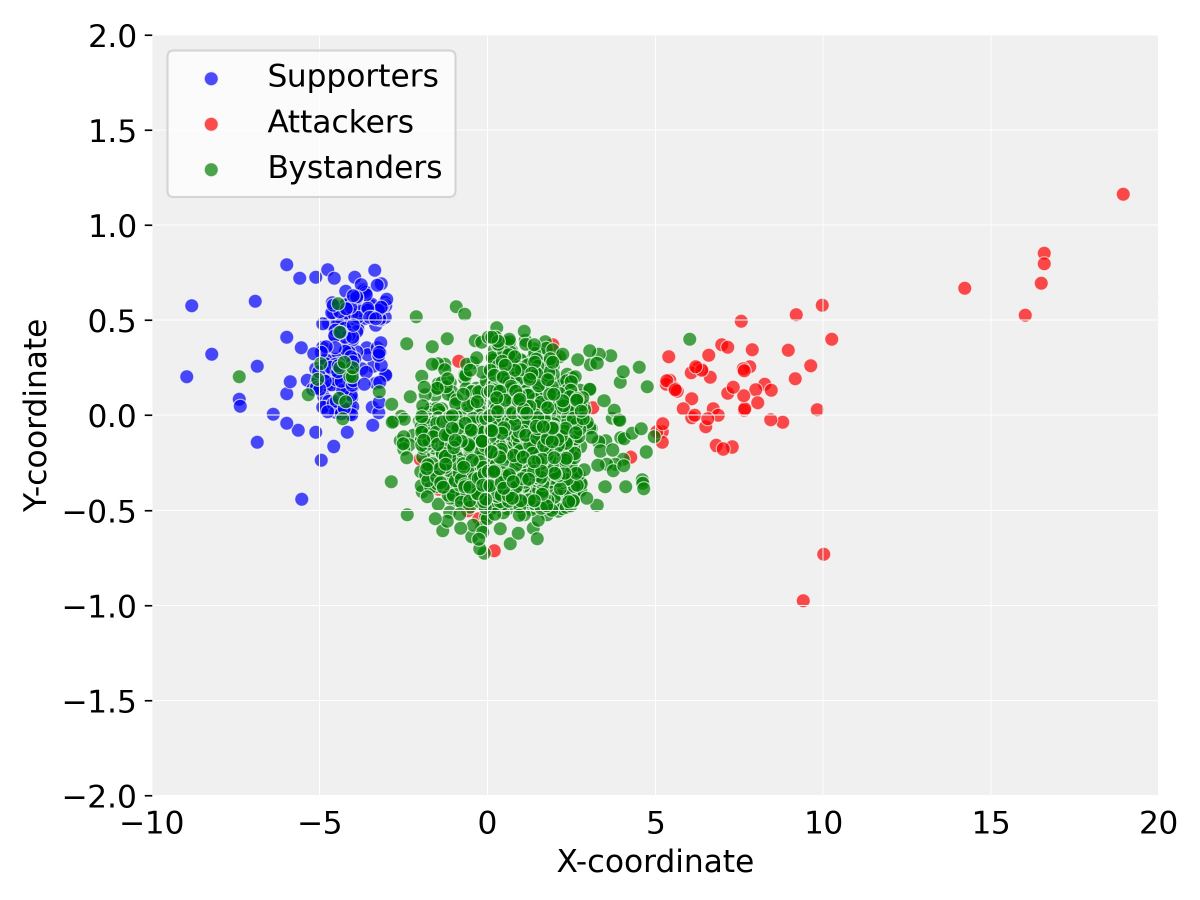}
        \caption{Clusters for Journalist C.}
        \label{fig:csub3}
    \end{subfigure}
    
    \caption{User clusters. For these three journalists, we see clear separations between attackers (red) and supporters (blue), but there are overlaps between bystanders (green) and attackers (red), indicating their similarity in behaviors. (See \Cref{sec:5.6}).}\label{fig:cluster} 
\end{figure}

In Figure~\ref{fig:cluster}, we provide a visualization of the clustering of the account embeddings derived via our \method\ technique. These embeddings are classified into three categories: attacker, supporter, and bystander accounts. We first average the embeddings for each user. Following this, we employ t-SNE~\cite{van2008visualizing} for dimensionality reduction. The final step involves visually presenting these 2-D coordinates.

From Figure~\ref{fig:csub1}, it is evident that the attacker group is significantly isolated from supporters. However, it is also noteworthy that there is some degree of overlap between the attackers and bystanders. This overlap suggests a certain level of similarity in the behaviors or characteristics of different groups, as captured by our model. In Figure~\ref{fig:csub2}, it is evident that there are distinct variations in the clustering visualizations between supporters and the other groups, with some overlap between attackers and bystanders. And the same result is shown in Figure~\ref{fig:csub3}. \textcolor{black}{While there is some overlap between attackers and bystanders in~\Cref{fig:csub2} and~\Cref{fig:csub3}, distinct clustering patterns are still present. The observed overlap is expected, as attackers and bystanders may share some behavioral similarities, especially in certain interactions.}

In this section, we introduced our group identification model architecture~\cref{sec:Classification Model Architecture}, discussed the account classification process~\cref{sec:account_classification}, and presented the results of our experiments~\cref{sec:5.6}. Our experiments indicate that \method~outperforms existing methods---including prior work on identifying coordinated attacks~\cite{sharma2021identifying}---in classifying users into attackers, supporters, and bystanders. 

These classification results provide a foundation for deeper analysis; in the following section, we explore the distinct strategic behaviors exhibited by attackers, supporters, and bystanders to further understand how these groups interact within conversations.

\section{Discovering Attacker, Supporter and Bystander Strategies (RQ2)}

In this section, we delve into the strategic behaviors of attackers, supporters, and bystanders. We present our model in~\cref{sub:The model} and discuss the strategy spaces in~\cref{sec:5.2}. We then evaluate the model's performance in identifying strategies in~\cref{sec:5.7}. Finally, we analyze the differences in strategic behaviors among the three groups in~\cref{sec:6.1}.

\subsection{The Strategic Behavior Model}
\label{sub:The model}

To analyze users' strategic reply behaviors, it is essential to determine the strategy distribution for each node $i$. We assume that strategy identification is influenced by two main elements: the content of the reply $z_i$ and its associated group $C_i$. Thus we formally express this strategy distribution as $P(s_j|z_i, C_{i})$, where $s_j$ represents the $j$-th strategy, $i$ signifies the reply $t_i$, $k$ identifies the group $k$ and We omit the index of layer for
simplicity and $z_i$ denotes the output of the self-attention. To integrate information about the group, for every node $i$, we concatenate the Transformer's output $z_i$ with the output from the classification module's second layer $f(z_i)$.
This combined vector is then processed through a two-layer linear model to derive $P(s_j|z_i, C_{i})$. Subsequently, we multiply this by $l(a_i|s_j, z_i, C_i)$, the likelihood of node $i$ responding to its parent node given strategy $s_j$ and the content information $z_i, C_i$. Details on the specific strategy space will be provided in Section~\ref{sec:5.2}.
Finally, the goal is to minimize the negative log-likelihood:
\begin{equation}
L_s = -\sum_{i=1}^N \text{log}\sum_{s_j \in S}l(a_i|s_{j}, z_i, C_i)P(s_{j}|z_i, C_i)
\label{eq: s_loss}
\end{equation}

Eq.~\ref{eq: s_loss} demonstrates that the generation of action is influenced by underlying strategy distributions, with each user possessing a unique strategy distribution shaped by their specific content and group affiliation. Then combining Eq.~\ref{eq: c_loss} and Eq.~\ref{eq: s_loss}, the total loss in our framework is:
\begin{equation}
    L = L_c + L_s
\label{eq:total_loss}
\end{equation}

\subsection{Strategy Spaces} \label{sec:5.2} 
In this section, we discuss three strategy spaces that influence the behavior of replies. Thus, consider a tweet $t_i$ that replies to another tweet $t_j$. We need to identify strategic considerations that explain the directed edges $(t_i, t_j)$. We identify three aspects based on time recency, similarity of topics and users. \textcolor{black}{Every time when user $i$ replies to user $j$, we need to identify the strategy considerations from user $i$ to user $j$.} When posting replies, attackers may be prone to compose content with similar topics to the original posts. They may also choose to reply right after the journalist creates a tweet.  
\begin{table}
    \centering
    \begin{tabular}{cl}
        \toprule
            Aspect & Strategy  \\
        \midrule
            \multirow{2}{*}{People} 
            & $s_{1,0}$, prefer to reply to people with similar interests \\
            & $s_{1,1}$, prefer to reply to people with dissimilar interests \\
            \multirow{2}{*}{Topic} & $s_{2,0}$, prefer similar topics \\
             & $s_{2,1}$, prefer dissimilar topics \\
            \multirow{2}{*}{Time} & $s_{3,0}$, prefer to respond quickly \\
            & $s_{3,1}$, prefer to respond at random time \\
        \bottomrule
    \end{tabular}
    \caption{Meaning of each pure strategy. Each composite citation strategy consists of one pure strategy from each of the four aspects. Each composite location strategy consists of one pure strategy from each of the first three aspects. (See \Cref{sec:5.2}).}
    \label{tab:pure_strategies}
\end{table}

\subsubsection{Topic}
For the topic strategy, \textcolor{black}{we use Twitter's tweet annotations and BERT embeddings~\cite{devlin2018bert} to assign each reply a topic vector $F_i$. In most cases, users are likely to reply to tweets with similar topics, or engage in conversations where the topics align with their own interests. However, in some cases, users may respond to tweets outside their usual topics, such as engaging in debates or discussions beyond their primary interests. Thus, we may explain an edge $(t_i, t_j)$ either through homophily (\ie, strategy $s_{1, 0}$) or by engaging with tweets on different topics (\ie, strategy $s_{1, 1}$), if the discussion spans multiple topics. Consequently, we set the likelihood of an edge $(t_i, t_j)$ based on strategy $s_{1, 0}$ to be} 
\begin{equation}
    l((t_i, t_j) | s_{1, 0}) \propto \exp {\frac{\langle F_i, F_j \rangle}{\|F_i\| \|F_j\|}}
\end{equation}
\textcolor{black}{where $F_i$ denotes the topic vector of tweet $t_i$. The likelihood of replying to a tweet with different topics is simply the complement of $l((t_i, t_j) | s_{1,0})$}
\subsubsection{People}
\textcolor{black}{We adopt a similar assumption to strategy $s_2$, where users are more likely t oreply to others who share similar interests. Accordingly, we define the likelihood of an edge $(t_i, t_j)$ based on strategy $s_{2, 0}$ as } 
\begin{equation}
    l((t_i, t_j) | s_{2, 0}) \propto \exp {\frac{\langle U_i, U_j \rangle}{\|U_i\| \|U_j\|}}
\end{equation}
\textcolor{black}{where $U_i$ denotes the topic vector for the author of reply $t_i$ indicating the range of topics that the user is likely to engage with. We derive this user-level topical interest by aggregating topic vectors from all their replies. The likelihood of replying to a user  with different interests $l((t_i, t_j) | s_{2, 1})$ is again the complement of $l((t_i, t_j) | s_{2, 0})$.}
\subsubsection{Time}
\textcolor{black}{Replying behavior on Twitter also exhibits recency bias, making time a crucial factor in explaining edge $(t_i, t_j)$. To model recency bias (\ie, strategy $s_{3,0}$), we first normalize the time gap between replies into a $24$-hour scheme and set a normalization scale of $240$ hours (10 days) using the transformation $\frac{t}{240}$. This choice is based on our observation that replies beyond 10 days are relatively rare, making it a reasonable upper bound for normalization.}

\textcolor{black}{Nexy, we define the normalized time difference between replies $t_i$ and $t_j$ as $0 \leq \delta \leq 1$. To incorporate timing into the model, we apply a Beta distribution to adjust the posterior probability of selecting tweets to reply to. Specifically, the likelihood of edge $(t_i, t_j)$ based on strategy $s_{3, 0}$ is given by}
\begin{equation}
    l((t_i, t_j)|s_{3, 0})\propto B(1-\delta|\alpha, \beta)
\end{equation}
\textcolor{black}{where $\alpha$ and$ \beta$ are the parameters of the Beta distribution. To reflect the tendency of users to respond more quickly to recent tweets, we select parameters that skew the distribution towards 1, favoring replies to newer tweets over older ones. For the alternative strategy (i.e., strategy $s_{3,1}$), a reply is chosen uniformly at random, ignoring the timestamps of the tweets.} 
\subsubsection{Composite Strategies}
We have discussed three different strategic behaviors to explain edge $(t_1,t_2)$: people, topic, and time recency. Thus the likelihood of the edge $(t_1,t_2)$ is a composite of each of the three strategies. Since each strategic consideration has two possibilities, we can enumerate $2^3 = 8$ composite strategies to explain edge $(t_1,t_2)$. For easy reference, we use a binary sequence to represent composite strategies with respect to pure strategies. For example, we can denote one replying strategy $S_0 = s_{1,0} \times s_{2,0} \times s_{3,0}$ as this certain user prefers similar people, topics and tend to reply right after the creation of the original post. So the distribution of forming an edge $(t_1, t_2))$ given a composite strategy $S_j$ is the normalized product of the likelihood of forming that edge given each of $S_i$'s pure strategies. \textcolor{black}{We acknowledge the potential value of incorporating additional factors such as sentiment, intent, or platform-specific dynamics into the strategic behavior model. While we considered these aspects, we did not include them in our current model because we found it challenging to systematically model them in a way that aligns with our framework. Our approach focuses on topic, user, and time as fundamental factors that directly influence replying behavior, and expanding the strategy space beyond these dimensions would require careful formulation and additional data.}

\subsection{Results for Strategy Identification} \label{sec:5.7}


To assess the efficacy of each model in terms of strategy discovery, we conduct an evaluation focused on predicting the probability of forming an edge between a reply, denoted as $t_i$, and its parent post in the conversation tree. This prediction task enables us to understand the underlying structural dynamics of conversational interactions. The detailed results of this evaluation are presented in Table~\ref{tab:strat_results_table}. \textcolor{black}{We selected Logistic Regression (LR)~\cite{nigam2000text} and Dirichlet Multinomial Mixture (DMM~\cite{yin2014dirichlet}) as baselines because they have been widely used in prior similar studies for modeling strategy tasks~\cite{xiao2020discovering}. These methods provide interpretable and well-established reference points for evaluating the effectiveness of our approach.} 
\begin{table}[htb]
\begin{tabular}{lccc}
\toprule
\textsc{Method} & \textsc{Accuracy} & \textsc{Recall} & F\textsubscript{1} \\ \midrule
LR~\cite{nigam2000text}      & 0.6342          & 0.6158          & 0.6906          \\
DMM~\cite{yin2014dirichlet} & 0.5526          & 0.5977          & 0.6219          \\
\method\   (Ours)         & \textbf{0.7715} & \textbf{0.7827} & \textbf{0.7770} \\ \bottomrule
\end{tabular}
\caption{Experiment results of strategy discovery across all journalists. \method\ achieves the highest scores.}
\label{tab:strat_results_table}
\end{table}
~\method\ consistently surpasses its counterparts in performance. This superior performance of \method\ suggests that our proposed method is more adept at capturing the formation of conversational threads. 

\subsection{How do Strategies Among Groups differ?}\label{sec:6.1}
We now investigate whether users employ strategies and if there are discernible differences in strategic behaviors exhibited by different groups. 

We first aggregate strategy distributions at the individual user level to unravel the strategic behaviors manifested by users across diverse groups. This process requires a classification of users based on their tendencies. During our examination, we observed some overlap of some users who post both attacking replies and supporting replies, after checking such examples, we found that they are false positive cases and labeled such users as the labels that appear the first in the whole conversation.

\begin{figure}[ht]
    \centering
        \includegraphics[width=0.9\textwidth]{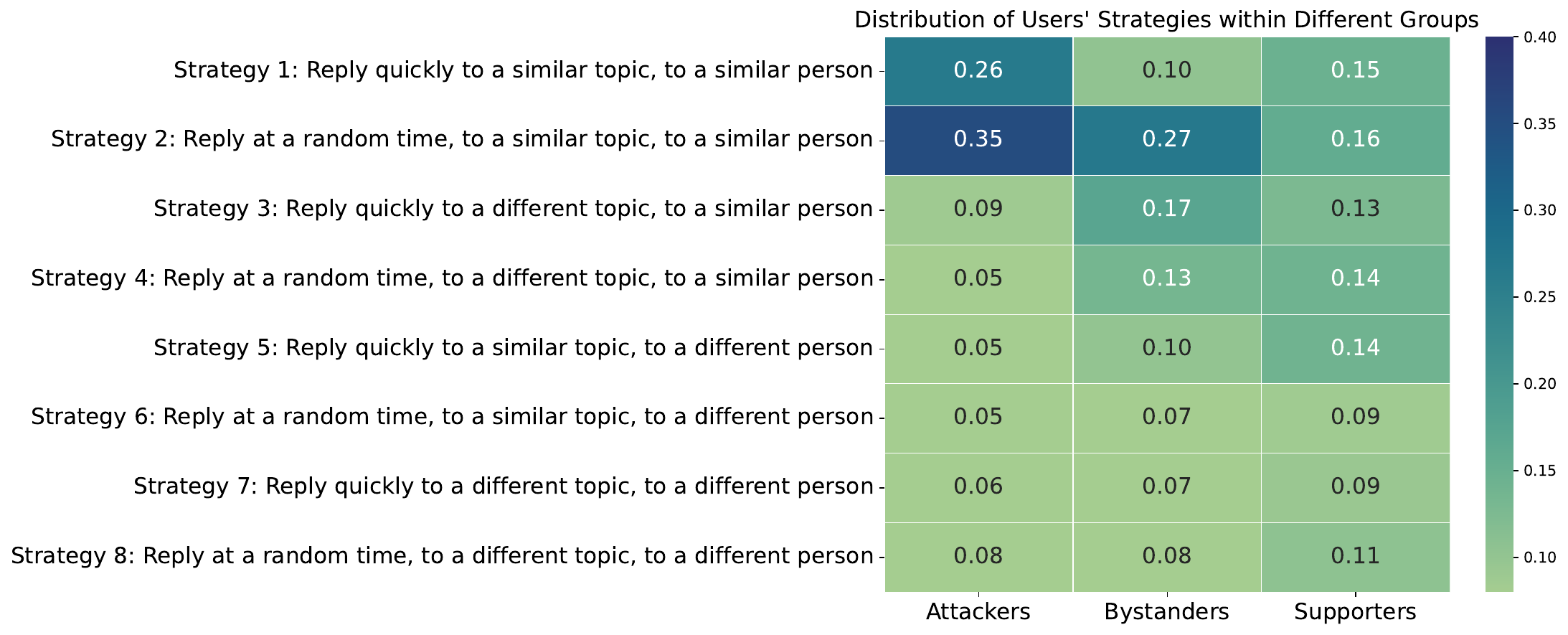} 
        \label{fig:overall_heat}
    \caption{Distributions of strategic behaviors among attackers, supporters, and bystanders with respect to the number of strategies (See \Cref{sec:6.1}). 
    Attackers seem to have clear strategies (strategies $s_1$ and $s_2$), often targeting users with similar interests or engaging in discussions on related topics while supporters do not appear to follow a specific strategy, showing a more varied or less coordinated approach to their replies. This suggests that attackers may operate in a more organized manner, while supporters exhibit more spontaneous or less structured behavior. 
    }
    \label{fig:combined}
\end{figure}

As a reminder, we distinguished between topic alignment with the original journalist post, and toxicity of the post. Thus, we categorize users with a tendency towards either attacking or supporting behaviors towards the journalist as non-neutral. Subsequently, we calculate the average strategy distribution for each user. By aggregating these individual distributions, we can compile a comprehensive overview of the average strategy distributions for each of the three identified groups: attackers, bystanders, and supporters.
By comparing these average distributions, we aim to uncover the differences in strategic behavior, understanding how various groups engage and interact within the platform. The more detailed results are shown in Figure~\ref{fig:combined}.

In our study, we identified distinct behavioral patterns among different user groups, particularly in how they engage in conversations. \textbf{Attackers}, in particular, tend to focus on posts and users that align with their areas of interest. This is reflected in their reliance on strategies such as $s_1$ and $s_2$, where they actively engage with content and individuals who share similar viewpoints or passions. This behavior suggests that attackers are not only motivated by the topic at hand but are also likely to seek support from like-minded individuals, forming more coordinated interactions and subsequently amplifying their attacks. Our findings indicate that attackers are more strategic in their approach, targeting specific users and topics to maximize their impact and influence the conversation. To contrast with prior work in coordinated attacks~\cite{keller2020political,weber2021amplifying} that focus on temporal coordination, our results suggest that while attackers do coordinate in time (part of strategy $s_1$), when we examine their behavior in a broader context (strategy $s_1$ and $s_2$), we notice attackers are more likely to respond to users with similar interests and to similar topics (part strategy $s_2$). This suggests that time coordination is not the only factor that attackers consider when coordinating their attacks. \textcolor{black}{We acknowledge that 'coordinated attacks' can imply pre-planned, strategic execution among attackers. However, in our study, we do not have direct evidence or data confirming that attackers explicitly plan their actions together. Instead, we observe behavioral patterns that resemble coordination where attackers engage in similar toxic behaviors, reinforce each other's actions, and create an amplified impact on the threads. Since this coordination emerges organically rather than through explicit planning, we refer to it as implicit coordination. This differs from the explicit coordination defined in~\cite{cai2023hate}, which often involve human-bot synchronization and explicit attack planning. 
In~\cite{sharma2021identifying}, there is no observable side channel of communication, but the ground truth labels were provided by Twitter.
However, our framing of implicit coordination is the norm, for example in~\cite{schoch2022coordination, keller2020political}, the authors identify time as the coordination strategy, but one cannot confirm this coordination, since the side-channel for communication is unobserved. By distinguishing these forms of coordination, we clarify that our findings focus on behavioral manifestations rather than verified pre-attack organization.}

\textbf{Bystanders}, on the other hand, exhibit a  different engagement pattern. While they also prefer interacting with users who share similar interests their participation appears to be less influenced by the specific content of the conversation (strategies $s_2$ and $s_3$). Additionally, their responses are not time-sensitive when responding to a similar person (strategies $s_1, s_2, s_3, s_4 $), indicating that bystanders might passively observe discussions for extended periods before engaging, often regardless of the current intensity or relevance of the topic. They are also somewhat likely to engage with a person different from themselves (strategy $s_5$), suggesting that they may be more open to engaging with diverse perspectives. 

In contrast, \textbf{supporters} do not demonstrate a clear strategic approach. Their engagement patterns are more diffused, suggesting a lack of structured or coordinated behavior. This absence of a specific strategy among supporters may imply that their involvement is more spontaneous or issue-driven, rather than being part of an organized effort to defend or promote their own opinions. This lack of cohesion could also explain why supporters are less influential in shaping the overall tone of a conversation compared to attackers.

In this section, we proposed a novel method to identify strategic behaviors among different user groups. Then, we explicated three particular strategies that users may employ (time, topic, people). We presented our model evaluation in~\cref{sec:5.7} and analyzed the differences in strategic behaviors among the three groups in~\cref{sec:6.1}. Our findings suggest that attackers are more strategic in their approach, targeting specific users and topics to maximize their impact and influence the conversation. In contrast, bystanders exhibit a more passive engagement pattern, while supporters do not demonstrate a clear strategic approach. 
\section{Change in Journalists' Behaviors After Attacks (RQ3)}\label{sec:6.2}
Now we explore whether the journalist's posting behavior is influenced by the volume of attacks they receive. Specifically, we aim to examine the relationship between the time intervals between consecutive posts and the proportion of replies from attackers that journalists receive in response to their posts.

For each conversation initiated by a journalist's post, we calculated the proportion of attackers in the thread. We further categorize a post as part of a ``non-toxic conversation'' if the proportion of attackers' replies is 0\%, as part of a ``low-toxic conversation'' if the proportion of attackers' replies is below 20\% but is higher than 0\%, and as part of a ``high toxic conversation'' if more than 80\% of the replies are from attackers. 

\begin{figure}[htbp]
    \centering
    \begin{subfigure}[b]{0.3\textwidth}
        \centering
        \includegraphics[width=\textwidth]{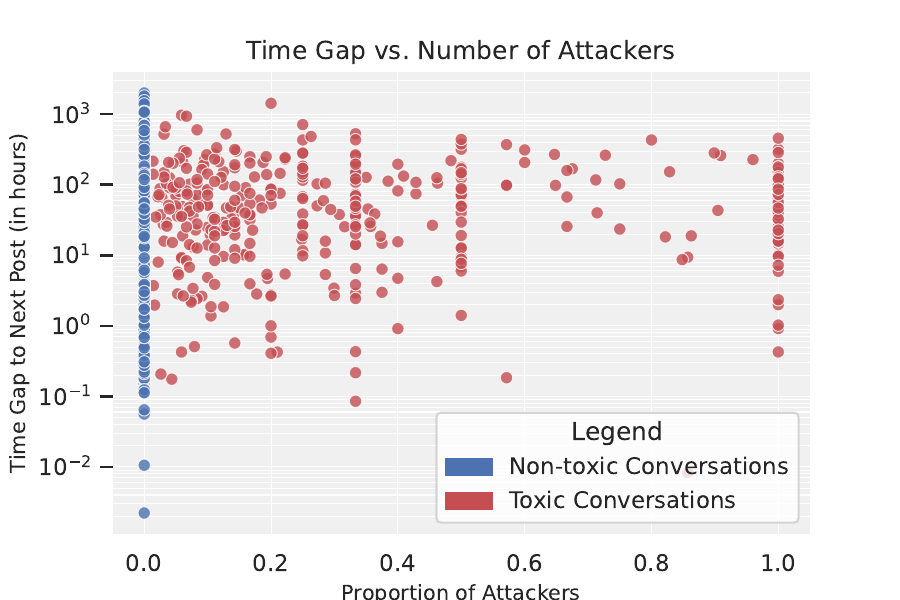} 
        \caption{Time Gap v.s. Attacker Ratios}
        \label{fig:agg_time_vs_ratio}
    \end{subfigure}
    \hfill
    \begin{subfigure}[b]{0.33\textwidth}
        \centering
        \includegraphics[width=\textwidth]{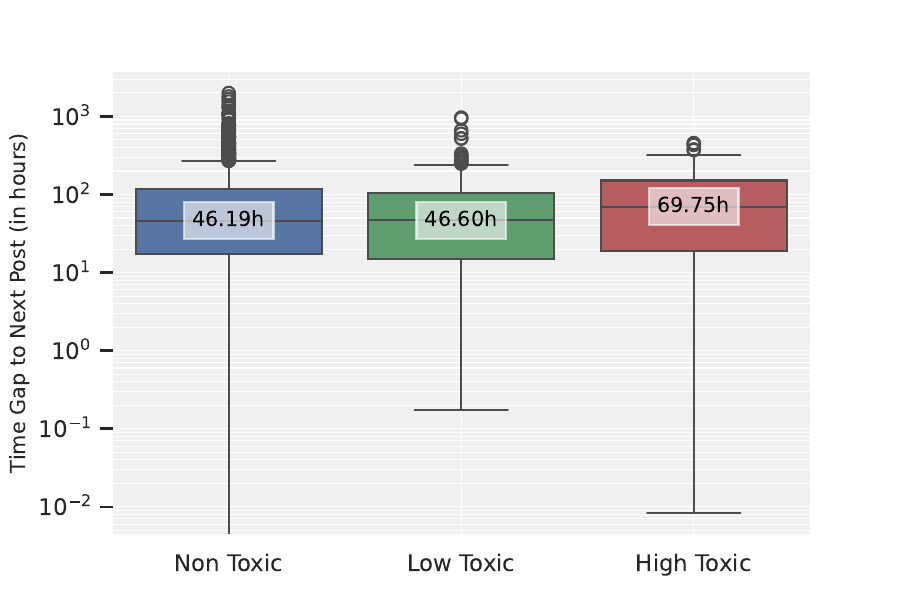} 
        \caption{\small{Time Gap Boxplots}}
        \label{fig:agg_time_gap_distribution}
    \end{subfigure}
    \hfill
    \begin{subfigure}[b]{0.3\textwidth}
        \centering
        \includegraphics[width=\textwidth]{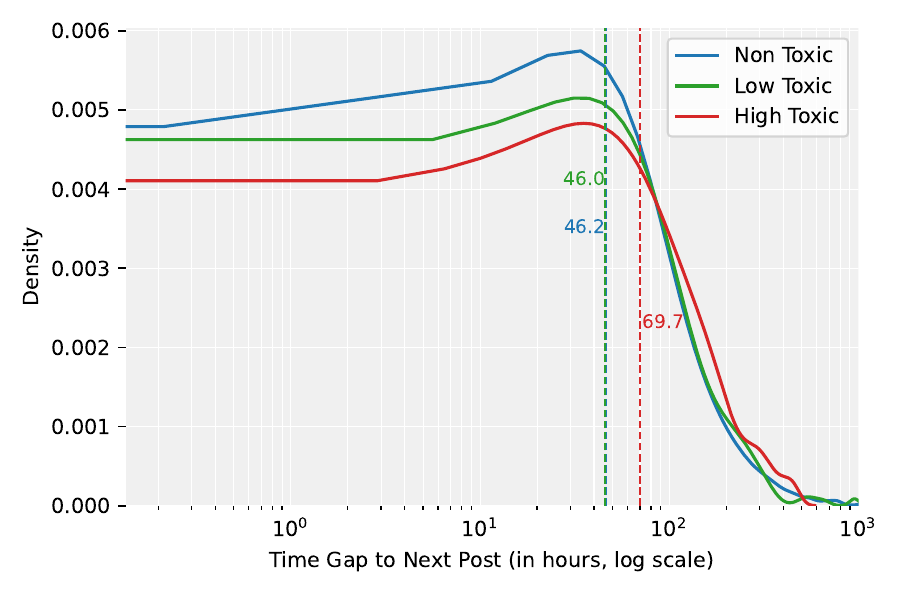} 
        \caption{Time Gap Distributions.}
        \label{fig:agg_kde}
    \end{subfigure}
    \caption{Journalists' Posting Frequency in Different Environments. Figure~\ref{fig:agg_time_vs_ratio} shows the distribution of time gaps (in hours, log scale) between consecutive posts, plotted against the proportion of attackers' replies in the previous post. Blue markers represent non-toxic conversations, while red markers indicate posts that received toxic replies. The figure suggests that journalists tend to delay or pause before posting again after receiving massive attacks (red), leading to longer time gaps. In contrast, when receiving no attacks (blue), the time gaps are more variable. Figure~\ref{fig:agg_time_gap_distribution} compares the time gaps in non-toxic, low-toxic, and high-toxic conversations through a box plot, showing that high-toxic conversations lead to longer time gaps till journalists' next posts. Lastly, Figure~\ref{fig:agg_kde} presents KDE plots, and median values are shown for non-toxic (blue number), low-toxic (green number) and high-toxic (red number) groups. The median line ($x=46.19$) for non-toxic group is covered by the median line ($x=45.96$) for the low-toxic group. We observe that journalists typically delay posting by about a day ($\approx$ 23.5 hours) after encountering more replies from attackers.  (See \Cref{sec:6.2})}
    \label{fig:agg_time}
\end{figure}

Figure~\ref{fig:agg_time_vs_ratio} shows that response times of the journalist to non-toxic conversations are more variable, while the response times in high-toxic conversations are more clustered. To assess whether there is a significant difference between non-toxic and high-toxic conversations, we conducted a t-test with unequal variance, which yields a P-value of $4.85 \times 10^{-8}$, and an F-test which yields a P-value of $2.22 \times 10 ^{-16}$, indicating these two distributions are significantly different. 
\textcolor{black}{This implies that the number of attackers in a conversation influences journalists' posting behavior, with a higher count associated with delays in their subsequent posts.}

This finding is further illustrated in Figure~\ref{fig:agg_time}. Figure~\ref{fig:agg_time_vs_ratio} illustrates the distribution of time gaps (in hours and log scale) between consecutive posts by journalists, plotted against the proportion of replies from attackers in the previous post. The blue markers indicate non-toxic conversations, where the previous post received no replies from attackers, while the red markers represent posts that received replies from attackers. It suggests that when the previous post gets attacks (red), journalists tend to delay or pause before making their next post, resulting in longer and more consistent time gaps. On the other hand, when no attacks are present (blue), the time between posts is more variable, spanning a broader range. \textcolor{black}{This highlights that the presence of attackers' interactions may be associated with "chilling effects," where journalists tend to slow their posting behavior.}
Figure~\ref{fig:agg_time_gap_distribution} presents a box plot comparing the time gaps among posts in non-toxic, low-toxic and high-toxic conversations, showing a slightly shorter time gap in high-toxic conversations. Meanwhile, Figure~\ref{fig:agg_kde}, which depicts KDE plots for the two environments, reveals that journalists tend to delay posting by a day after experiencing more attacks.

One possible explanation for this behavior is that journalists may feel the need to pause after receiving a high volume of attacks, potentially to avoid further hostility. Conversely, when they experience fewer attacks, they may post more frequently to extend their influence. This indicates that the conversational environment can influence journalists' willingness to engage. To counter this, platforms could offer psychological support mechanisms, such as introducing cooldown periods in threads that are identified as toxic. \textcolor{black}{It's worth noticing that different journalists may have different strategies to deal with the attackers, and the combination aggregation across all journalists cannot capture the distinct patterns for all journalists. Therefore, we present the each journalist's posting frequency in~\ref{append:freq}.}

\textcolor{black}{While our findings suggest a chilling effect on the posting behavior of journalists, we recognize that other factors could also contribute to variations in response time. For example, political events, editorial schedules, or personal preferences might influence when journalists choose to post. Additionally, certain journalists may already have lower baseline activity levels, making it difficult to isolate the direct effect of online attacks. A more comprehensive analysis that incorporates other potential confounding factors, \eg, overall news cycle intensity, or individual posting habits, would further clarify the extent to which attacks lead to these behavioral changes.}

\section{Change in Users' Behaviors After Attacks (RQ4)}\label{sec:6.3}
In this section, we examine how users adjust their posting behavior within a conversation thread. Specifically, we ask: In~\cref{sec:6.3.1}, \textit{how does the reply environment change as conversations progress}? \textit{Does the behavior of the attackers change the posting timing of other users}~\cref{sec:6.3.2}? \textit{How do attackers, supporters, and bystanders adjust their tone as the conversation depth increases}~\cref{sec:6.3.3}? Finally, in~\cref{sec:6.3.4}, we ask \textit{do people change their opinions}?

\subsection{How Does the Reply Environment Change as Conversations Progress?}\label{sec:6.3.1}
\begin{figure}[ht]
    \centering
    \begin{subfigure}[b]{0.3\textwidth}
        \centering
        \includegraphics[width=\textwidth]{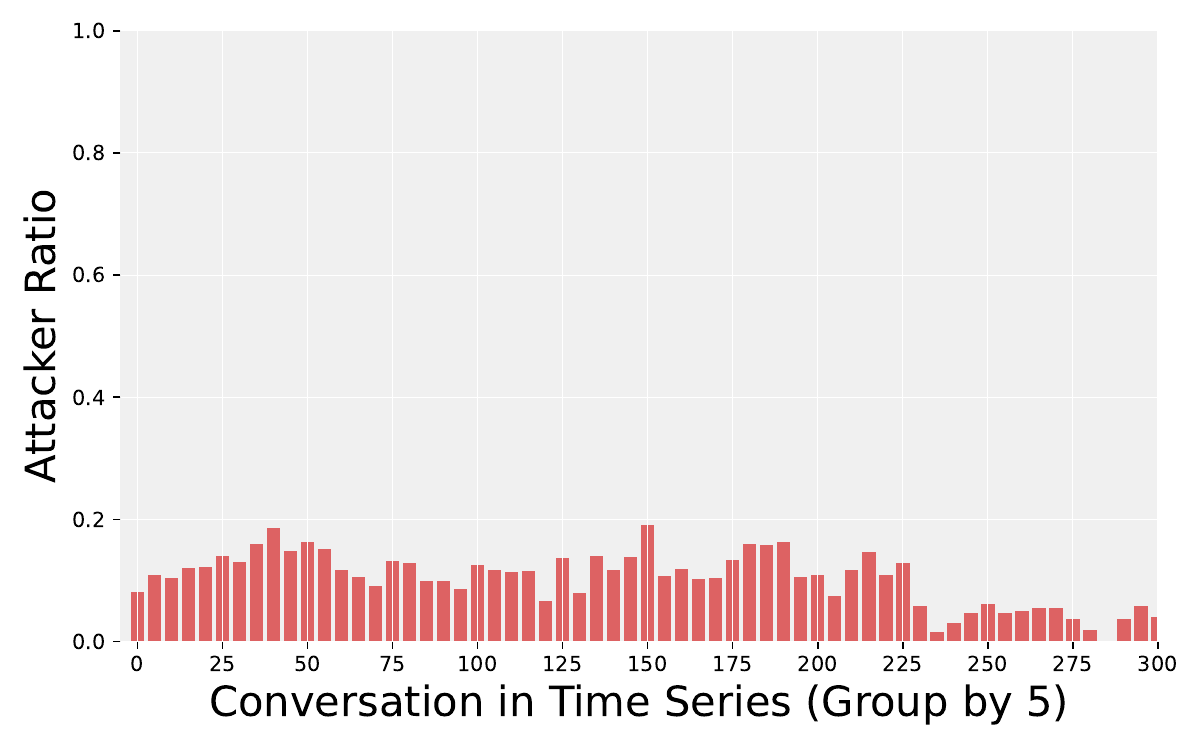} 
        \caption{Attacker fraction over time.}
        \label{fig:agg_attacker_long}
    \end{subfigure}
    \hfill
    \begin{subfigure}[b]{0.3\textwidth}
        \centering
        \includegraphics[width=\textwidth]{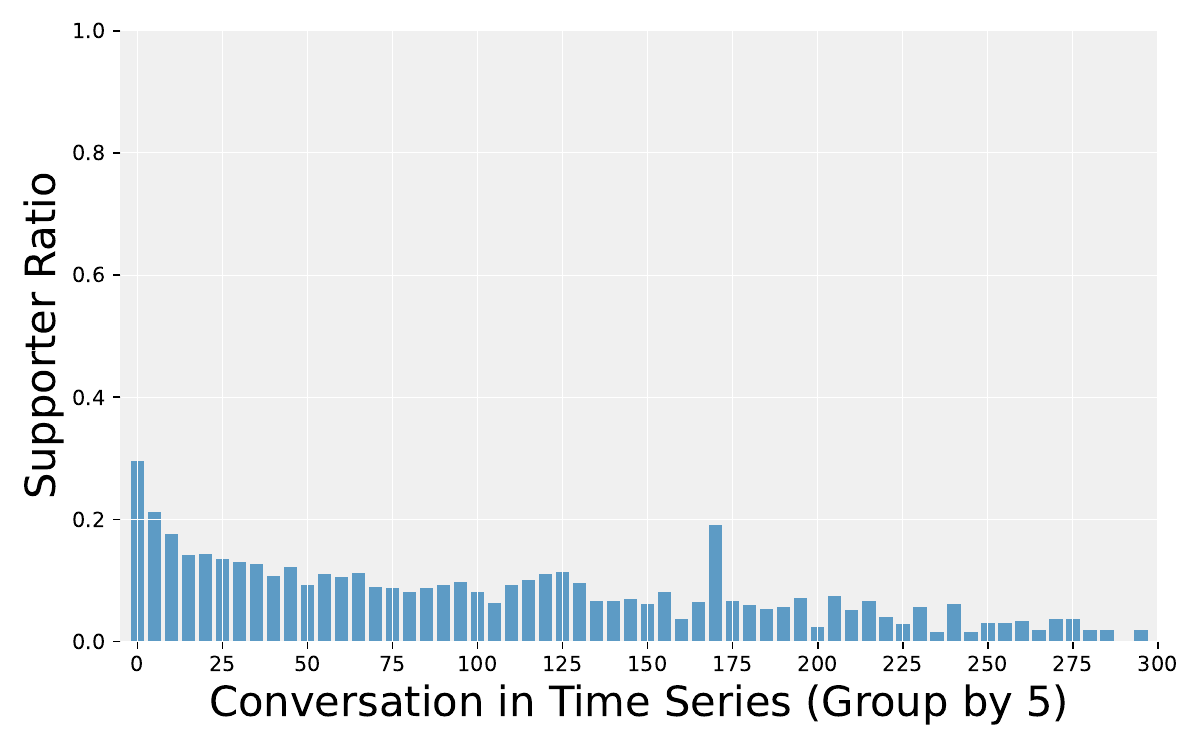} 
        \caption{Supporter fraction over time.}
        \label{fig:agg_sup_long}
    \end{subfigure}
    \hfill
    \begin{subfigure}[b]{0.3\textwidth}
        \centering
        \includegraphics[width=\textwidth]{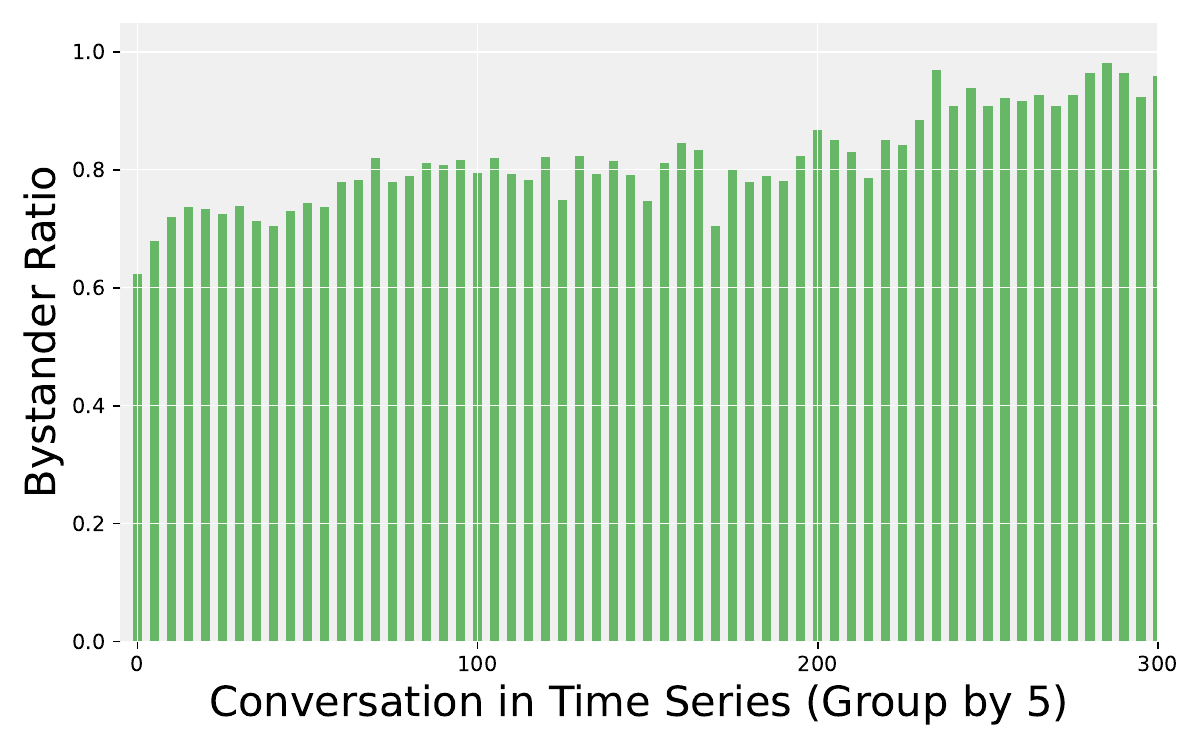} 
        \caption{Bystander fraction over time.}
        \label{fig:agg_by_long}
    \end{subfigure}
    \caption{Ratio of attackers (Figure~\ref{fig:agg_attacker_long}), supporters (Figure~\ref{fig:agg_sup_long}), and bystanders (Figure~\ref{fig:agg_by_long}) over the length of conversations, grouped in time series (grouped by 5 levels). 
    As conversations progress, the supporter ratio consistently decreases, while the bystander ratio steadily increases, suggesting a shift in focus away from the original post and a decline in active participation by supporters. (See \Cref{sec:6.3.1}). 
    }
    \label{fig:agg_ratio_long}
\end{figure}

For each conversation, we track user labels and toxicity scores over time, then average the results across all conversations across all journalists to generate Figures~\ref{fig:agg_ratio_long}. To smooth the data, we apply a sliding window with a size of 5 after averaging the conversations by their length. This allows us to better observe trends.

Our analysis reveals that as conversations progress, the number of supporters decreases, while the number of bystanders increases. One possible explanation is that as discussions continue, they tend to deviate from the original topic or focus of the journalist’s post. Instead, the conversations shift to interactions between other users, moving away from the initial subject matter. This observation also suggests that supporters may become discouraged from posting replies as the conversation progresses. The increasing presence of bystanders and the decreasing engagement from supporters imply that the evolving tone of the discussion may deter active participation from those who initially support journalists while prompting more activities from bystanders.

\begin{figure}[ht]
    \centering
    \begin{subfigure}[b]{0.3\textwidth}
        \centering
        \includegraphics[width=\textwidth]{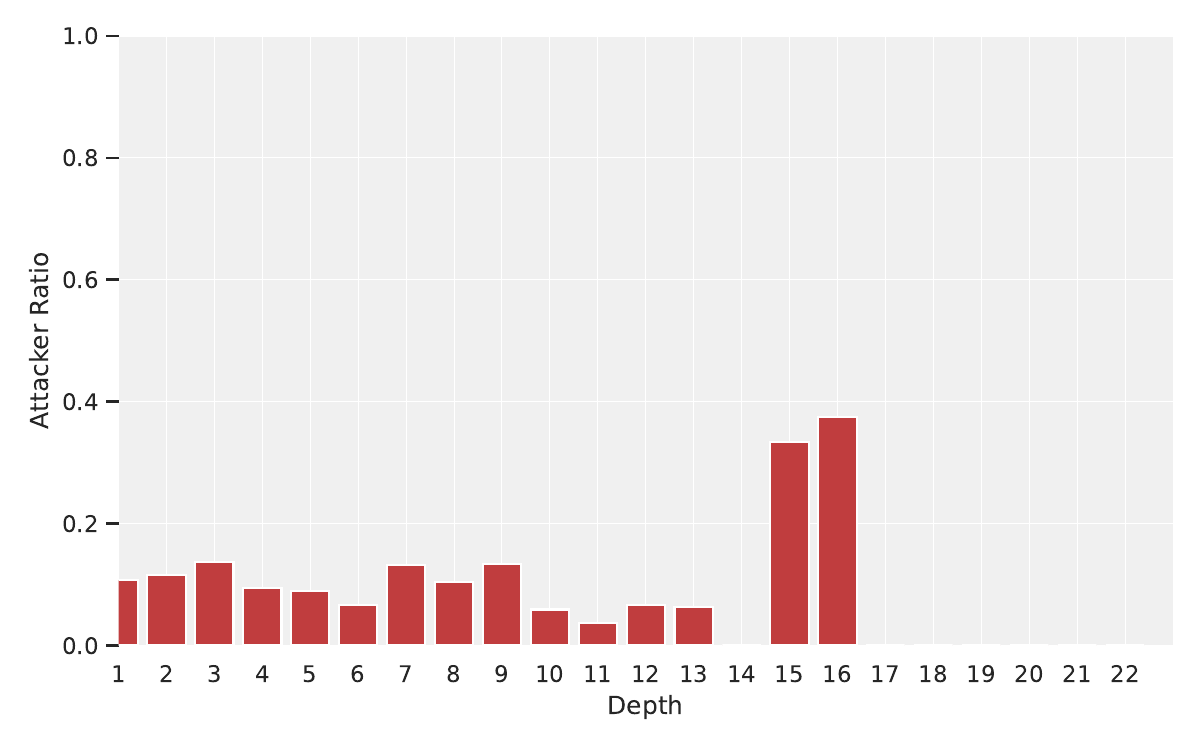} 
        \caption{Attacker by conv. depth.}
        \label{fig:agg_attacker_depth}
    \end{subfigure}
    \hfill
    \begin{subfigure}[b]{0.3\textwidth}
        \centering
        \includegraphics[width=\textwidth]{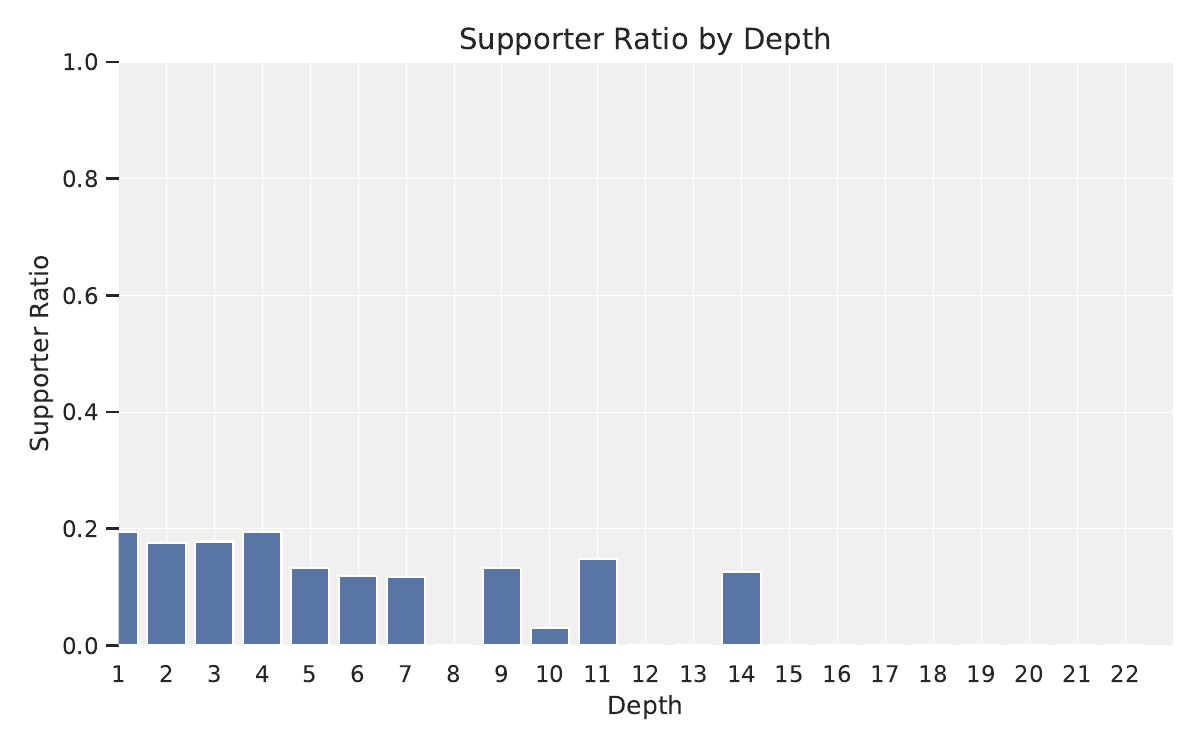} 
        \caption{Supporter by conv. depth.}
        \label{fig:agg_sup_depth}
    \end{subfigure}
    \hfill
    \begin{subfigure}[b]{0.3\textwidth}
        \centering
        \includegraphics[width=\textwidth]{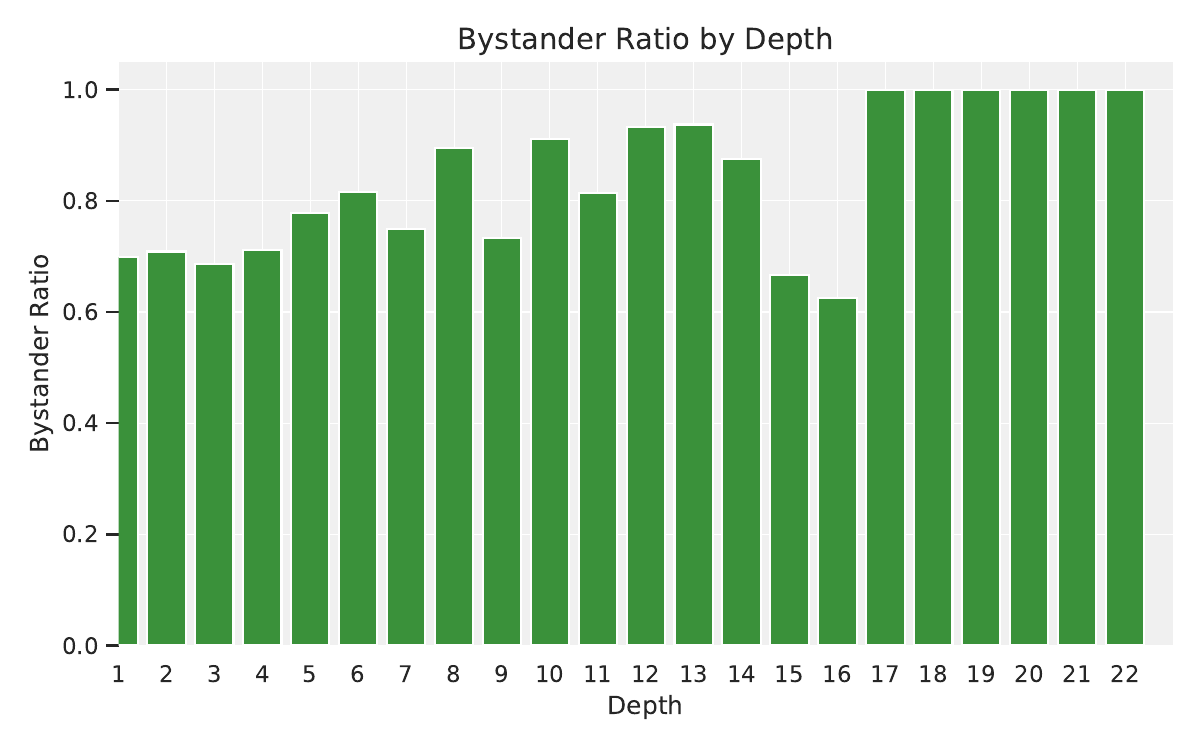} 
        \caption{Bystander by conv. depth.}
        \label{fig:agg_by_depth}
    \end{subfigure}
    \caption{Ratio of attackers, supporters, and bystanders based on conversation depth. Figure~\ref{fig:agg_attacker_depth} shows the attacker ratio, Figure~\ref{fig:agg_sup_depth} shows the supporter ratio, and Figure~\ref{fig:agg_by_depth} shows the bystander ratio. The attacker ratio shows a peak at greater depths, particularly around depth 15, while the supporter ratio declines as depth increases. In contrast, the bystander ratio steadily grows with conversation depth, suggesting that deeper conversations tend to attract more passive observers and fewer active supporters. (See \Cref{sec:6.3.1})}
    \label{fig:agg_ratio_depth}
\end{figure}

We also calculated the ratios of different user groups based on conversation depth, as shown in Figure~\ref{fig:agg_ratio_depth}. In our case, conversation depth is a structural measure derived from the conversation's tree-like hierarchy. Each reply in a conversation is assigned a depth based on its position in the thread relative to the initial post. The results are consistent with our previous observations, reinforcing the trend of decreasing supporter participation and increasing bystander presence as conversations deepen.
The rise in bystander participation, coupled with their toxicity, indicates that while more users are present in the discussion, they are less likely to contribute positively. This dynamic can exacerbate the problem, as fewer constructive voices are heard, and the focus of the conversation moves away from the original topic.


\subsection{How Do Attackers Change the Posting Timing of Other Users?}\label{sec:6.3.2}
\begin{figure}[ht]
    \centering
        \includegraphics[width=0.6\textwidth]{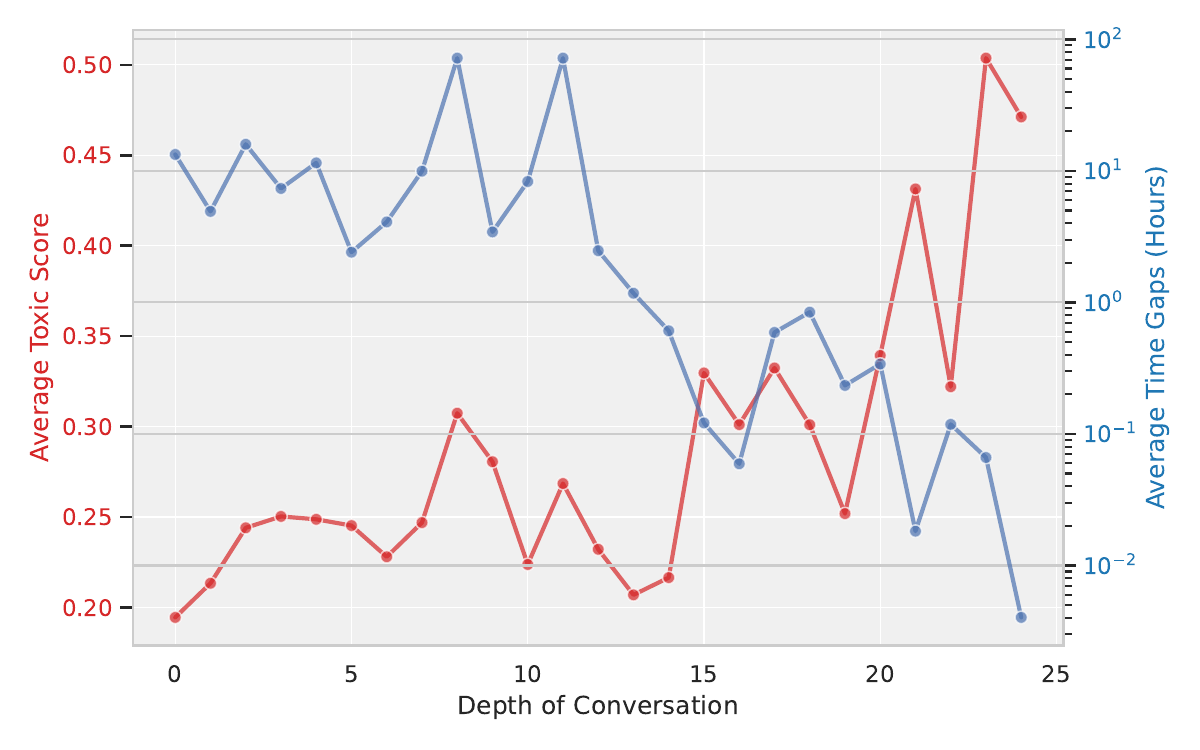} 
        
    \caption{Trends in post toxicity, time gaps, and their combined effects as conversations deepen. 
    The red line shows the average post toxicity, and the blue line indicates the average time gap. The results show a strong correlation between rising toxicity and shorter time gaps, highlighting how discussions are getting more aggressive.}
    \label{fig:agg_overall_vs_depth_combined}
\end{figure}

We aim to investigate whether attackers influence the posting behavior of targeted users. As illustrated in Figure~\ref{fig:agg_overall_vs_depth_combined}, we observe that as conversations deepen, the toxicity scores tend to increase, while the time gaps between replies decrease. This pattern suggests that as discussions become longer, they also become more intense, with escalating hostility and a quicker pace of replies, likely driven by the increasingly toxic environment. These indicate a positive correlation between the depth of conversations and the level of aggression. However, as we point out in~\cref{fig:agg_ratio_long}, and in~\cref{sec:6.3.4}, the increase in toxicity is partly explained by \textit{attackers and supporters attacking each other}. This leads to a neutral stance towards the journalist's post, thus classified as bystanders by our framework. This observation suggests that the presence of attackers can influence the behavior of other users, leading to more aggressive and faster-paced discussions.

\subsection{How Do Users Adjust Their Tone as the Conversation Depth Increases?}\label{sec:6.3.3}
\begin{figure}[ht]
    \centering
    \includegraphics[width=0.6\textwidth]{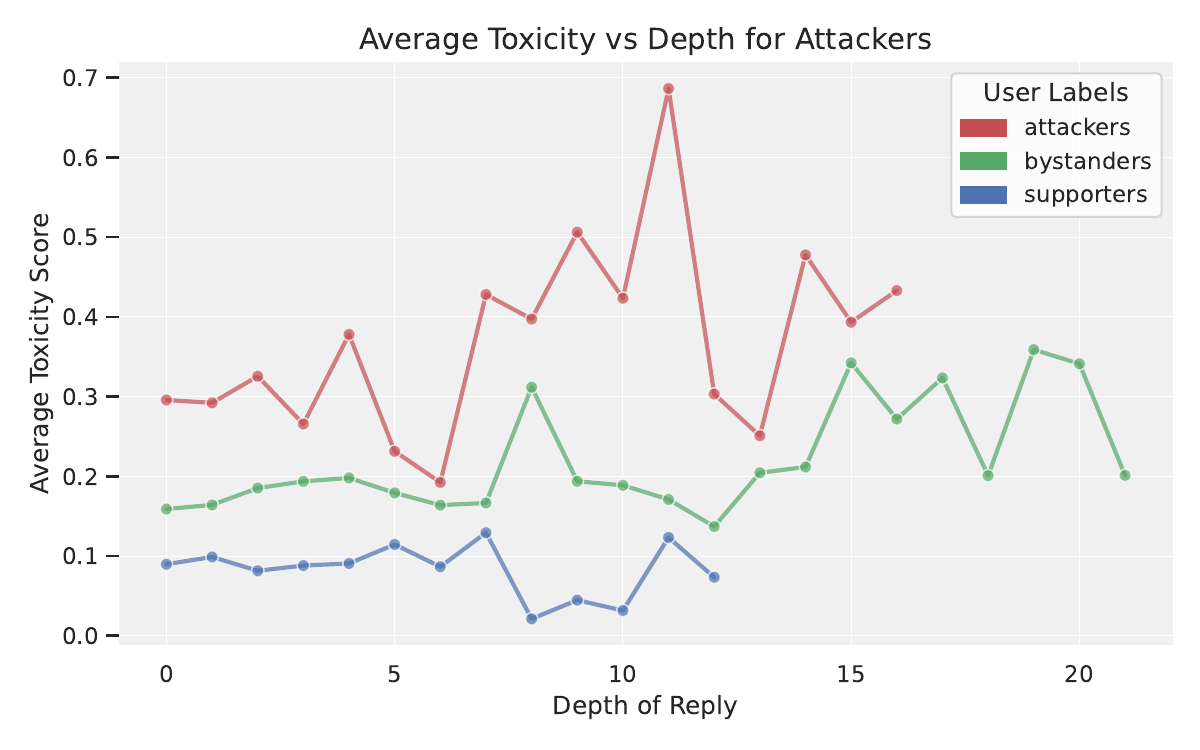} 
    \caption{Toxicity Scores of Different Groups by Depth (See \Cref{sec:6.3.3}). We observe that (1) Supporters are present only during the early stages of the conversations, suggesting that the increasingly toxic environment and the presence of attackers discourage them from participating in the discussions; (2) Both attackers and bystanders are getting more toxic as the conversation progresses, indicating that attackers may incite more toxic responses from bystanders.}
    \label{fig:agg_toxic_vs_depth_012}
\end{figure}

To further investigate the behaviors of attackers, supporters, and bystanders, we analyze their toxicity scores across different conversation depths. As shown in Figure~\ref{fig:agg_toxic_vs_depth_012}, we find that in the early stages, attackers’ replies exhibit the highest toxicity. However, as the conversations progress, both attackers and bystanders become increasingly toxic. This trend implies that as discussions deviate from the original posts, the posts from attackers can incite more toxic messages from bystanders.

\subsection{Do Users Change Their Opinions?}\label{sec:6.3.4}
In this section, we explore how users’ opinions evolve during the evolvement of a conversation.

\begin{figure}[ht]
        \centering
        \includegraphics[width=0.7\textwidth]{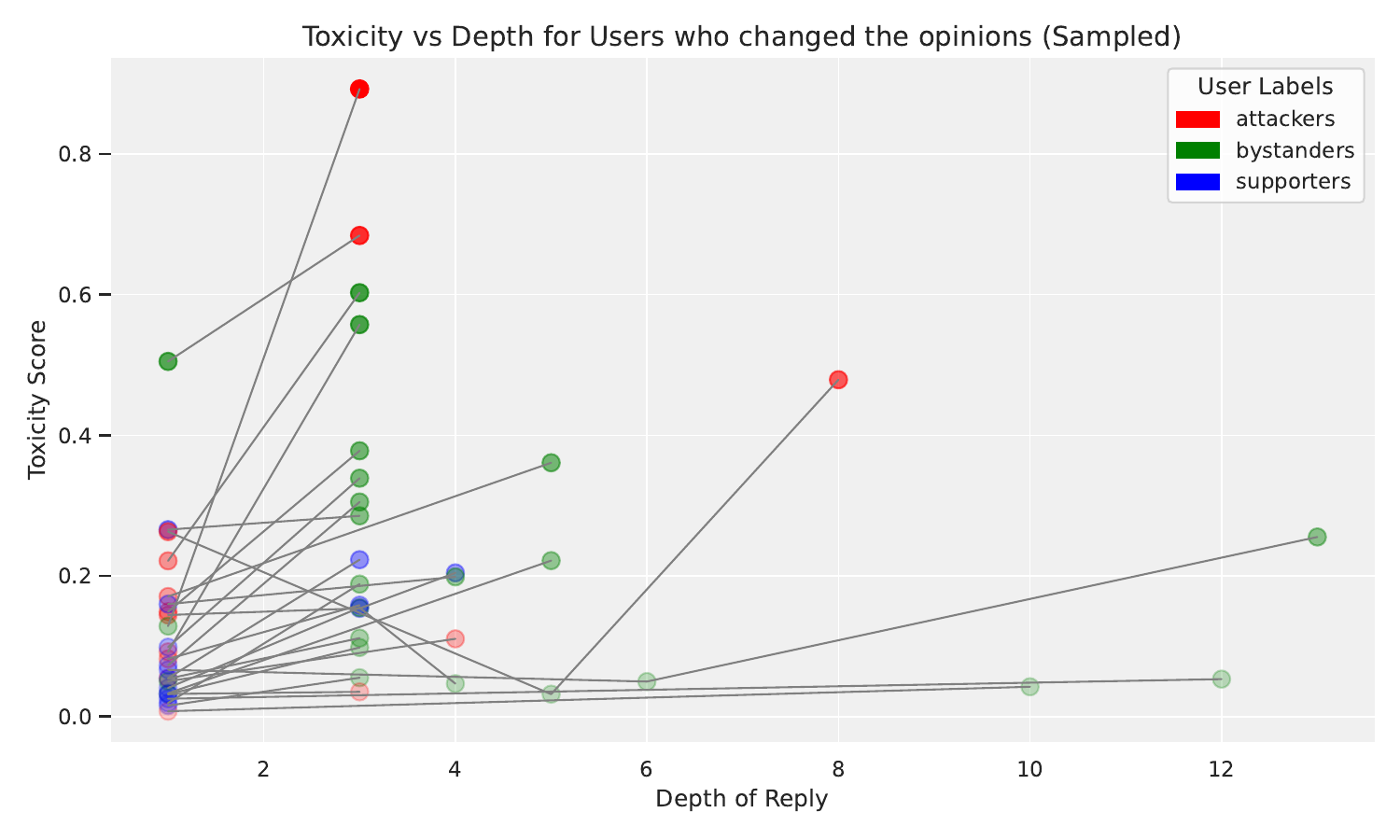} 
    \caption{Toxicity Change Examples. Most of the users change to bystanders and are getting more toxic. There are also users changing from bystanders to attackers, possibly related to the interactions with other attackers. (See~\cref{sec:6.3.4})}
    \label{fig:agg_toxic}
\end{figure}

In Figure~\ref{fig:agg_toxic}, we highlight several users who change their stance throughout the discussion. Notably, most users shift to becoming bystanders, suggesting that they are straying from the original post. Additionally, we observe a rise in toxicity scores, which is consistent with our earlier findings, indicating that as users' focus shifts away from the main topic, the overall tone of the conversation becomes more toxic.

\begin{figure}[htbp]
  \centering
\begin{tikzpicture}[
  node distance=0.9cm and 6cm,
  every node/.style={font=\footnotesize\sffamily},
  class/.style={circle, draw, minimum size=0.9cm, font=\bfseries, text=black},
  attacker/.style={class, fill=red!20},
  bystander/.style={class, fill=green!20},
  supporter/.style={class, fill=myblue}
]

\node[attacker] (A1) {A};
\node[bystander, below=of A1] (B1) {B};
\node[supporter, below=of B1] (S1) {S};

\node[attacker, right=6cm of A1] (A2) {A};
\node[bystander, below=of A2] (B2) {B};
\node[supporter, below=of B2] (S2) {S};

\node[left=0.2cm of A1] {\scriptsize 23\%};
\node[left=0.2cm of B1] {\scriptsize 52.4\%};
\node[left=0.2cm of S1] {\scriptsize 24.6\%};

\node[right=0.2cm of A2] {\scriptsize 18.5\%};
\node[right=0.2cm of B2] {\scriptsize 61.8\%};
\node[right=0.2cm of S2] {\scriptsize 19.7\%};

\draw[->, draw=yellow!90!black, line width=2.31pt] (A1) -- (A2) node[pos=0.3, above] {\scriptsize 57.7\%};
\draw[->, draw=green!50!black, line width=1.49pt] (A1) -- (B2) node[pos=0.3, above, sloped] {\scriptsize 37.2\%};
\draw[->, draw=black, dashed, line width=0.20pt] (A1) -- (S2) node[pos=0.2, below, sloped] {\scriptsize 5.1\%};

\draw[->, draw=black, line width=0.224pt] (B1) -- (A2) node[pos=0.8, above, sloped] {\scriptsize 5.6\%};
\draw[->, draw=yellow!90!black, line width=3.448pt] (B1) -- (B2) node[pos=0.7, above] {\scriptsize 86.2\%};
\draw[->, draw=black, line width=0.328pt] (B1) -- (S2) node[pos=0.7, below, sloped] {\scriptsize 8.2\%};

\draw[->, draw=black, dashed, line width=0.168pt] (S1) -- (A2) node[pos=0.2, above, sloped] {\scriptsize 4.2\%};
\draw[->, draw=green!50!black, line width=1.456pt] (S1) -- (B2) node[pos=0.3, above, sloped] {\scriptsize 36.4\%};
\draw[->, draw=yellow!90!black, line width=2.376pt] (S1) -- (S2) node[pos=0.3, below] {\scriptsize 59.4\%};

\node[below=0.6cm of S1, font=\bfseries\small] {Initial};
\node[below=0.6cm of S2, font=\bfseries\small] {Final};

\node[attacker, minimum size=0.5cm, right=1.2cm of A2] (LA) {};
\node[anchor=west] at ([xshift=0.15cm]LA.east) {\tiny Attacker (A)};
\node[bystander, minimum size=0.5cm, below=0.4cm of LA] (LB) {};
\node[anchor=west] at ([xshift=0.15cm]LB.east) {\tiny Bystander (B)};
\node[supporter, minimum size=0.5cm, below=0.4cm of LB] (LS) {};
\node[anchor=west] at ([xshift=0.15cm]LS.east) {\tiny Supporter (S)};

\end{tikzpicture}
\caption{The probability of change of opinions. Yellow arrows represent no change of opinions, meaning the majority of users maintain their initial positions. Besides, a great number of supporters and attackers change to bystanders (green arrows), suggesting that users are engaging in discussions irrelevant to the original posts. Black dashed arrows denote transit from attackers to supporters or supporters to attackers, after checking some cases, we find that these transitions are false positive cases.}
    \label{fig:agg_user_label_change}
\end{figure}
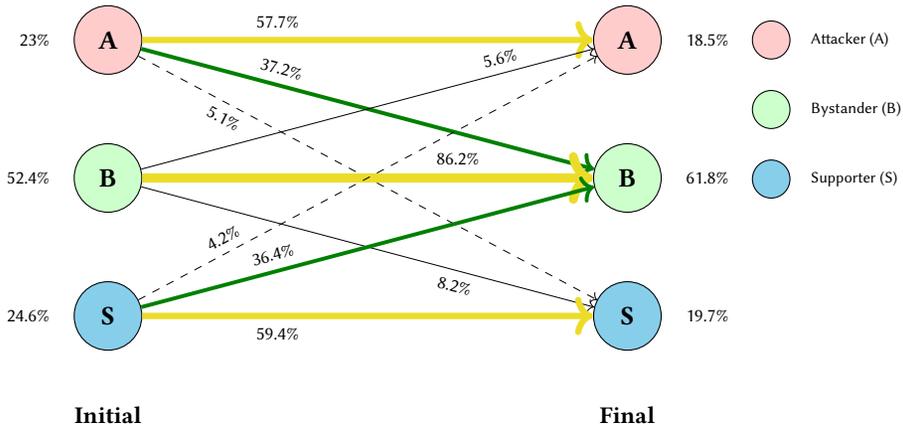

To provide a statistical representation of users' stance changes, we plot the conditional probability of users changing their stance given their initial opinions throughout the conversation in Figure~\ref{fig:agg_user_label_change}. The results show that the majority of users maintain their initial positions (yellow arrows). However, we also observe a notable number of both supporters and attackers transitioning to bystanders (green arrows), which suggests that these users are engaging in off-topic discussions or becoming less actively involved in the core debate. This shift further underscores the trend of conversations drifting away from the original subject as they progress. 
There are also edges transit from attackers to supporters or supporters (black dashed arrows) to attackers, after checking some cases, we find that these transitions are False positive cases.
We also notice that compared to the initial ratio of bystanders, the overall ratio of bystanders increases while the numbers of attackers and bystanders decrease.

In this section, we observed that as conversations increase (either over time or depth), there are fewer supporters and attackers, and more bystanders~\cref{sec:6.3.1}. Furthermore, we found that the toxicity of the conversation increases, and the time between replies decreases (\cref{sec:6.3.2,sec:6.3.3}). Finally, we discovered in \cref{sec:6.3.4} that approximately a third of the attackers and supporters change to bystanders, suggesting that they are attacking each other (since they are classified as bystanders by our framework) and engaged in toxic conversations.

\subsection{Are There Attention Seekers?}
\textcolor{black}{Studying groups who actively seek conflict, such as attention seekers, is intriguing, and at the same time challenging as their motivations are complex and not always clear. Furthermore, their interactions and engagement patterns can shift over time, influenced by external factors such as audience reactions or responses from the original poster. Here is how we find attention seekers in our work: \textit{they are users who reply multiple times within one conversation, with at least one reply directed at the journalist and exhibiting toxic behavior (with toxic scores greater than 0.6).} While it is beyond the scope of our current work, we have some interesting findings based on our definition of attention seekers.}

\begin{figure}[htbp]
    \centering
    \includegraphics[width=1.0\linewidth]{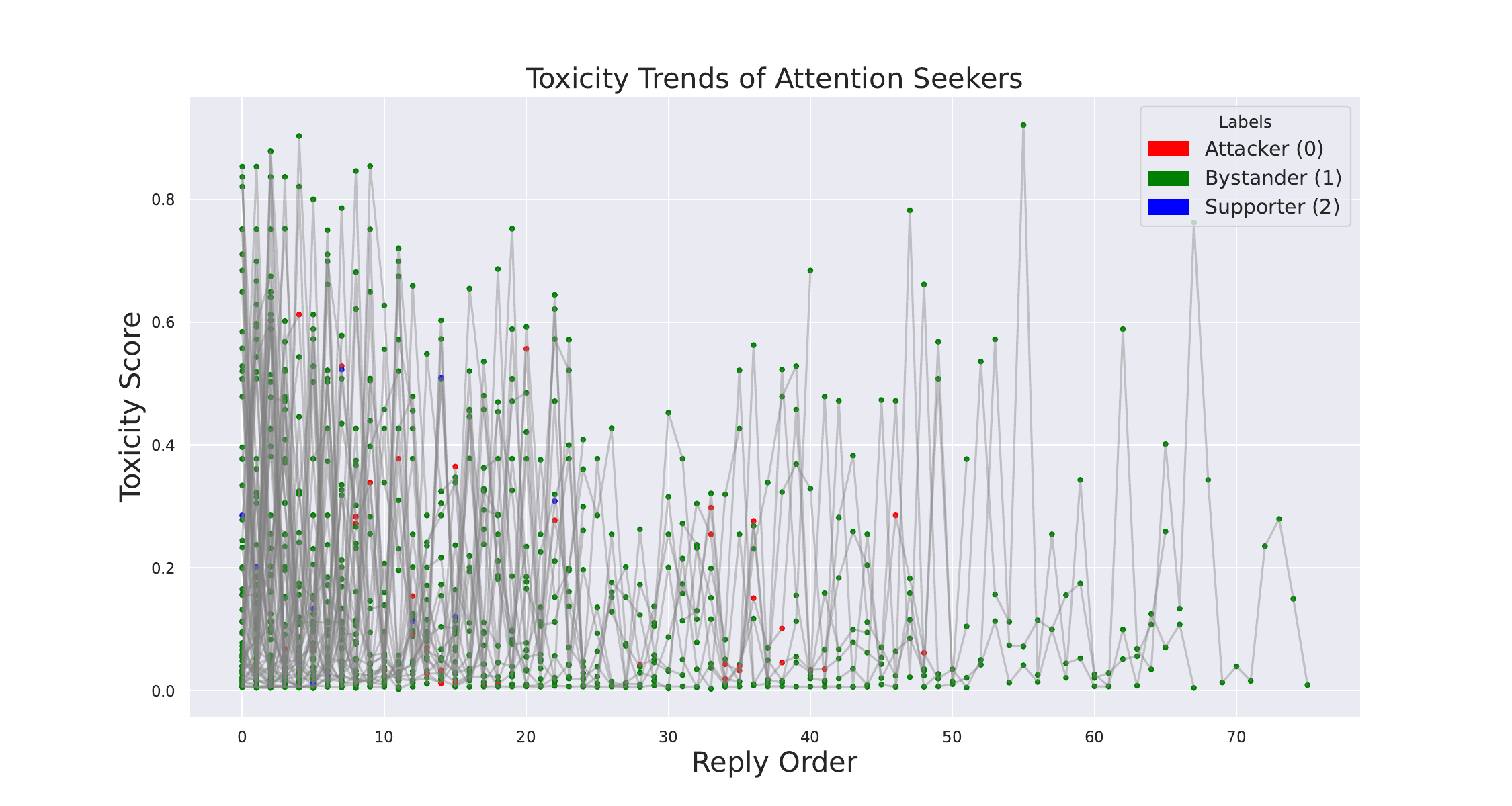}
    \caption{Toxicity Change Examples of Attention Seekers. We cut the length of replies to 100 for clarity.}
    \label{fig:toxic_attention}
\end{figure}

\noindent \textcolor{black}{\textbf{Toxicity Change Examples within Attention Seekers.} In Figure~\ref{fig:toxic_attention}, we illustrate how the toxicity scores and labels of attention seekers evolve over time. We observe that supporters can participate in a conversation multiple times with some toxic content. With closer inspection, most supporters exhibit toxicity when responding to attackers or expressing strong agreement with journalists. Additionally, some attackers continuously reply to journalists or other users within the same conversation, aiming to provoke conflict or directly attack the journalist. Furthermore, we notice that toxic bystanders emerge at the later stages of conversations, suggesting they may be discussing off-topic subjects with a toxic tone, diverging from the original topic.}

\noindent \textcolor{black}{\textbf{Bomb Replying.} We also observe that some users engage in "bomb replying" within conversations, \ie, repeatedly replying to journalists or other users within a short period. Most of these users are attackers attempting to overwhelm or provoke the target. Additionally, some attackers repeatedly reply with identical toxic content. Below in~\cref{tab:bomb}, we provide an example of an attacker bomb replying to a journalist with the same message:}
\begin{table}[h]
    \centering
    \begin{tabular}{lp{5cm}p{4cm}}
        \toprule
        \textbf{Users} & \textbf{Replies} & \textbf{Time} \\
        \hline
        \textbf{User 1} & "Propagandist. \#NewNuremburg"
        & 2021-06-20 19:02:24 \\ 
        & & 2021-06-20 19:02:33 \\ 
        & & 2021-06-20 19:02:43 \\ 
        & & 2021-06-20 19:03:55 \\ 
        & & 2021-06-20 19:04:02 \\ 
        & & 2021-06-20 19:05:12 \\ \hline
        
        \textbf{User 2} & "What is the difference between you and a thief? To defraud money is no different from a thief."
        & 2022-12-27 02:33:38 \\ 
        & & 2022-12-27 03:07:38 \\ 
        
        & "The golden master behind you has begun to have trouble again, it seems that it is short of money."
        & 2022-12-09 02:05:32 \\ \hline
        
        \textbf{User 3} & "The golden master behind you has begun to have trouble again, it seems that it is short of money."
        & 2022-12-09 15:23:15 \\
        \hline
    \end{tabular}
    \caption{\textcolor{black}{Examples of false positive and false negative edges.}}
    \label{tab:bomb}
\end{table}

\textcolor{black}{We also find cases where different users reply with identical toxic content (user 2 and user 3), suggesting the possibility of coordinated attacks or automated bot activity. Such repetition of the exact same message could indicate attempts to amplify harmful narratives or harass the journalist. This pattern is particularly concerning when the duplicated messages are aimed at the journalist or spreading misinformation. Further analysis would be needed to determine whether these users are part of an organized effort or simply engaging in copy-paste behavior.}

\section{Discussion}
\label{sec:Discussion}
Our findings indicate the presence of clear coordination among attackers, a lack of coordination among supporters, and increase in the frequency of toxic behavior (though not at the journalist; about a third of the attackers and supporters end up attacking each other) as conversations deepen, and 
that journalists experience ``chilling effects''---delaying their responses significantly---due to toxic, attacker responses.
Now we discuss the implications for the governance of social media platforms, and the design of tools for journalists and end-users.

\subsection{Fostering Healthier Online Environments and Resilient Conversations}
Our findings have implications for constructing healthier conversation environments on social platforms. By identifying patterns of coordinated attacks early, platforms can intervene by reducing the visibility of toxic content or preventing its spread. Besides, the passive engagement of bystanders, who are not influenced by topic relevance or timing, implies that platforms could encourage more positive interaction by prompting users to engage with constructive content. Additionally, combining Figure~\ref{fig:combined} and Figure~\ref{fig:utility} (See~\cref{sec:5.3}), we observe that the primary strategies used by attackers are associated with higher utility scores. This suggests that attackers may have to an extent inferred the relationship between strategies and payoffs and employ strategies that attract more attention, providing further evidence of coordinated behavior. 

The observation that bystanders become more toxic than attackers in the later stages of a conversation has significant implications. It suggests that passive participants may be influenced by the toxic replies and shift toward more aggressive behavior over time. This deviation from the original topic and the rise in bystander toxicity implies that conversations are not only becoming more hostile but are also moving away from constructive discourse.

To foster healthier environments, platforms need to implement proactive strategies to interrupt this cycle of escalating toxicity~\cite{zhang2018conversations, hessel2019something} and promote resilience~\cite{lambert2022conversational} in the face of adverse events within online conversations. Potential interventions like slowing the pace of replies in highly toxic conversations, reducing the visibility of inflammatory content, or providing users with tools to de-escalate hostile interactions could help curb the aggression.

\subsection{Implications for  Online Moderation}
\label{sub:Moderation Implications}

Platforms face significant challenges in moderation, and while this is well documented~\cite{Gibson2019,Jhaver2021,Jhaver2019,Srinivasan2019a,Chandrasekharan2017,Chandrasekharan2017a,Chandrasekharan2019,Koshy2023}, much of the work deals with what we could classify as instance-based moderation. That is, much of the research focuses on what is to be done when a \textit{single} post is toxic, including content removal~\cite{Srinivasan2019a}, de-platforming~\cite{Chandrasekharan2017}, design of moderation tools~\cite{Chandrasekharan2019} for platforms such as Reddit that rely on volunteer moderators or decentralized moderation schemes~\cite{Gordon2022}. Our work, suggests that moderation tools dealing with groups of coordinated attackers are urgently needed. Furthermore, our work suggests that the attackers are not necessarily attacking the journalist, but are disrupting the conversation (and conversational outcomes).

The lack of coordination among supporters (in contrast to coordination among attackers) may partly explain why toxic behavior predominates. This suggests that the platform ought to be proactive in identifying instances of coordination among attackers. And when the platform can identify coordinated groups, take measures to ensure that the behavior of these groups is not disruptive. For example, platforms could consider first warning members of the group that their behavior is disruptive, and if the behavior persists, the platform could consider mechanisms that disrupt the coordination strategy employed among the members of the group. If the platform discovers that the group is using a particular strategy where the members respond similar topics, or by members similar to themselves, that platform could consider mechanisms that disrupt this strategy (\eg delaying receiving or posting messages on this topic or from similar individuals).

\subsection{Design Implications: Tools for Journalists and Users}
\label{sub:Tools for Journalists and Users}

Journalists on social media platform deserve enhanced protection. As~\citet{Posetti2021} mention in their report to UNESCO, the journalists face significant harassment based on their identity, with a significant number reporting threats of violence. Our findings show that journalists delay their responses due to toxic responses, suggesting that journalists who are subject to online attacks experience chilling effects. Sustained toxic environments may lead these journalists to potentially withdraw from the platform, or discontinue engaging with their audience, resulting in a critical loss to maintaining a healthy democracy. Platforms could consider mechanisms that protect journalists from toxic behavior, such as having enhanced individual-level moderation tools, including the ability to block or delay responses from users, or ensuring that only users who meet a sufficient threshold of civility (\eg have posted a sufficient number of messages, with a toxicity score below a threshold) can engage with the journalist's audience. They could also support journalists by having a mechanism that allows journalists to report coordinated attacks.

We discovered that as conversations deepen, toxicity increases (not necessarily directed at the journalist; about a third of the attackers and supporters attack each other), and participants begin to respond quickly. To address this, the platform could consider providing user facing tools (\eg a button) to highlight this conversation out to the platform, so that the platform can take action to slow down the conversation, or to provide additional moderation. This could help reduce the toxicity of the conversation, as well as give users time to reflect on their responses.

\section{Limitations and Future Work}
\label{sec:limitations}
In this section, we discuss the limitations of our work and possible future directions. 

\begin{description}
    \item[Limited Dataset] First, our dataset is limited to interactions involving 13 female journalists. While female journalists are particularly vulnerable to online attacks, a broader analysis comparing the impact of attacks on both female and male journalists would provide a more comprehensive understanding. Additionally, exploring attacks across multiple platforms could help us discover the distinct dynamics of different conversations. Future work should therefore aim to include a larger and more diverse sample of journalists across different social media platforms. \textcolor{black}{Testing our model on other datasets or user groups, such as different professions, genders, or platforms, would indeed be valuable to assess the generalizability of our findings. However, extending this analysis to broader contexts is beyond the scope of this study due to inherent methodological constraints. Specifically, patterns of interactions observed on Twitter may not simply transfer to platforms like Reddit—for example, Reddit's community-driven moderation (e.g., upvote/downvote mechanisms, subreddit-specific rules) and nested thread structures differ substantially from Twitter's follower-based interaction model and reply structure, requiring different analytical approaches and data preprocessing methods.}
    \item[Restricted Strategy Space] Second, our analysis of strategies is constrained to a limited set. Other potential strategies or \textcolor{black}{confounding factors}, such as considering a user’s number of friends and followers, could reveal strategic behaviors with finer granularity. Expanding the strategy space in future research would enable a deeper understanding of user behaviors.
    \item[Lack of Private Message Data] Third, we lack access to private message data that could offer insights into why certain users form clusters and how they coordinate. Including private messaging information would enhance our ability to analyze coordinated behaviors more accurately.
\end{description}
\section{Conclusion} \label{sec:8}
In this paper, we analyze how various user groups strategically interact with female journalists on Twitter. A key challenge in this research was identifying the latent strategic behaviors behind users' actions. To tackle this problem, we introduce a novel tree-structured Transformer approach, \method, to classify response groups and represent strategy distributions as hidden vectors that influence subsequent action distributions. Via extensive testing, we find that our approach surpasses current state-of-the-art methods by 8\% for the task of identifying strategies and by 3\% for the task of classifying user accounts \textcolor{black}{and is robust to journalists across diverse cultures}. 
We also demonstrate that the proportion of attackers' replies affects journalists' posting behaviors, with higher toxicity leading to longer delays between posts. Additionally, we show that as conversations deepen, toxicity intensifies, with both attackers and bystanders becoming more toxic over time. We also find that attackers tend to use coordinated strategies to incite toxic messages from bystanders, while supporters lack organization, making their impact less effective. Our findings suggest that platforms should implement tools to detect and mitigate coordinated toxic behavior, support positive contributors, and encourage constructive bystander involvement to foster resilient and healthier online environments.

\bibliographystyle{ACM-Reference-Format}
\bibliography{reference}
\clearpage
\appendix
\section{Appendix}
\label{sec:append}
\subsection{Data Description} \label{append:data}
\textcolor{black}{Here we will describe the time distribution of the original posts of 13 journalists and those of their replies.}
\begin{figure*}[htbp]
  \centering
  \begin{subfigure}[b]{0.3\textwidth}
    \includegraphics[width=\linewidth]{figure/timeline_aliceysu.pdf}
  \end{subfigure}
  \begin{subfigure}[b]{0.3\textwidth}
    \includegraphics[width=\linewidth]{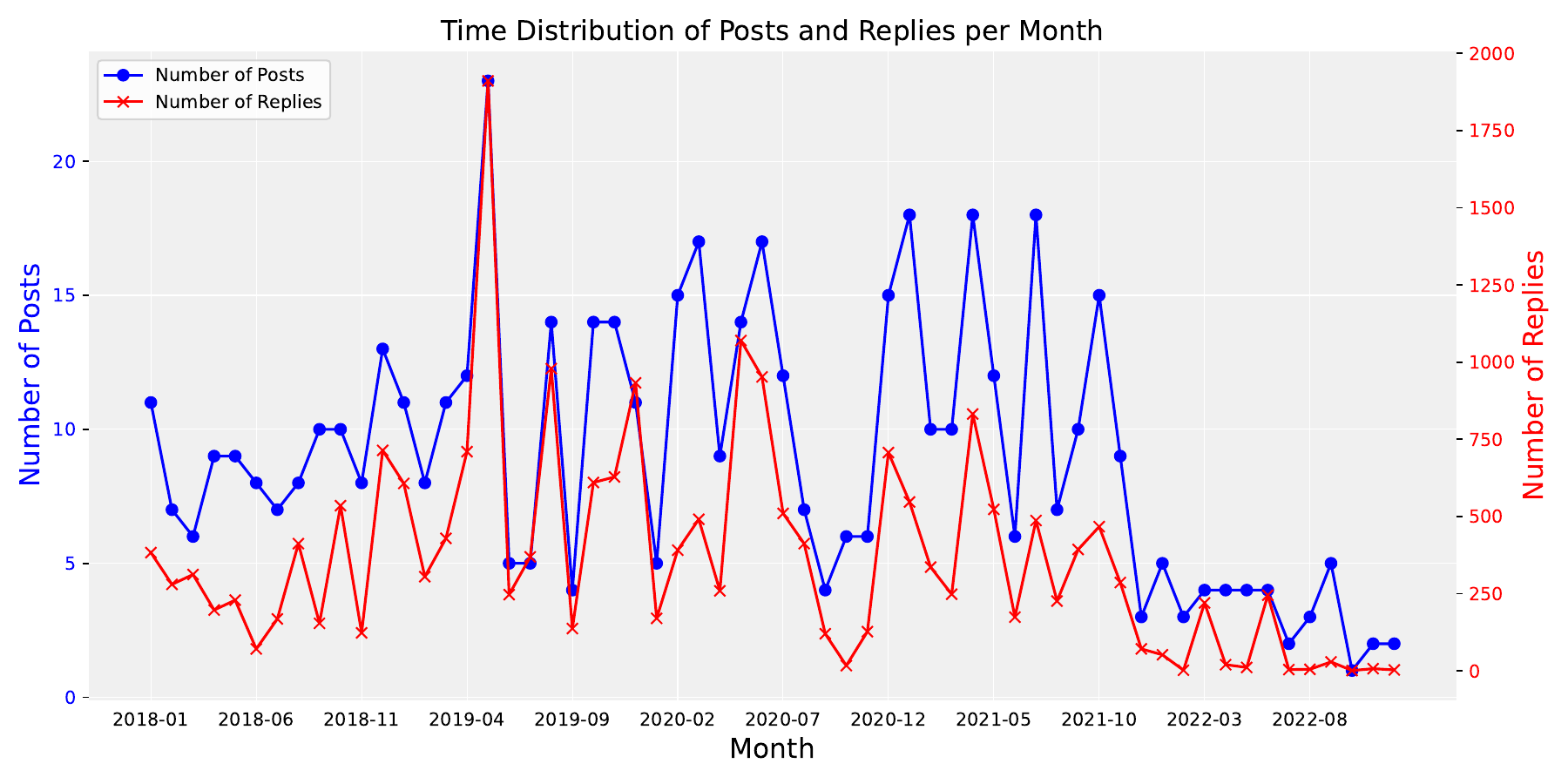}
  \end{subfigure}
  \begin{subfigure}[b]{0.3\textwidth}
    \includegraphics[width=\linewidth]{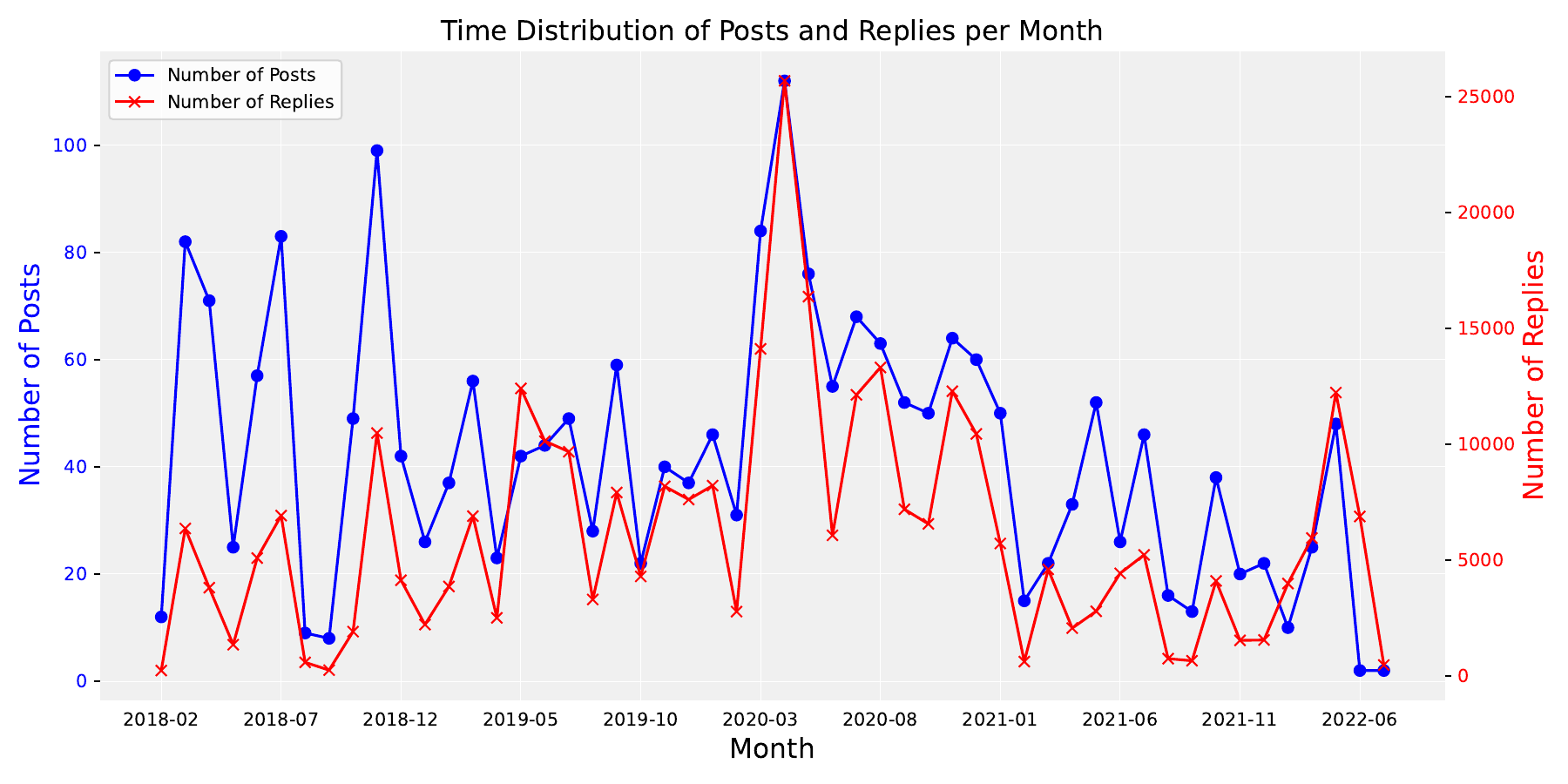}
  \end{subfigure}

  \vspace{1em}
  
  \begin{subfigure}[b]{0.3\textwidth}
    \includegraphics[width=\linewidth]{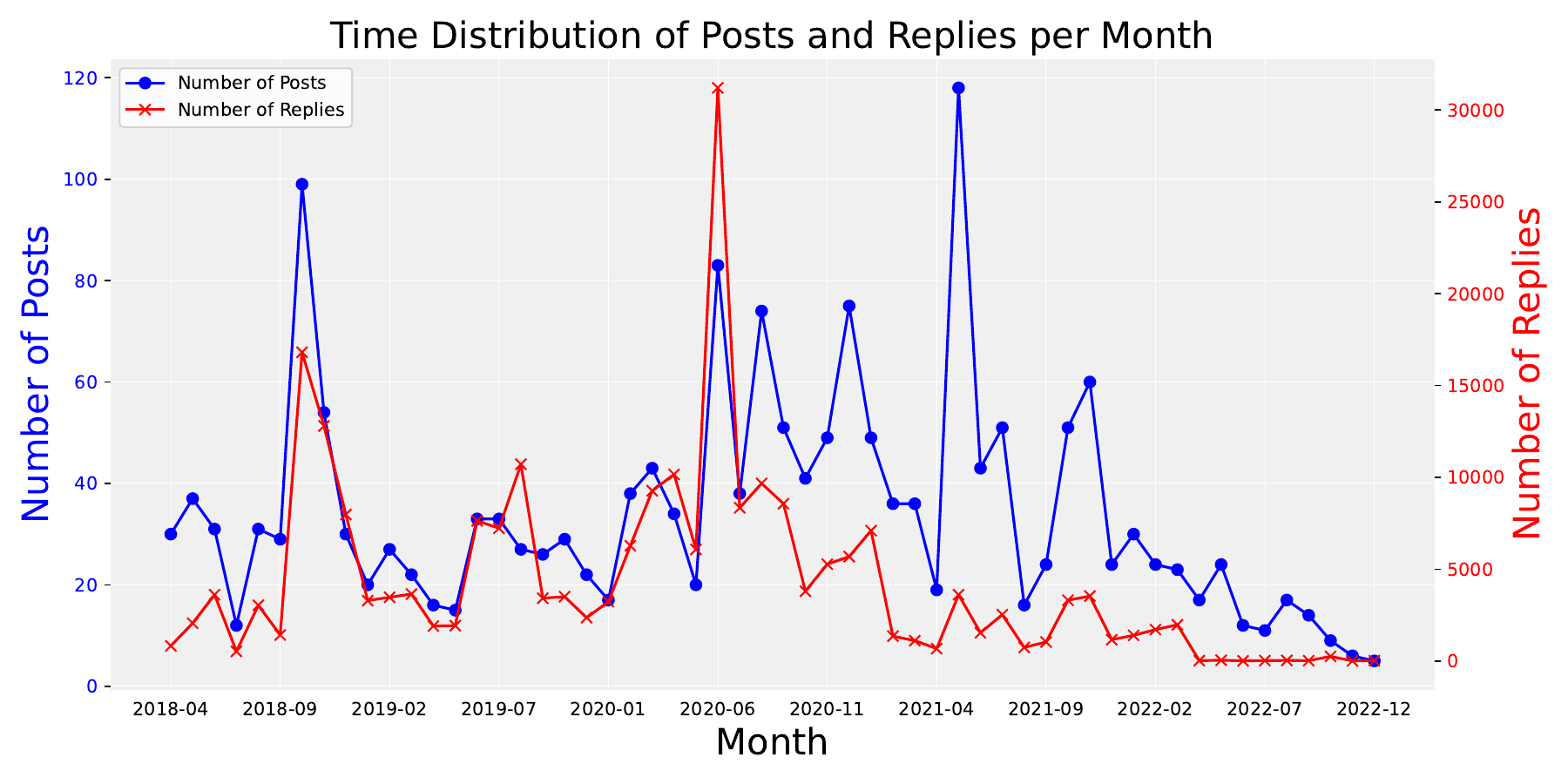}
  \end{subfigure}
  \begin{subfigure}[b]{0.3\textwidth}
    \includegraphics[width=\linewidth]{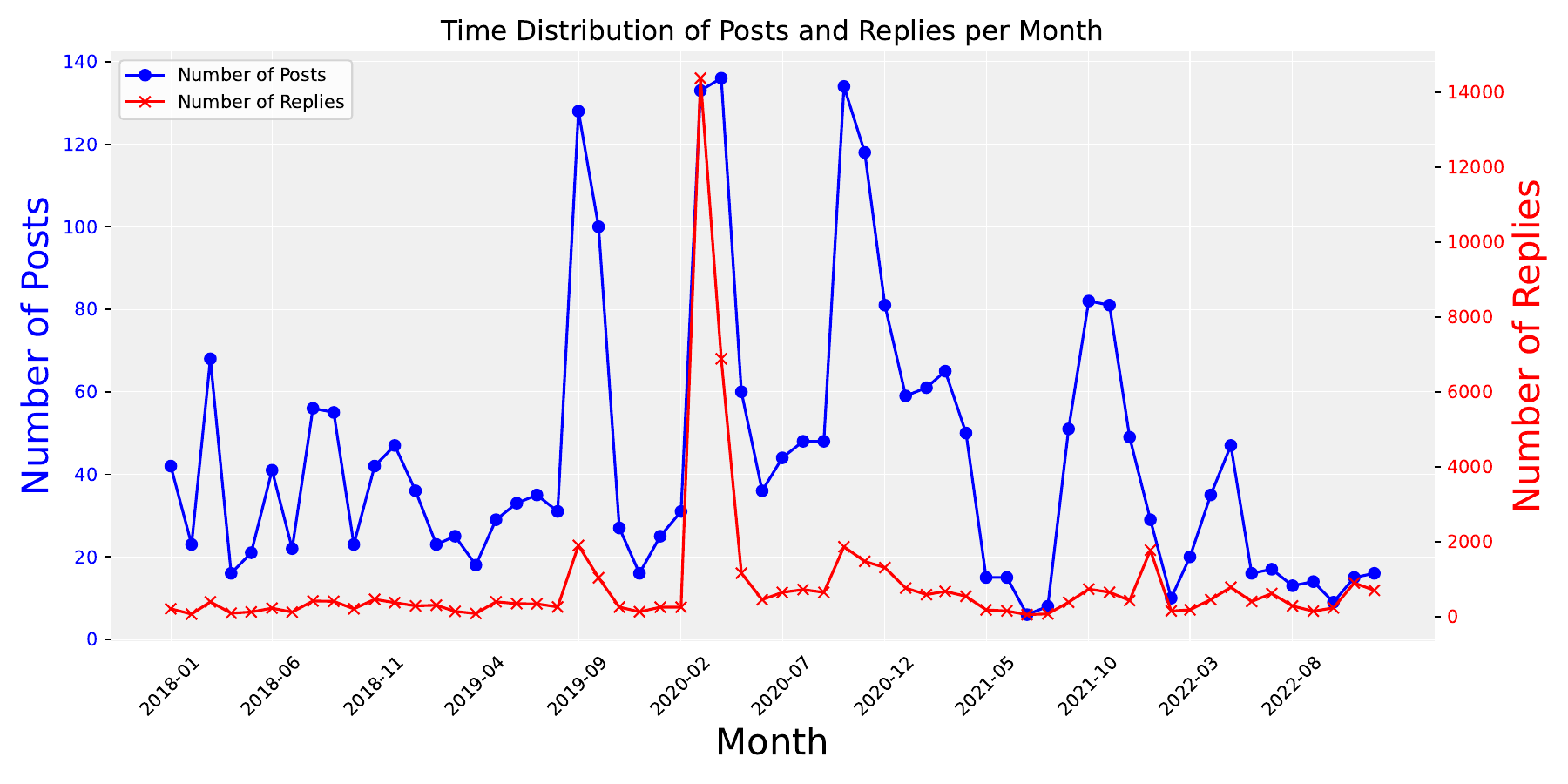}
  \end{subfigure}
  \begin{subfigure}[b]{0.3\textwidth}
    \includegraphics[width=\linewidth]{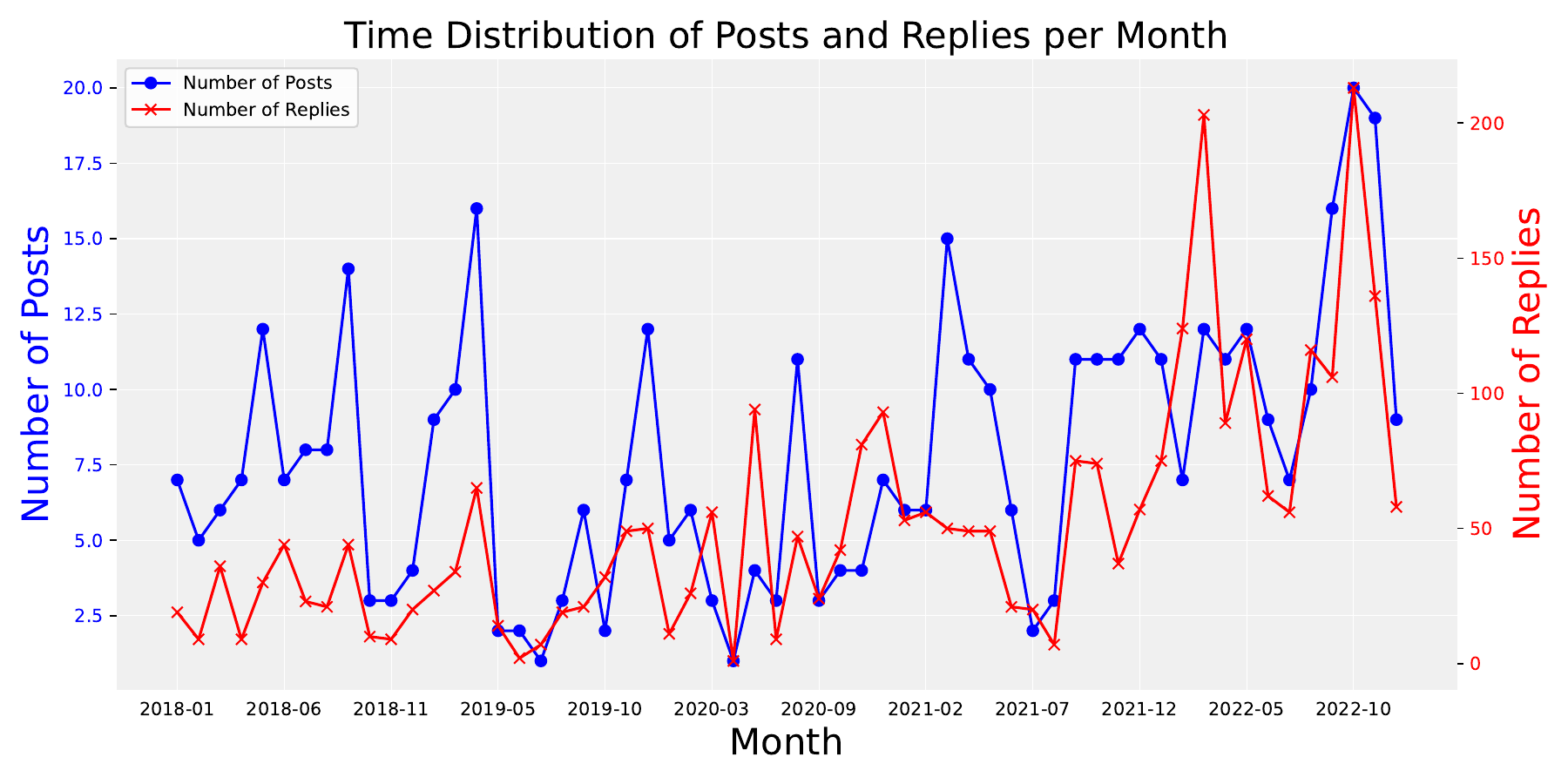}
  \end{subfigure}
  
  \vspace{1em}
  
  \begin{subfigure}[b]{0.3\textwidth}
    \includegraphics[width=\linewidth]{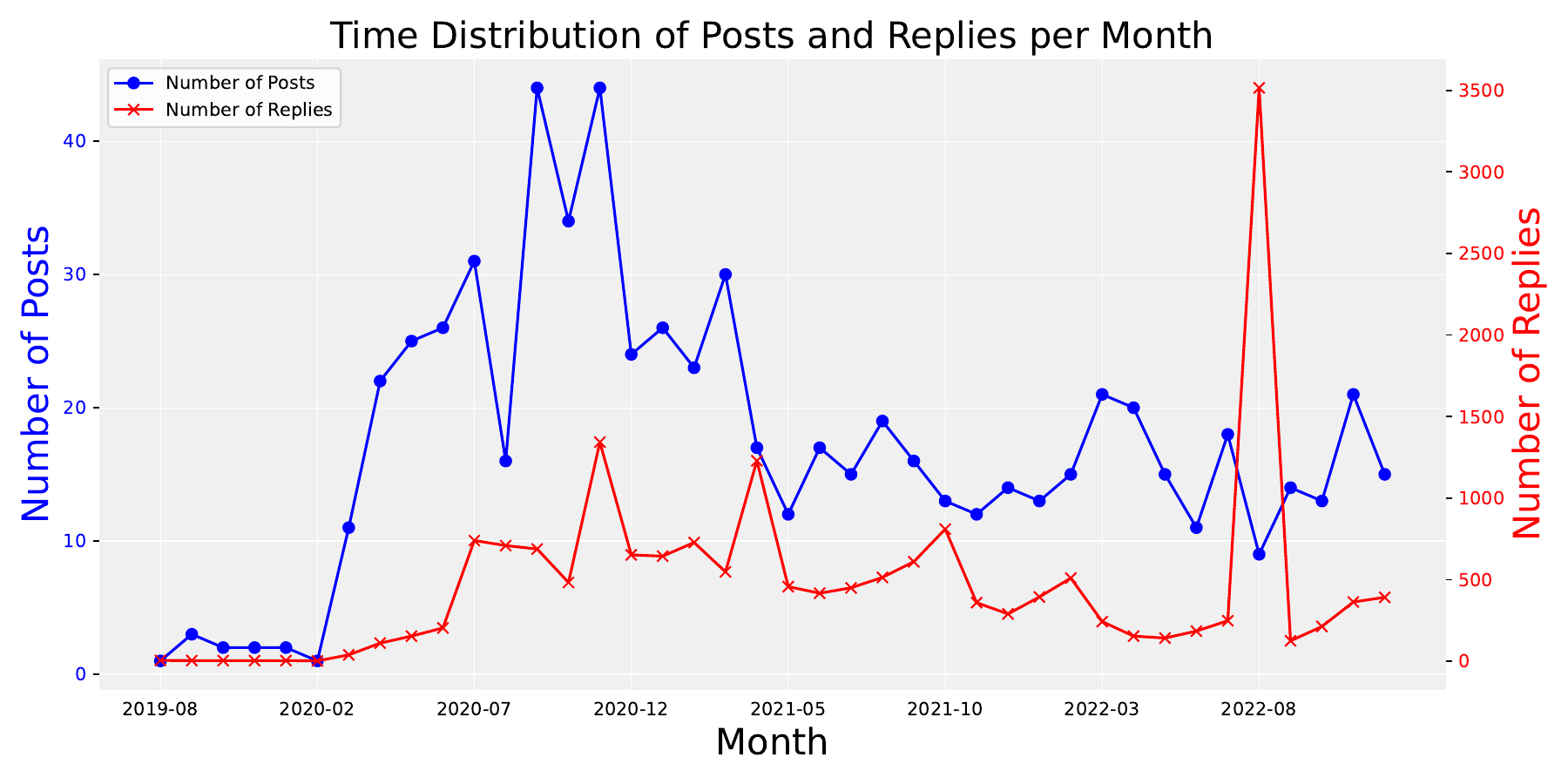}
  \end{subfigure}
  \begin{subfigure}[b]{0.3\textwidth}
    \includegraphics[width=\linewidth]{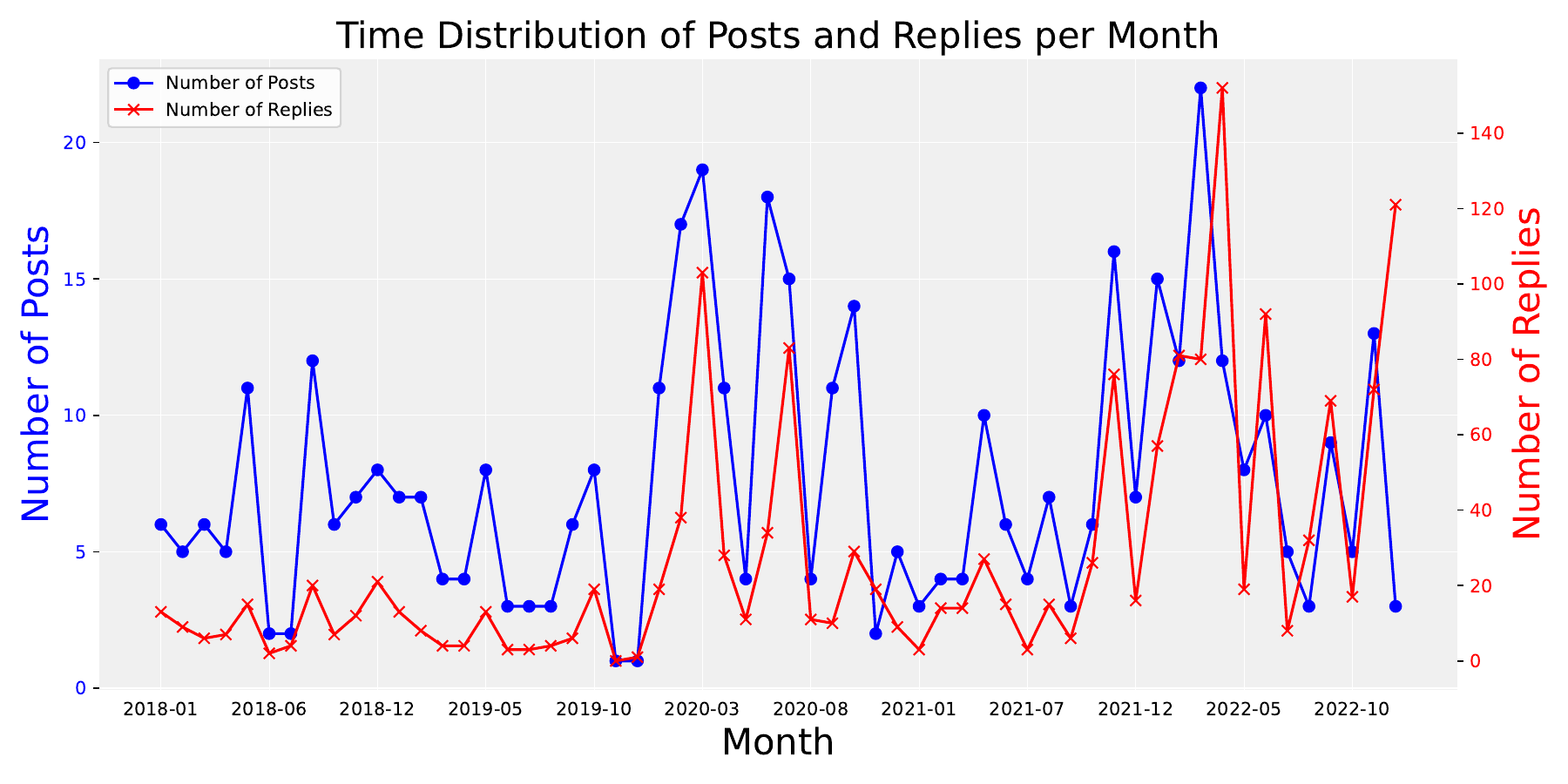}
  \end{subfigure}
  \begin{subfigure}[b]{0.3\textwidth}
    \includegraphics[width=\linewidth]{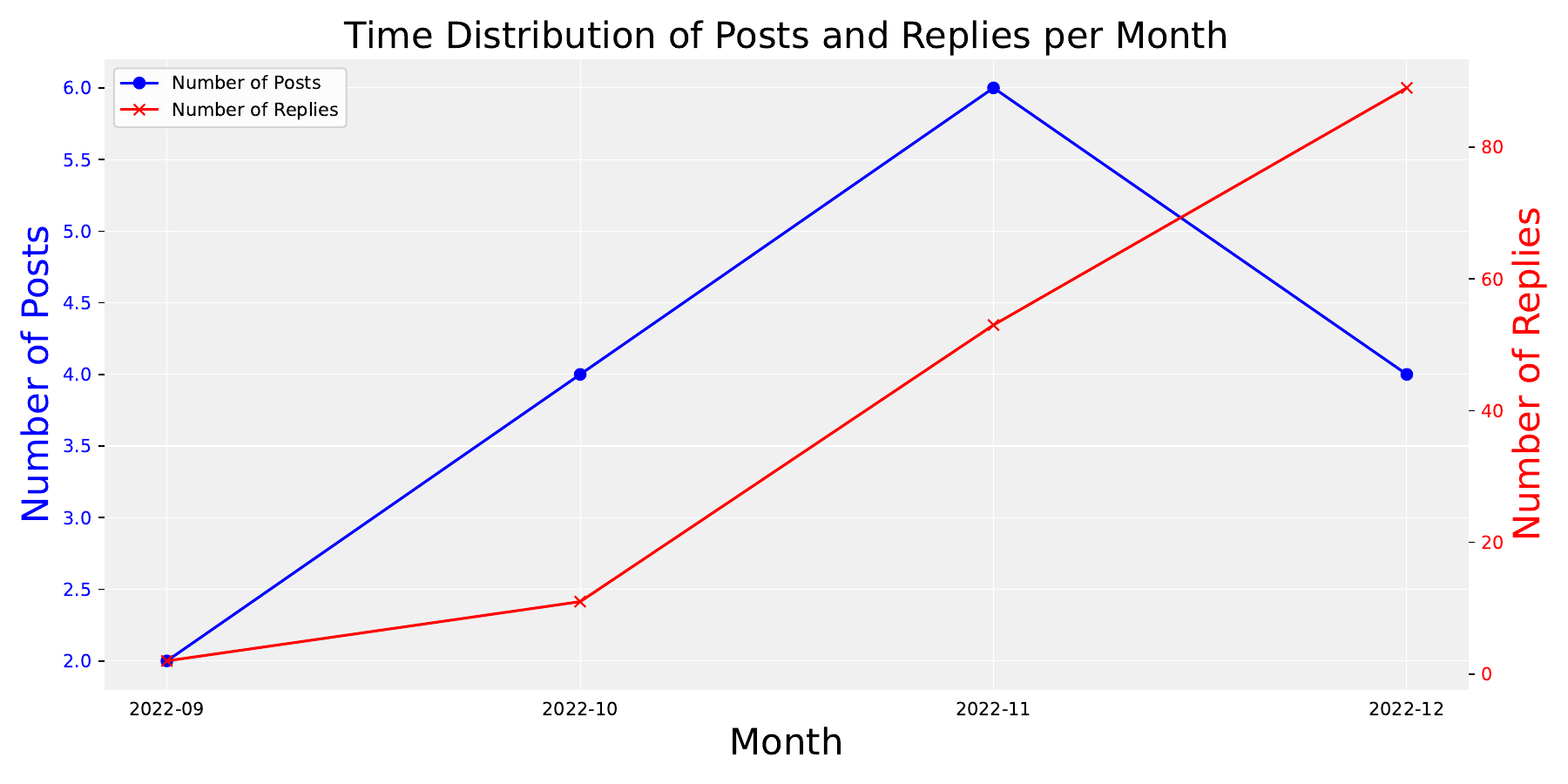}
  \end{subfigure}
  \vspace{1em}
  
  \begin{subfigure}[b]{0.3\textwidth}
    \includegraphics[width=\linewidth]{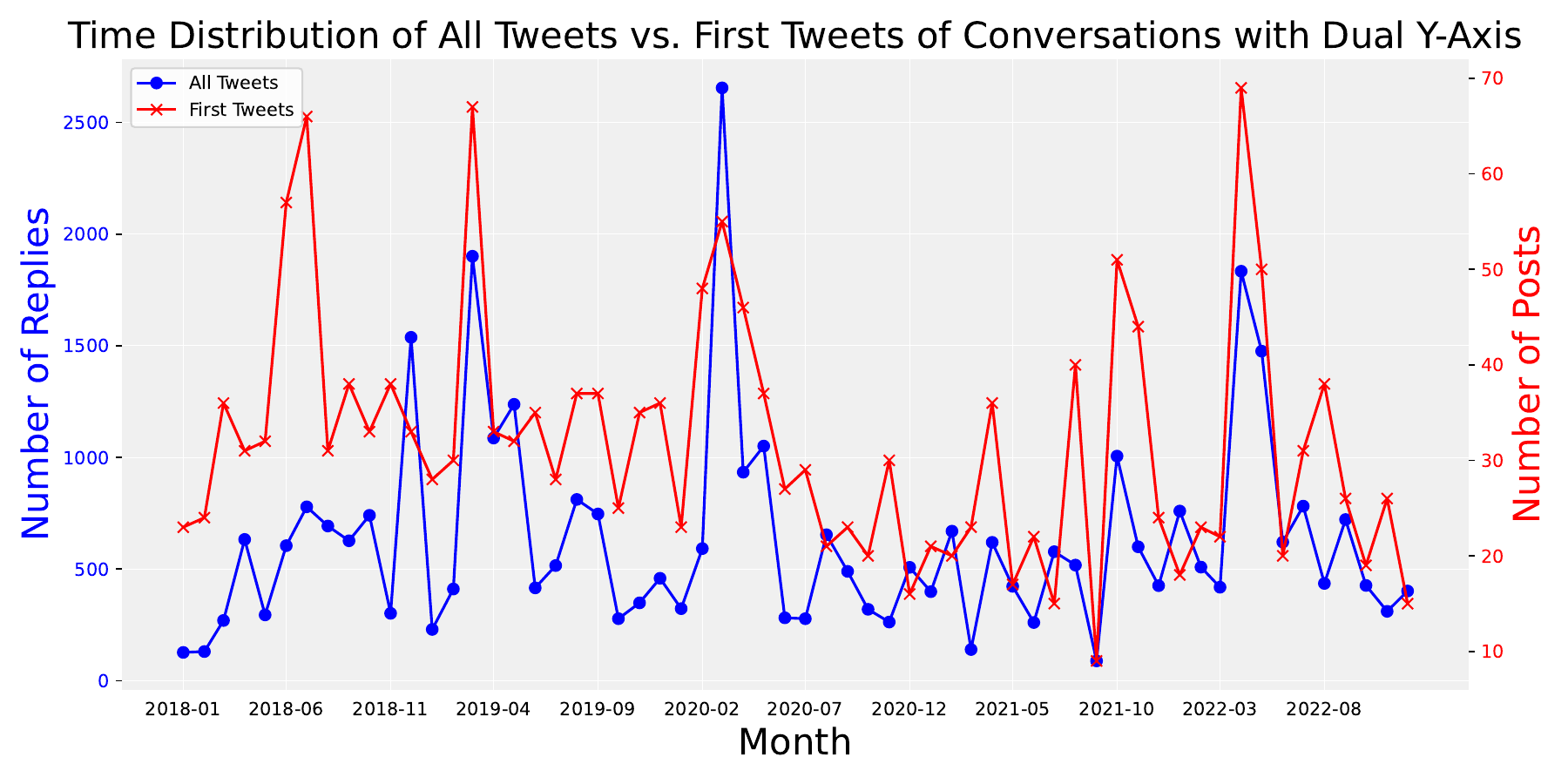}
  \end{subfigure}
  \begin{subfigure}[b]{0.3\textwidth}
    \includegraphics[width=\linewidth]{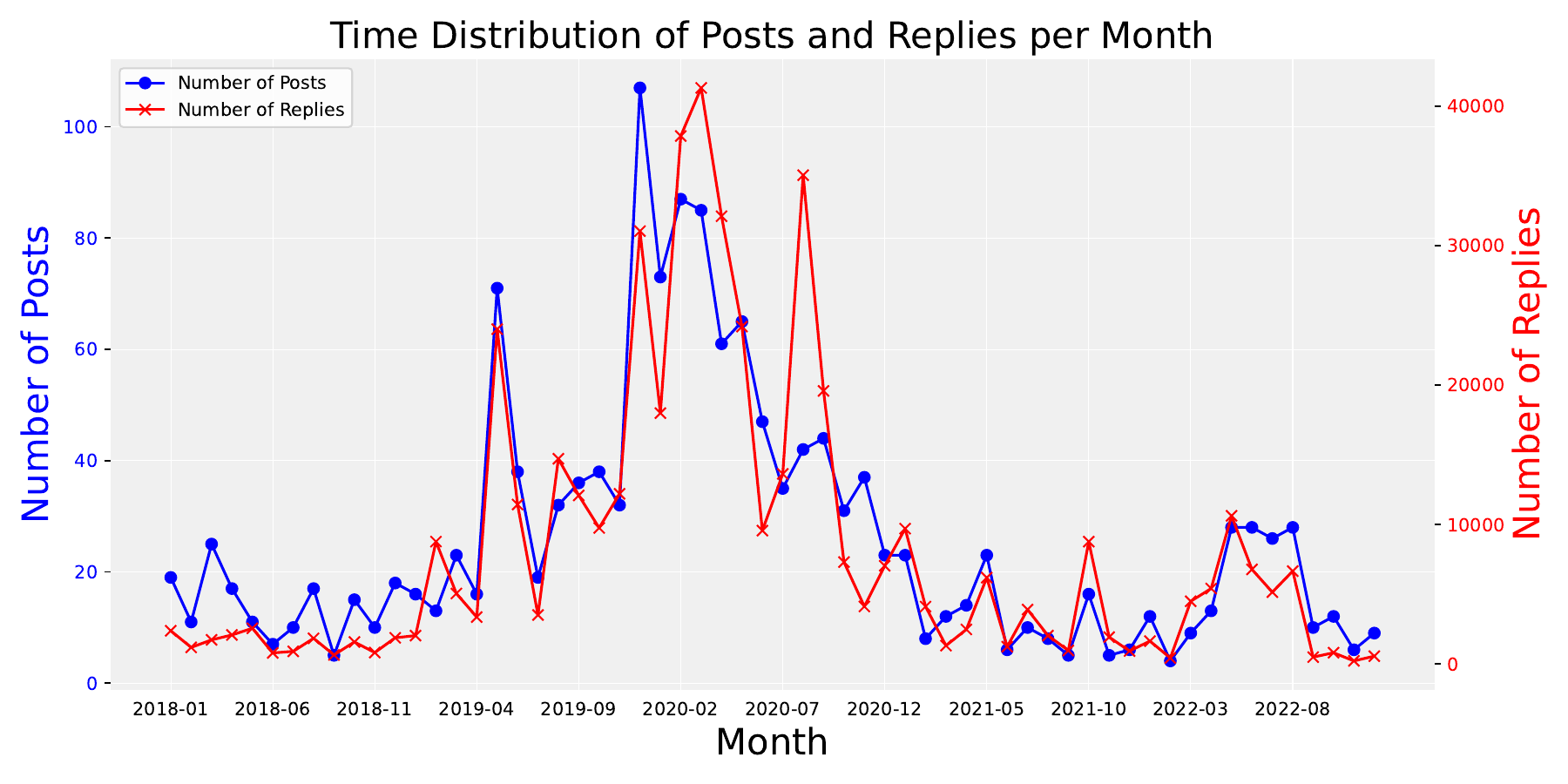}
  \end{subfigure}
  \begin{subfigure}[b]{0.3\textwidth}
    \includegraphics[width=\linewidth]{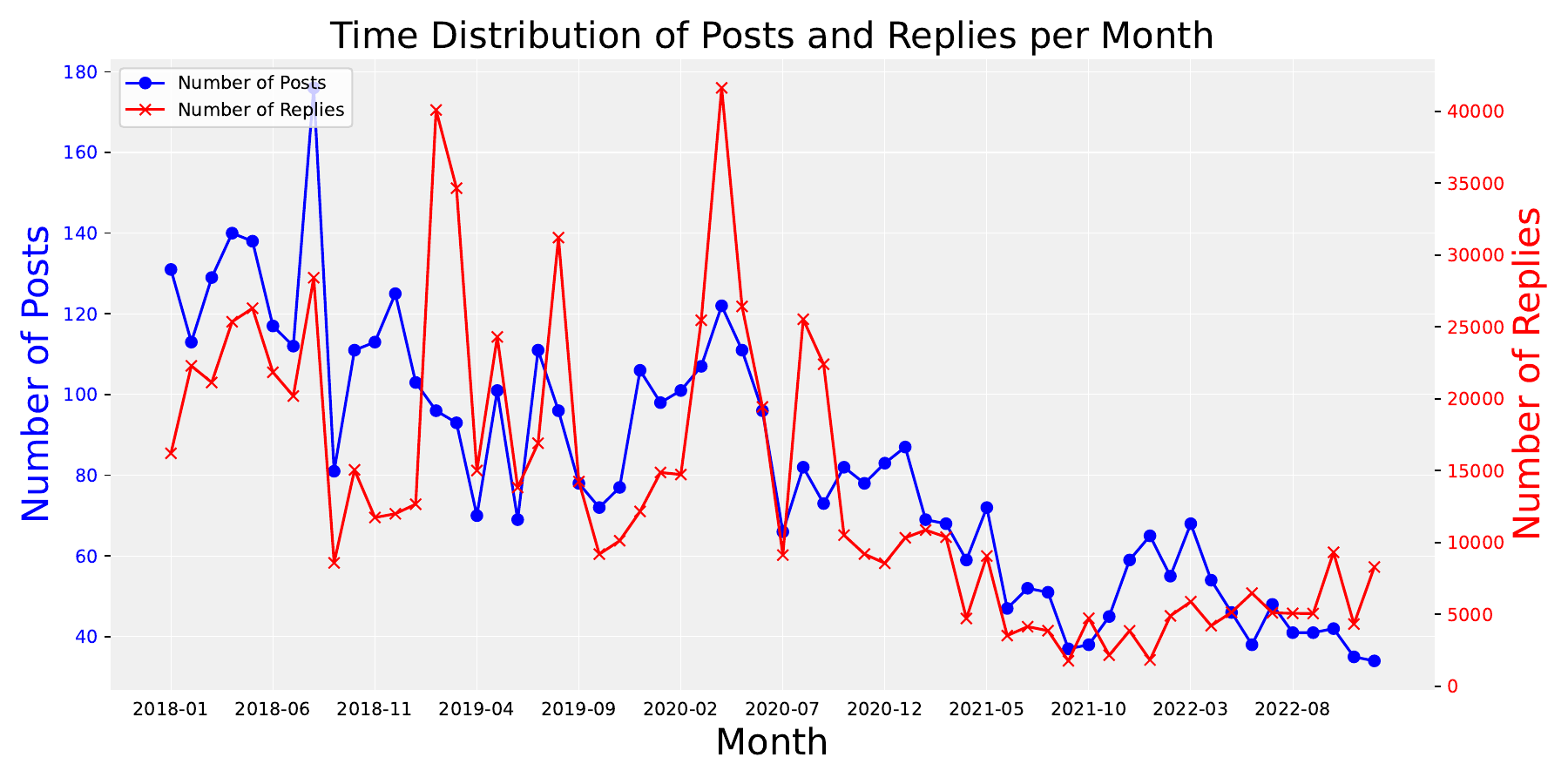}
  \end{subfigure}

  \vspace{1em}

  \begin{subfigure}[b]{0.3\textwidth}
    \includegraphics[width=\linewidth]{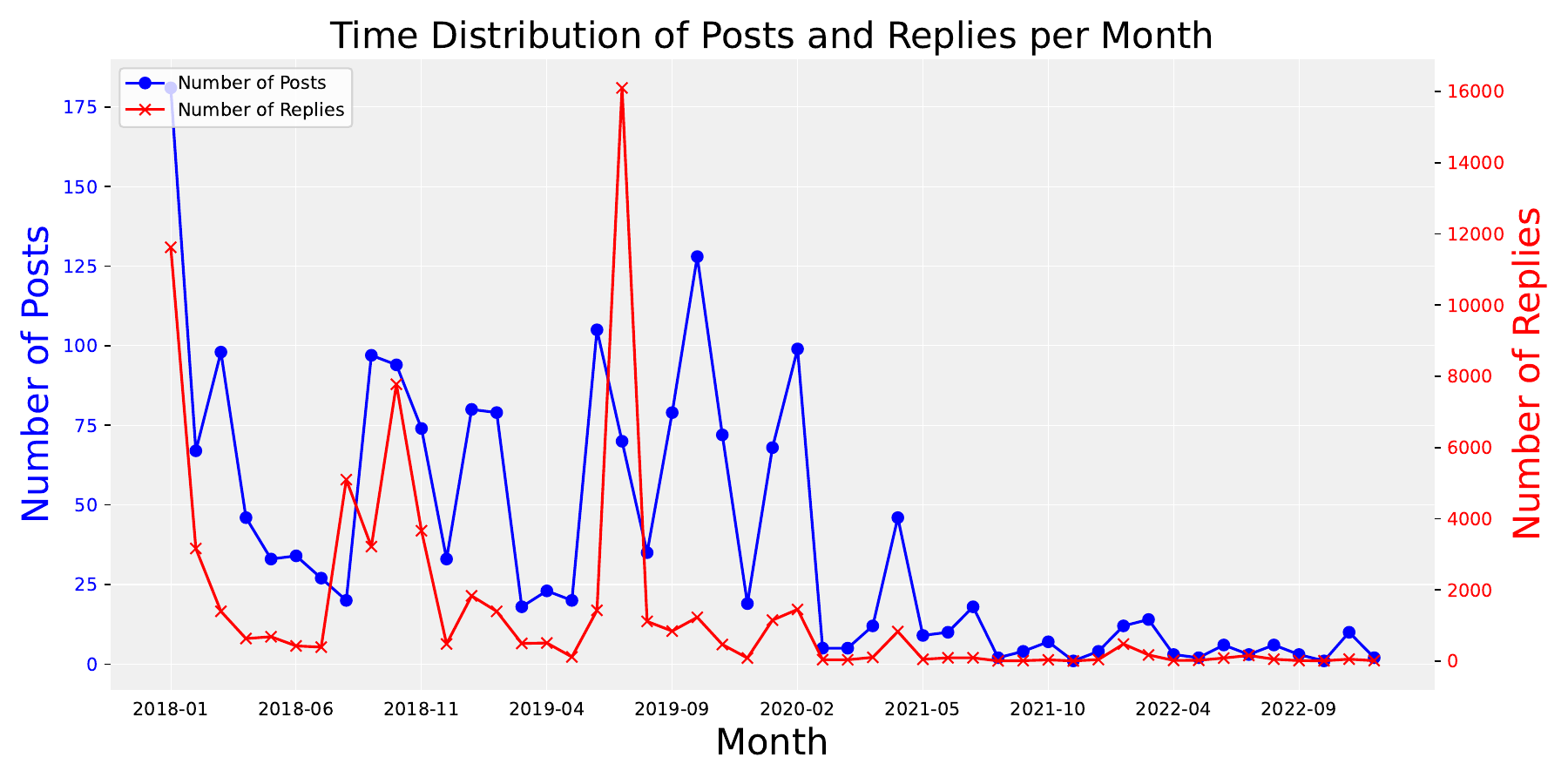}
  \end{subfigure}

  \caption{\textcolor{black}{The activity timelines for 13 journalists.}}
  \label{fig:all_journalist_timelines}
\end{figure*}
\textcolor{black}{Figure~\ref{fig:all_journalist_timelines} illustrates the monthly distribution of posts and replies by journalists. Posting behavior exhibits distinct peaks, often aligning with major events like Covid, \eg, Covid, followed by periods of reduced activity. Across most timelines, the number of replies vastly exceeds the number of posts—while posts typically number in the hundreds per month, replies often reach into the tens of thousands. This reflects the central role journalists play in initiating conversations that attract substantial audience engagement. And in most cases, posting and replying spike concurrently, suggesting real-time engagement with ongoing events. Some timelines show isolated bursts of activity, while others reflect sustained posting over time. Overall, journalists demonstrate a interactive and multi-turn communication pattern across different time periods.}

\subsection{Payoff Maximization} \label{sec:5.3}
When responding to a post, a user exhibiting rational behavior ought to assess the benefits of all possible strategies to select the optimal one. Nevertheless, identifying rational behavior presents challenges; although the actions (like a reply) are visible, the underlying strategies guiding these actions remain concealed. Consequently, users must associate these hidden strategies with their outcomes. In this section, we explore methods to reveal these connections.

We define three utility scores to quantify the benefits of responding to a post as follows:
\begin{itemize}
    \item $\mu^a_i = \frac{1}{k} \sum_{r\in \mathcal{R}_i} (retweets(i) + replies(i) + quotes(i) + likes(i) + \log(views(i)))$, representing the aggregated interaction metrics for reply $i$, 
    to account for their relative importance.
    \item $\mu^b_i = \frac{1}{k} \sum_{r\in \mathcal{R}_i} \mathbf{1}_{C_i \neq C_r}$, indicating the count of responses from users from a different group.
    \item $\mu^c_i = \frac{1}{k} \sum_{r\in \mathcal{R}_i} \mathbf{1}_{au(r)=A}$, reflecting the number of replies from journalist A.
\end{itemize}
The term $k$ denotes the elapsed time since the original post, normalizing the utility scores. The whole utility score for each user $i$ is the sum of all three utility scores, i.e. $\mu_i = \mu_i^a + \mu_i^b + \mu_i^c$. The utility of each reply is then normalized by the total number of replies generated by the user. In our model, we associate each utility score with the most likely strategy as determined by Eq.~\ref{eq: s_loss}. A rational user aims to maximize their expected global utility $\hat{\mu}_u$ across all strategies $s\in S$. Discovering this rational behavior offers a baseline for assessing whether users are strategic in their responses to posts. 

\begin{figure}[ht]
    \centering
        \includegraphics[width=0.75\textwidth]{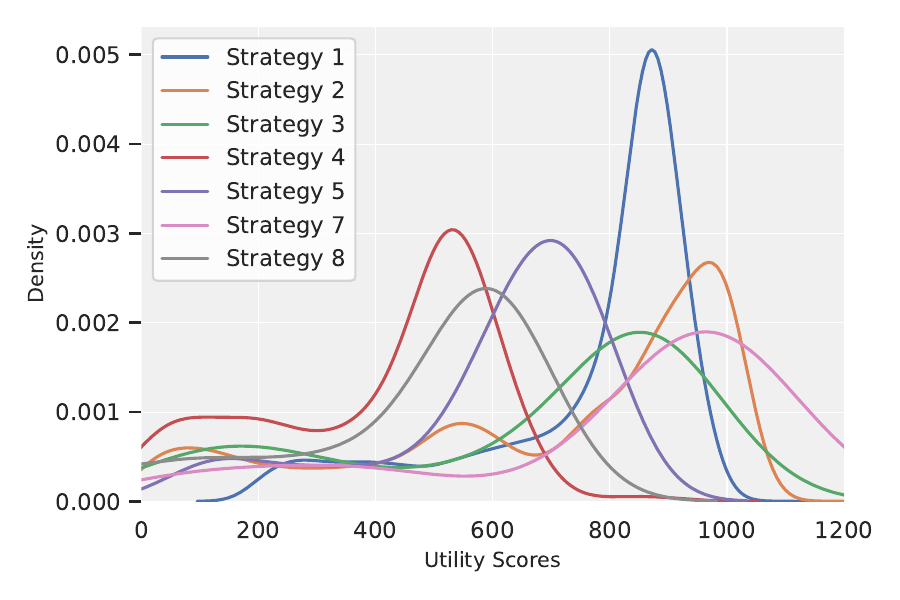} 
        \label{fig:overall_utility}
    \caption{Distributions of utility scores across different strategies. Using $s_2$ (orange) results in the highest median utility score, while using $s_1$ (blue), $s_7$ (pink) and $s_3$ (green) also lead to gaining more attention. In contrast, employing $s_4$ (red), $s_8$ (grey) results in lower utility scores, indicating timely replies play a key role in gaining more attention. (See \Cref{sec:6.1})}
    \label{fig:utility}
\end{figure}

We examine the utility scores (\ie anticipated payoffs) across various strategies discussed in  \Cref{sec:5.3}. To identify each user's predominant strategy, we consider the strategy with the highest probability as their primary strategy. Figure~\ref{fig:utility} shows that using strategies $s_1$ and $s_2$ results in higher utility scores, meaning these approaches attract more attention from others. Additionally, we see that $s_3$ and $s_7$ also lead to relatively higher attention, suggesting that timely replies play a key role in gaining more attention. Combining Figure~\ref{fig:combined} and Figure~\ref{fig:utility}, we observe that the primary strategies used by attackers are associated with higher utility scores. This suggests that attackers may have to an extent inferred the relationship between strategies and payoffs and employ strategies that attract more attention, providing further evidence of coordinated behavior. 

\subsection{Baselines} \label{sec:5.4}
We have two sets of baselines to verify the effectiveness of both our group classification model and the identification of strategic behaviors.
\subsubsection{Group Classification (RQ1)}
\label{Group Classification Baselines}
We compare our proposed model with existing approaches that utilize tweet metadata to identify attackers, supporters and bystanders.
\begin{enumerate}
    \item \textbf{THP}~\cite{zuo2020transformer}: The Transformer Hawkes Process (THP) model leverages the self-attention mechanism to capture long-term dependencies and analyze event streams. We evaluate its performance on assigning replies to one of three groups.
    \item \textbf{\textbf{AMDN-HAGE}}~\cite{sharma2021identifying}: This work developed a method for identifying coordinated accounts solely based on their activities on social media platforms. This approach is unique as it does not rely on linguistic analysis, metadata, or platform-specific features, but instead focuses on the patterns and behaviors exhibited by these accounts in their interactions.
    \item \textbf{\textbf{TreeTransformer}}~\cite{peng2022rethinking}: In this work, the authors introduce an innovative approach with a tree Transformer that assigns positions to each node using a two-dimensional representation of tree structures. 

\end{enumerate}

\subsubsection{Strategy Discovery (RQ2)} Following the work of~\cite{xiao2020discovering}, we compare our proposed framework with one topic model~\cite{yin2014dirichlet} and one traditional regression model \cite{hosmer2013applied}. 
\begin{enumerate}
    \item \textbf{Logistic Regression (LR)}~\cite{nigam2000text}: When using the logistic regression model for each reply separately, we consider the probability of edge formation as the predictive variable and the actual observed connections as the response variable. The parameters within the regression model are restricted to be non-negative and their total must equal 1. This constraint ensures that these coefficients can be understood as distributions of different strategies.
    \item \textbf{Dirichlet Multinomial Mixture Model (DMM)}~\cite{yin2014dirichlet}: Words become strategies and topics over words become distributions over strategies. Users need to pick one strategy from their strategy distributions to form an edge. We set the number of topics to be the same as the number of strategies so that each topic is initialized with a maximum likelihood strategy. 

\end{enumerate}

\textcolor{black}{To better understand the performance of our strategy modeling, we checked the false positive and false negative edges and here we present some examples in Table~\ref{tab:error-analysis}.}
\begin{table}[h]
    \centering
    \begin{tabular}{lp{5cm}p{5cm}}
        \toprule
        \textbf{Label} & \textbf{Parent} & \textbf{Child}\\
        \hline
        \textbf{False Positive} & "I contributed to this by @BBCFergusWalsh with more details on those false rumours about the vaccine trial spreading on social media So here’s a quick thread on how to spot fake news like this" 
        9:08 AM · Apr 26, 2020 
        & "Ok thanks for the info !!!" 
        1:42 PM · Apr 26, 2020\\
        \textbf{False Positive} & "China's domestic issues will always trump financial concerns. Wall Street bankers are finding that out firsthand as they go hungry in Shanghai" 
        8:31 PM · Apr 26, 2022
        & "Data can always be made up easily in this evil country."
        3:27 AM · Apr 27, 2022\\
        \textbf{False Negative} & "Pappu will always be Pappu" 9:53 PM · Dec 23, 2022 
        & "English....Hindi....ukhad....lehra...."
        10:14 PM · Dec 23, 2022 \\
        \textbf{False Negative} & "In Hong Kong, remembering June4  is now a crime. It's 33 years after TiananmenSquareMassacre and 26 activists--inc. Apply Daily's Jimmy Lai--have been arrested for trying to commemorate the event. Universities have removed memorials."
        8:57 AM · Jun 3, 2022 
        & "“Don't wait until my father is martyred to start talking about him..” A heart wrenching message from the daughter of prisoner Khalil Awawdeh who is on hunger strike for 93 days.. Support him from now!
        1:07 PM · Jun 3, 2022\\
        \hline
    \end{tabular}
    \caption{\textcolor{black}{Examples of false positive and false negative edges.}}
    \label{tab:error-analysis}
\end{table}
\textcolor{black}{Through checking misclassified cases, we found that most false negative cases occur between a journalist’s original post and replies that are part of an extended thread initiated by the journalist. This suggests that the model struggles to accurately recognize contextual continuity when replies are indirectly related to the original post, especially if they are part of a longer conversation thread rather than direct replies. 
And most false positive cases arise when replies consist of content that is unlikely to be genuine responses, including replies that are just GIFs or media responses without textual content, posts containing advertisements or promotional material, which do not contribute to the conversation meaningfully, and messages made up of random or nonsensical words, likely generated by bots or spam accounts.}
\subsection{\textcolor{black}{Results for Different Journalists}}\label{append:each}
\textcolor{black}{In~\cref{tab:each}, we show classification results for journalists from different cultures. We can see that there is minimal variation across journalists from different cultural backgrounds.}
\begin{table}[]
\begin{tabular}{l|l|l|l}
\hline
Coutry   & Accuracy & Recall & F1     \\ \hline
China    & 0.8212   & 0.8358 & 0.8284 \\
India    & 0.8265   & 0.8396 & 0.8329 \\
US       & 0.8245   & 0.8383 & 0.8313 \\
UK       & 0.8201   & 0.8277 & 0.8238 \\
Pakistan & 0.8219   & 0.8320 & 0.8283 \\
Lebanon  & 0.8123   & 0.8175 & 0.8148 \\ \hline
\end{tabular}
\label{tab:each}
\caption{\textcolor{black}{Evaluation of classification results across journalists from different cultures.}}
\end{table}
\subsection{Journalists Frequency}
\label{append:freq}
\begin{figure*}[htbp]
  \centering

  \begin{subfigure}[b]{0.3\textwidth}
    \includegraphics[width=\linewidth]{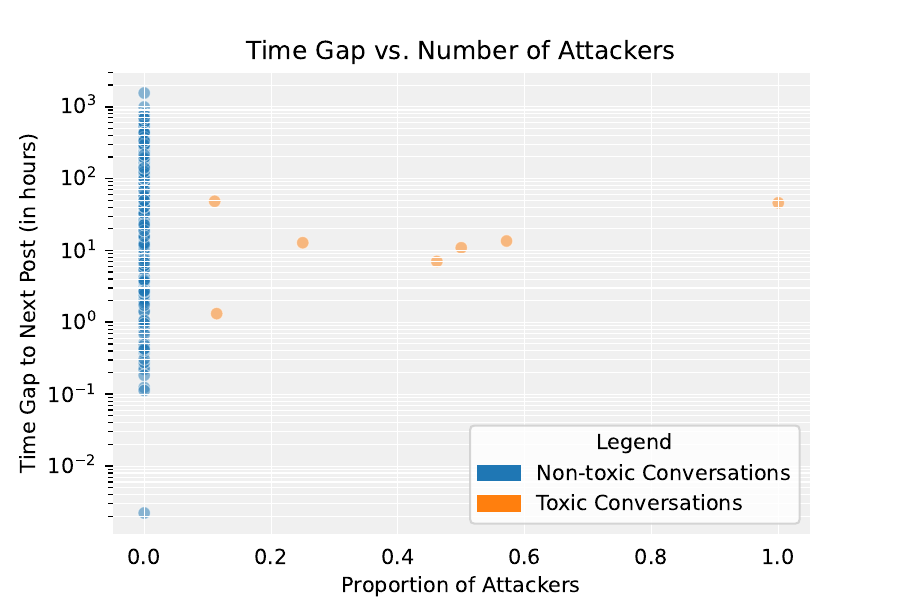}
  \end{subfigure}
  \begin{subfigure}[b]{0.3\textwidth}
    \includegraphics[width=\linewidth]{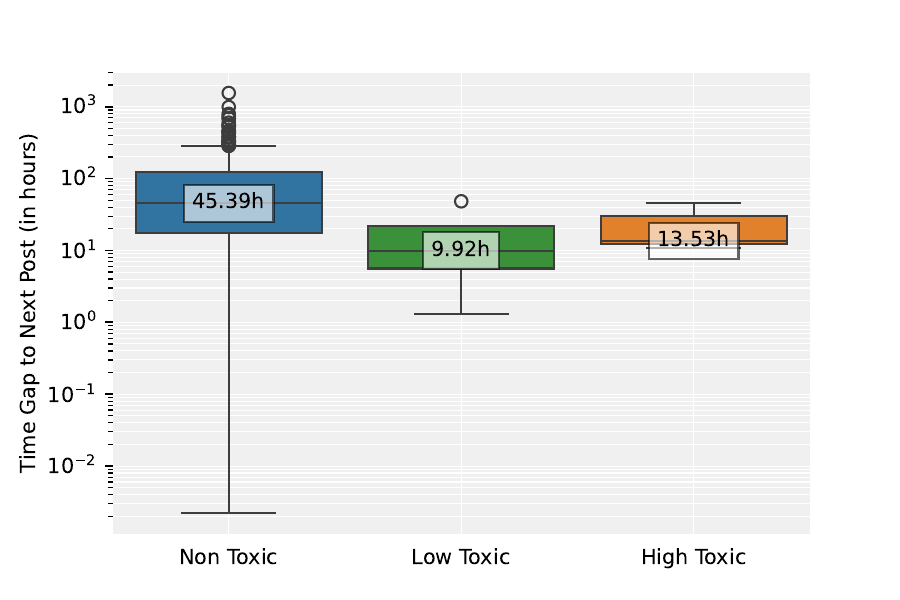}
  \end{subfigure}
  \begin{subfigure}[b]{0.3\textwidth}
    \includegraphics[width=\linewidth]{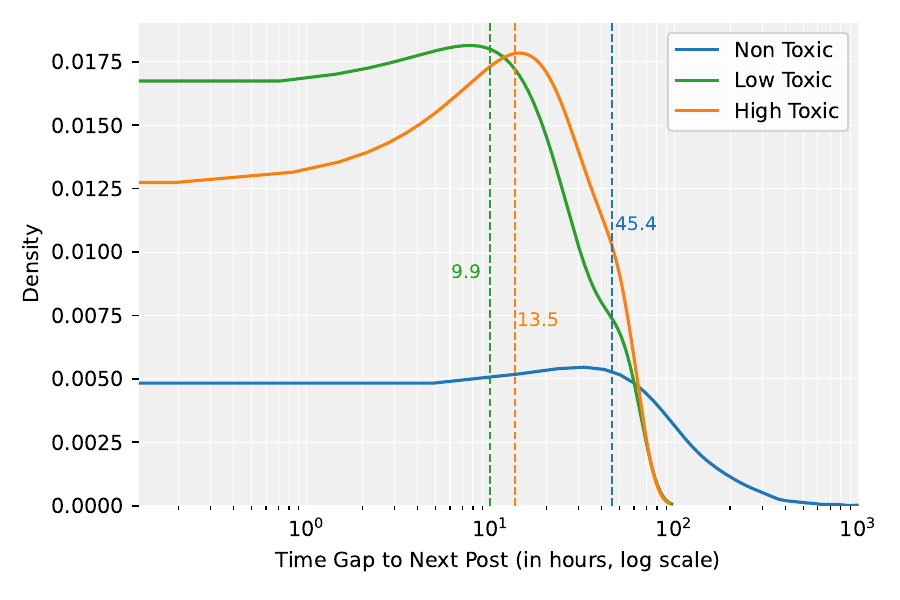}
  \end{subfigure}


  \begin{subfigure}[b]{0.3\textwidth}
    \includegraphics[width=\linewidth]{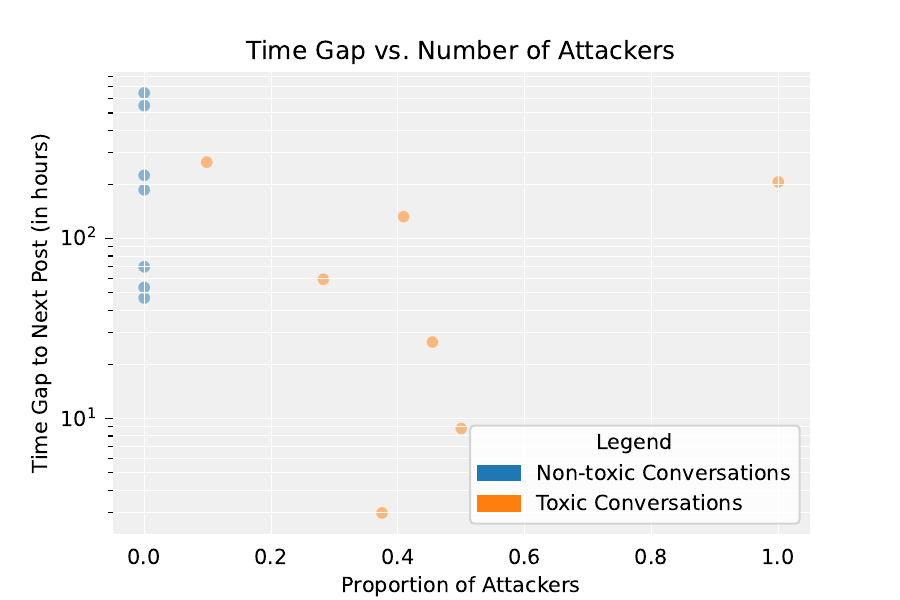}
  \end{subfigure}
  \begin{subfigure}[b]{0.3\textwidth}
    \includegraphics[width=\linewidth]{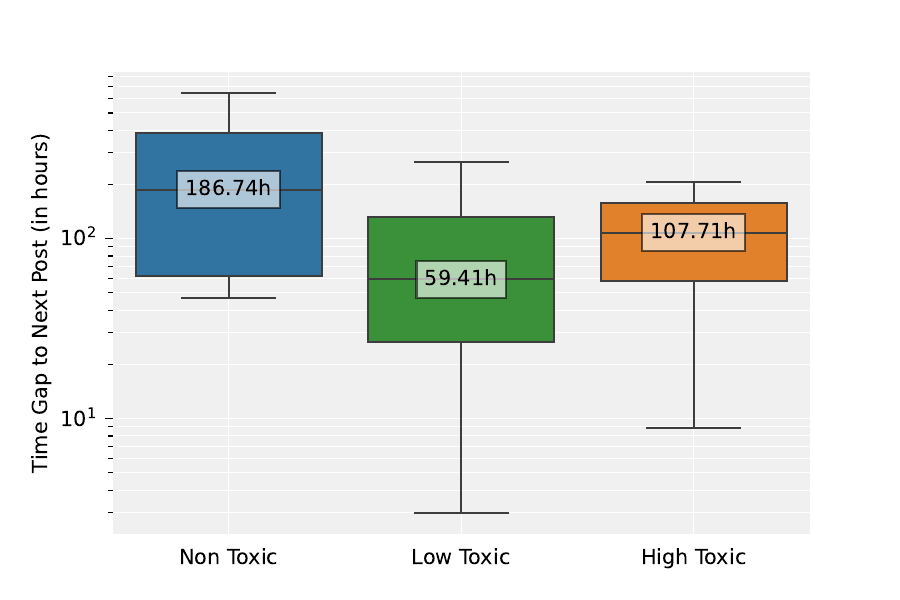}
  \end{subfigure}
  \begin{subfigure}[b]{0.3\textwidth}
    \includegraphics[width=\linewidth]{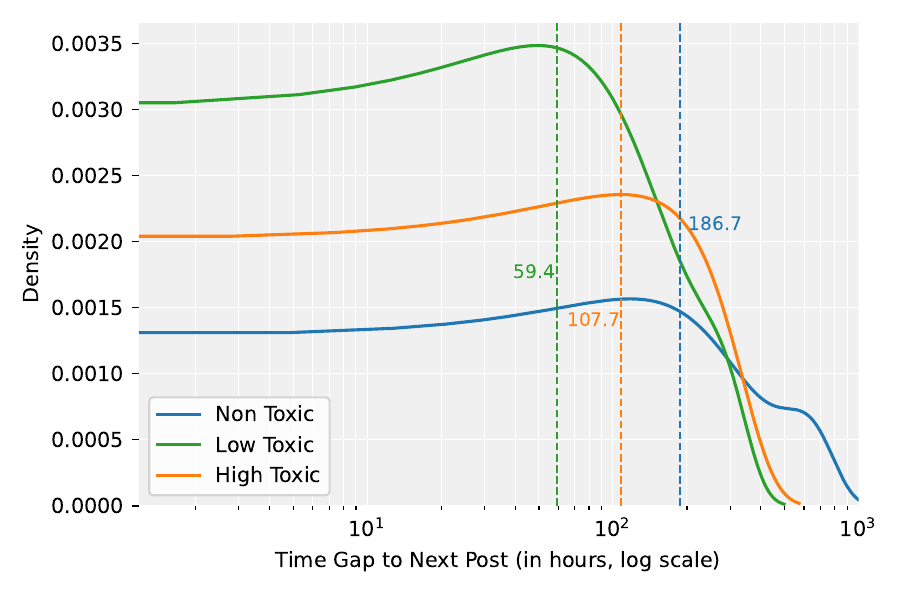}
  \end{subfigure}


  \begin{subfigure}[b]{0.3\textwidth}
    \includegraphics[width=\linewidth]{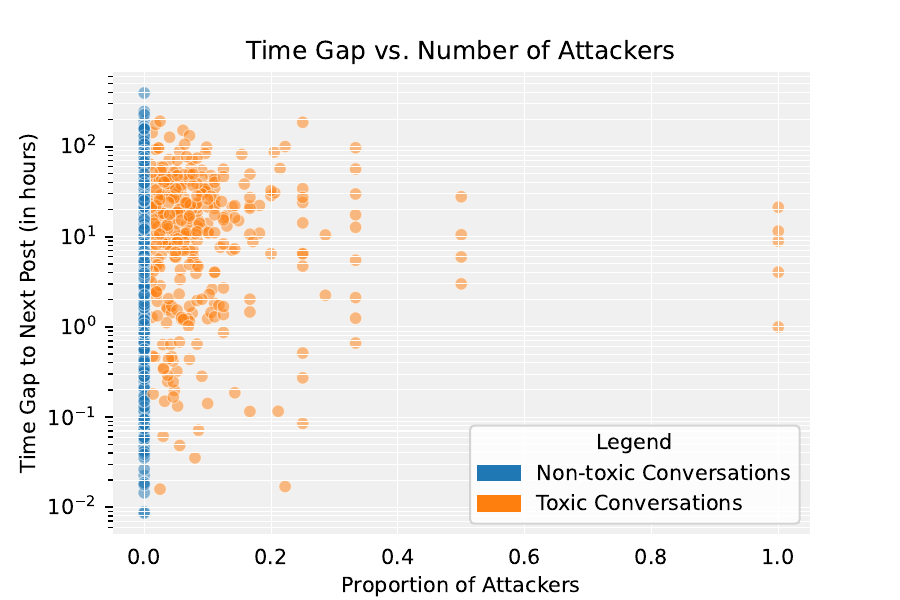}
  \end{subfigure}
  \begin{subfigure}[b]{0.3\textwidth}
    \includegraphics[width=\linewidth]{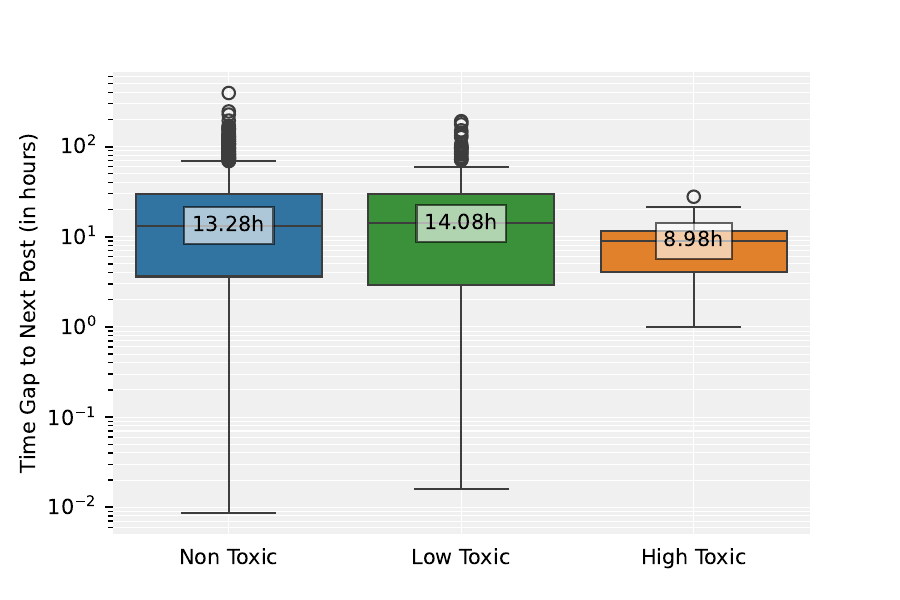}
  \end{subfigure}
  \begin{subfigure}[b]{0.3\textwidth}
    \includegraphics[width=\linewidth]{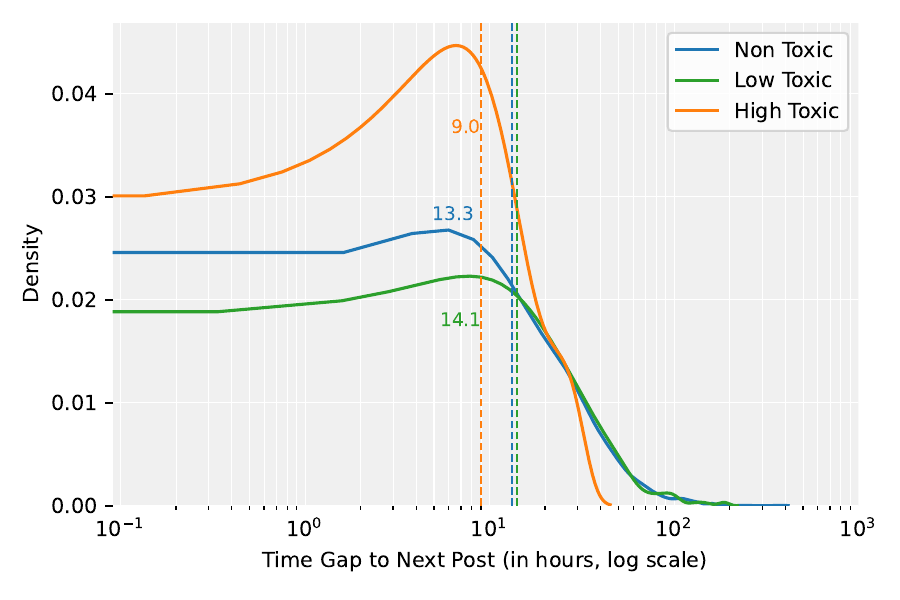}
  \end{subfigure}


  \begin{subfigure}[b]{0.3\textwidth}
    \includegraphics[width=\linewidth]{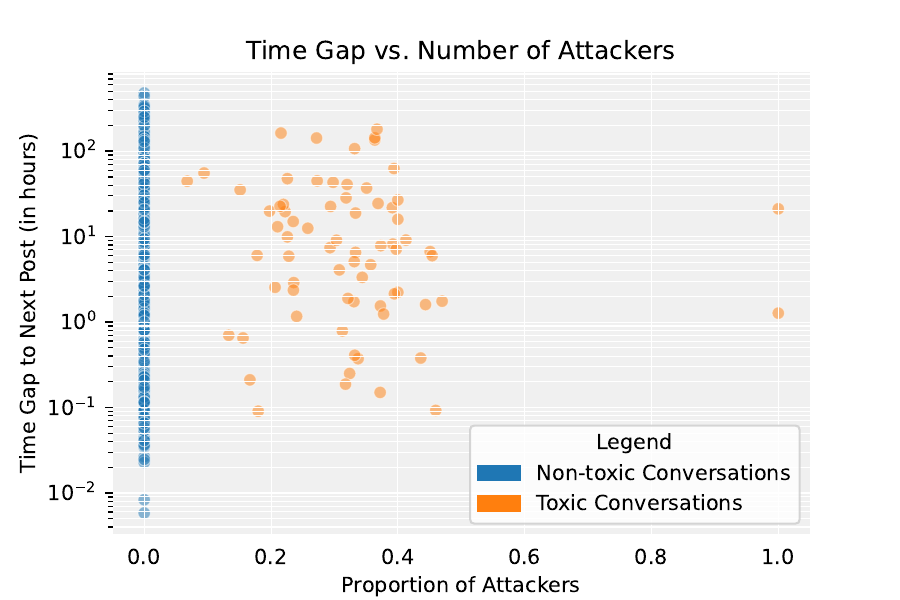}
  \end{subfigure}
  \begin{subfigure}[b]{0.3\textwidth}
    \includegraphics[width=\linewidth]{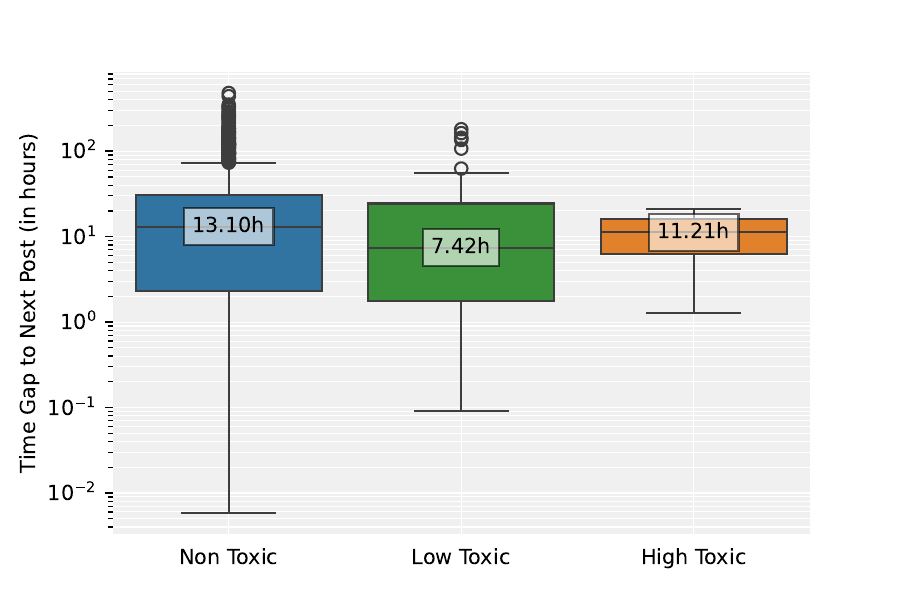}
  \end{subfigure}
  \begin{subfigure}[b]{0.3\textwidth}
    \includegraphics[width=\linewidth]{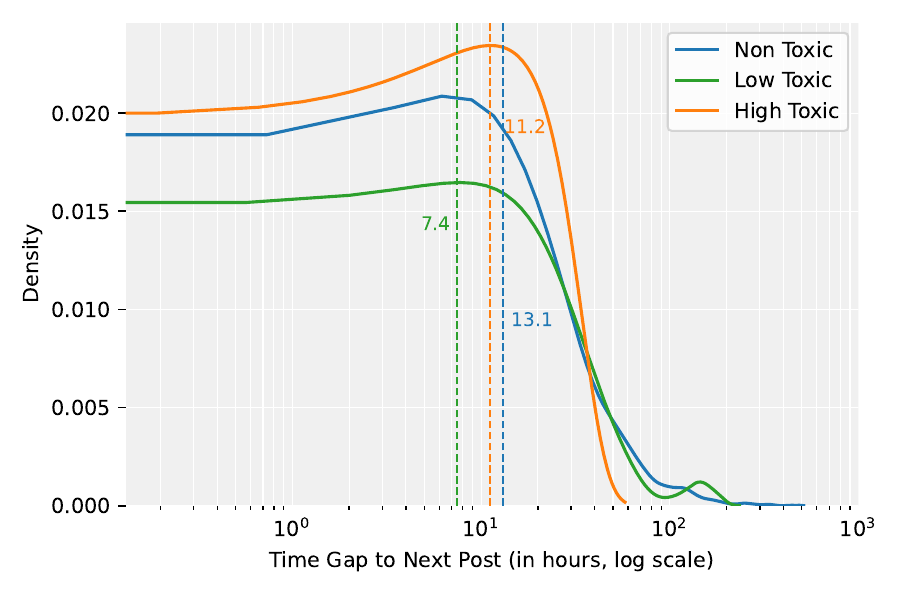}
  \end{subfigure}


  \begin{subfigure}[b]{0.3\textwidth}
    \includegraphics[width=\linewidth]{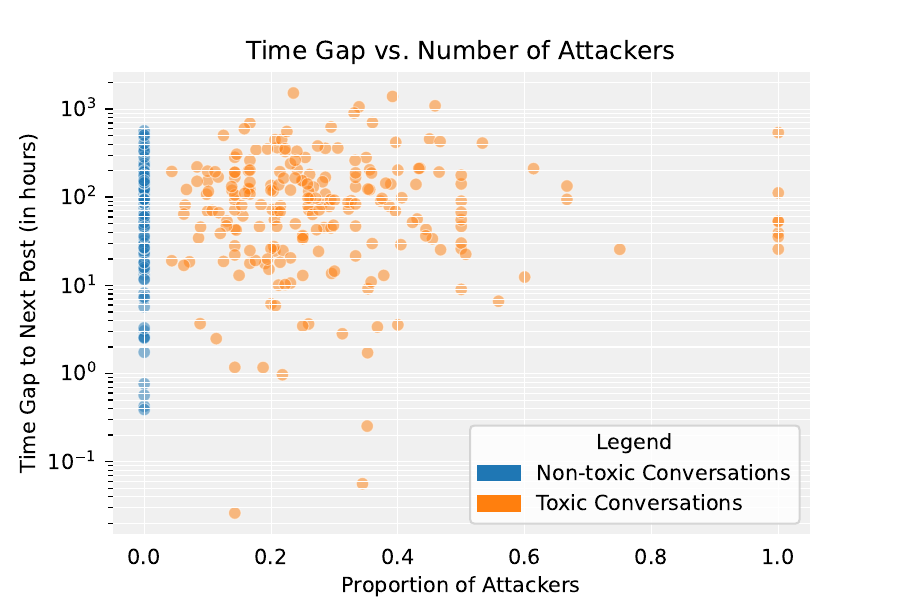}
  \end{subfigure}
  \begin{subfigure}[b]{0.3\textwidth}
    \includegraphics[width=\linewidth]{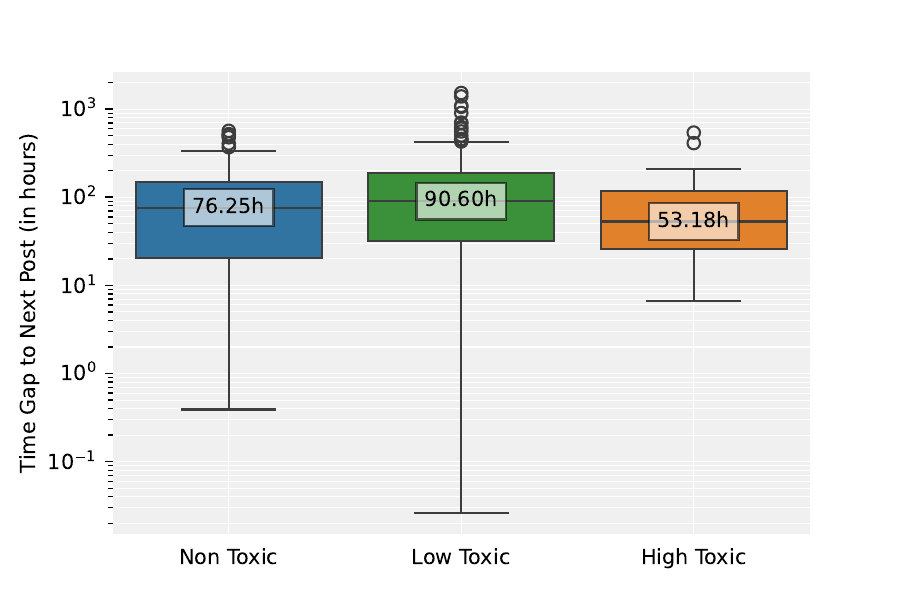}
  \end{subfigure}
  \begin{subfigure}[b]{0.3\textwidth}
    \includegraphics[width=\linewidth]{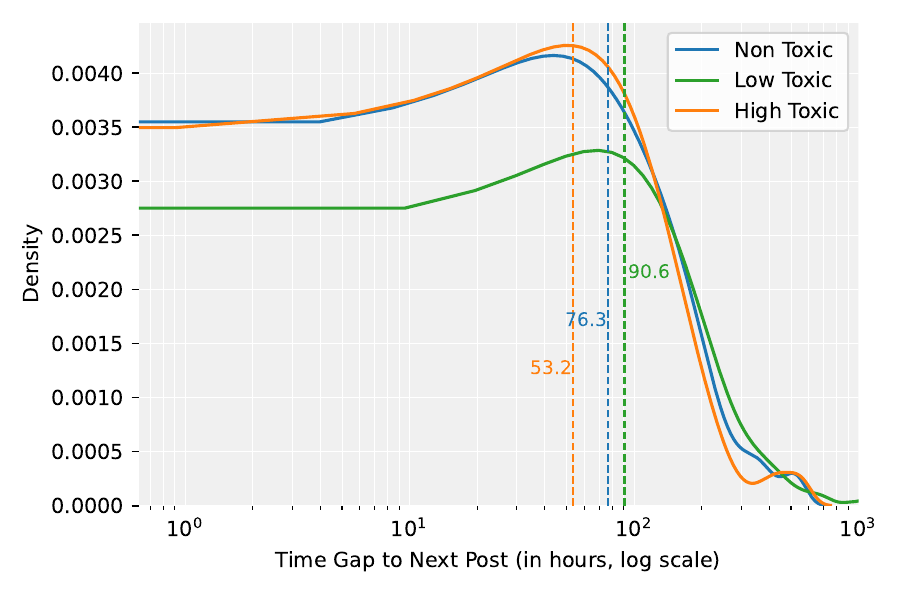}
  \end{subfigure}

  \vspace{1em}

  \begin{subfigure}[b]{0.3\textwidth}
    \includegraphics[width=\linewidth]{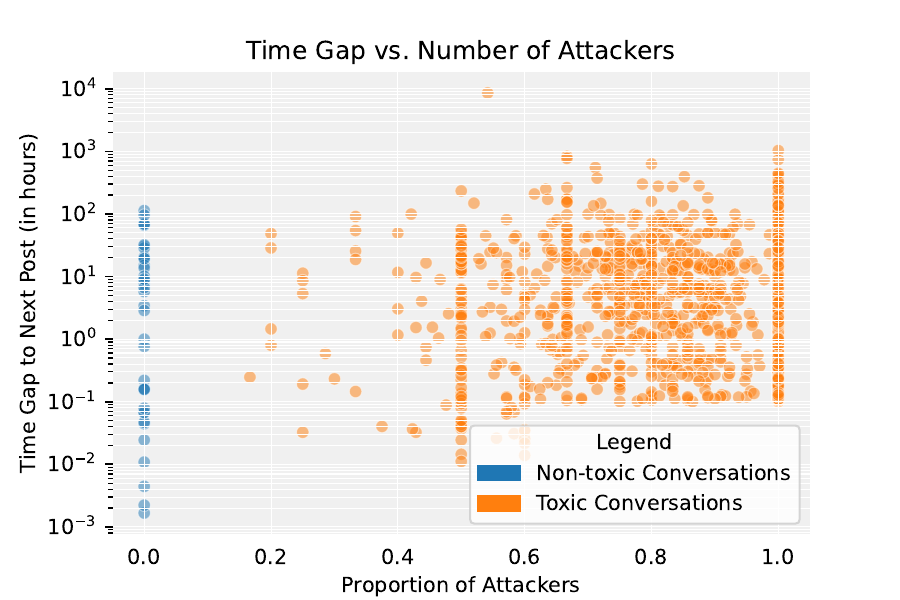}
    \caption*{Time Gap v.s. Attacker Ratios}
  \end{subfigure}
  \begin{subfigure}[b]{0.3\textwidth}
    \includegraphics[width=\linewidth]{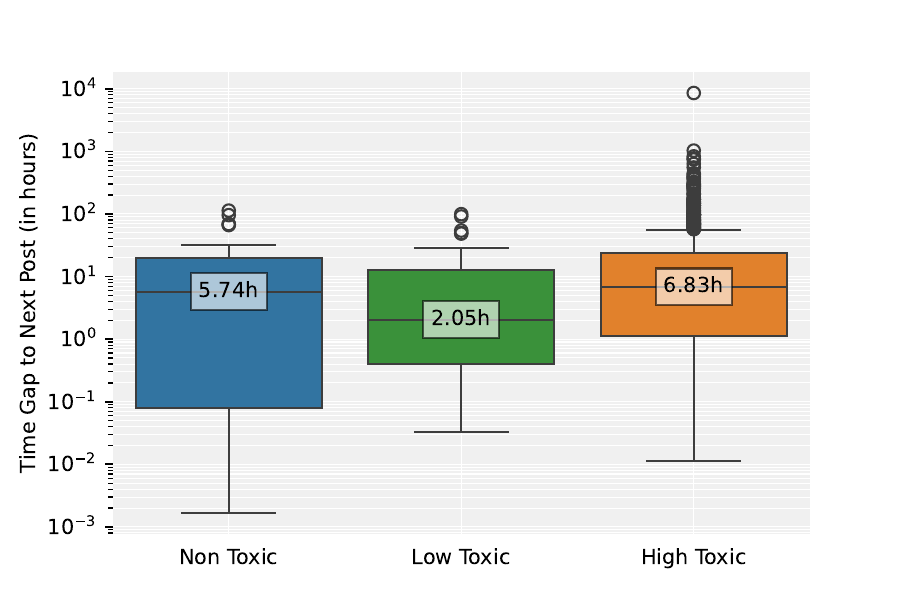}
    \caption*{Time Gap Boxplots}
  \end{subfigure}
  \begin{subfigure}[b]{0.3\textwidth}
    \includegraphics[width=\linewidth]{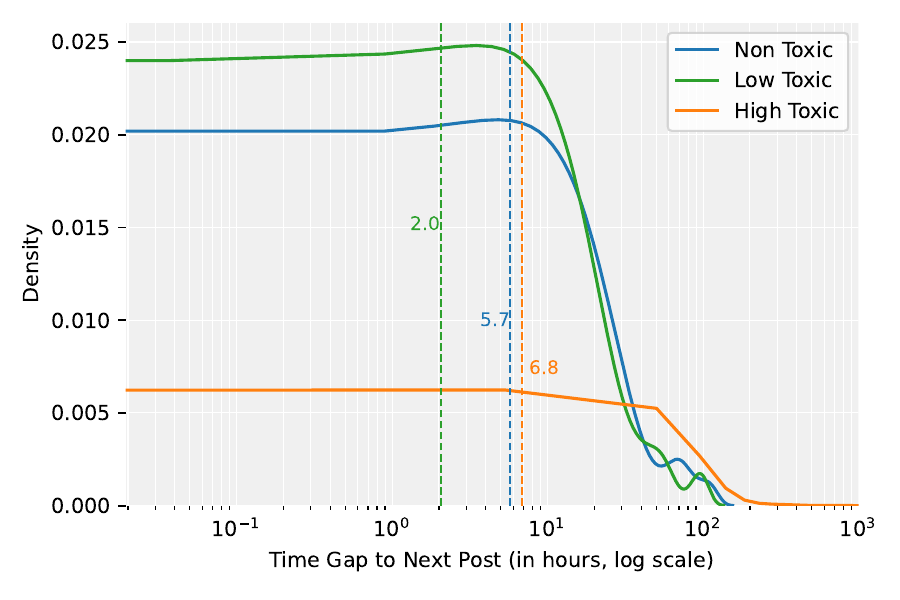}
    \caption*{Time Gap Distributions}
  \end{subfigure}

  \vspace{1em}

  \caption{\textcolor{black}{Posting behavior of 13 journalists across different environments. Each row corresponds to one journalist, showing their posting frequencies.}}
  \label{fig:journalist_overview}
\end{figure*}

\textcolor{black}{To capture the diverse strategies employed by different journalists in responding to attackers, and to uncover any distinct patterns that may emerge, we present each journalist’s posting frequency across different environments in Figure~\ref{fig:journalist_overview}.}
\newpage
\begin{figure*}[htbp]
  \centering
  \begin{subfigure}[b]{0.3\textwidth}
    \includegraphics[width=\linewidth]{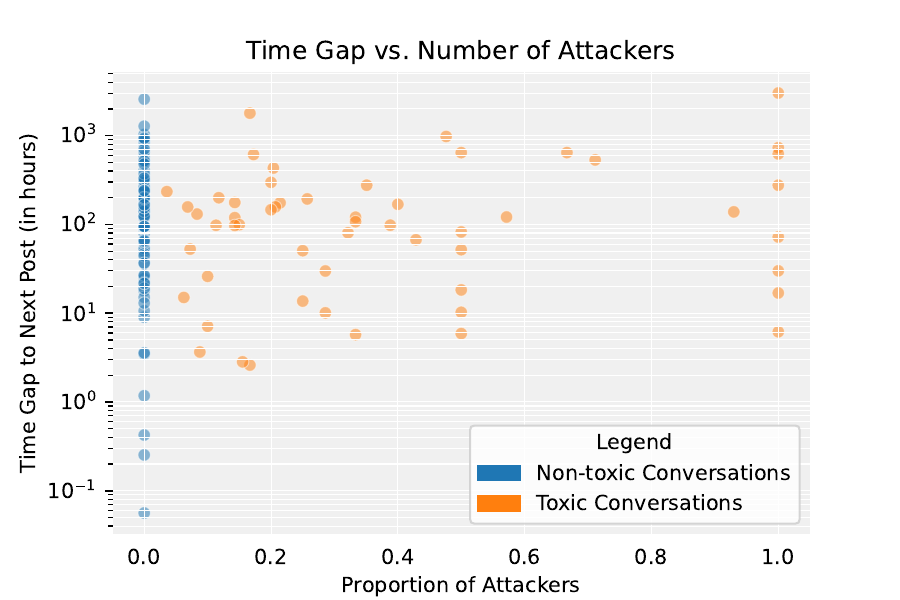}
  \end{subfigure}
  \begin{subfigure}[b]{0.3\textwidth}
    \includegraphics[width=\linewidth]{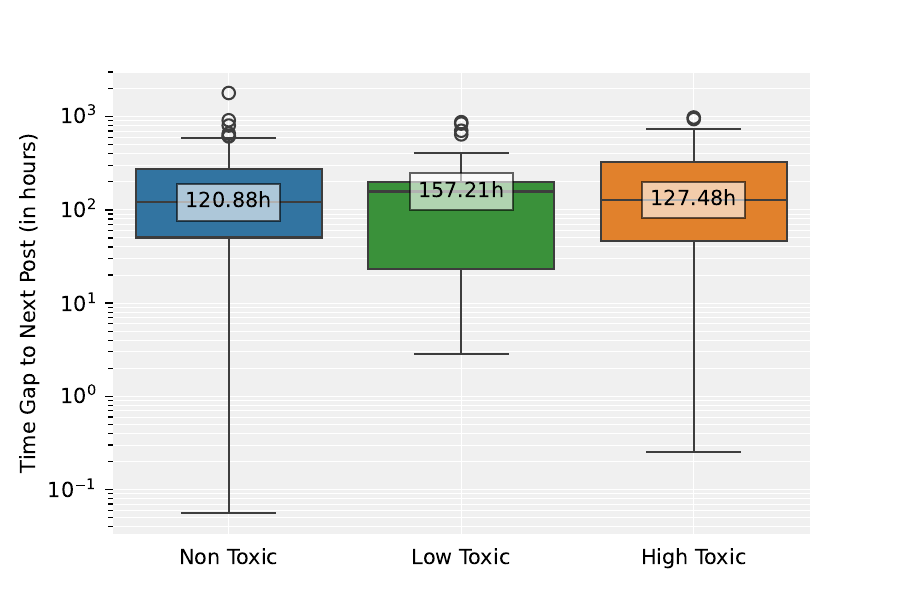}
  \end{subfigure}
  \begin{subfigure}[b]{0.3\textwidth}
    \includegraphics[width=\linewidth]{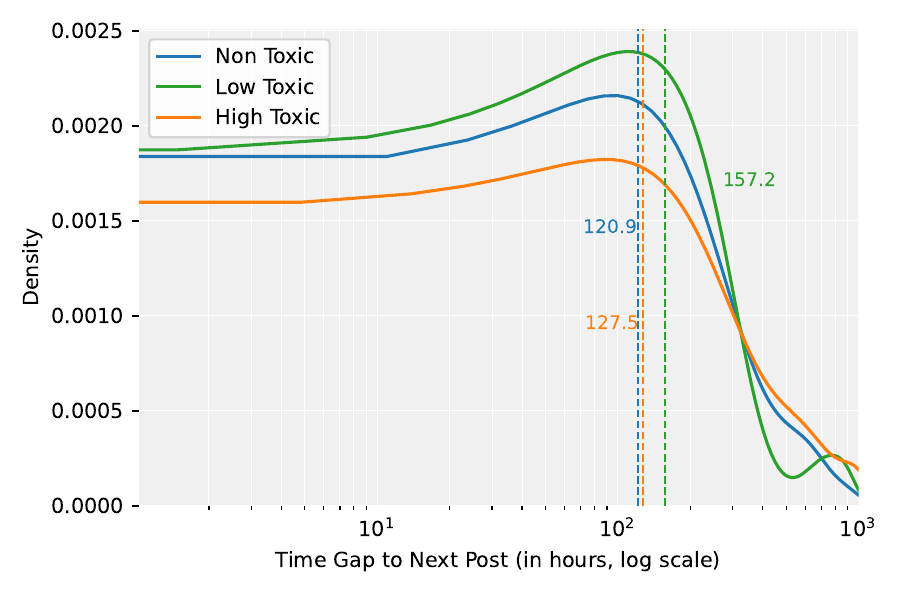}
  \end{subfigure}


  \begin{subfigure}[b]{0.3\textwidth}
    \includegraphics[width=\linewidth]{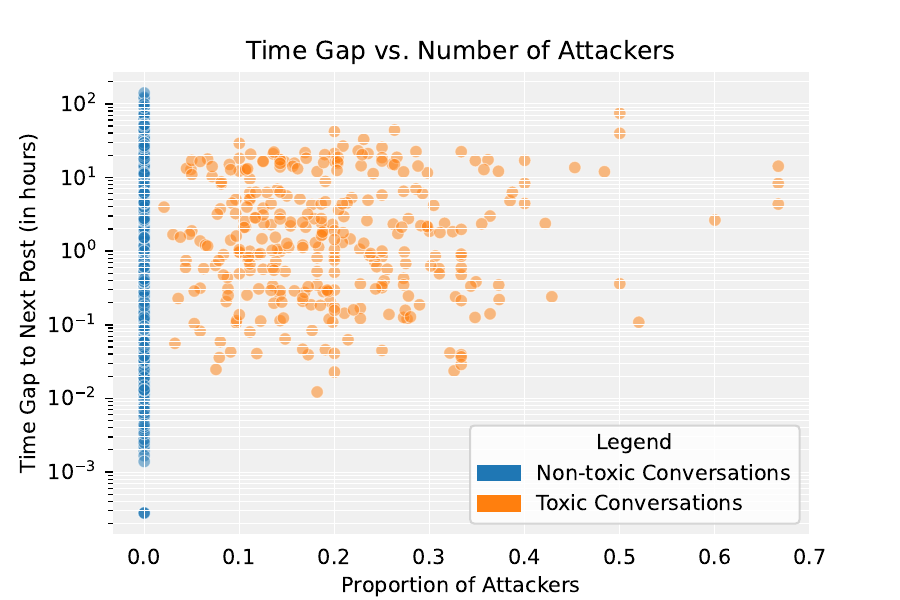}
  \end{subfigure}
  \begin{subfigure}[b]{0.3\textwidth}
    \includegraphics[width=\linewidth]{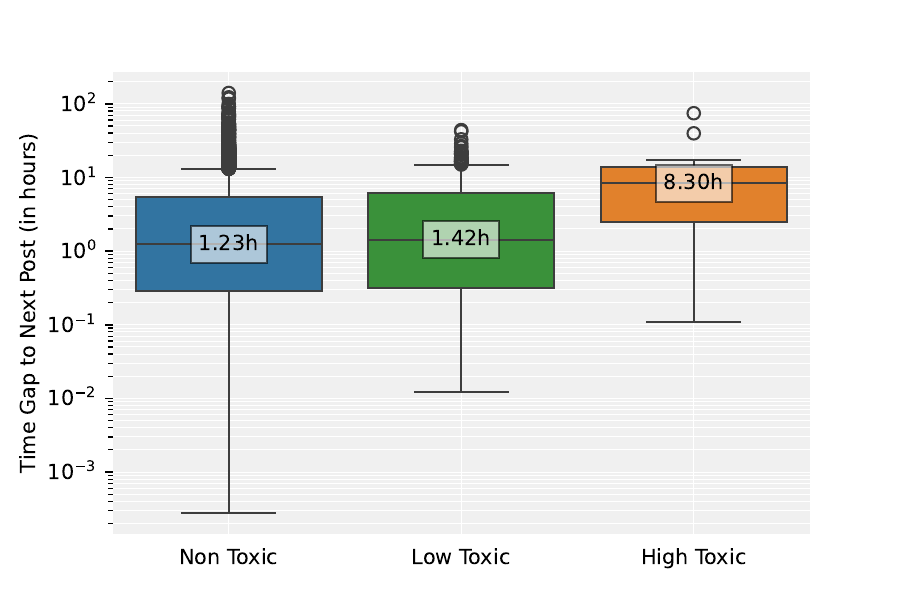}
  \end{subfigure}
  \begin{subfigure}[b]{0.3\textwidth}
    \includegraphics[width=\linewidth]{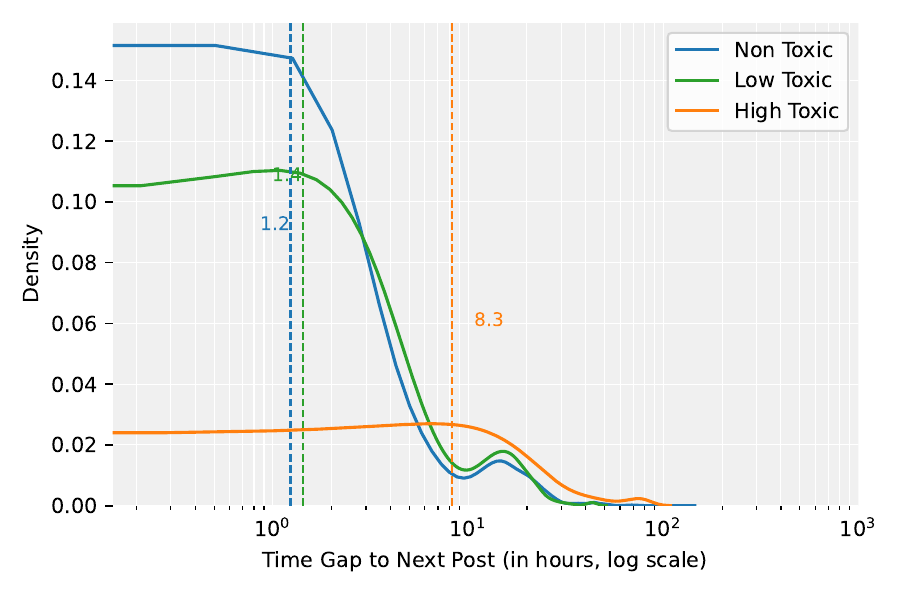}
  \end{subfigure}


  \begin{subfigure}[b]{0.3\textwidth}
    \includegraphics[width=\linewidth]{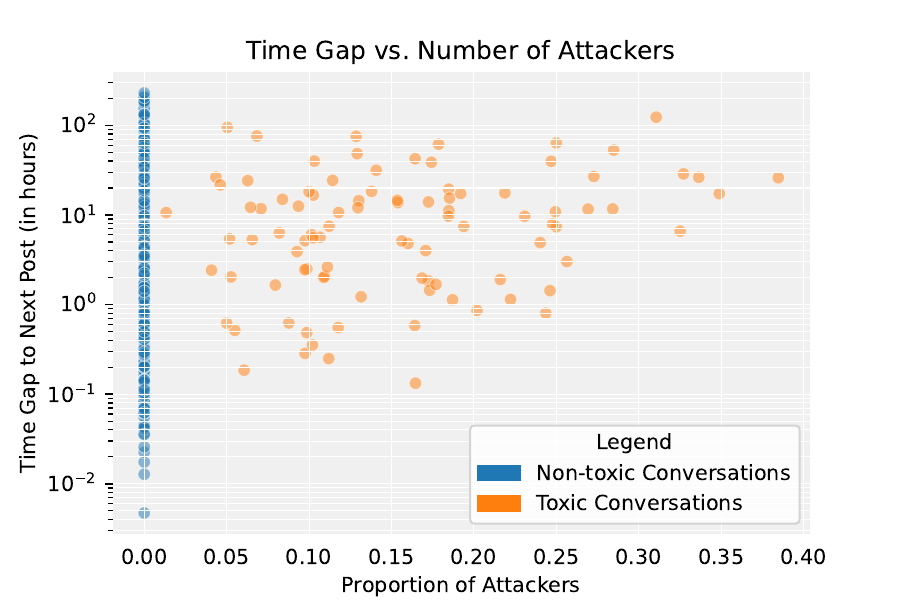}
  \end{subfigure}
  \begin{subfigure}[b]{0.3\textwidth}
    \includegraphics[width=\linewidth]{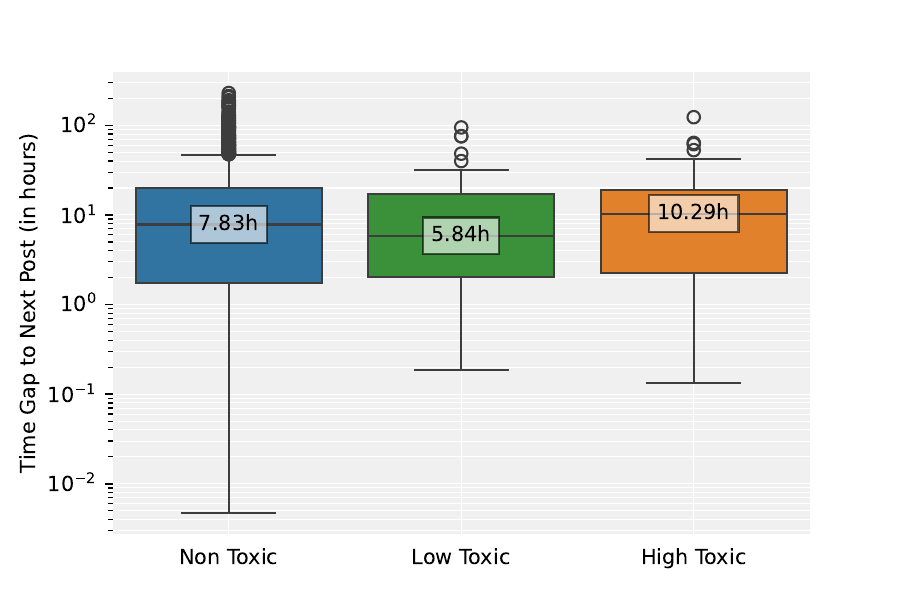}
  \end{subfigure}
  \begin{subfigure}[b]{0.3\textwidth}
    \includegraphics[width=\linewidth]{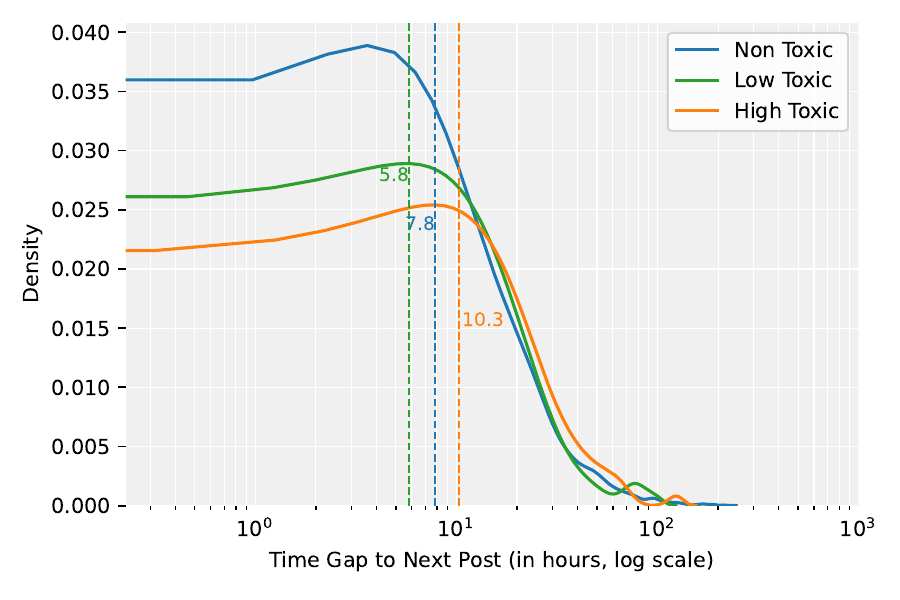}
  \end{subfigure}


  \begin{subfigure}[b]{0.3\textwidth}
    \includegraphics[width=\linewidth]{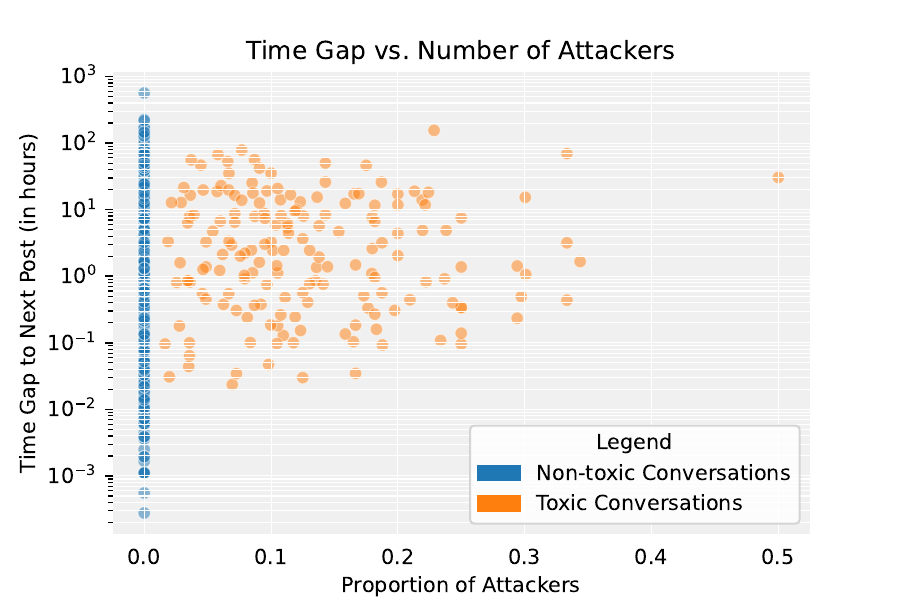}
  \end{subfigure}
  \begin{subfigure}[b]{0.3\textwidth}
    \includegraphics[width=\linewidth]{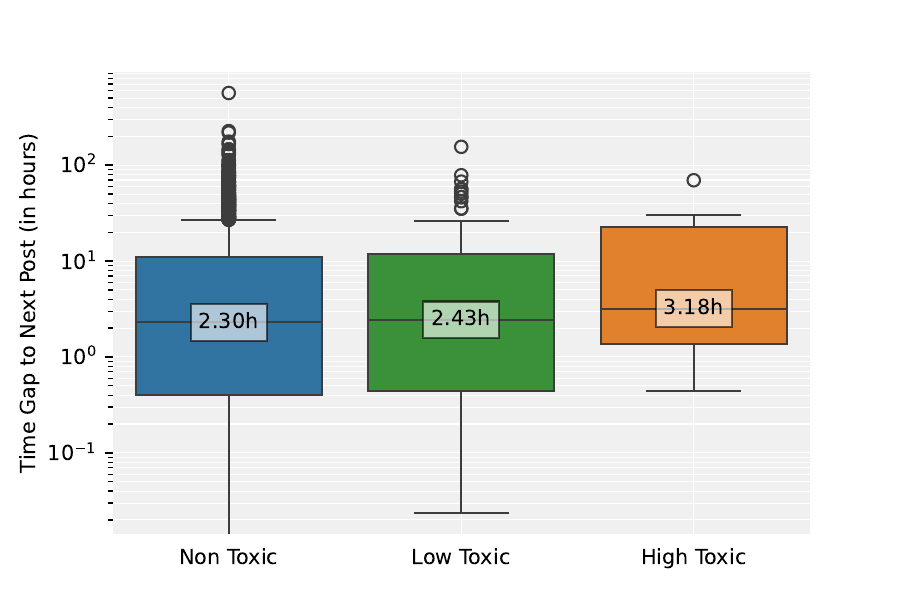}
  \end{subfigure}
  \begin{subfigure}[b]{0.3\textwidth}
    \includegraphics[width=\linewidth]{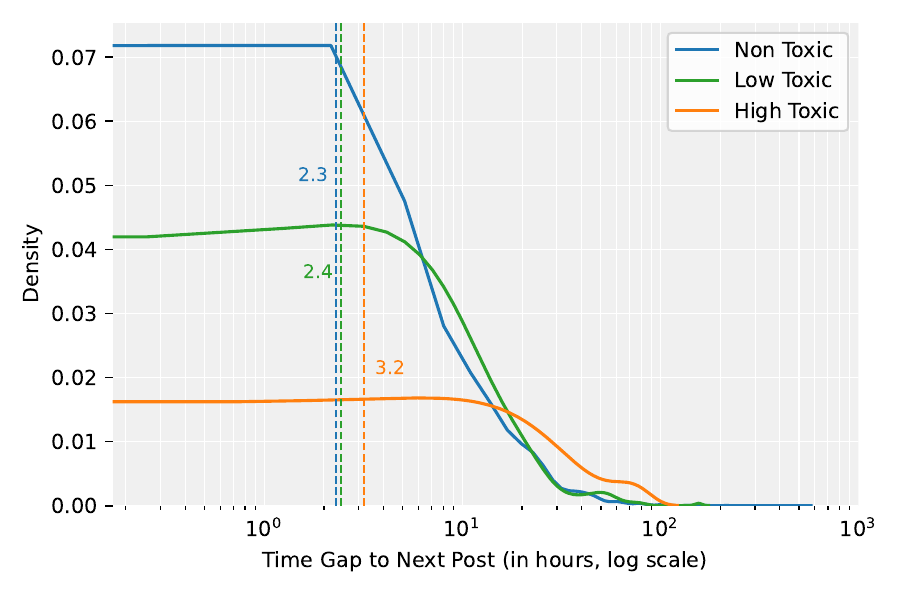}
  \end{subfigure}


  \begin{subfigure}[b]{0.3\textwidth}
    \includegraphics[width=\linewidth]{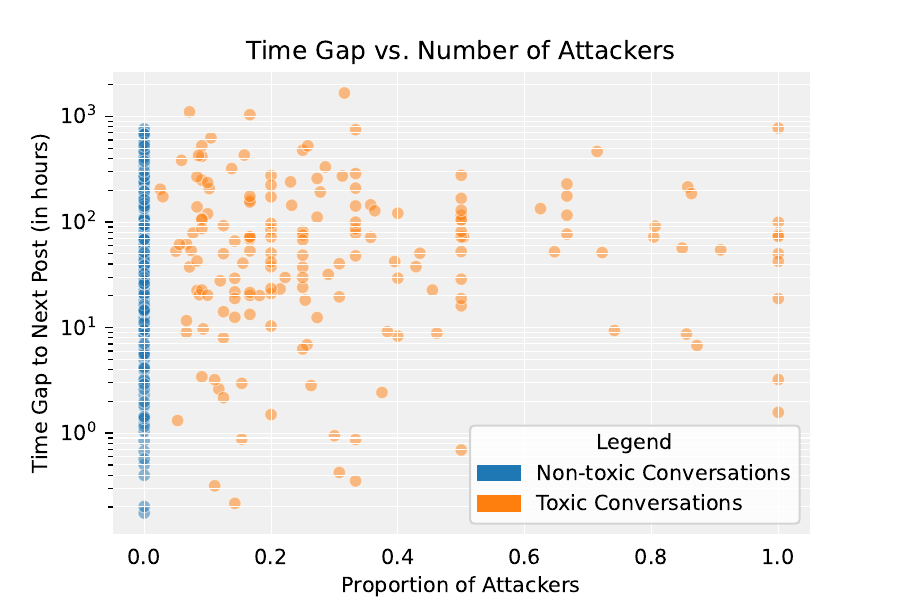}
  \end{subfigure}
  \begin{subfigure}[b]{0.3\textwidth}
    \includegraphics[width=\linewidth]{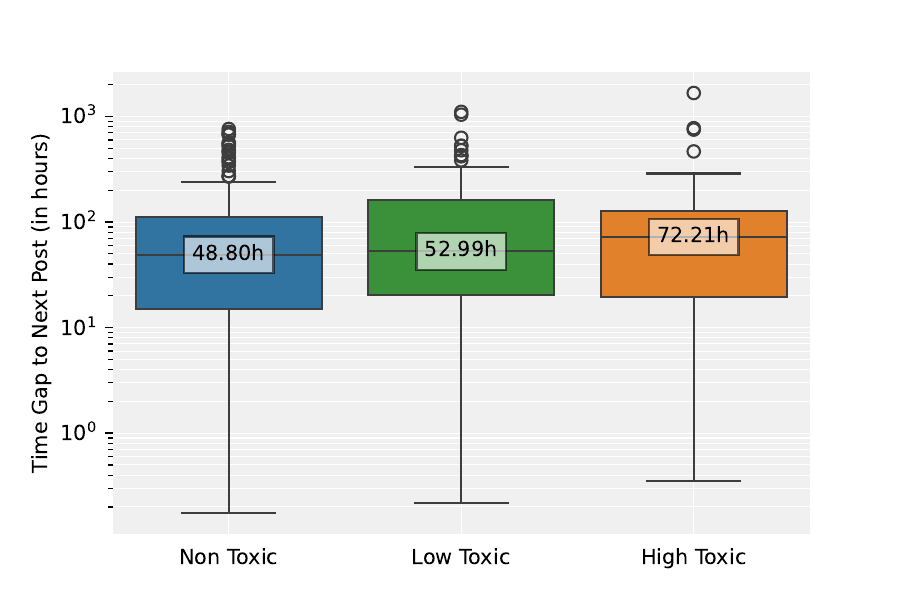}
  \end{subfigure}
  \begin{subfigure}[b]{0.3\textwidth}
    \includegraphics[width=\linewidth]{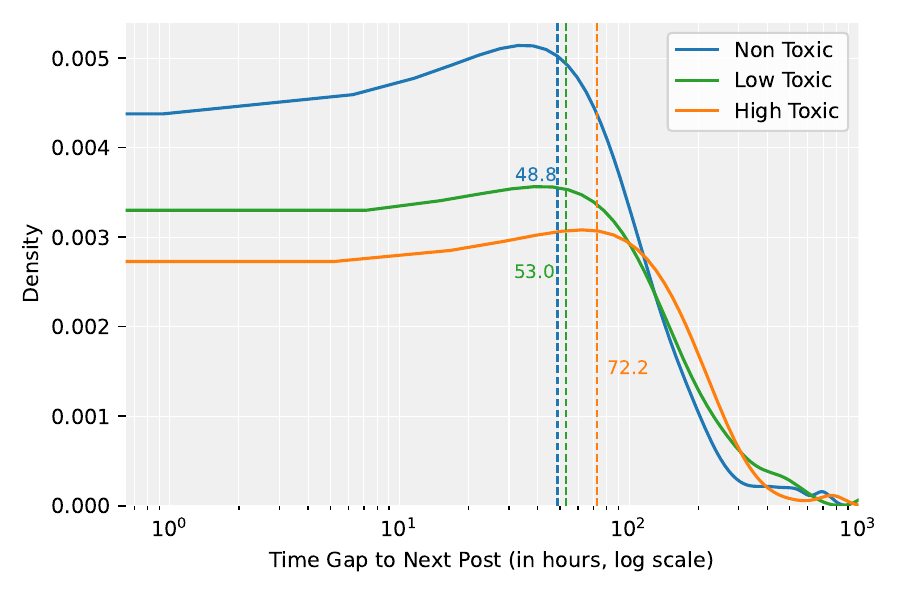}
  \end{subfigure}


  \begin{subfigure}[b]{0.3\textwidth}
    \includegraphics[width=\linewidth]{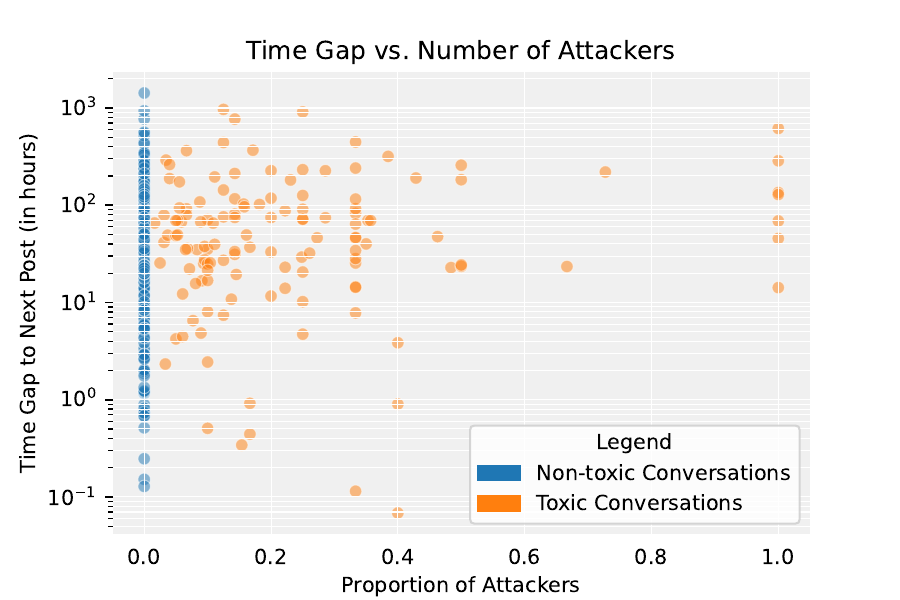}
  \end{subfigure}
  \begin{subfigure}[b]{0.3\textwidth}
    \includegraphics[width=\linewidth]{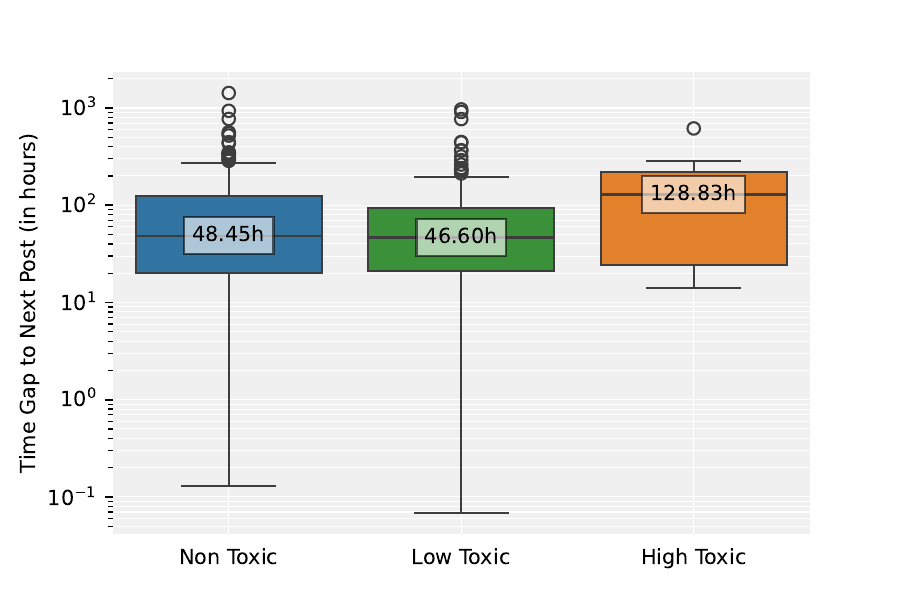}
  \end{subfigure}
  \begin{subfigure}[b]{0.3\textwidth}
    \includegraphics[width=\linewidth]{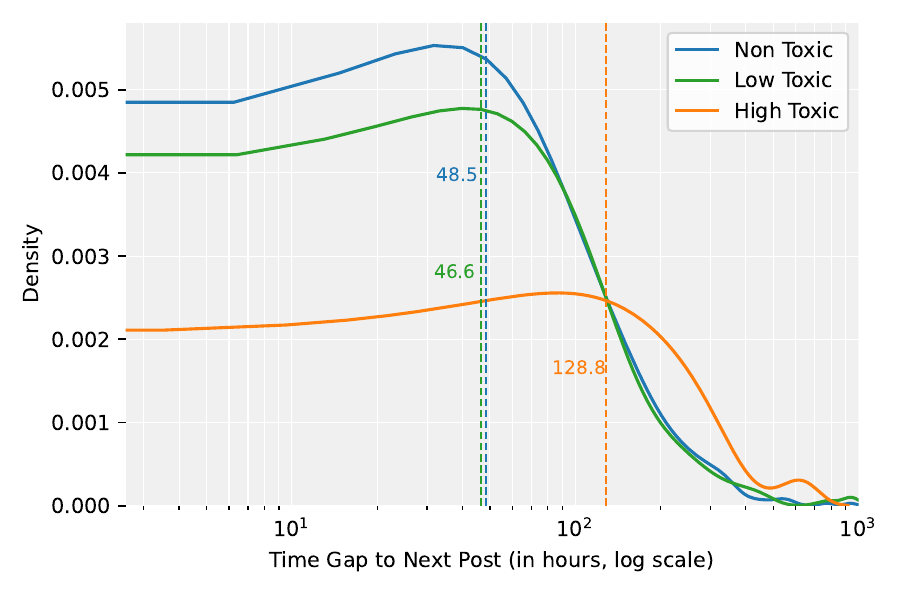}
  \end{subfigure}

  \begin{subfigure}[b]{0.3\textwidth}
    \includegraphics[width=\linewidth]{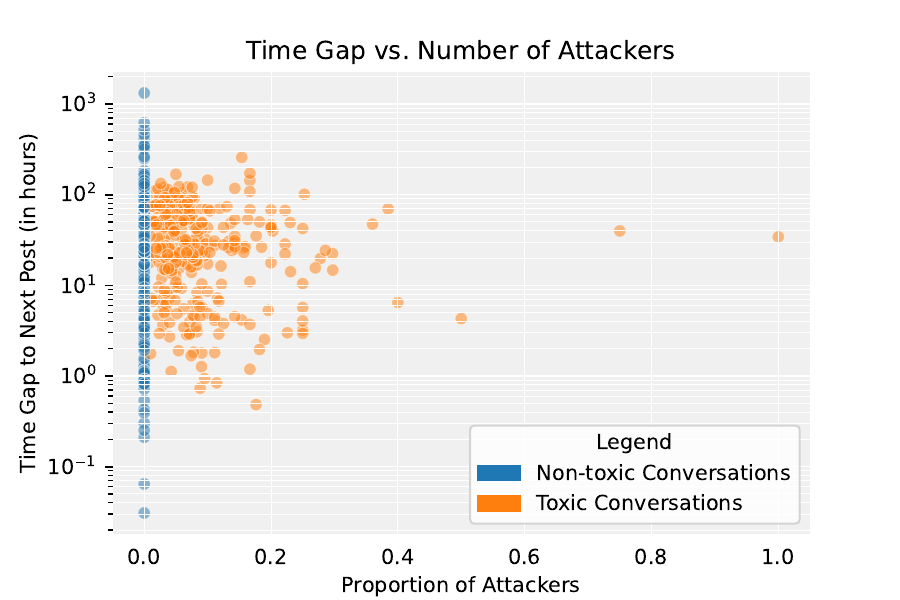}
    \caption*{Time Gap v.s. Attacker Ratios}
  \end{subfigure}
  \begin{subfigure}[b]{0.3\textwidth}
    \includegraphics[width=\linewidth]{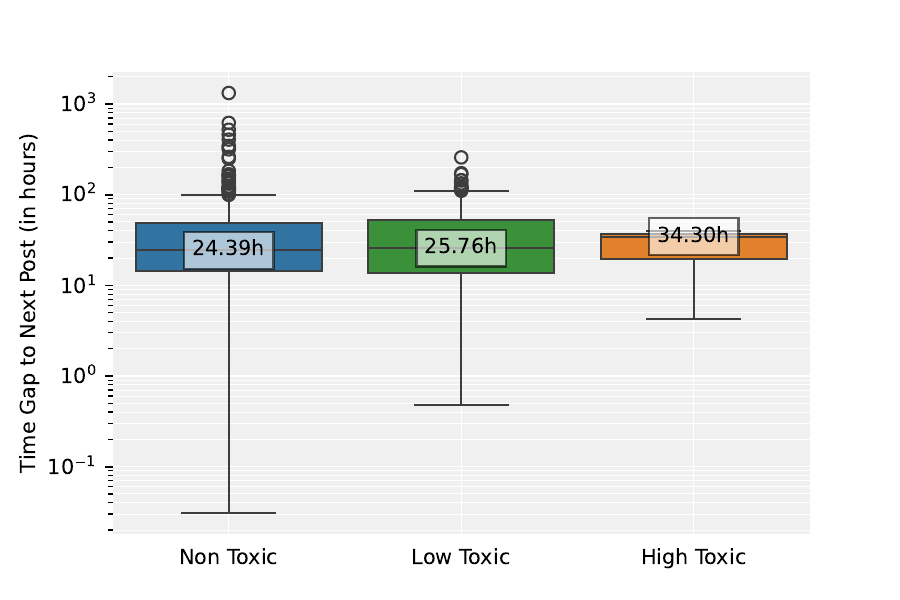}
    \caption*{Time Gap Boxplots}
  \end{subfigure}
  \begin{subfigure}[b]{0.3\textwidth}
    \includegraphics[width=\linewidth]{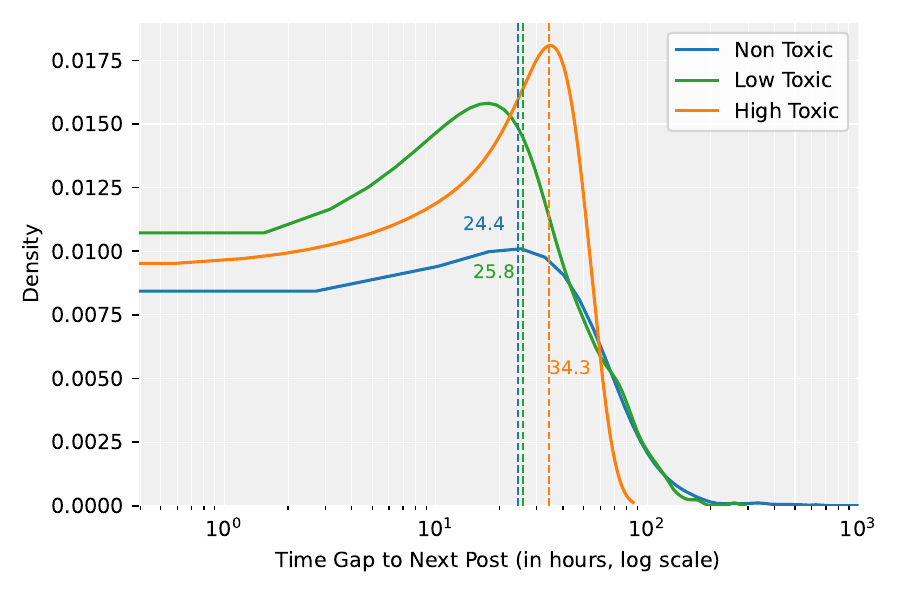}
    \caption*{Time Gap Distributions}
  \end{subfigure}
\end{figure*}

\textcolor{black}{We observe distinct posting patterns across different journalists. While most tend to delay posting after experiencing a high volume of attacks (attacker ratio above 50\%), some appear indifferent when facing such hostility. This indifference may arise when conversations contain few attackers or when the overall tone of the discussions is not particularly toxic. However, these individual variations are often masked when the data is aggregated, giving the appearance of a more uniform pattern.}

\subsection{Direct Supporters' Toxicity Level}
\begin{figure}[ht]
  \centering
  \begin{subfigure}[t]{0.5\textwidth}
    \includegraphics[width=\linewidth]{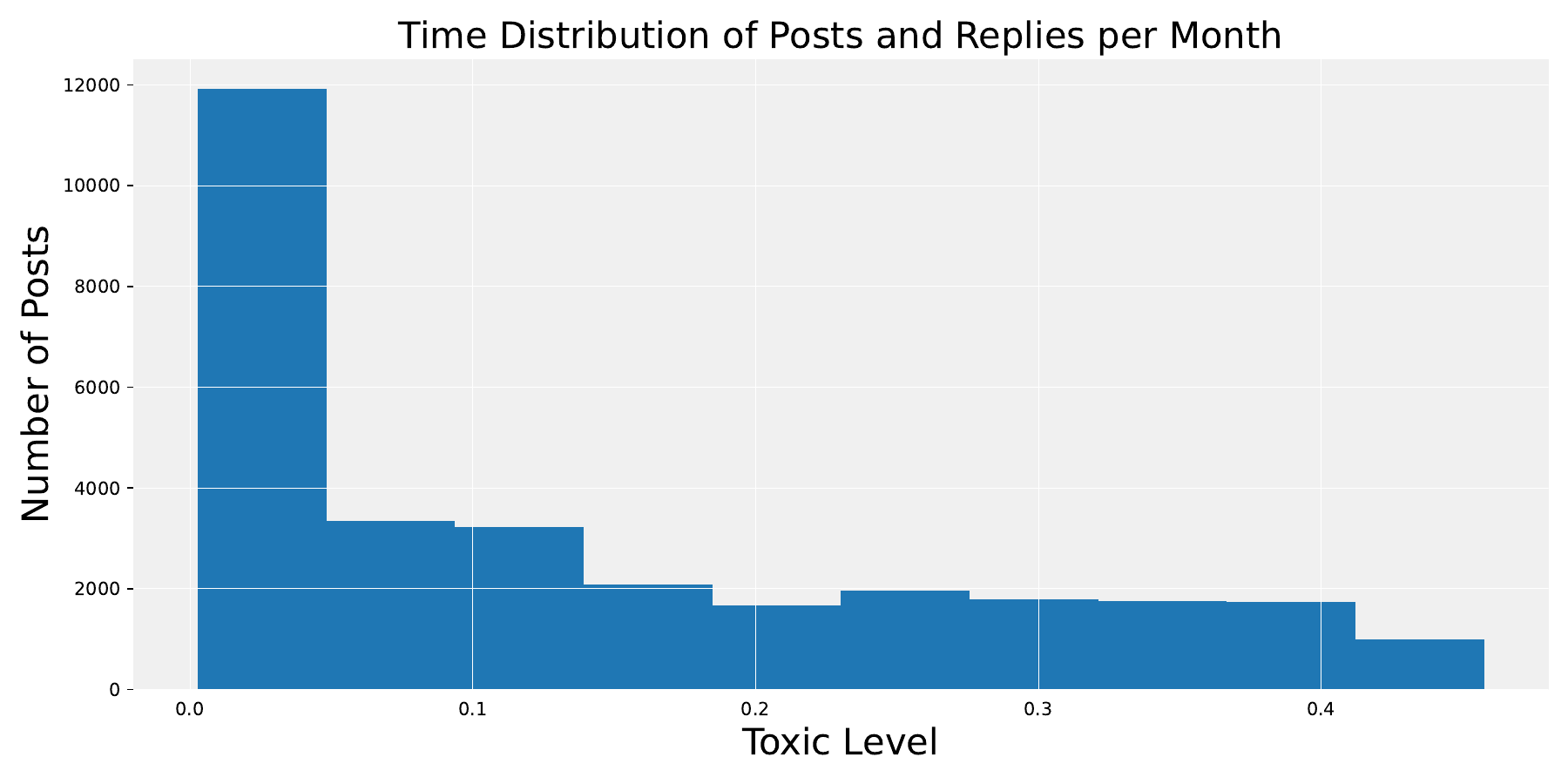}
    \label{fig:sub1}
  \end{subfigure}
  \hfill
  \begin{subfigure}[t]{0.5\textwidth}
    \includegraphics[width=\linewidth]{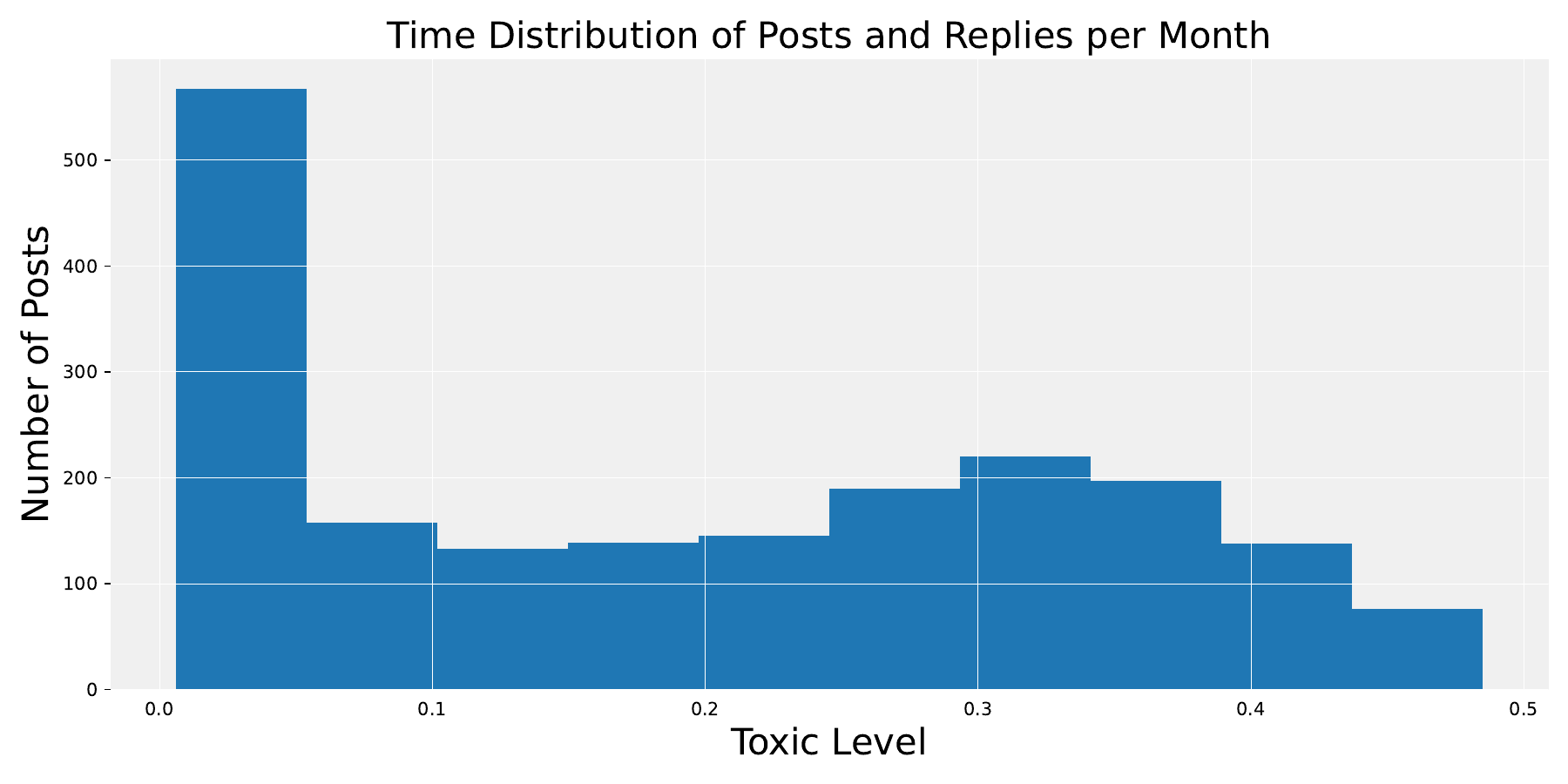}
    \label{fig:sub2}
  \end{subfigure}
  \hfill
  \begin{subfigure}[t]{0.5\textwidth}
    \includegraphics[width=\linewidth]{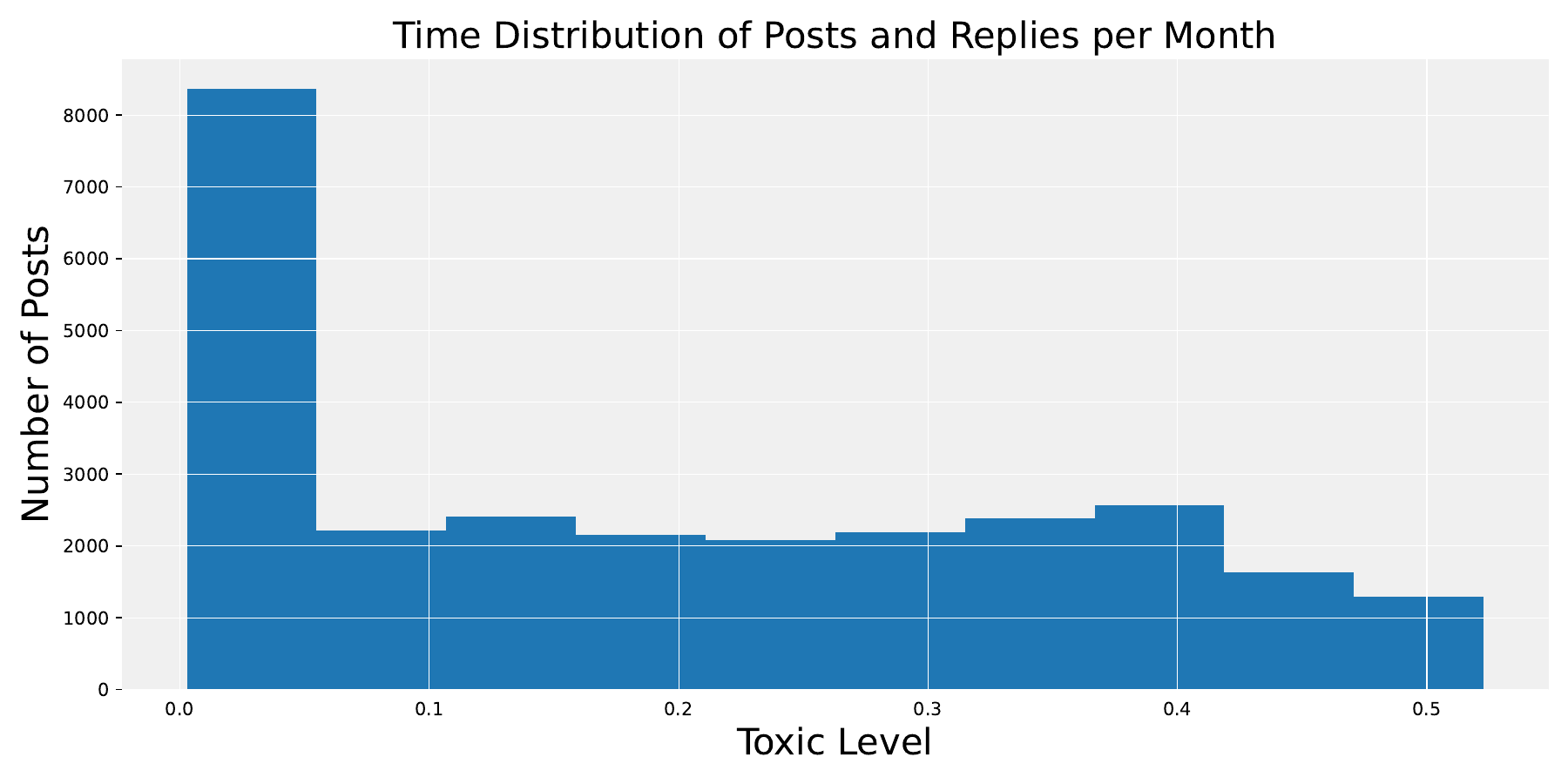}
    \label{fig:sub3}
  \end{subfigure}
  \caption{Toxicity Distribution of Direct Supporters's Replies}
  \label{fig:three_figs}
\end{figure}
\subsection{Macro Structures of Conversation Trees} \label{sec:12.3}
In this section, we shift focus toward understanding the interactions between attackers, bystanders, supporters, and journalists in different conversation environments. Specifically, we define a tweet's conversation environment using three features: the label of the parent tweet, the label of the grandparent tweet, and their position within the conversation tree (depth). We expect the label of the child tweet to be correlated to the conversation environment. Formally, it is represented in Eq.~\ref{eq:probability_model}.

\begin{equation}
\text{P(ChildLabel | ParentLabel, GrandParentLabel, Depth)}
\label{eq:probability_model}
\end{equation}

To this end, we developed a Bayesian model (details in subsection~\cref{sec:12.4}) to predict the label of the child tweet conditioned on the grandparent label, parent label, and tweet depth. 

\begin{figure}[htbp]
    \begin{subfigure}[b]{\textwidth}
        \centering
        \includegraphics[width=\textwidth]{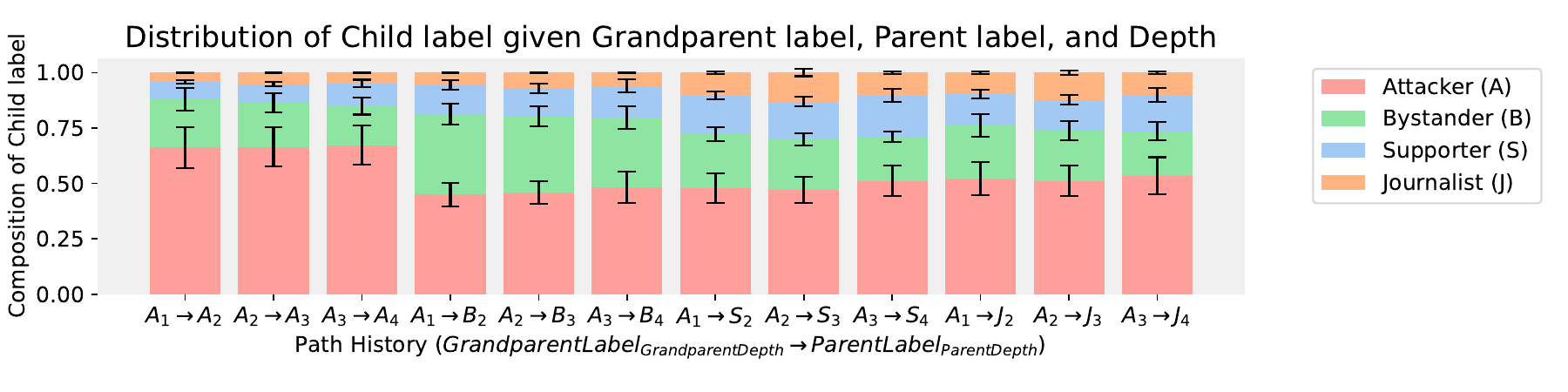} 
        \caption{Aggregated effect of attacker grandparent.}
        \label{fig:aggregated_attacker_grandparent}
    \end{subfigure}
    \begin{subfigure}[b]{\textwidth}
        \centering
        \includegraphics[width=\textwidth]{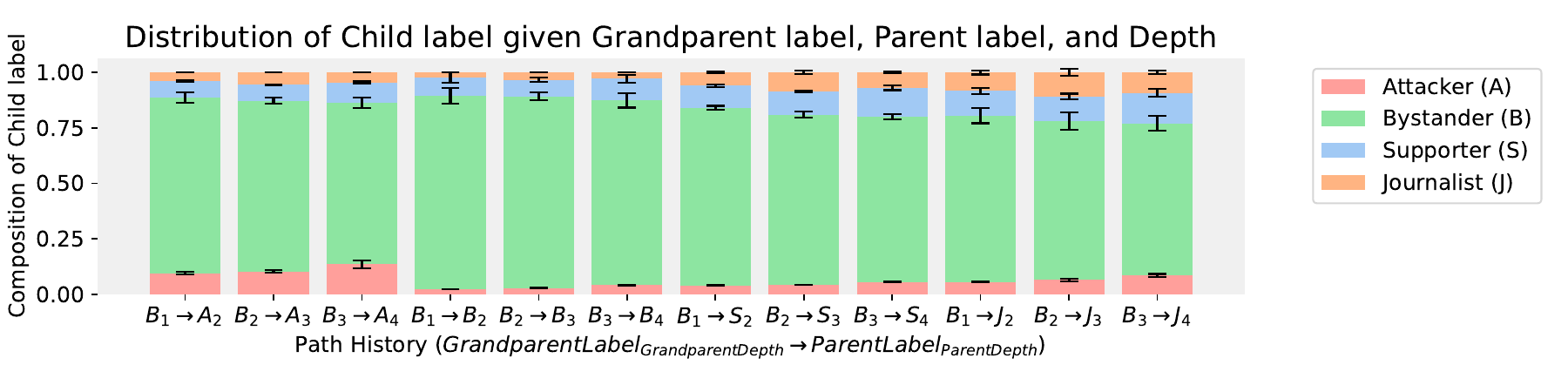} 
        \caption{Aggregated effect of bystander grandparent.}
        \label{fig:aggregated_bystander_grandparent}
    \end{subfigure}
    \begin{subfigure}[b]{\textwidth}
        \centering
        \includegraphics[width=\textwidth]{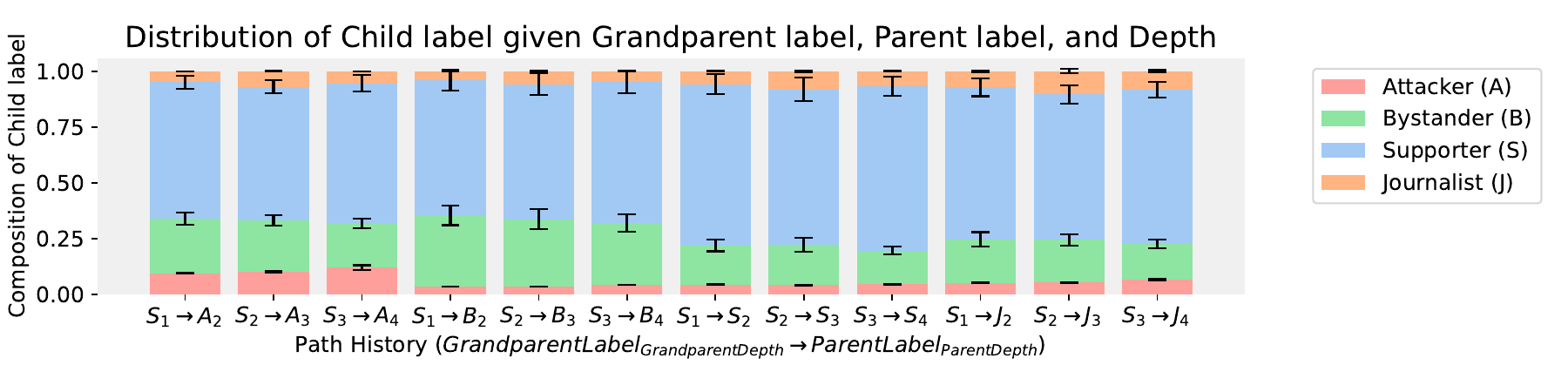} 
        \caption{Aggregated effect of supporter grandparent.}
        \label{fig:aggregated_supporter_grandparent}
    \end{subfigure}
    \begin{subfigure}[b]{\textwidth}
        \centering
        \includegraphics[width=\textwidth]{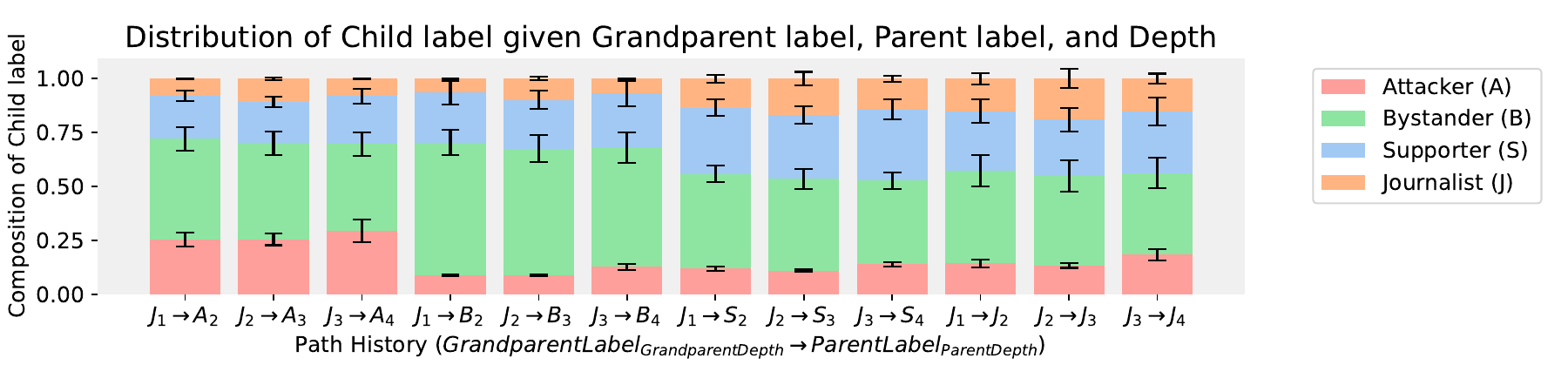} 
        \caption{Aggregated effect of journalist grandparent.}
        \label{fig:aggregated_journalist_grandparent}
    \end{subfigure}
    \caption{Aggregated effect of grandparent label on child distribution (See \Cref{sec:12.3}). 
    Journalists prefer conversing with supporters while having comparable engagement with attackers and bystanders. The notion of social affirmation is reinforced through the preference of user group members to engage with people from the same group. Attackers appear active in responding to journalists and supporters }
    \label{fig:aggregated_grandparent_label_effect}
\end{figure}

We chose to use a Bayesian model for two reasons. First, the priors and the likelihood function make the model's assumptions fully transparent, eliminating the need to validate distributional assumptions against the data. Second, priors within this model regularize and make the model less prone to overfitting. At the same time, the priors are chosen to be weakly informative to ensure they do not dominate inference. 
To understand generic dialogue characteristics that exist across journalists, we aggregate the results from our Bayesian model across 13 journalists to smoothen out individual differences and highlight fundamental patterns. The aggregation is done by taking the mean composition of the four groups for each conversation environment. The results are shown in figure~\ref{fig:aggregated_grandparent_label_effect}.  The variance of each stack is plotted to illustrate that the spread of the stacked bar plots is reasonable to support the idea that generic dialogue patterns exist. 

The results in Figure~\ref{fig:aggregated_grandparent_label_effect} indicate that attackers across 13 journalists tend to pile on each other's replies while having a preference to interact with supporters and journalists as opposed to bystanders. Attackers appear to interact more with journalists than supporters, hinting at an incentive to directly impact the journalists as opposed to countering supporters. These interaction patterns are consistent across all parent labels, which is inferred by examining the height of the red stack across the x-axis of each sub-figure in Figure~\ref{fig:aggregated_grandparent_label_effect}, and grandparent labels, which is inferred by examining the difference in the sequence of heights of red stacks in figures~\ref{fig:aggregated_attacker_grandparent}, ~\ref{fig:aggregated_bystander_grandparent}, ~\ref{fig:aggregated_supporter_grandparent}, and ~\ref{fig:aggregated_journalist_grandparent}. The analysis is conducted for each interest group to understand its activity in the presence of other groups. Here, the "presence" of another group is defined by its appearance in the path history, specifically as a parent or grandparent tweet.


Bystanders are much more likely to interact with other bystanders. They also interact more with journalists than with attackers and supporters. These activity trends are inferred by examining the height of the green stack across the x-axis for each sub-figure in Figure~\ref{fig:aggregated_grandparent_label_effect} and by examining the difference in sequence of heights of green stacks in sub-figures~\ref{fig:aggregated_attacker_grandparent}, ~\ref{fig:aggregated_bystander_grandparent}, ~\ref{fig:aggregated_supporter_grandparent}, and ~\ref{fig:aggregated_journalist_grandparent}. The activity trends of bystanders are in line with the idea that bystanders interact with each other or the journalist and refrain from participating in conversations that express opinions about the journalist.

A similar examination of the blue stacks in sub-figures~\ref{fig:aggregated_attacker_grandparent}, ~\ref{fig:aggregated_bystander_grandparent}, ~\ref{fig:aggregated_supporter_grandparent}, and ~\ref{fig:aggregated_journalist_grandparent}, suggests that supporters are less likely to engage with attackers and bystanders than with other supporters or the journalist. They appear to be focused on supporting the journalist directly rather than countering attackers or engaging in neutral conversations with bystanders. 

Journalist trends in Figure~\ref{fig:aggregated_grandparent_label_effect} indicate that they participate actively in the presence of supporters, while they are more reserved with attackers and bystanders. This suggests that journalists make more active efforts to participate in a positive environment as opposed to defending themselves from attackers. Although attacker tweets justify a response from the journalists, while bystander tweets do not, it is noteworthy that the journalists participate equally in both environments.

\subsection{Model Details for estimating the Macro Structures of Conversation Trees}
\label{sec:12.4}
In this section, we discuss the details of the proposed Bayesian model. The model's key components are the likelihood function, linear models, and prior distributions. The outcome variable, the child label, is a multi-class categorical variable, justifying our use of a categorical likelihood function. The values passed to the likelihood function are derived by applying softmax to the output of four linear models. The ouputs $s_{m,i}$ in Equations \eqref{eq:bayesian_model_eq12b}, \eqref{eq:bayesian_model_eq12c}, \eqref{eq:bayesian_model_eq12d}, and  \eqref{eq:bayesian_model_eq12e} represent the unnormalized probability of data point 'i' belonging to class m. There are a set of parameters associated with each class. The $\alpha$ parameters in Equations \eqref{eq:bayesian_model_eq12f} - \eqref{eq:bayesian_model_eq12i} determine $s_{0,i}$; the $\beta$ parameters of Equations \eqref{eq:bayesian_model_eq12f} - \eqref{eq:bayesian_model_eq12i} define $s_{1,i}$; the gamma parameters of Equations \eqref{eq:bayesian_model_eq12f} - \eqref{eq:bayesian_model_eq12i} determine $s_{2,i}$. A reference class, $s_{3,i}$, is used as baseline category for multinomial models. The log-odds of the other classes are computed relative to this category.

The parameters $\alpha_I$, $\beta_I$, and $\gamma_I$ are intercepts for the corresponding unnormalized class probability. The other $\alpha$, $\beta$, and $\gamma$ parameters are indexed by the value of input variables: grandparent label, parent label, and depth. The grandparent label indexes into Equation \eqref{eq:bayesian_model_eq12g} taking on values $\{0,1,2,3\}$, while the parent label indexes into Equation \eqref{eq:bayesian_model_eq12h} taking on the same values. The input index variable of depth in \eqref{eq:bayesian_model_eq12i} takes on values from 0 to maximum depth, which is used to obtain the corresponding $\alpha_D$, $\beta_D$, and $\gamma_D$ parameters. Each input value of the grandparent label, parent label, and depth is associated with a parameter for each output class. The priors of these parameters are distributed as a StudentT distribution as this distribution handles data with heavier tails better, thereby being robust to outliers. 

\begin{subequations}
\label{eq:bayesian_model}
\begin{align}
    \text{ChildLabel}_{i} & \sim \text{Categorical}\left(\text{Softmax}\left[s_{0,i}, s_{1,i}, s_{2,i}, s_{3,i}\right]\right) \label{eq:bayesian_model_eq12a}\\
    s_{0,i} & = \alpha_{I[i]} + \alpha_{G[i]} + \alpha_{P[i]} + \alpha_{D[i]} \quad \quad \quad \quad \text{Linear model for P(attacker)} \label{eq:bayesian_model_eq12b}\\
    s_{1,i} & = \beta_{I[i]} + \beta_{G[i]} + \beta_{P[i]} + \beta_{D[i]} \quad \quad \quad \quad \text{Linear model for P(bystander)}\label{eq:bayesian_model_eq12c}\\
    s_{2,i} & = \gamma_{I[i]} + \gamma_{G[i]} + \gamma_{P[i]} + \gamma_{D[i]}  \quad \quad \quad \quad \ \ \text{Linear model for P(supporter)}\label{eq:bayesian_model_eq12d}\\
    s_{3,i} & = 0 \hspace{5cm} \  \text{Reference class}\label{eq:bayesian_model_eq12e}\\
    \alpha_{I}, \beta_{I}, \gamma_{I} & \sim \text{StudentT}(\nu = 2, \mu = 0, \sigma = 1) \hspace{1cm} \ \ \text{Prior for Intercept} \label{eq:bayesian_model_eq12f}\\
    \alpha_{G[i]}, \beta_{G[i]}, \gamma_{G[i]} & \sim \text{StudentT}(\nu = 2, \mu = 0, \sigma = 1)  \hspace{1cm} \ \ \text{Prior for grandparent label} \label{eq:bayesian_model_eq12g}\\
    \alpha_{P[i]}, \beta_{P[i]}, \gamma_{P[i]} & \sim \text{StudentT}(\nu = 2, \mu = 0, \sigma = 1) \hspace{1cm} \ \ \text{Prior for parent label}\label{eq:bayesian_model_eq12h}\\
    \alpha_{D[i]}, \beta_{D[i]}, \gamma_{D[i]} & \sim \text{StudentT}(\nu = 2, \mu = 0, \sigma = 1)\hspace{1cm} \ \ \text{Prior for depth}\label{eq:bayesian_model_eq12i}
\end{align}
\end{subequations}

\newpage
The results of the model for the 13 journalists along with the histograms of the posterior predictive outcomes and ground truth distribution are displayed below.

\begin{figure}[htbp]
    \begin{subfigure}[b]{\textwidth}
        \centering
        \includegraphics[width=\textwidth]{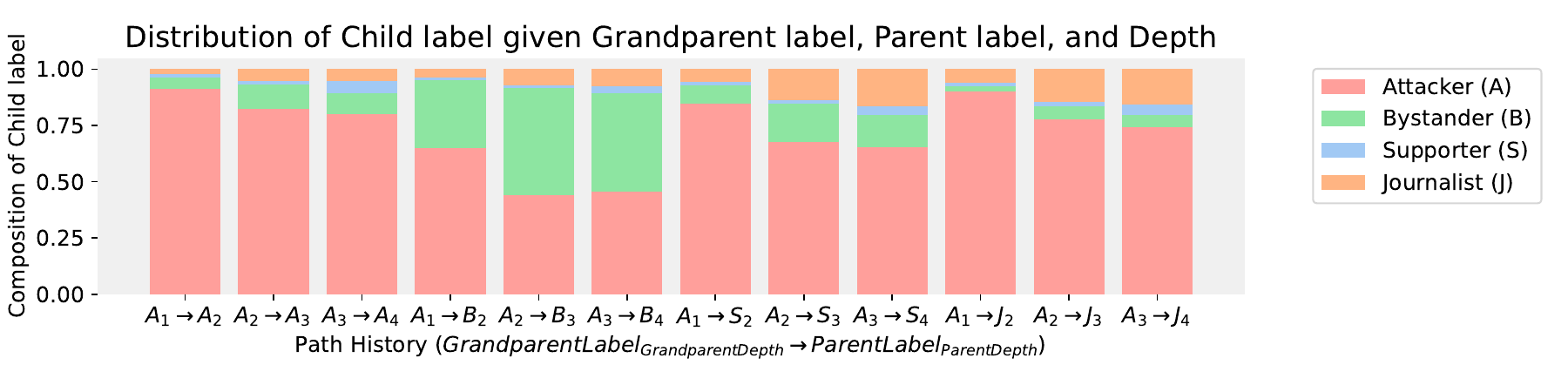} 
        \caption{Journalist A: attacker grandparent.}
    \end{subfigure}
    \begin{subfigure}[b]{\textwidth}
        \centering
        \includegraphics[width=\textwidth]{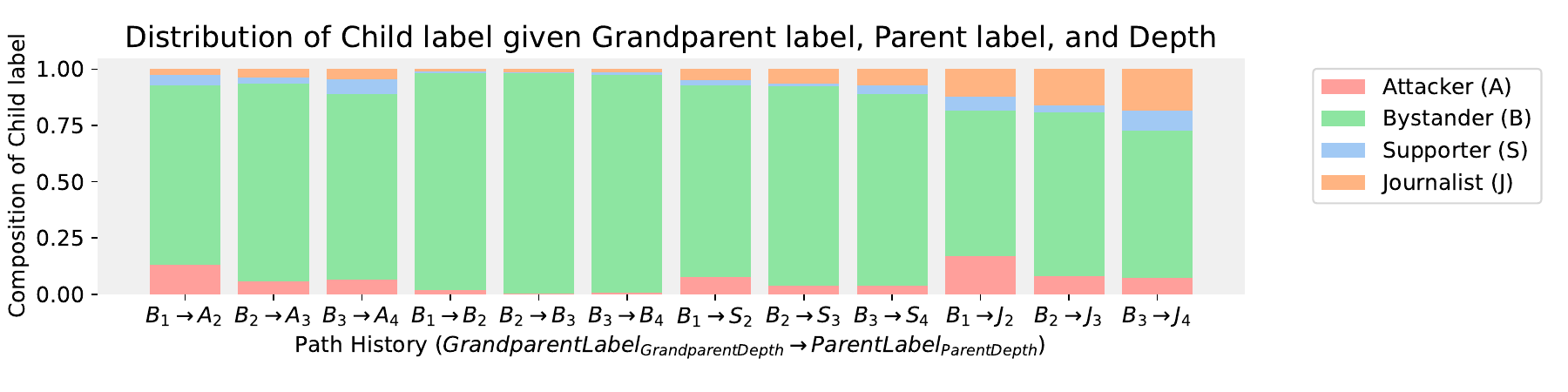} 
        \caption{Journalist A: bystander grandparent.}
    \end{subfigure}
    \begin{subfigure}[b]{\textwidth}
        \centering
        \includegraphics[width=\textwidth]{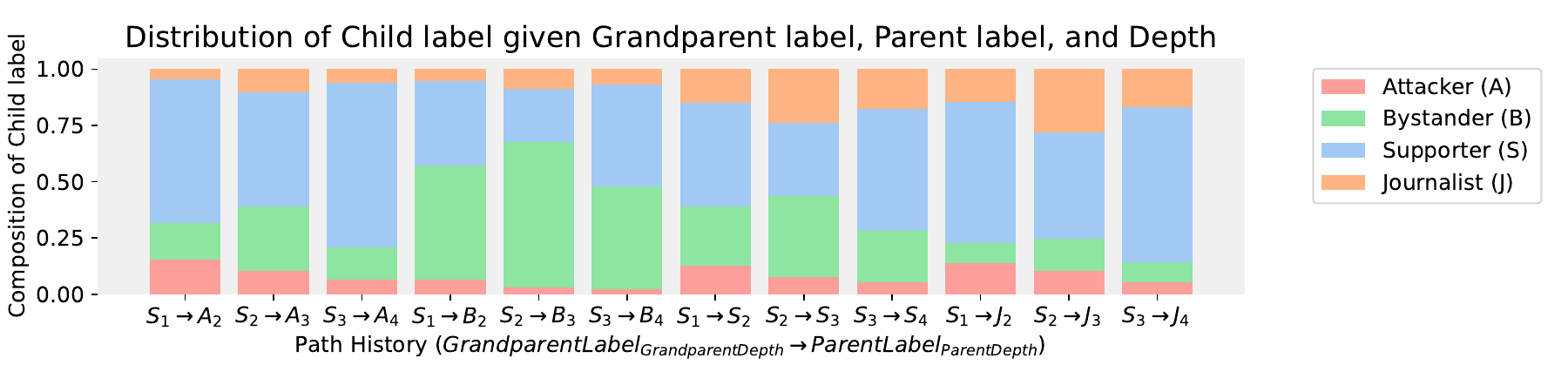} 
        \caption{Journalist A: supporter grandparent.}
    \end{subfigure}
    \begin{subfigure}[b]{\textwidth}
        \centering
        \includegraphics[width=\textwidth]{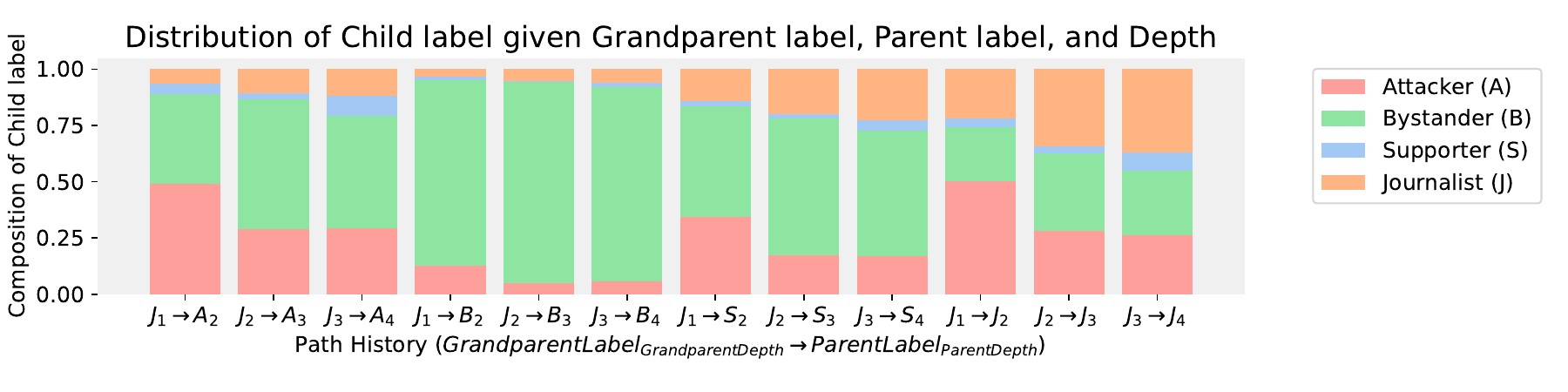} 
        \caption{Journalist A: journalist grandparent.}
    \end{subfigure}
    \caption{Journalist A: Effect of grandparent and parent label on child label distribution}
\end{figure}

\begin{figure}[ht]
    \centering
    \begin{subfigure}[b]{0.45\textwidth}
        \centering
        \includegraphics[width=\textwidth]{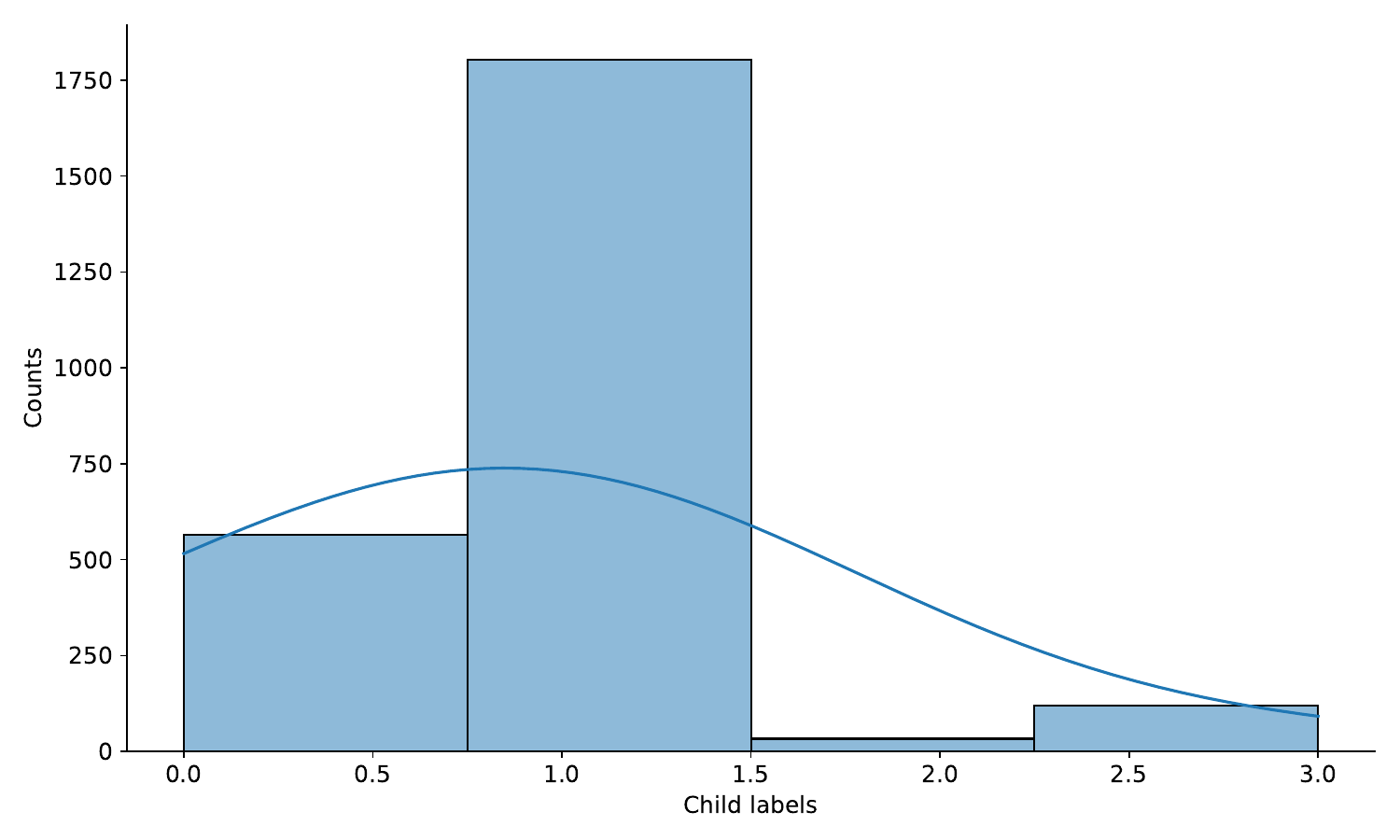} 
        \caption{Journalist A: Histogram of posterior predictive outcome}
    \end{subfigure}
    \hfill
    \begin{subfigure}[b]{0.45\textwidth}
        \centering
        \includegraphics[width=\textwidth]{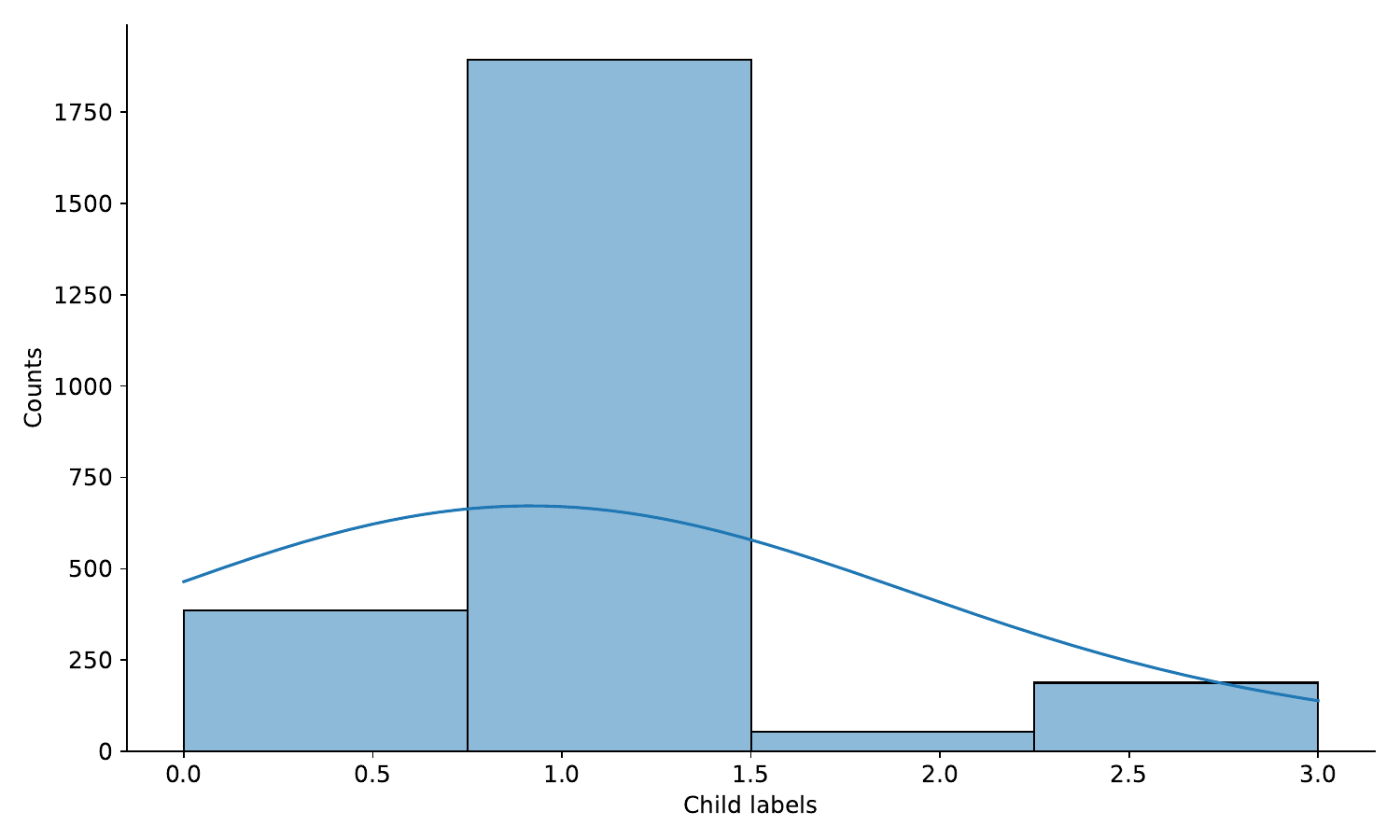} 
        \caption{Journalist A: Histogram of ground truth outcome distribution}
    \end{subfigure}
    \caption{Journalist A: Posterior predictive outcome distribution vs. ground truth outcome distribution}
\end{figure}

\begin{figure}[htbp]
    \begin{subfigure}[b]{\textwidth}
        \centering
        \includegraphics[width=\textwidth]{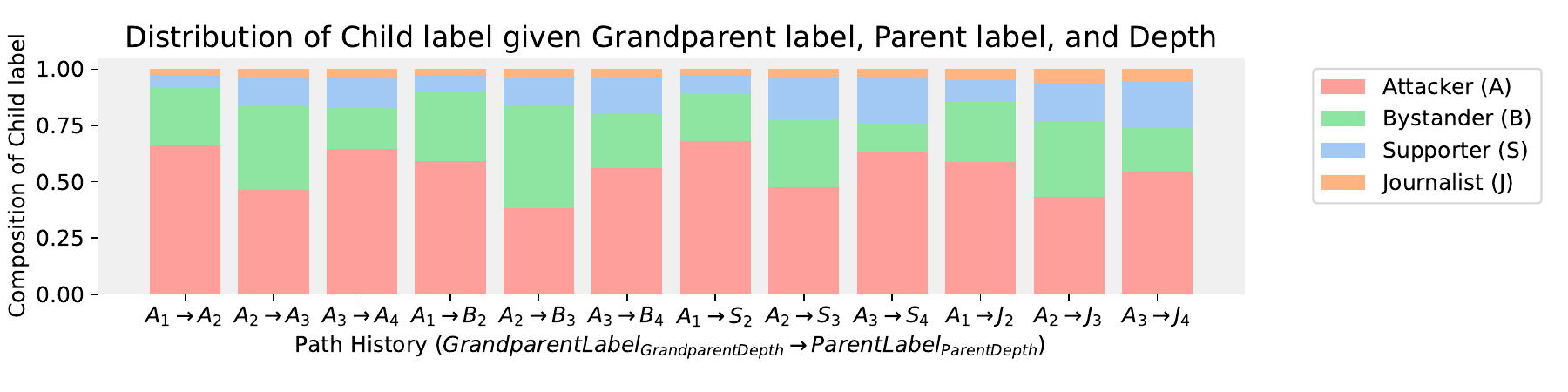} 
        \caption{Journalist B: attacker grandparent.}
    \end{subfigure}
    \begin{subfigure}[b]{\textwidth}
        \centering
        \includegraphics[width=\textwidth]{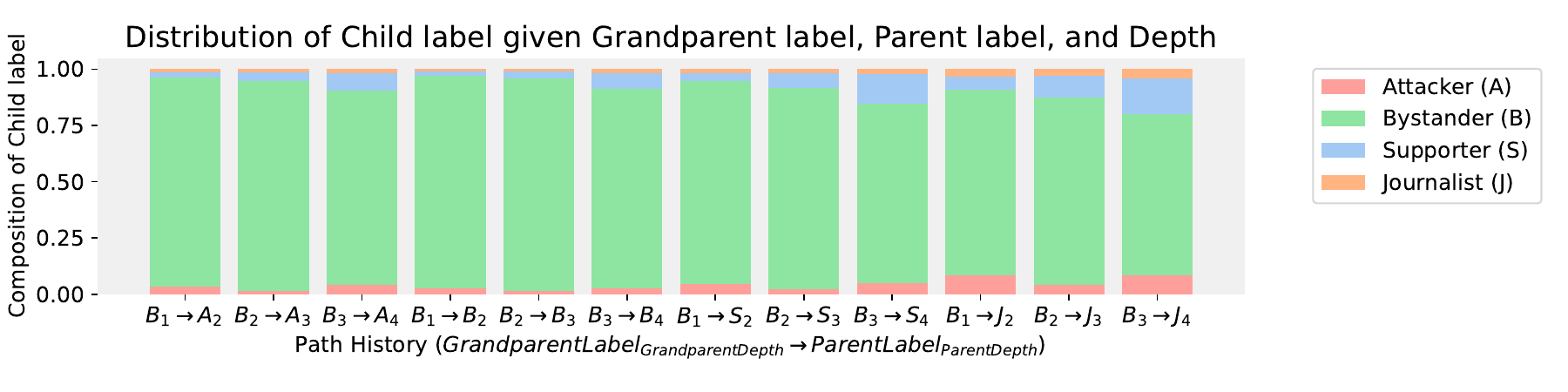} 
        \caption{Journalist B: bystander grandparent.}
    \end{subfigure}
    \begin{subfigure}[b]{\textwidth}
        \centering
        \includegraphics[width=\textwidth]{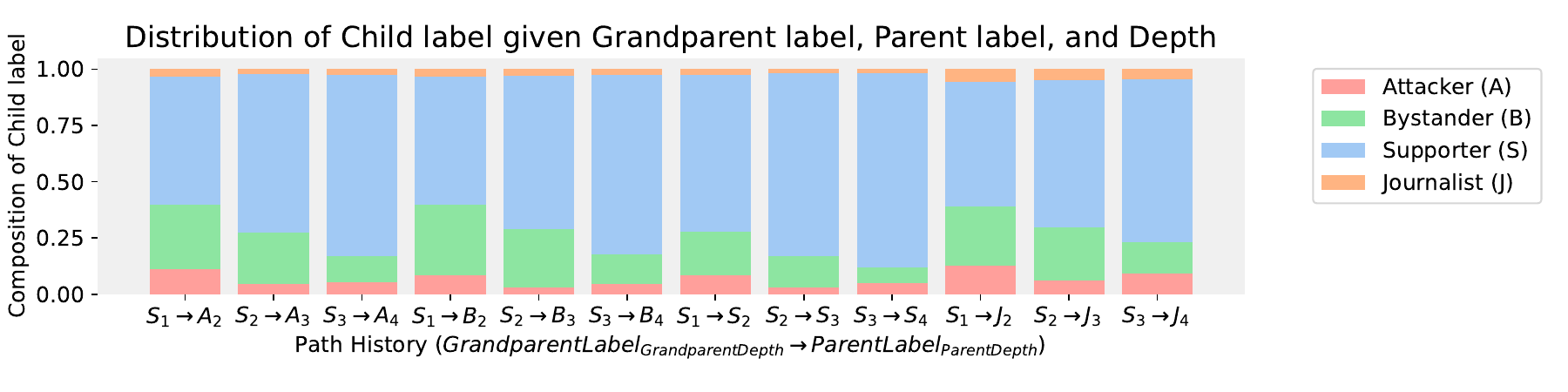} 
        \caption{Journalist B: supporter grandparent.}
    \end{subfigure}
    \begin{subfigure}[b]{\textwidth}
        \centering
        \includegraphics[width=\textwidth]{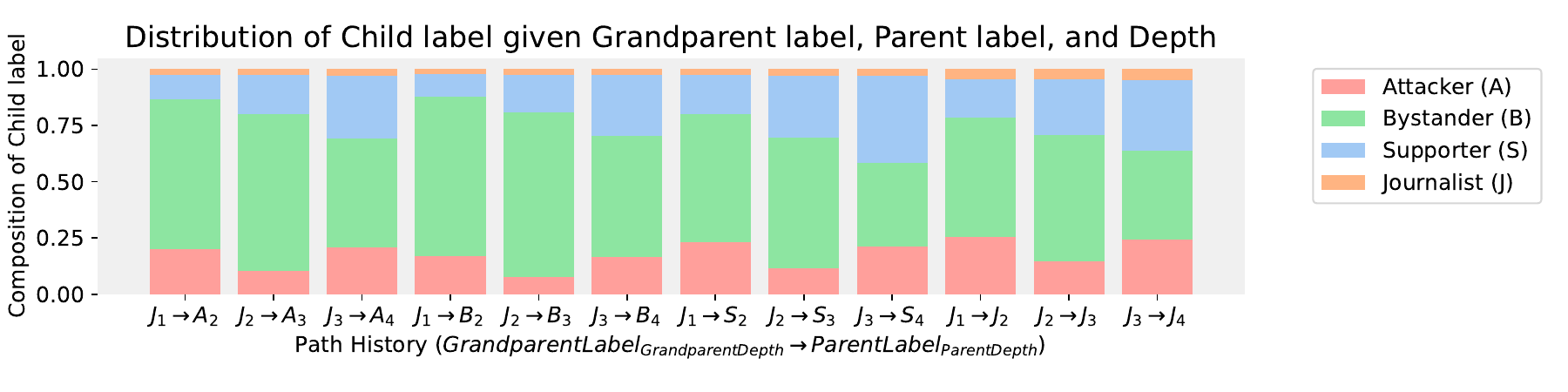} 
        \caption{Journalist B: journalist grandparent.}
    \end{subfigure}
    \caption{Journalist B: Effect of grandparent and parent label on child label distribution}
\end{figure}

\begin{figure}[ht]
    \centering
    \begin{subfigure}[b]{0.45\textwidth}
        \centering
        \includegraphics[width=\textwidth]{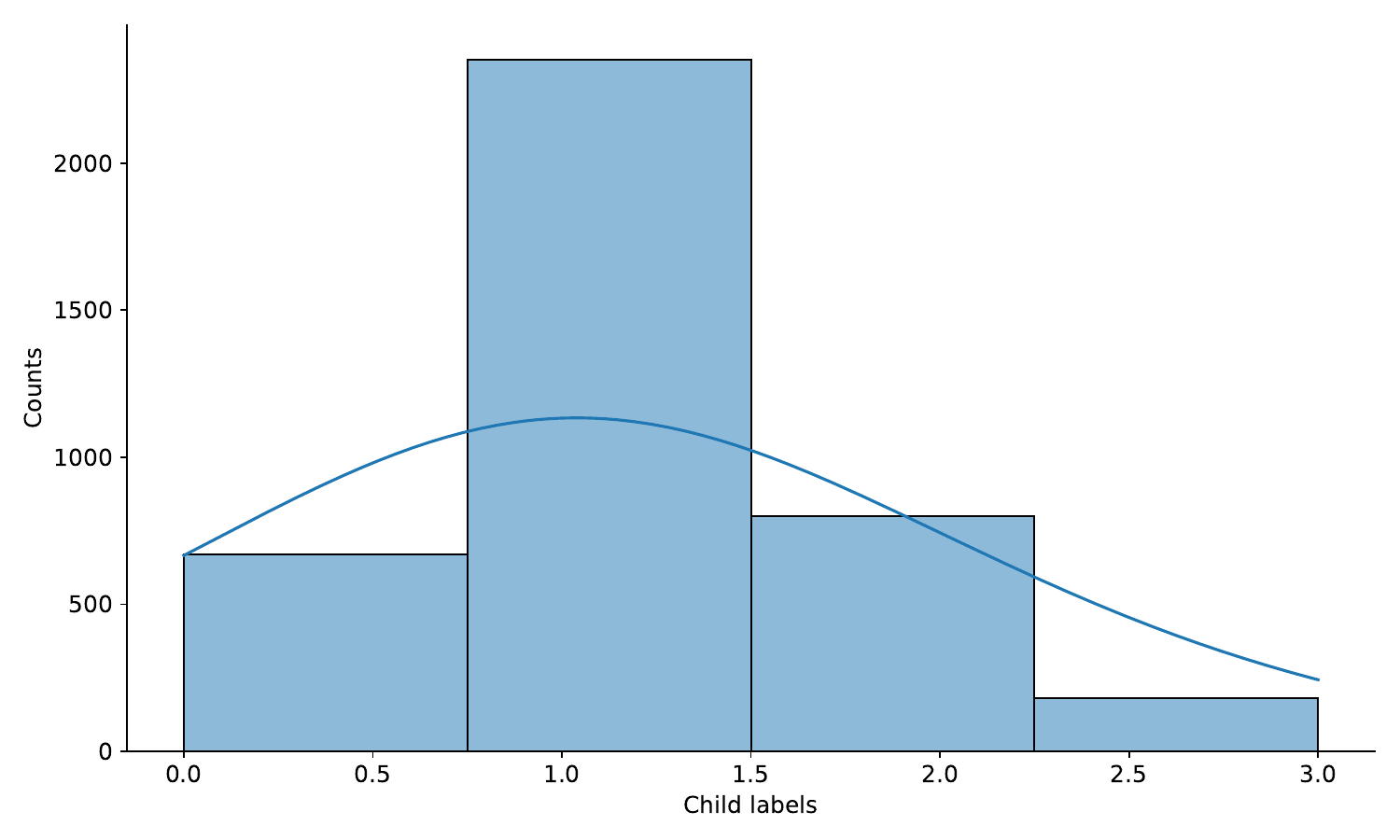} 
        \caption{Journalist B: Histogram of posterior predictive outcome}
    \end{subfigure}
    \hfill
    \begin{subfigure}[b]{0.45\textwidth}
        \centering
        \includegraphics[width=\textwidth]{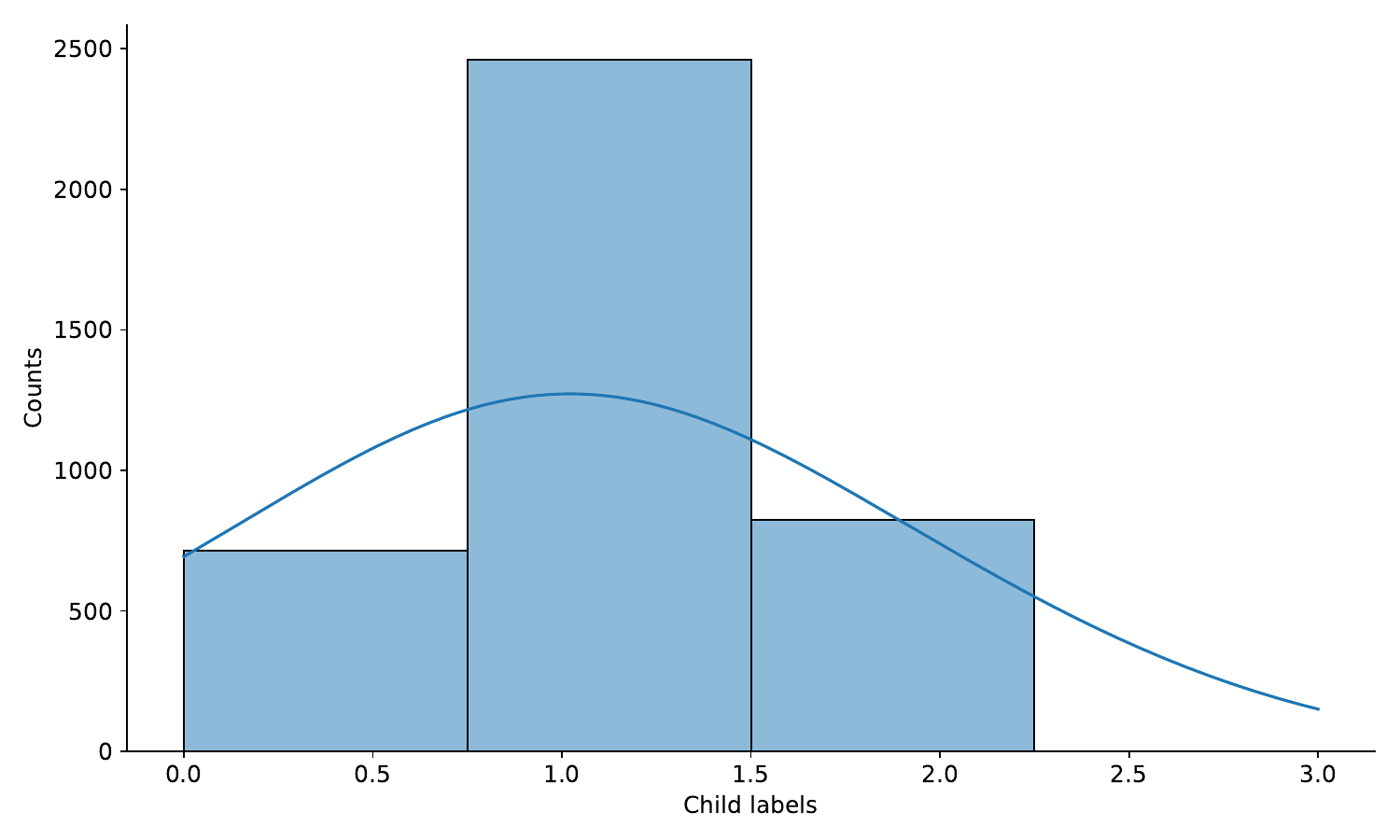} 
        \caption{Journalist B: Histogram of ground truth outcome distribution}
    \end{subfigure}
    \caption{Journalist B: Posterior predictive outcome distribution vs. ground truth outcome distribution}
\end{figure}

\begin{figure}[htbp]
    \begin{subfigure}[b]{\textwidth}
        \centering
        \includegraphics[width=\textwidth]{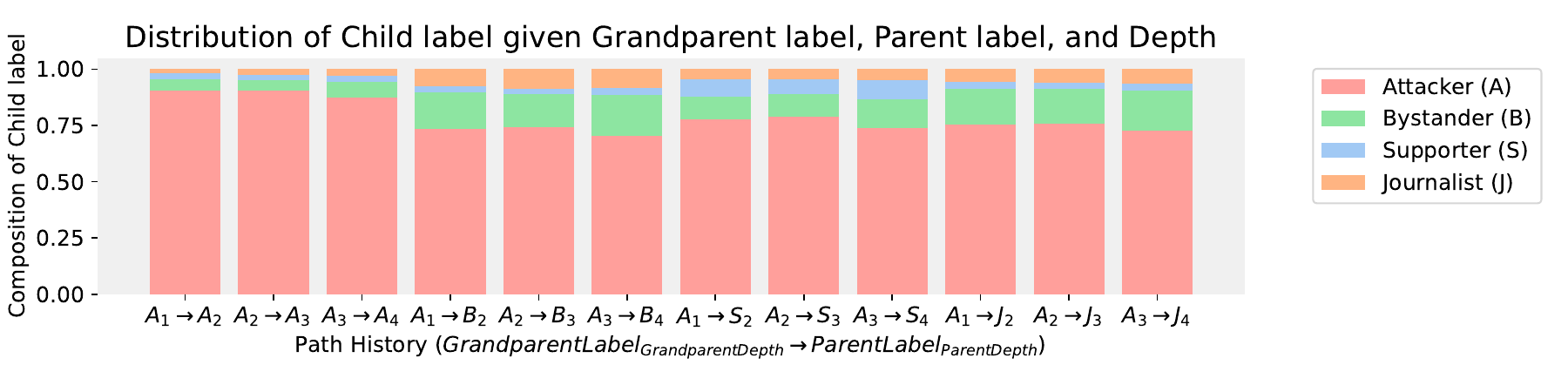} 
        \caption{Journalist C: attacker grandparent.}
    \end{subfigure}
    \begin{subfigure}[b]{\textwidth}
        \centering
        \includegraphics[width=\textwidth]{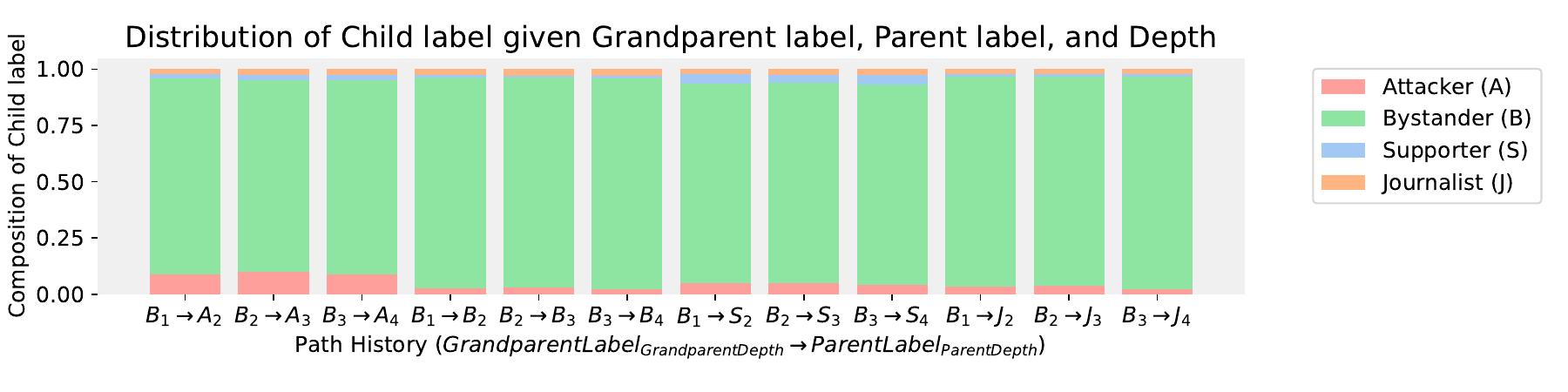} 
        \caption{Journalist C: bystander grandparent.}
    \end{subfigure}
    \begin{subfigure}[b]{\textwidth}
        \centering
        \includegraphics[width=\textwidth]{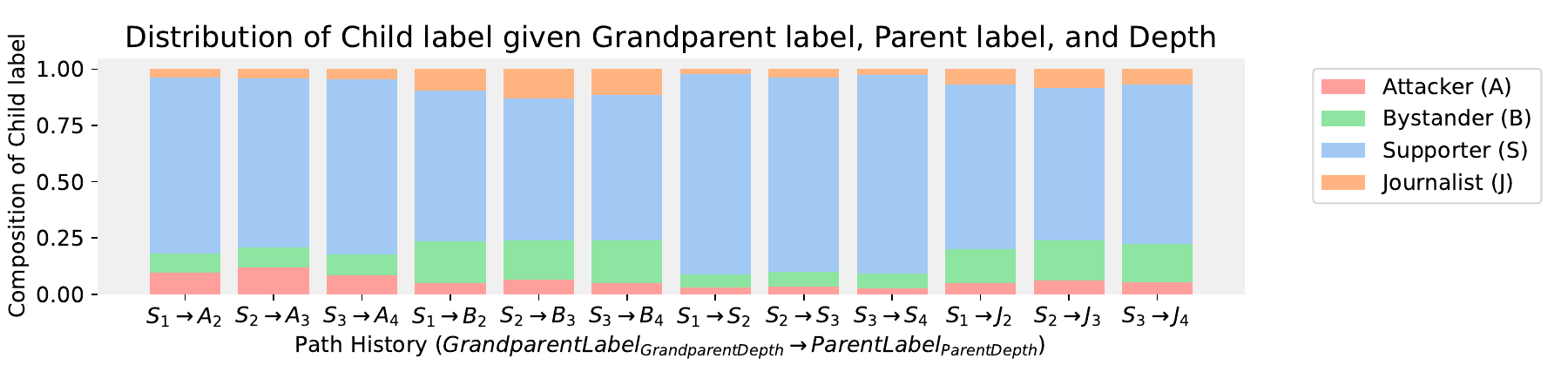} 
        \caption{Journalist C: supporter grandparent.}
    \end{subfigure}
    \begin{subfigure}[b]{\textwidth}
        \centering
        \includegraphics[width=\textwidth]{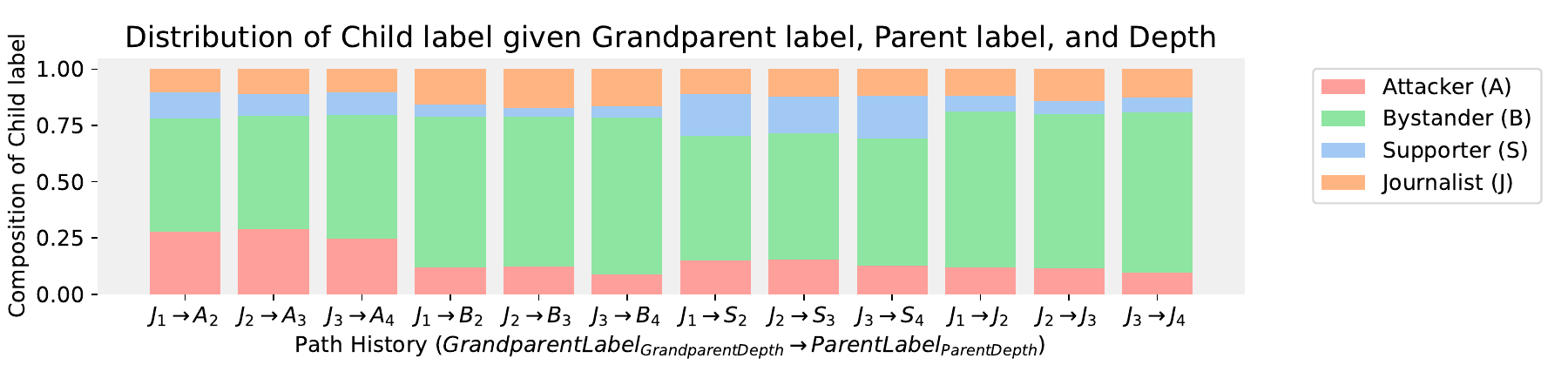} 
        \caption{Journalist C: journalist grandparent.}
    \end{subfigure}
    \caption{Journalist C: Effect of grandparent and parent label on child label distribution}
\end{figure}

\begin{figure}[ht]
    \centering
    \begin{subfigure}[b]{0.45\textwidth}
        \centering
        \includegraphics[width=\textwidth]{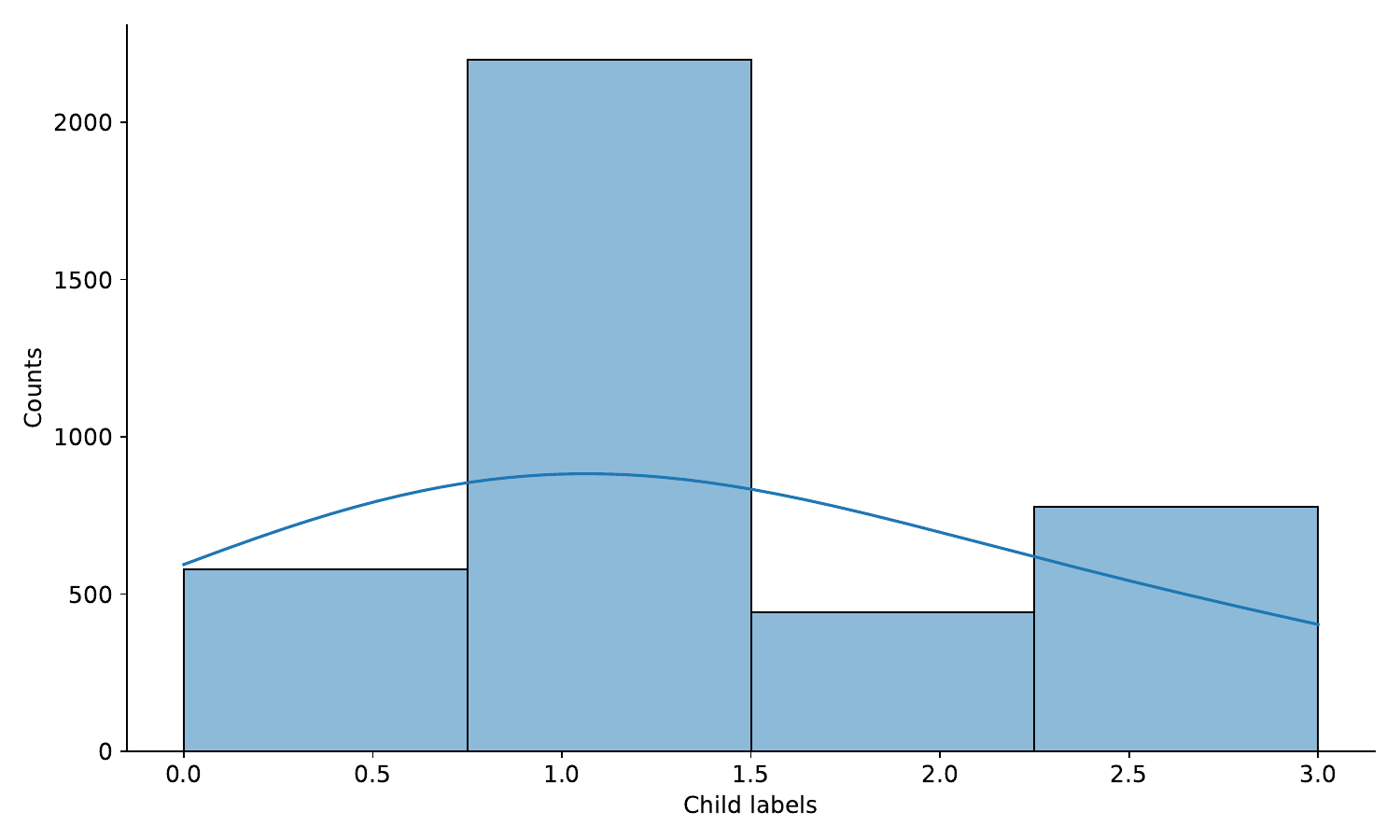} 
        \caption{Journalist C: Histogram of posterior predictive outcome}
    \end{subfigure}
    \hfill
    \begin{subfigure}[b]{0.45\textwidth}
        \centering
        \includegraphics[width=\textwidth]{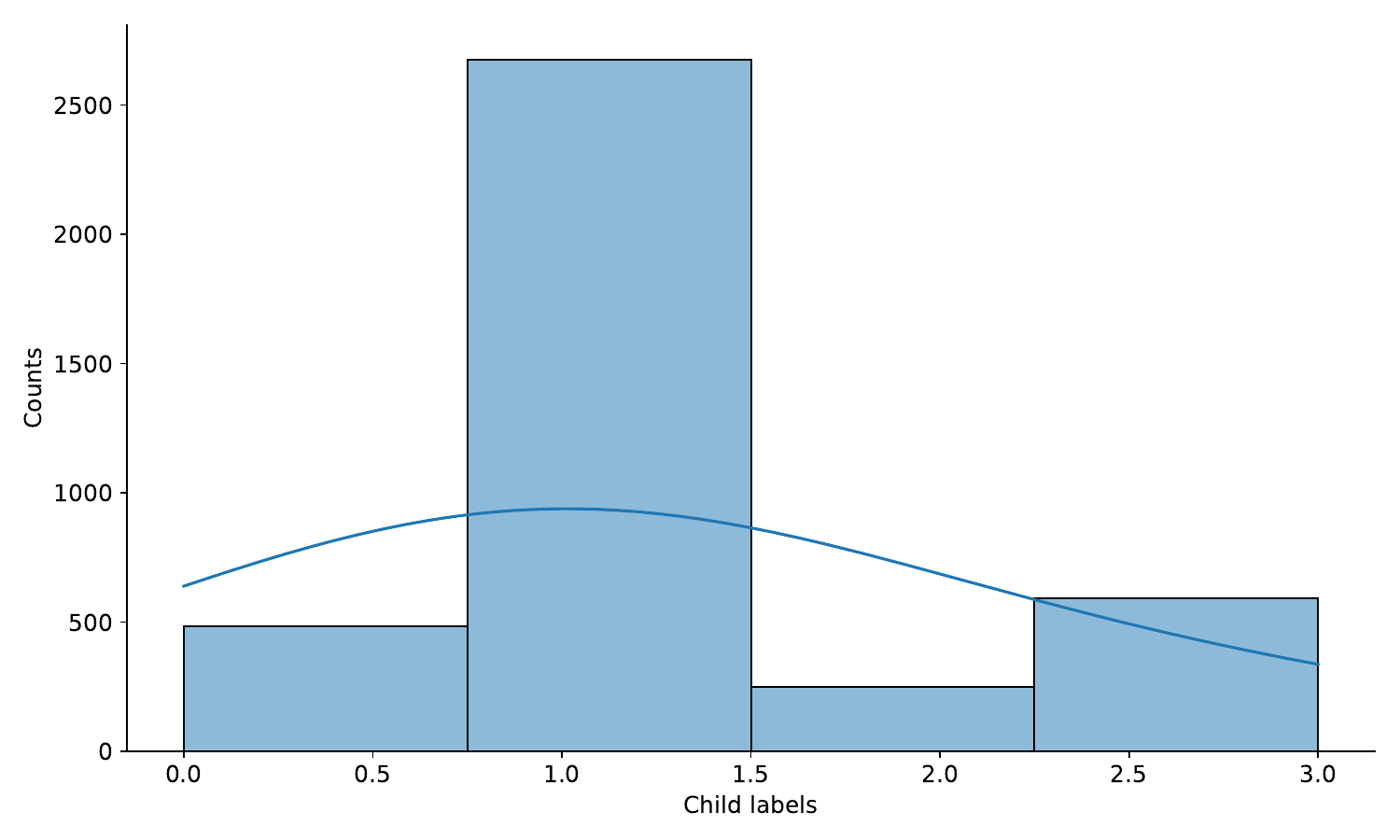} 
        \caption{Journalist C: Histogram of ground truth outcome distribution}
    \end{subfigure}
    \caption{Journalist C: Posterior predictive outcome distribution vs. ground truth outcome distribution}
\end{figure}

\begin{figure}[htbp]
    \begin{subfigure}[b]{\textwidth}
        \centering
        \includegraphics[width=\textwidth]{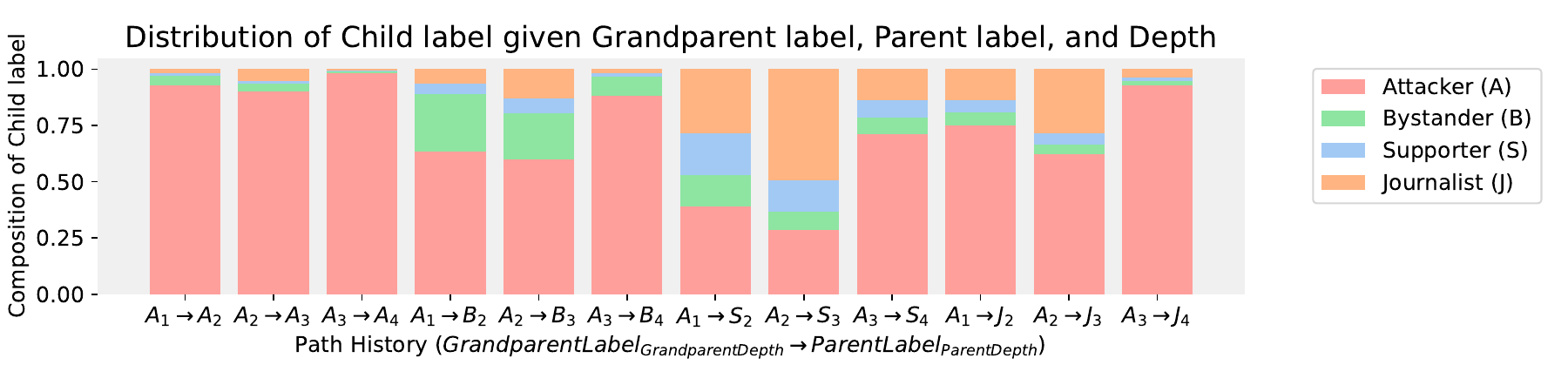} 
        \caption{Journalist D: attacker grandparent.}
    \end{subfigure}
    \begin{subfigure}[b]{\textwidth}
        \centering
        \includegraphics[width=\textwidth]{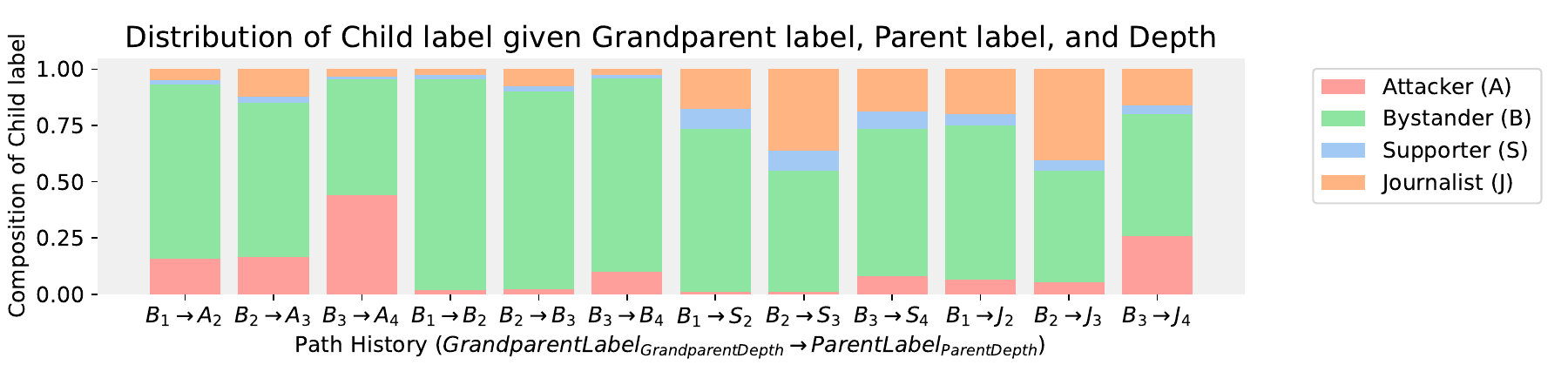} 
        \caption{Journalist D: bystander grandparent.}
    \end{subfigure}
    \begin{subfigure}[b]{\textwidth}
        \centering
        \includegraphics[width=\textwidth]{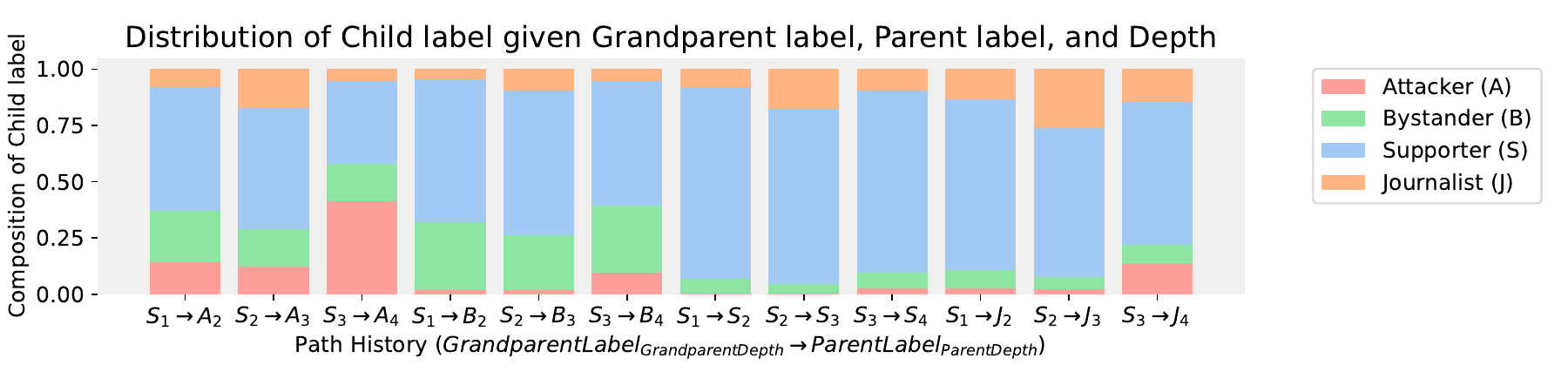} 
        \caption{Journalist D: supporter grandparent.}
    \end{subfigure}
    \begin{subfigure}[b]{\textwidth}
        \centering
        \includegraphics[width=\textwidth]{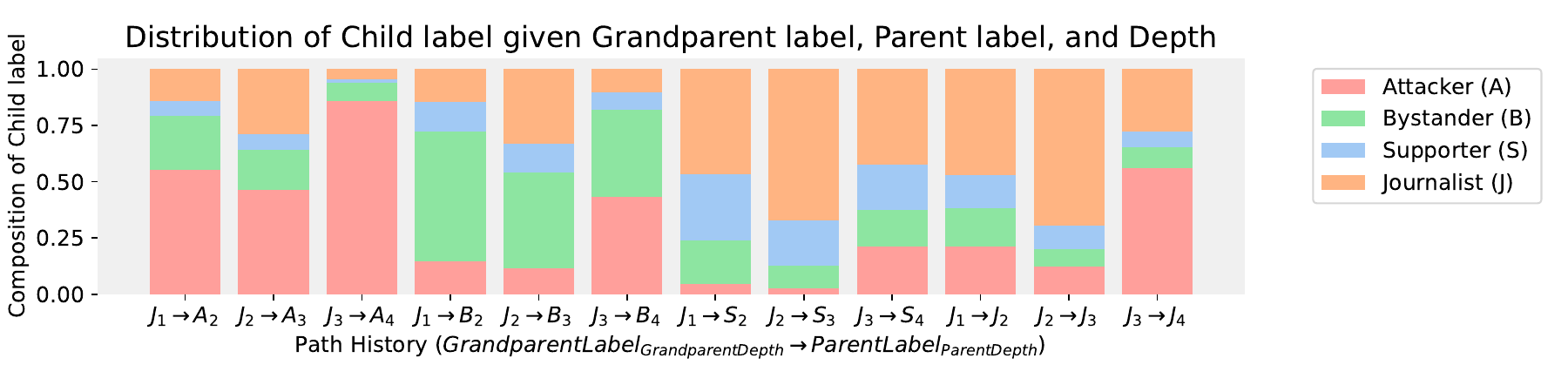} 
        \caption{Journalist D: journalist grandparent.}
    \end{subfigure}
    \caption{Journalist D: Effect of grandparent and parent label on child label distribution}
\end{figure}

\begin{figure}[ht]
    \centering
    \begin{subfigure}[b]{0.45\textwidth}
        \centering
        \includegraphics[width=\textwidth]{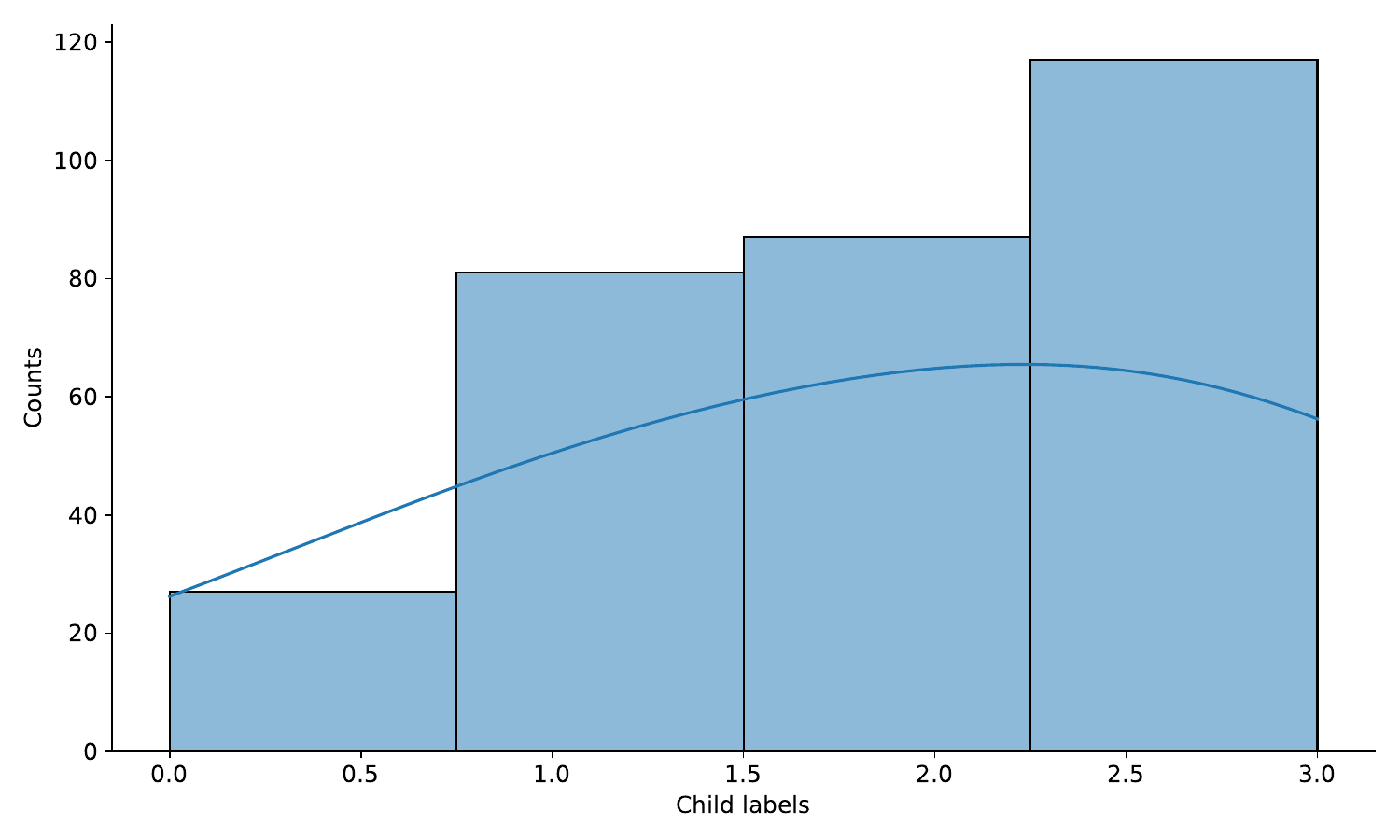} 
        \caption{Journalist D: Histogram of posterior predictive outcome}
    \end{subfigure}
    \hfill
    \begin{subfigure}[b]{0.45\textwidth}
        \centering
        \includegraphics[width=\textwidth]{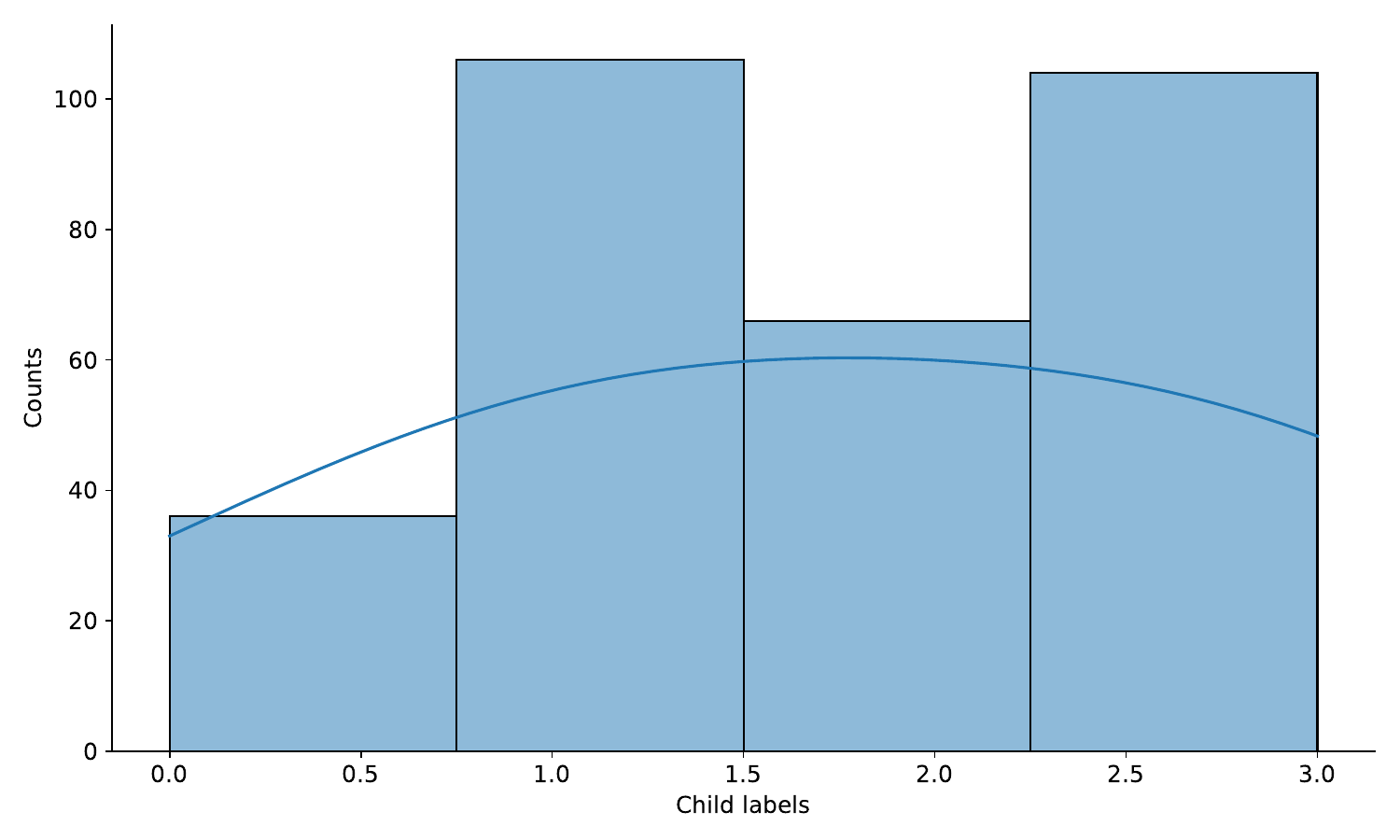} 
        \caption{Journalist D: Histogram of ground truth outcome distribution}
    \end{subfigure}
    \caption{Journalist D: Posterior predictive outcome distribution vs. ground truth outcome distribution}
\end{figure}

\begin{figure}[htbp]
    \begin{subfigure}[b]{\textwidth}
        \centering
        \includegraphics[width=\textwidth]{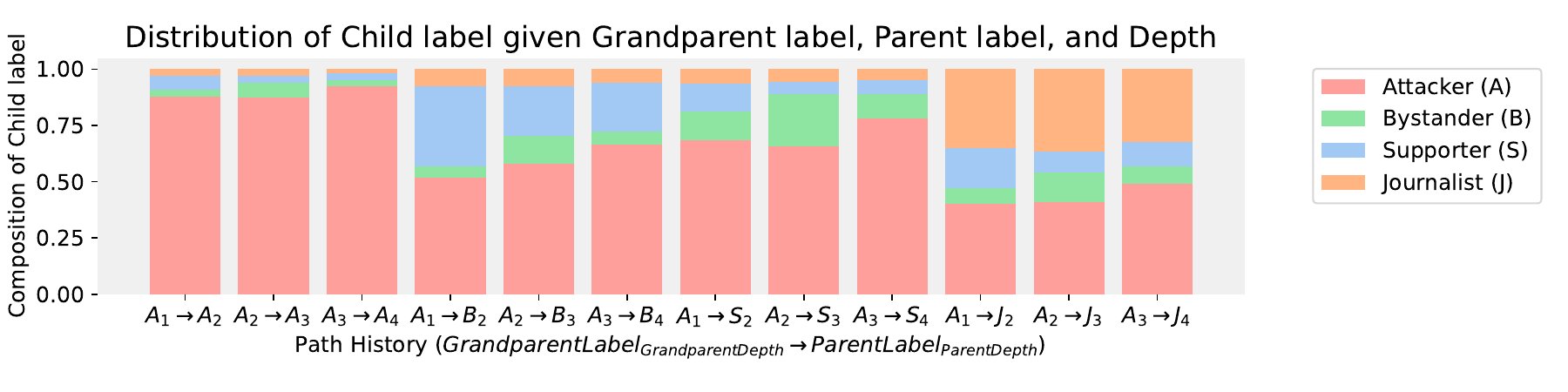} 
        \caption{Journalist E: attacker grandparent.}
    \end{subfigure}
    \begin{subfigure}[b]{\textwidth}
        \centering
        \includegraphics[width=\textwidth]{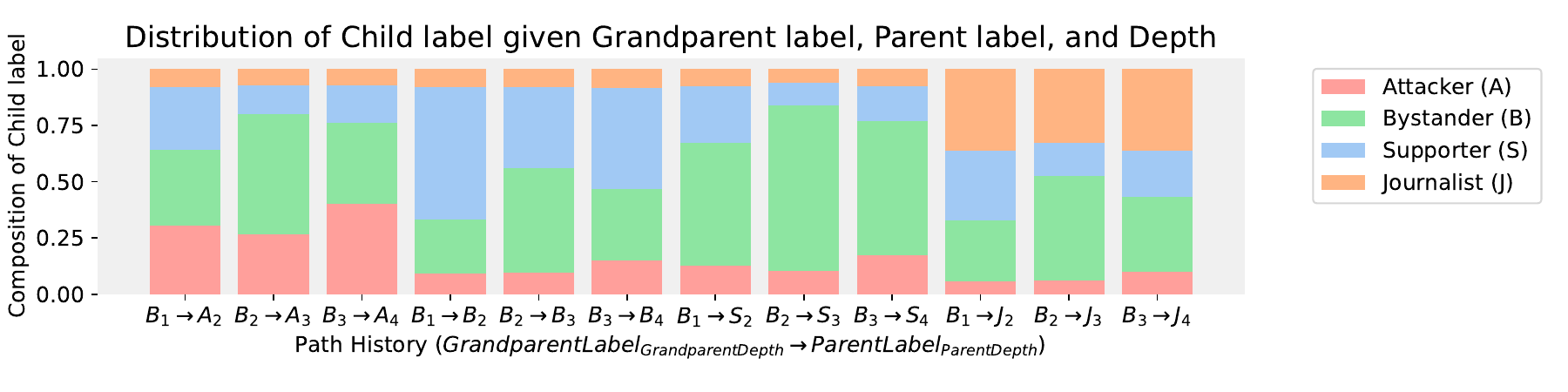} 
        \caption{Journalist E: bystander grandparent.}
    \end{subfigure}
    \begin{subfigure}[b]{\textwidth}
        \centering
        \includegraphics[width=\textwidth]{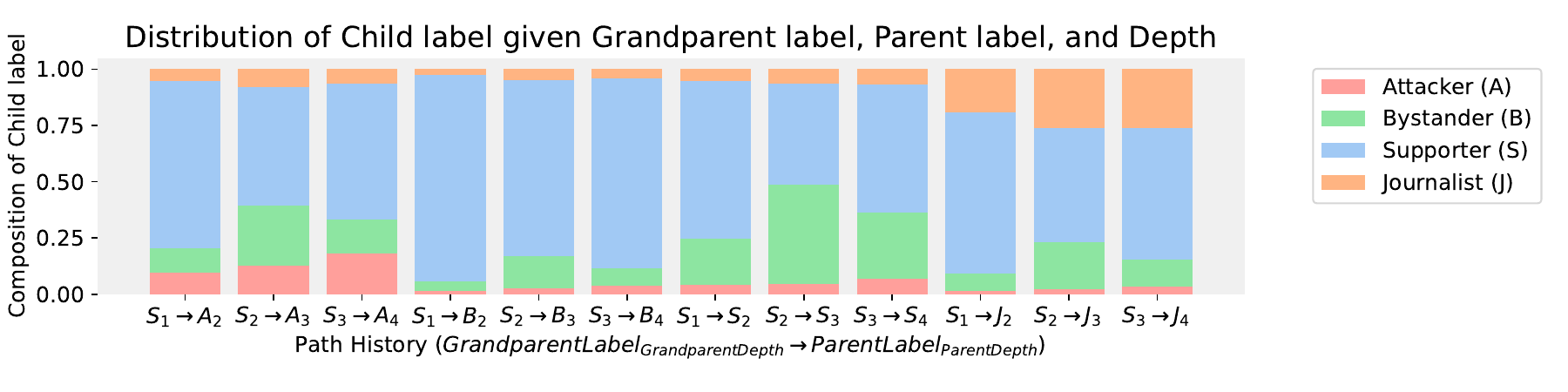} 
        \caption{Journalist E: supporter grandparent.}
    \end{subfigure}
    \begin{subfigure}[b]{\textwidth}
        \centering
        \includegraphics[width=\textwidth]{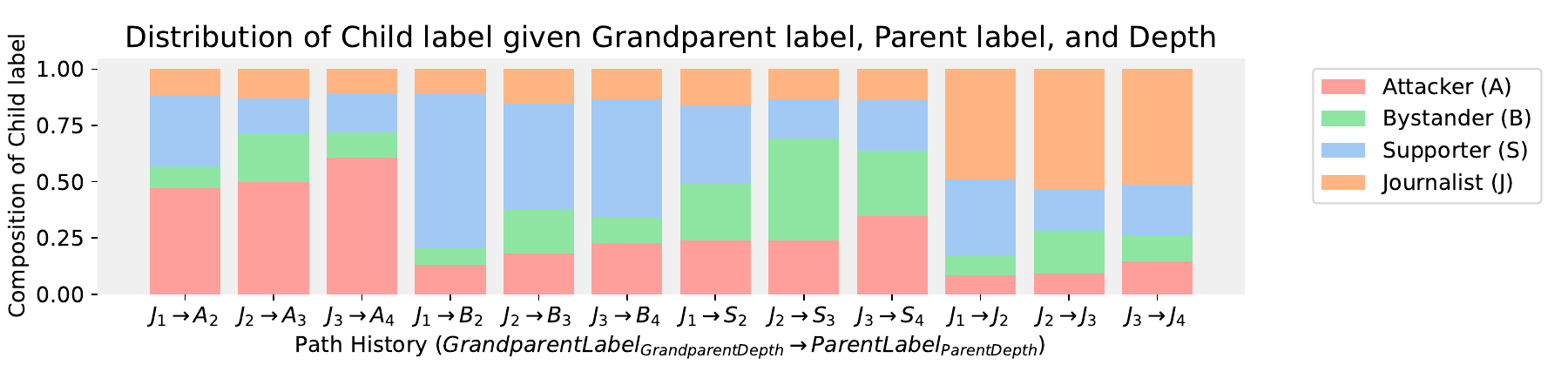} 
        \caption{Journalist E: journalist grandparent.}
    \end{subfigure}
    \caption{Journalist E: Effect of grandparent and parent label on child label distribution}
\end{figure}

\begin{figure}[ht]
    \centering
    \begin{subfigure}[b]{0.45\textwidth}
        \centering
        \includegraphics[width=\textwidth]{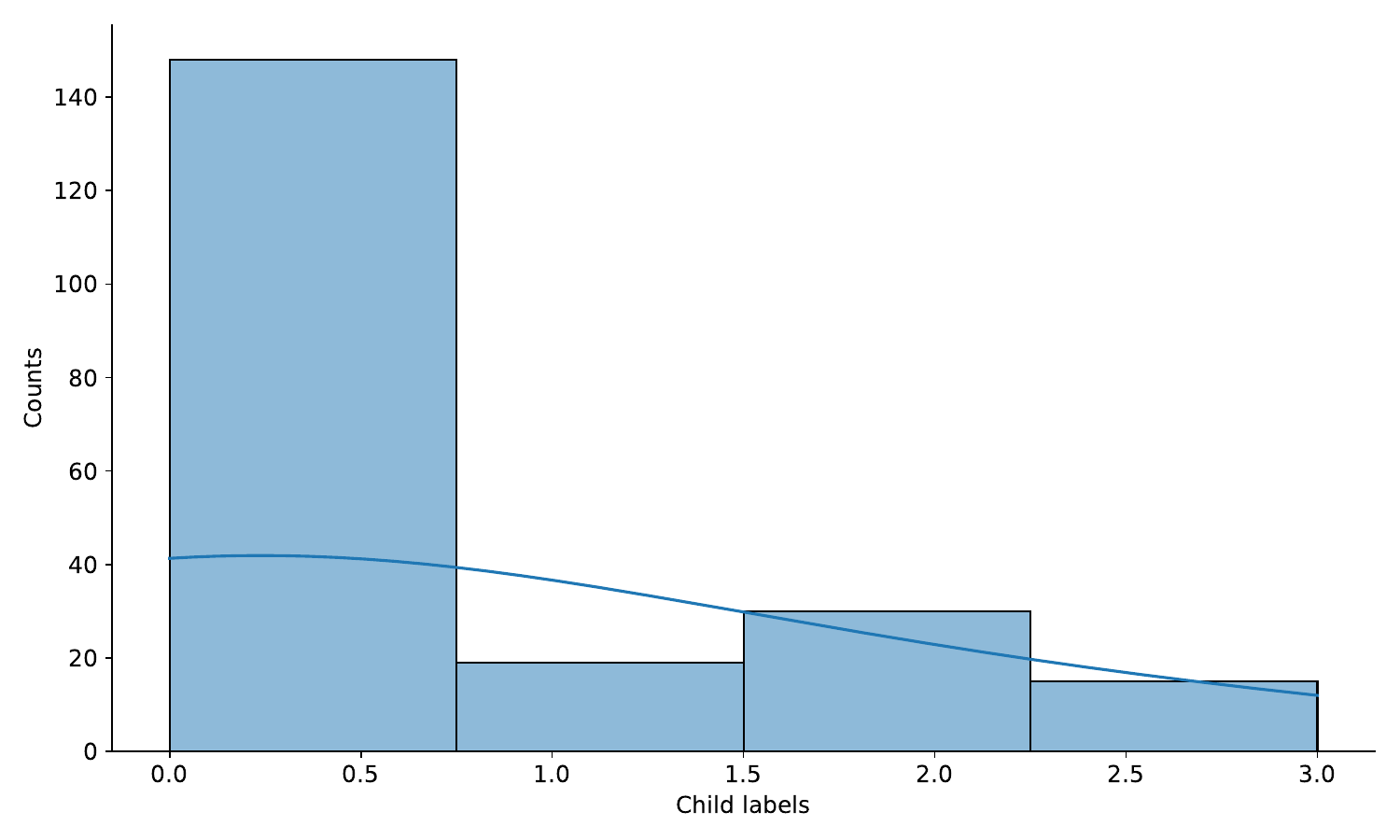} 
        \caption{Journalist E: Histogram of posterior predictive outcome}
    \end{subfigure}
    \hfill
    \begin{subfigure}[b]{0.45\textwidth}
        \centering
        \includegraphics[width=\textwidth]{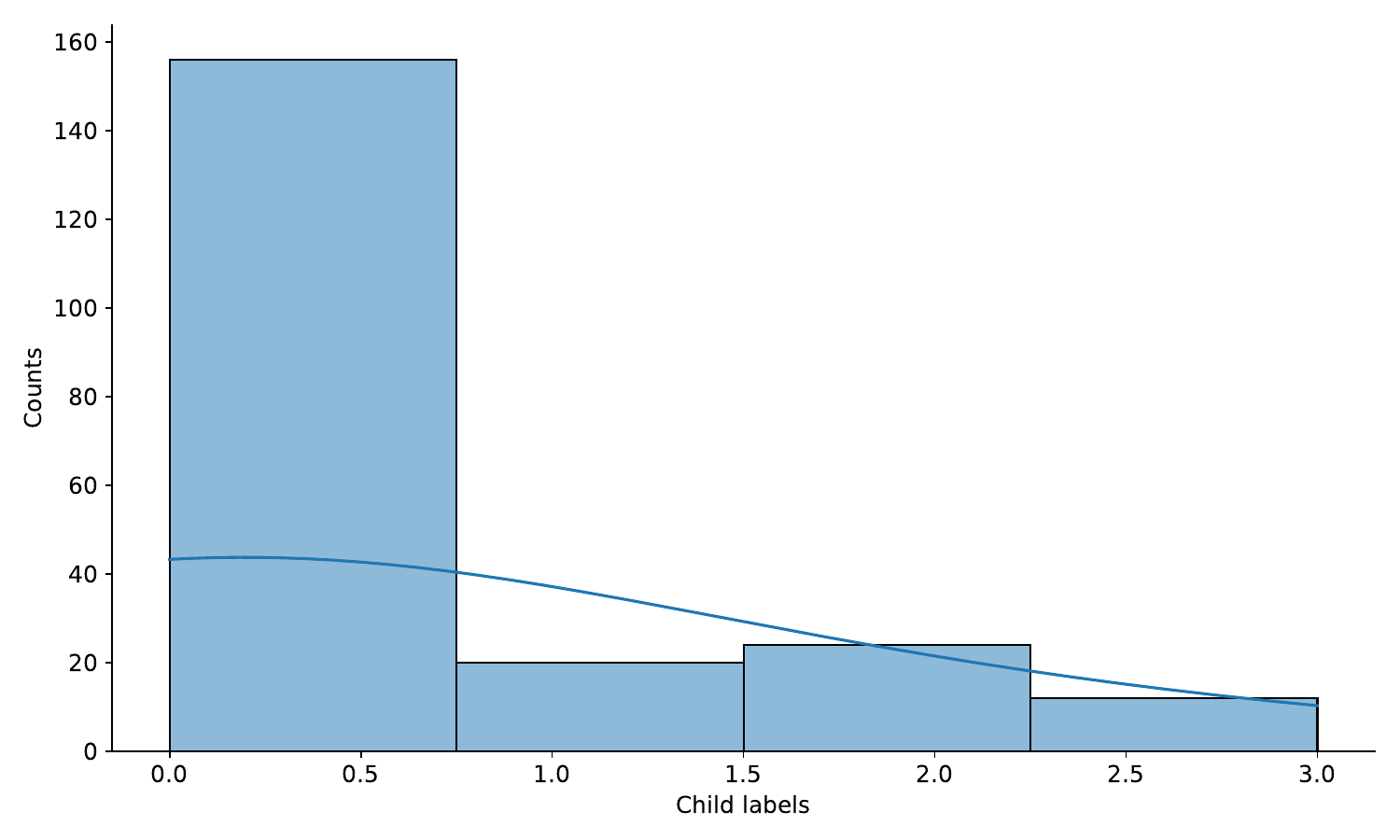} 
        \caption{Journalist E: Histogram of ground truth outcome distribution}
    \end{subfigure}
    \caption{Journalist E: Posterior predictive outcome distribution vs. ground truth outcome distribution}
\end{figure}

\begin{figure}[htbp]
    \begin{subfigure}[b]{\textwidth}
        \centering
        \includegraphics[width=\textwidth]{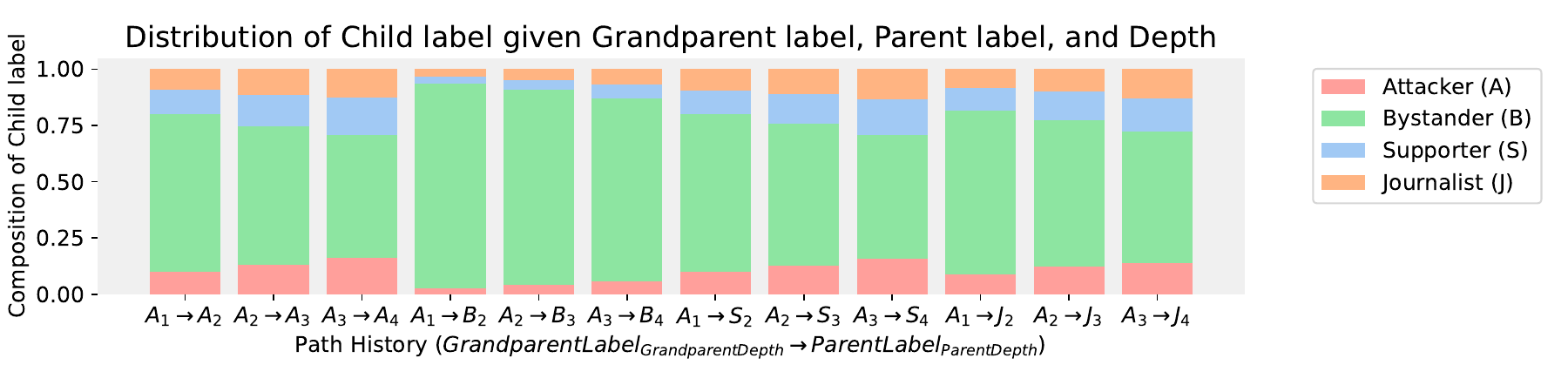} 
        \caption{Journalist F: attacker grandparent.}
    \end{subfigure}
    \begin{subfigure}[b]{\textwidth}
        \centering
        \includegraphics[width=\textwidth]{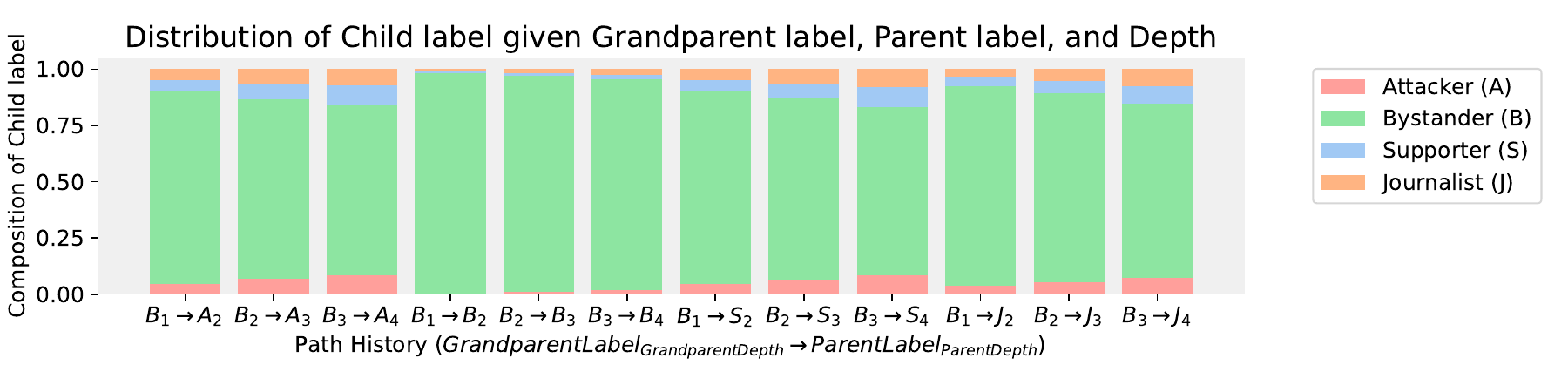} 
        \caption{Journalist F: bystander grandparent.}
    \end{subfigure}
    \begin{subfigure}[b]{\textwidth}
        \centering
        \includegraphics[width=\textwidth]{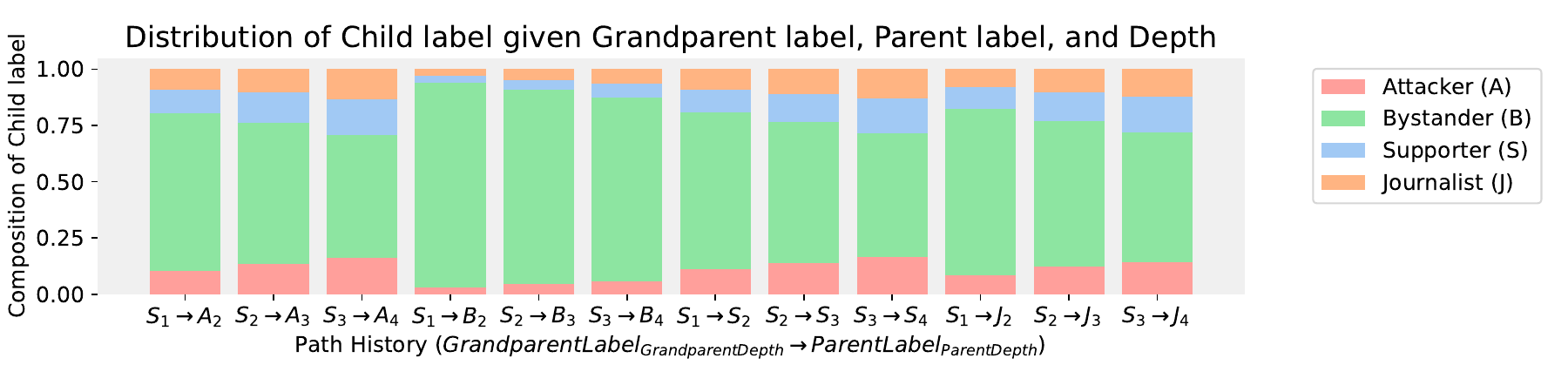} 
        \caption{Journalist F: supporter grandparent.}
    \end{subfigure}
    \begin{subfigure}[b]{\textwidth}
        \centering
        \includegraphics[width=\textwidth]{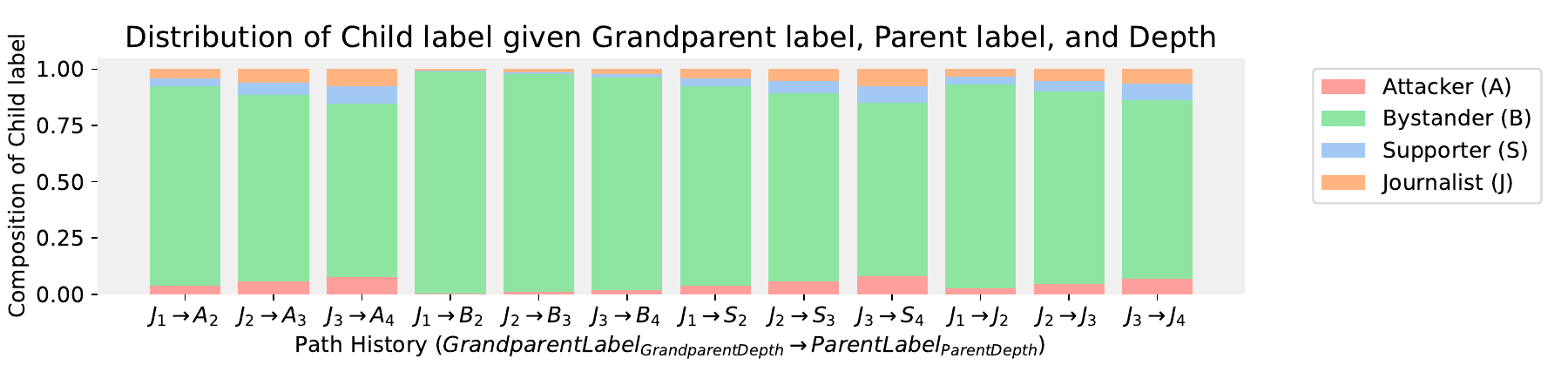} 
        \caption{Journalist F: journalist grandparent.}
    \end{subfigure}
    \caption{Journalist F: Effect of grandparent and parent label on child label distribution}
\end{figure}

\begin{figure}[ht]
    \centering
    \begin{subfigure}[b]{0.45\textwidth}
        \centering
        \includegraphics[width=\textwidth]{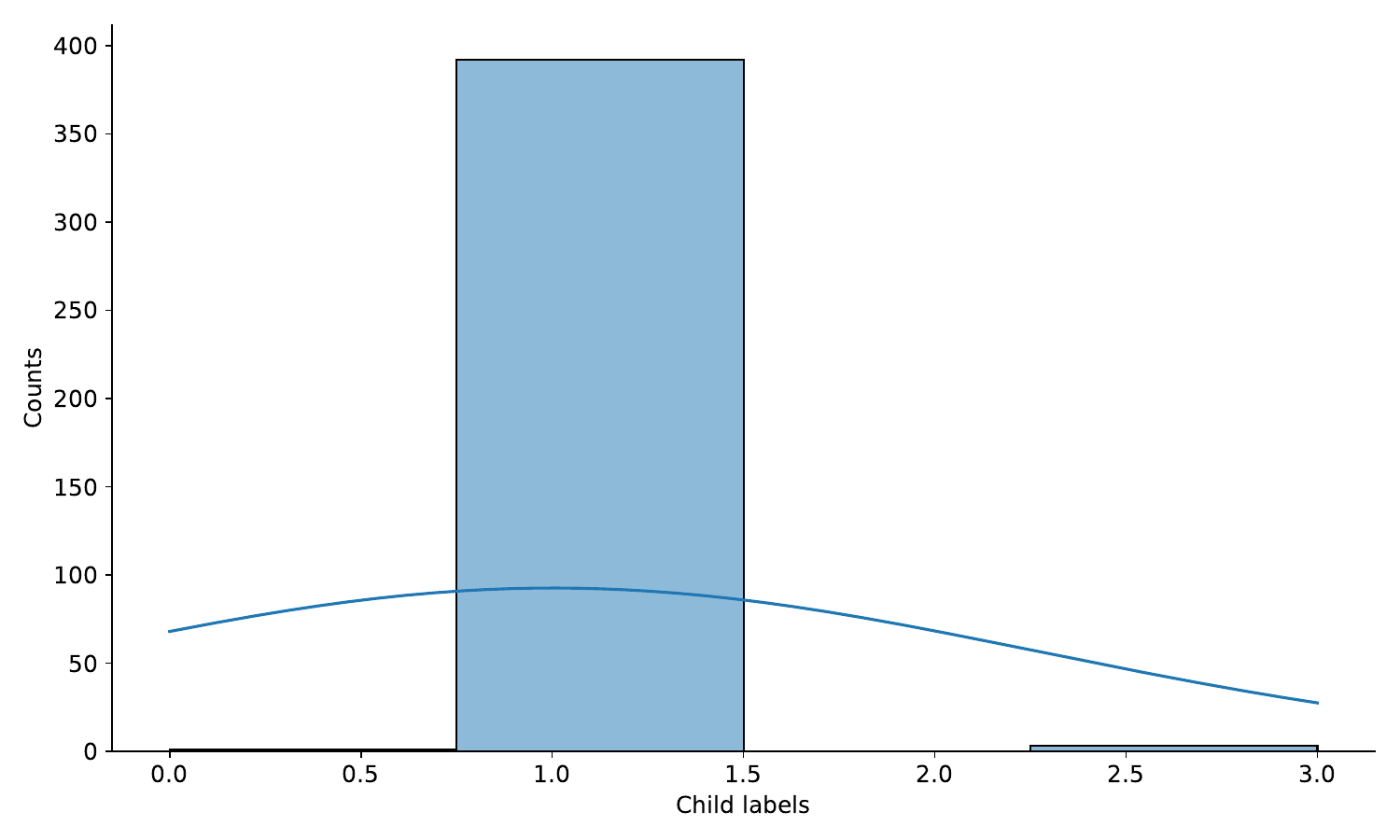} 
        \caption{Journalist F: Histogram of posterior predictive outcome}
    \end{subfigure}
    \hfill
    \begin{subfigure}[b]{0.45\textwidth}
        \centering
        \includegraphics[width=\textwidth]{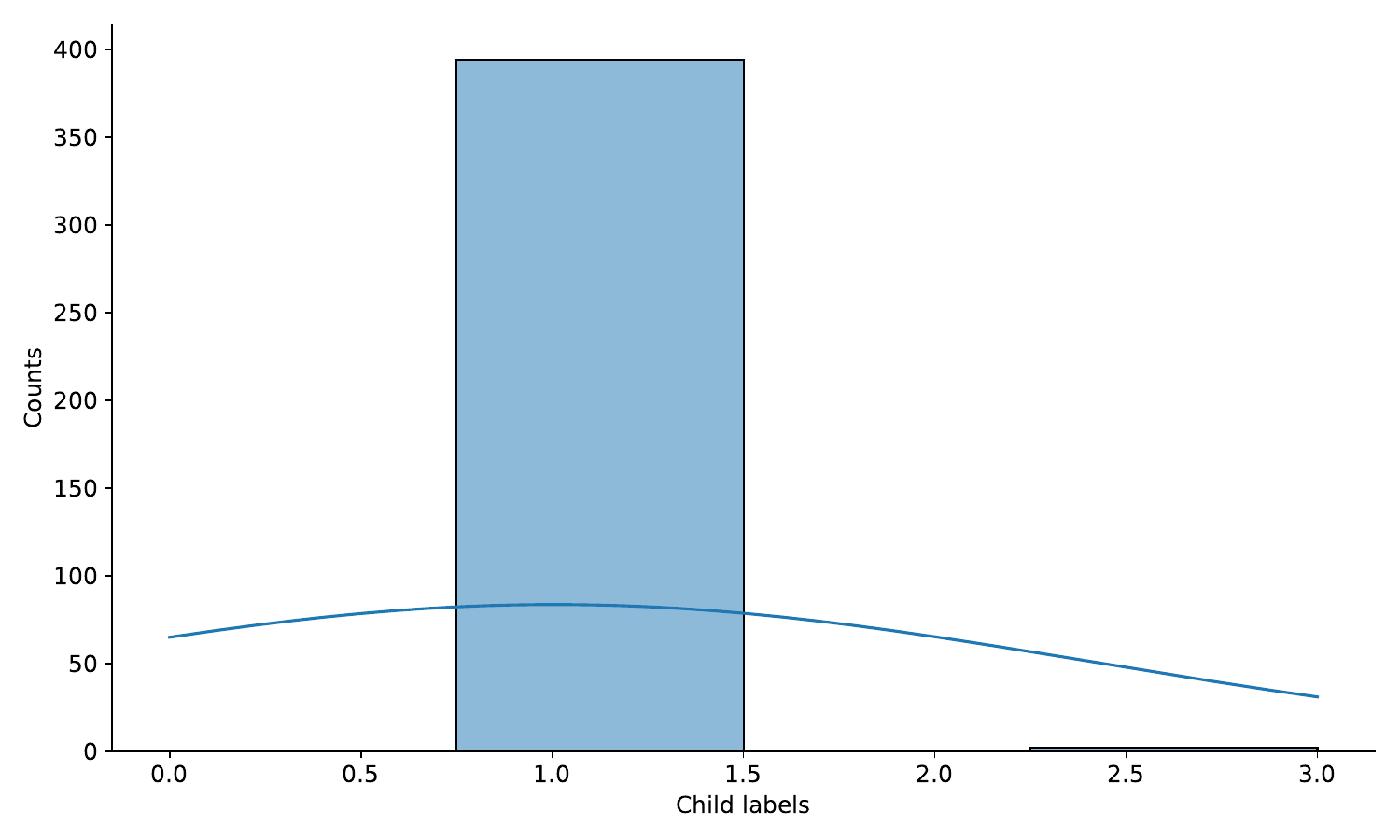} 
        \caption{Journalist F: Histogram of ground truth outcome distribution}
    \end{subfigure}
    \caption{Journalist F: Posterior predictive outcome distribution vs. ground truth outcome distribution}
\end{figure}

\begin{figure}[htbp]
    \begin{subfigure}[b]{\textwidth}
        \centering
        \includegraphics[width=\textwidth]{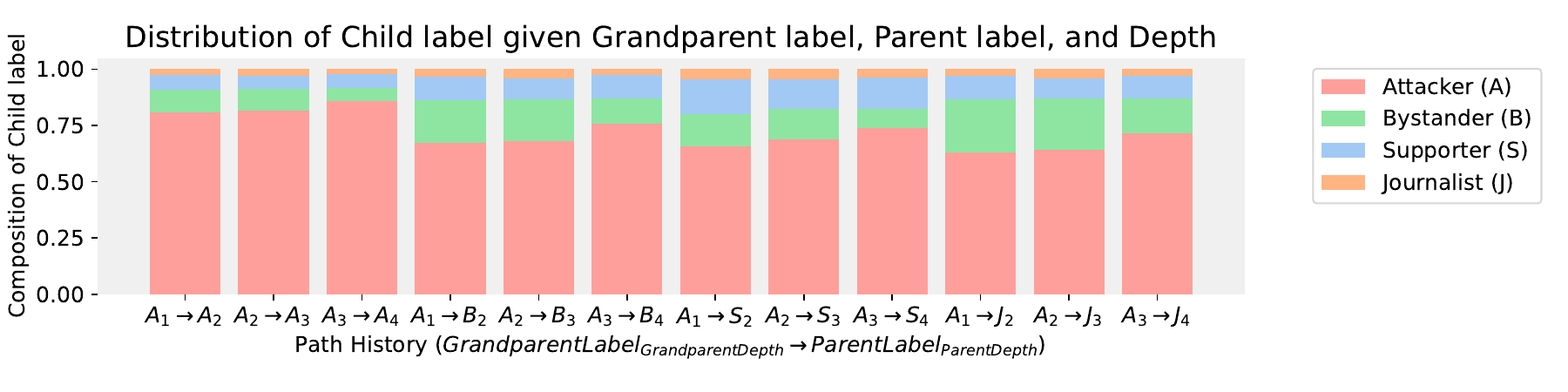} 
        \caption{Journalist G: attacker grandparent.}
    \end{subfigure}
    \begin{subfigure}[b]{\textwidth}
        \centering
        \includegraphics[width=\textwidth]{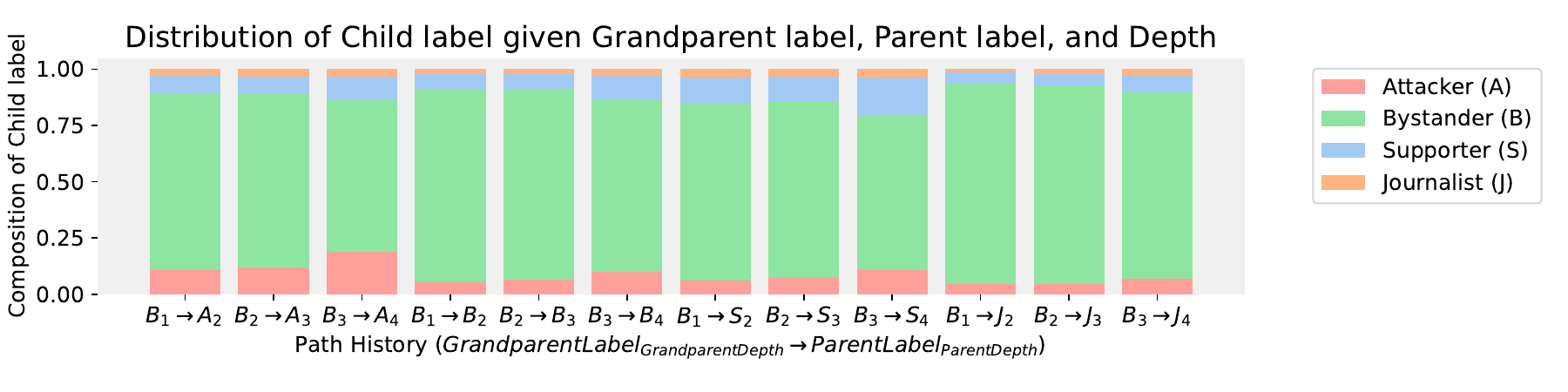} 
        \caption{Journalist G: bystander grandparent.}
    \end{subfigure}
    \begin{subfigure}[b]{\textwidth}
        \centering
        \includegraphics[width=\textwidth]{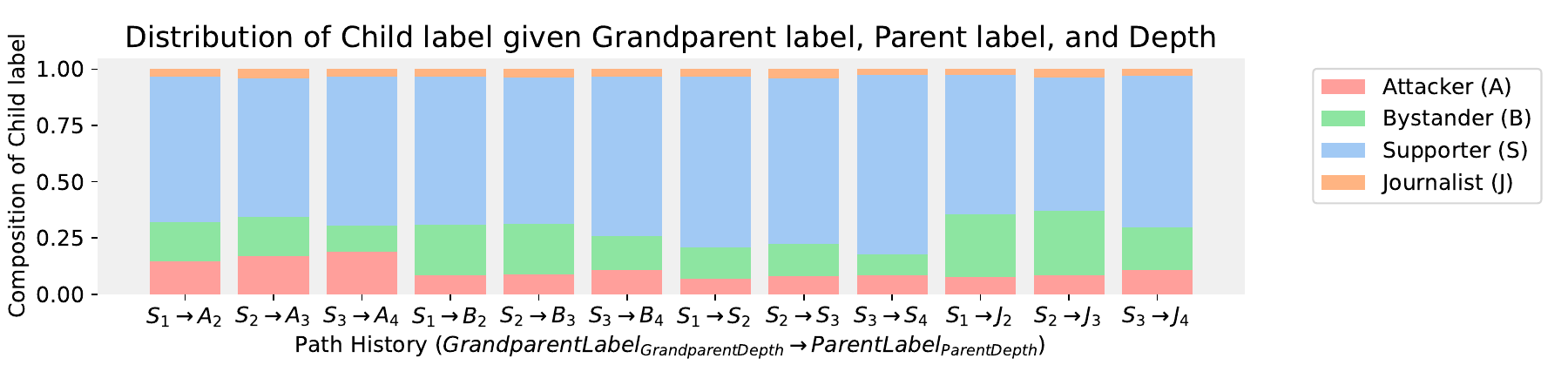} 
        \caption{Journalist G: supporter grandparent.}
    \end{subfigure}
    \begin{subfigure}[b]{\textwidth}
        \centering
        \includegraphics[width=\textwidth]{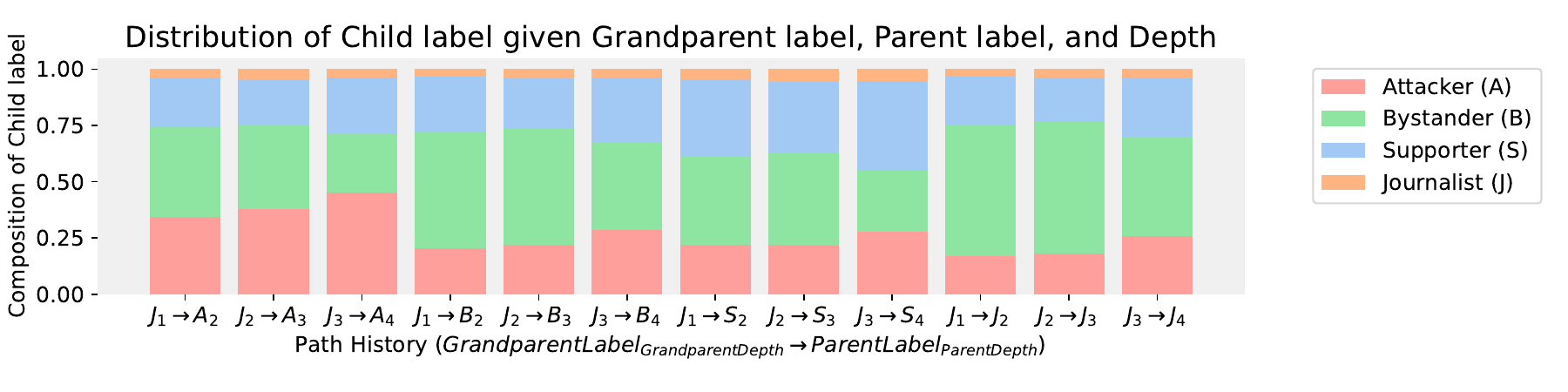} 
        \caption{Journalist G: journalist grandparent.}
    \end{subfigure}
    \caption{Journalist G: Effect of grandparent and parent label on child label distribution}
\end{figure}

\begin{figure}[ht]
    \centering
    \begin{subfigure}[b]{0.45\textwidth}
        \centering
        \includegraphics[width=\textwidth]{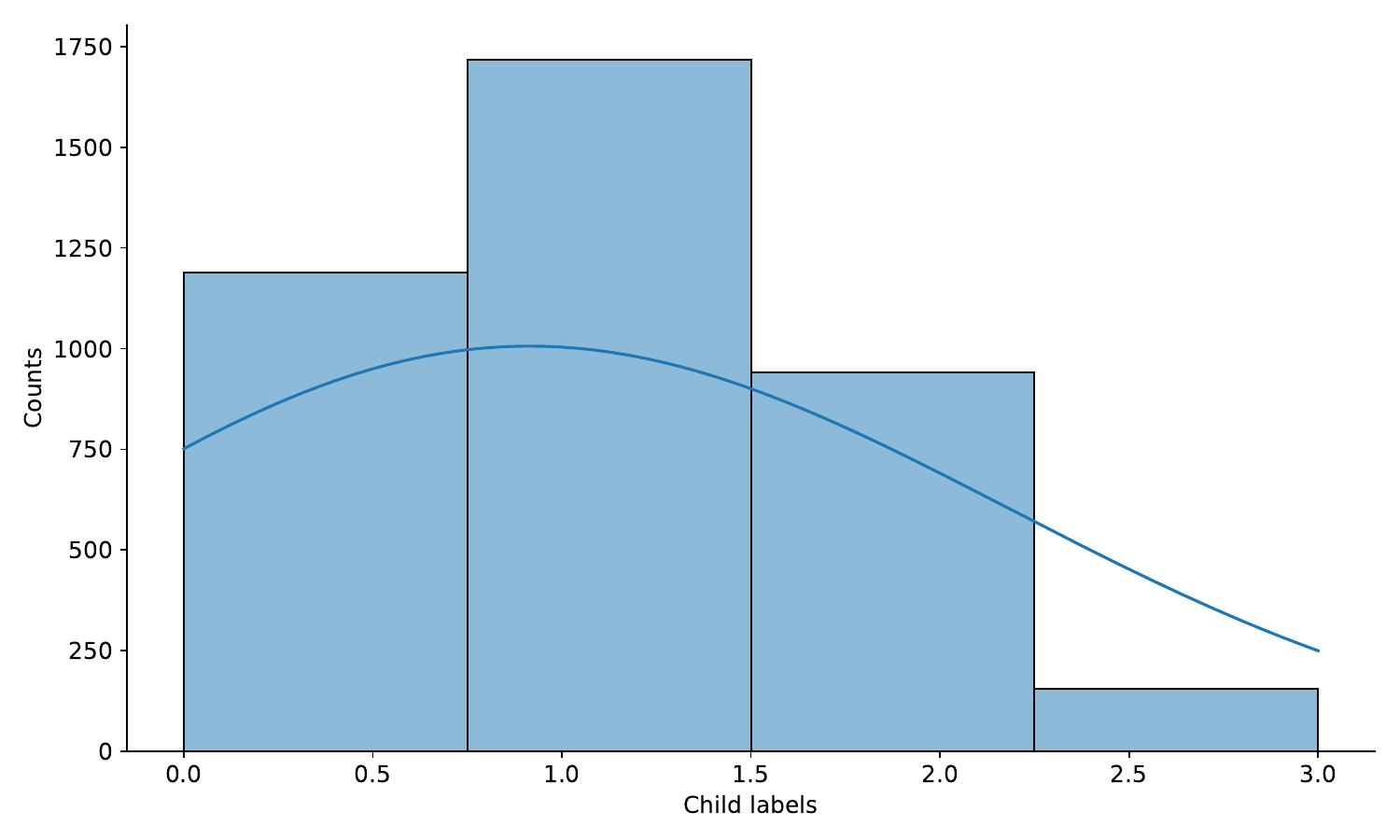} 
        \caption{Journalist G: Histogram of posterior predictive outcome}
    \end{subfigure}
    \hfill
    \begin{subfigure}[b]{0.45\textwidth}
        \centering
        \includegraphics[width=\textwidth]{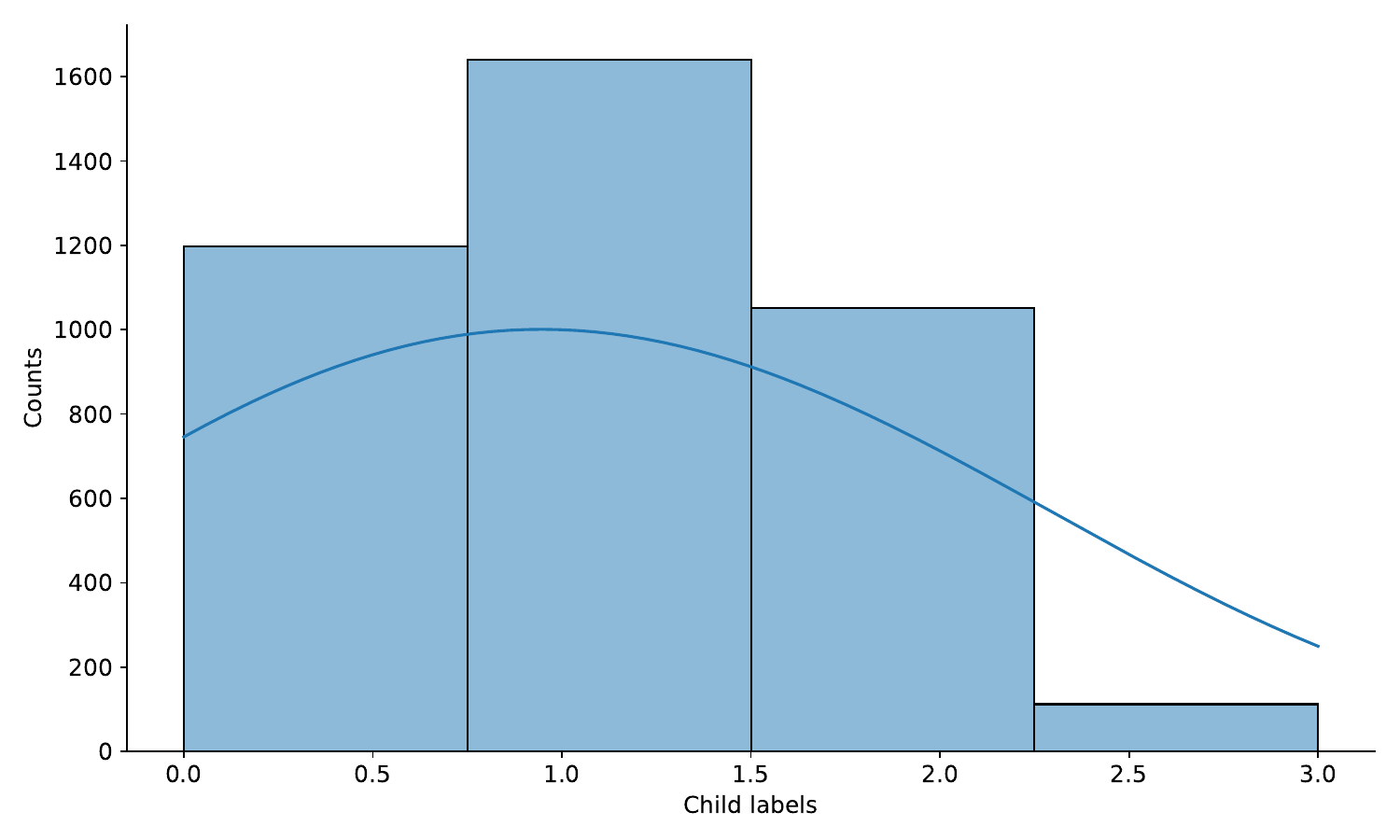} 
        \caption{Journalist G: Histogram of ground truth outcome distribution}
    \end{subfigure}
    \caption{Journalist G: Posterior predictive outcome distribution vs. ground truth outcome distribution}
\end{figure}

\begin{figure}[htbp]
    \begin{subfigure}[b]{\textwidth}
        \centering
        \includegraphics[width=\textwidth]{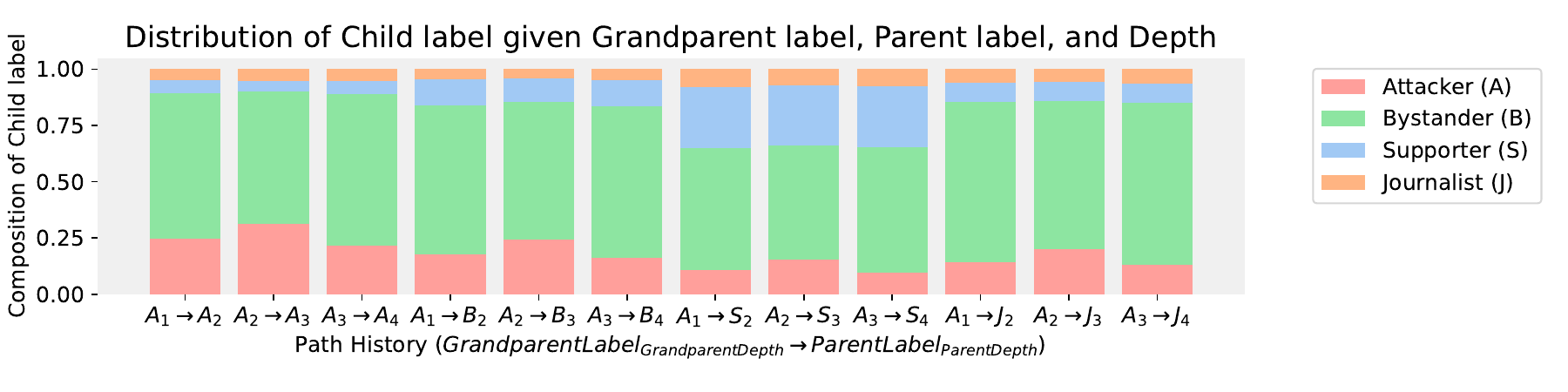} 
        \caption{Journalist H: attacker grandparent.}
    \end{subfigure}
    \begin{subfigure}[b]{\textwidth}
        \centering
        \includegraphics[width=\textwidth]{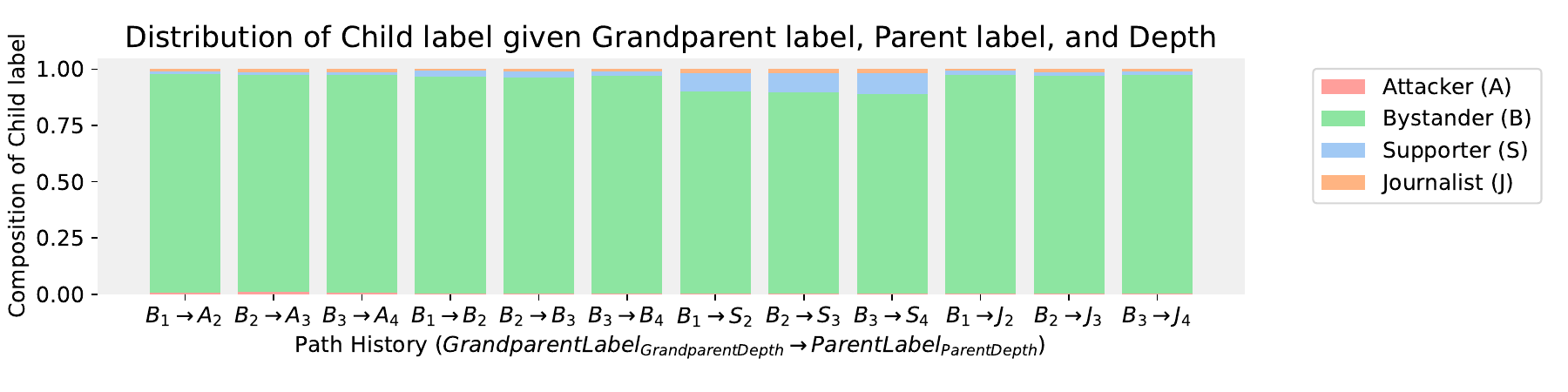} 
        \caption{Journalist H: bystander grandparent.}
    \end{subfigure}
    \begin{subfigure}[b]{\textwidth}
        \centering
        \includegraphics[width=\textwidth]{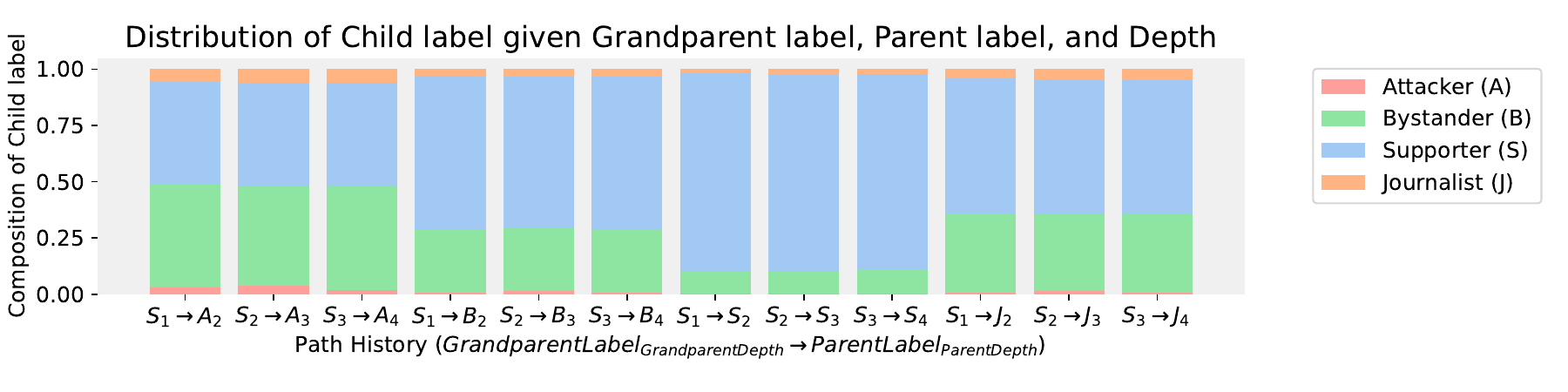} 
        \caption{Journalist H: supporter grandparent.}
    \end{subfigure}
    \begin{subfigure}[b]{\textwidth}
        \centering
        \includegraphics[width=\textwidth]{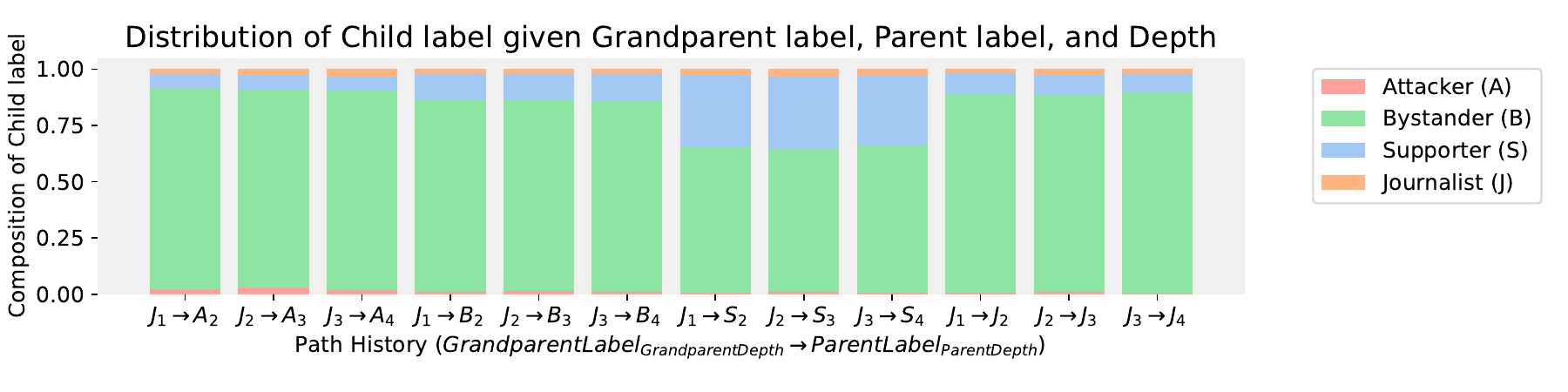} 
        \caption{Journalist H: journalist grandparent.}
    \end{subfigure}
    \caption{Journalist H: Effect of grandparent and parent label on child label distribution}
\end{figure}

\begin{figure}[ht]
    \centering
    \begin{subfigure}[b]{0.45\textwidth}
        \centering
        \includegraphics[width=\textwidth]{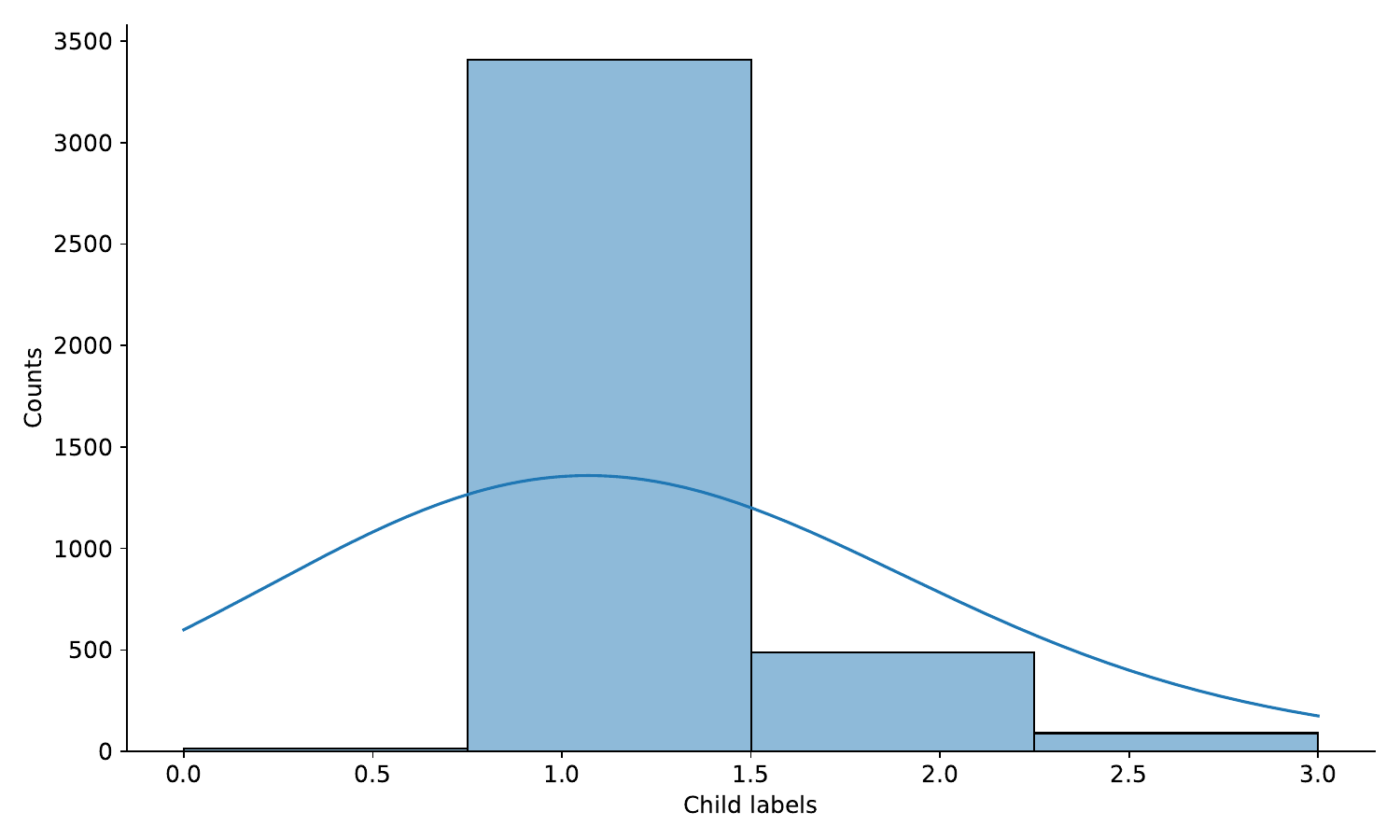} 
        \caption{Journalist H: Histogram of posterior predictive outcome}
    \end{subfigure}
    \hfill
    \begin{subfigure}[b]{0.45\textwidth}
        \centering
        \includegraphics[width=\textwidth]{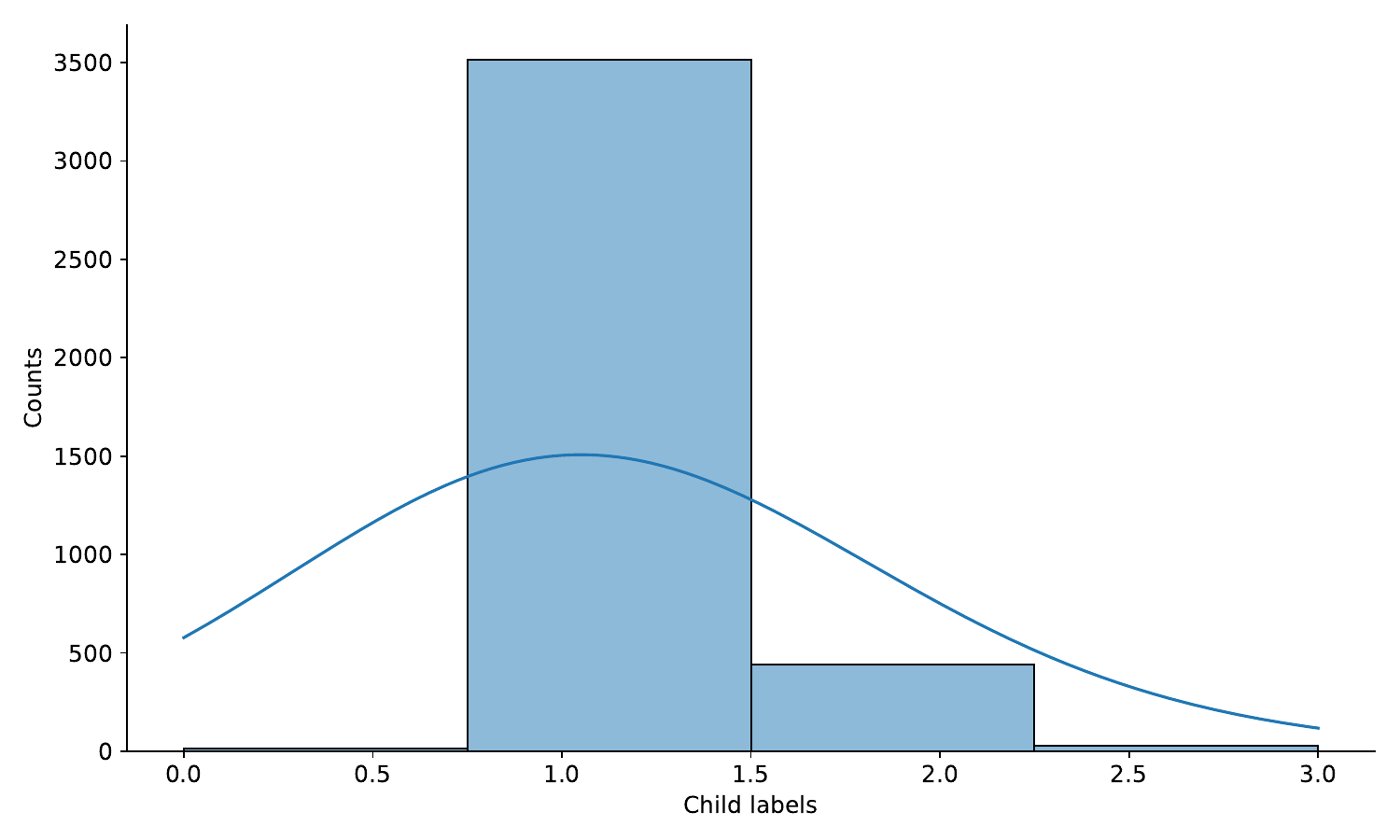} 
        \caption{Journalist H: Histogram of ground truth outcome distribution}
    \end{subfigure}
    \caption{Journalist H: Posterior predictive outcome distribution vs. ground truth outcome distribution}
\end{figure}

\begin{figure}[htbp]
    \begin{subfigure}[b]{\textwidth}
        \centering
        \includegraphics[width=\textwidth]{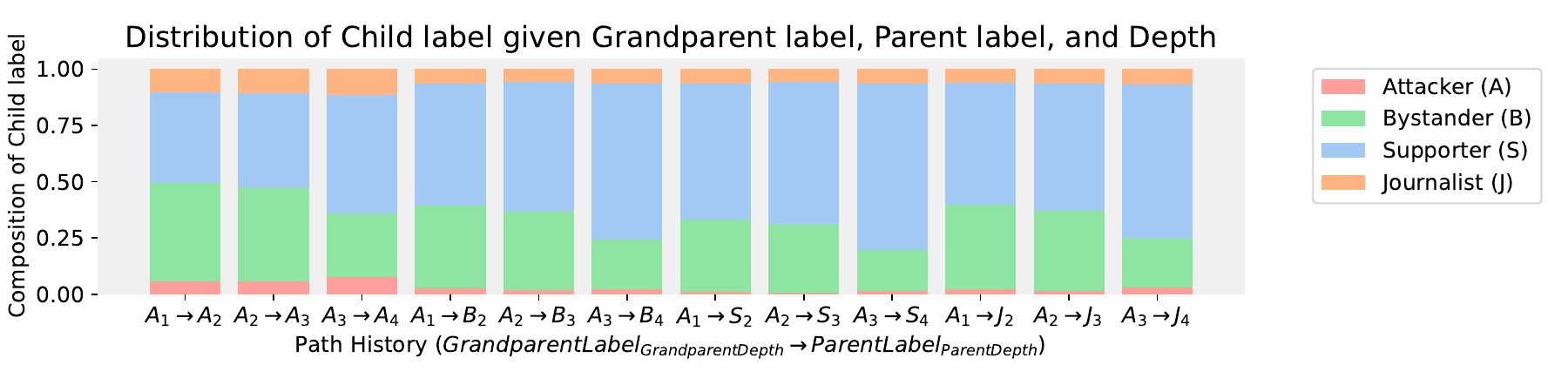} 
        \caption{Journalist I: attacker grandparent.}
    \end{subfigure}
    \begin{subfigure}[b]{\textwidth}
        \centering
        \includegraphics[width=\textwidth]{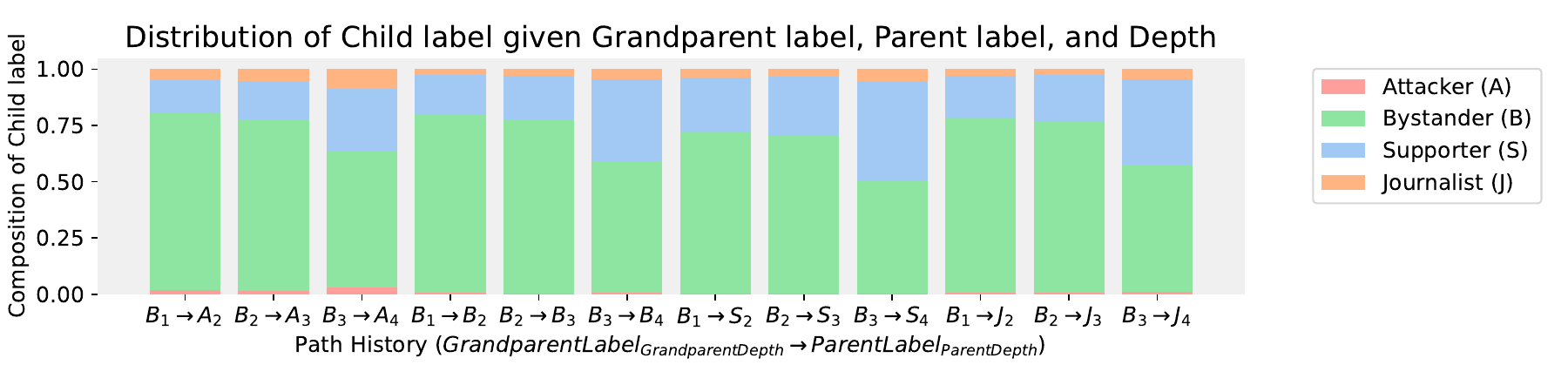} 
        \caption{Journalist I: bystander grandparent.}
    \end{subfigure}
    \begin{subfigure}[b]{\textwidth}
        \centering
        \includegraphics[width=\textwidth]{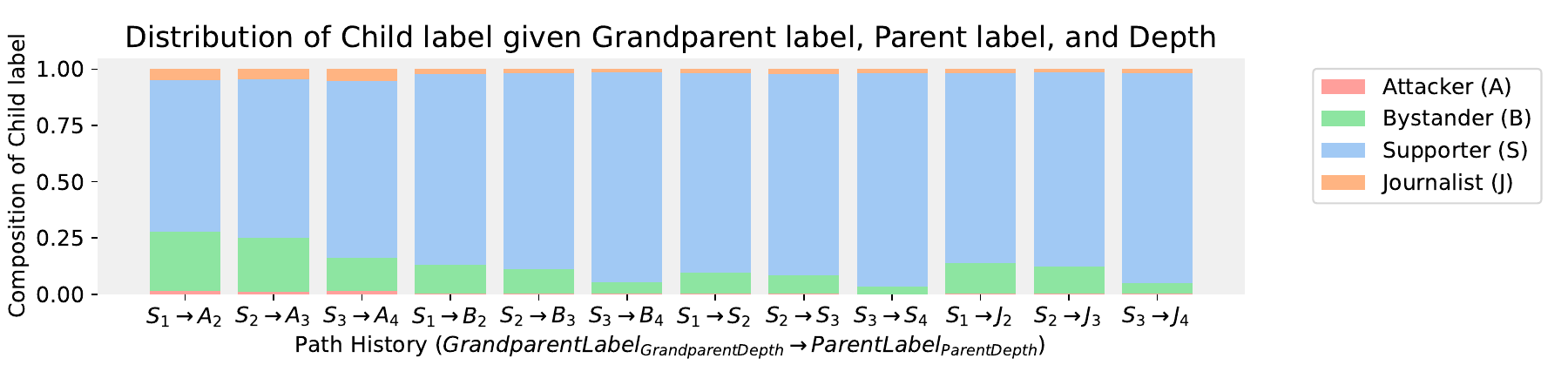} 
        \caption{Journalist I: supporter grandparent.}
    \end{subfigure}
    \begin{subfigure}[b]{\textwidth}
        \centering
        \includegraphics[width=\textwidth]{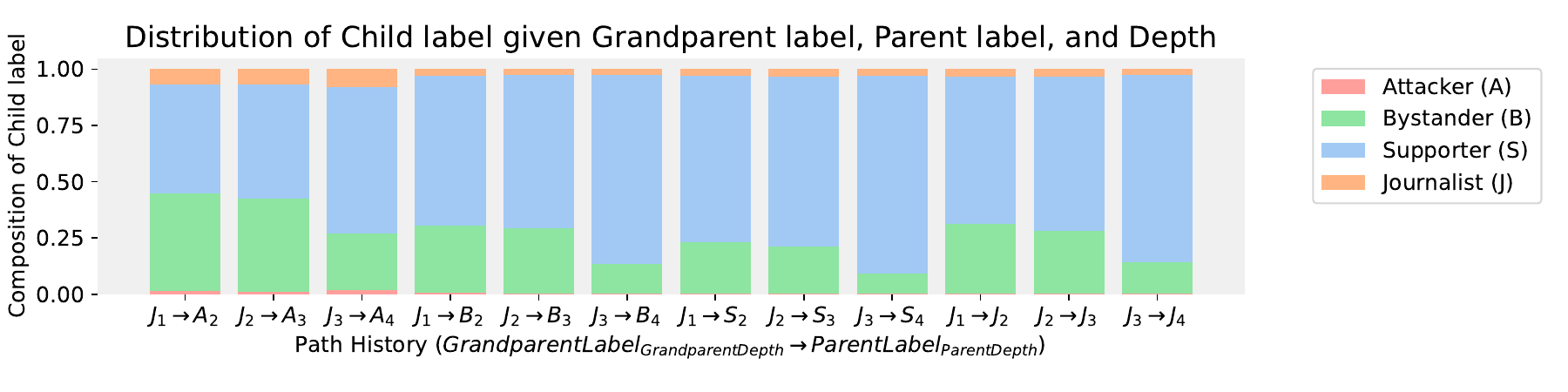} 
        \caption{Journalist I: journalist grandparent.}
    \end{subfigure}
    \caption{Journalist I: Effect of grandparent and parent label on child label distribution}
\end{figure}

\begin{figure}[ht]
    \centering
    \begin{subfigure}[b]{0.45\textwidth}
        \centering
        \includegraphics[width=\textwidth]{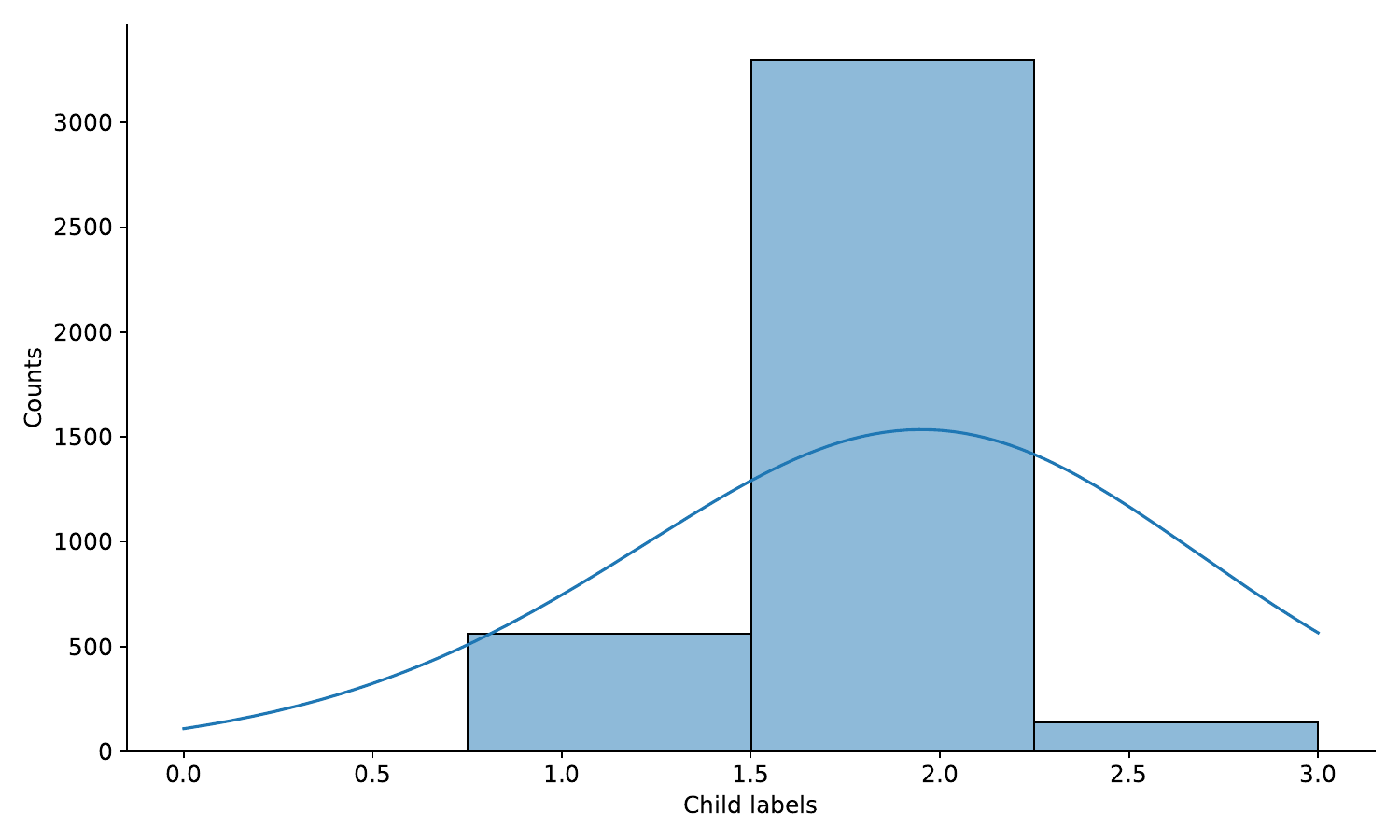} 
        \caption{Journalist I: Histogram of posterior predictive outcome}
    \end{subfigure}
    \hfill
    \begin{subfigure}[b]{0.45\textwidth}
        \centering
        \includegraphics[width=\textwidth]{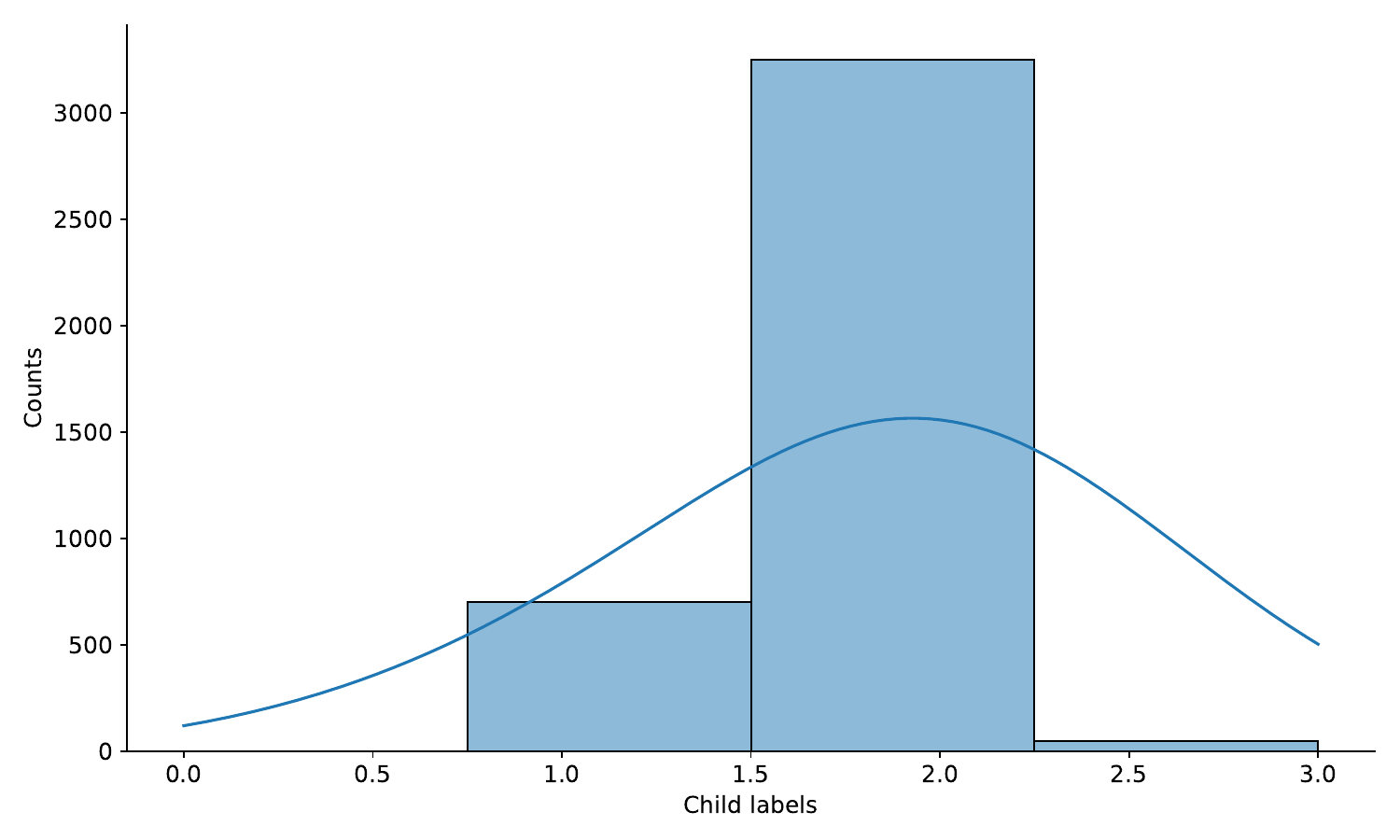} 
        \caption{Journalist I: Histogram of ground truth outcome distribution}
    \end{subfigure}
    \caption{Journalist I: Posterior predictive outcome distribution vs. ground truth outcome distribution}
\end{figure}

\begin{figure}[htbp]
    \begin{subfigure}[b]{\textwidth}
        \centering
        \includegraphics[width=\textwidth]{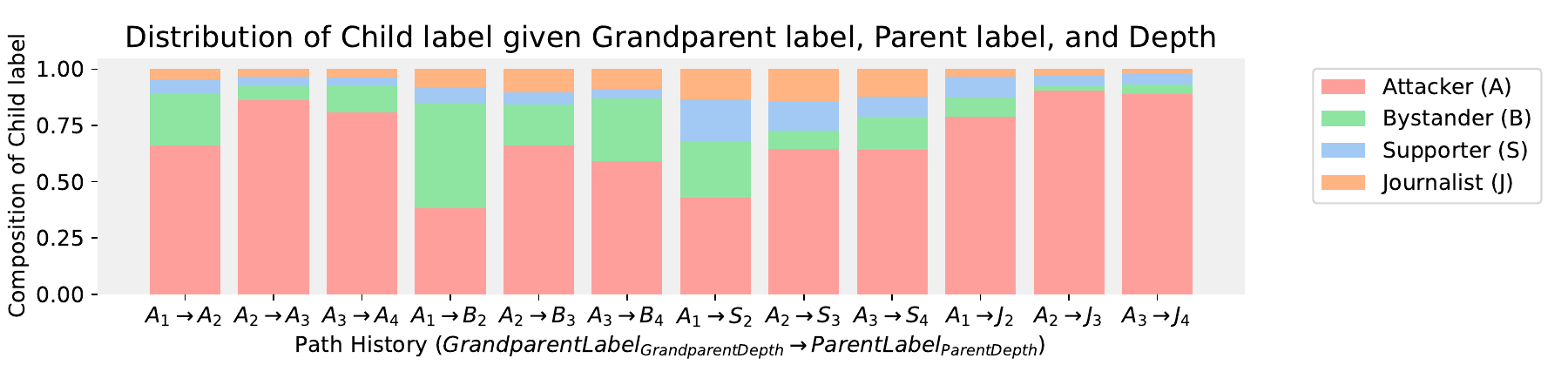} 
        \caption{Journalist J: attacker grandparent.}
    \end{subfigure}
    \begin{subfigure}[b]{\textwidth}
        \centering
        \includegraphics[width=\textwidth]{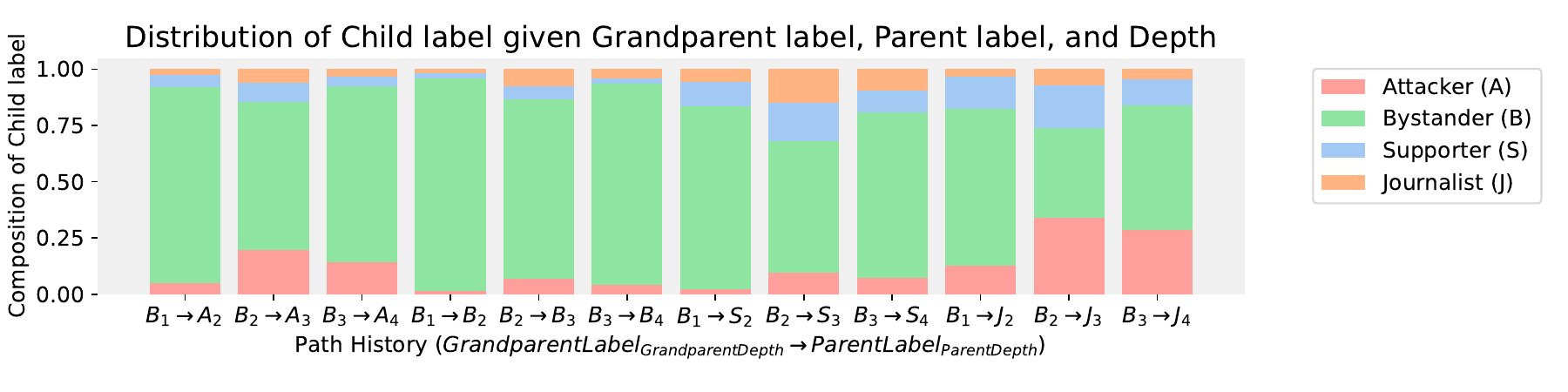} 
        \caption{Journalist J: bystander grandparent.}
    \end{subfigure}
    \begin{subfigure}[b]{\textwidth}
        \centering
        \includegraphics[width=\textwidth]{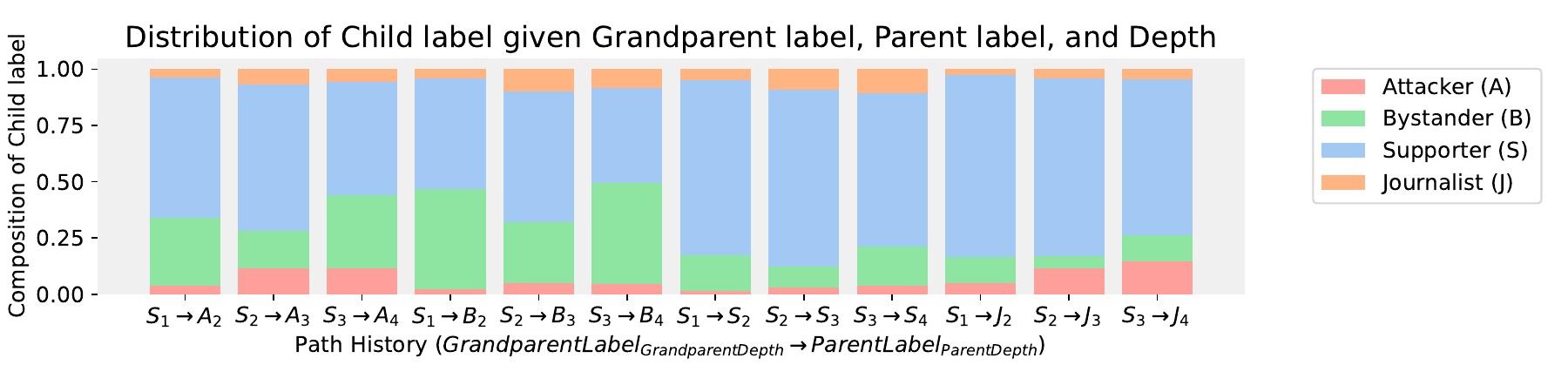} 
        \caption{Journalist J: supporter grandparent.}
    \end{subfigure}
    \begin{subfigure}[b]{\textwidth}
        \centering
        \includegraphics[width=\textwidth]{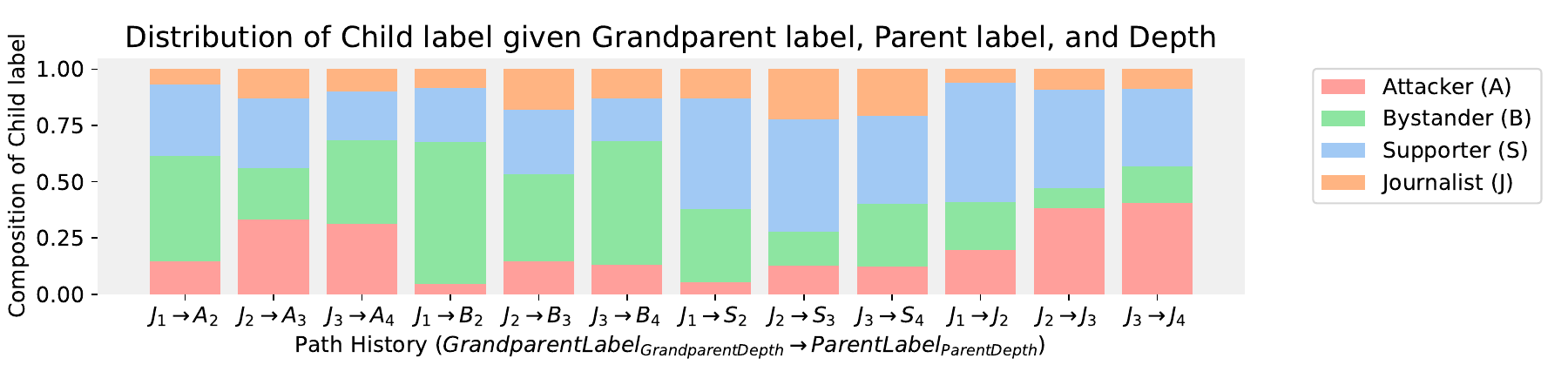} 
        \caption{Journalist J: journalist grandparent.}
    \end{subfigure}
    \caption{Journalist J: Effect of grandparent and parent label on child label distribution}
\end{figure}

\begin{figure}[ht]
    \centering
    \begin{subfigure}[b]{0.45\textwidth}
        \centering
        \includegraphics[width=\textwidth]{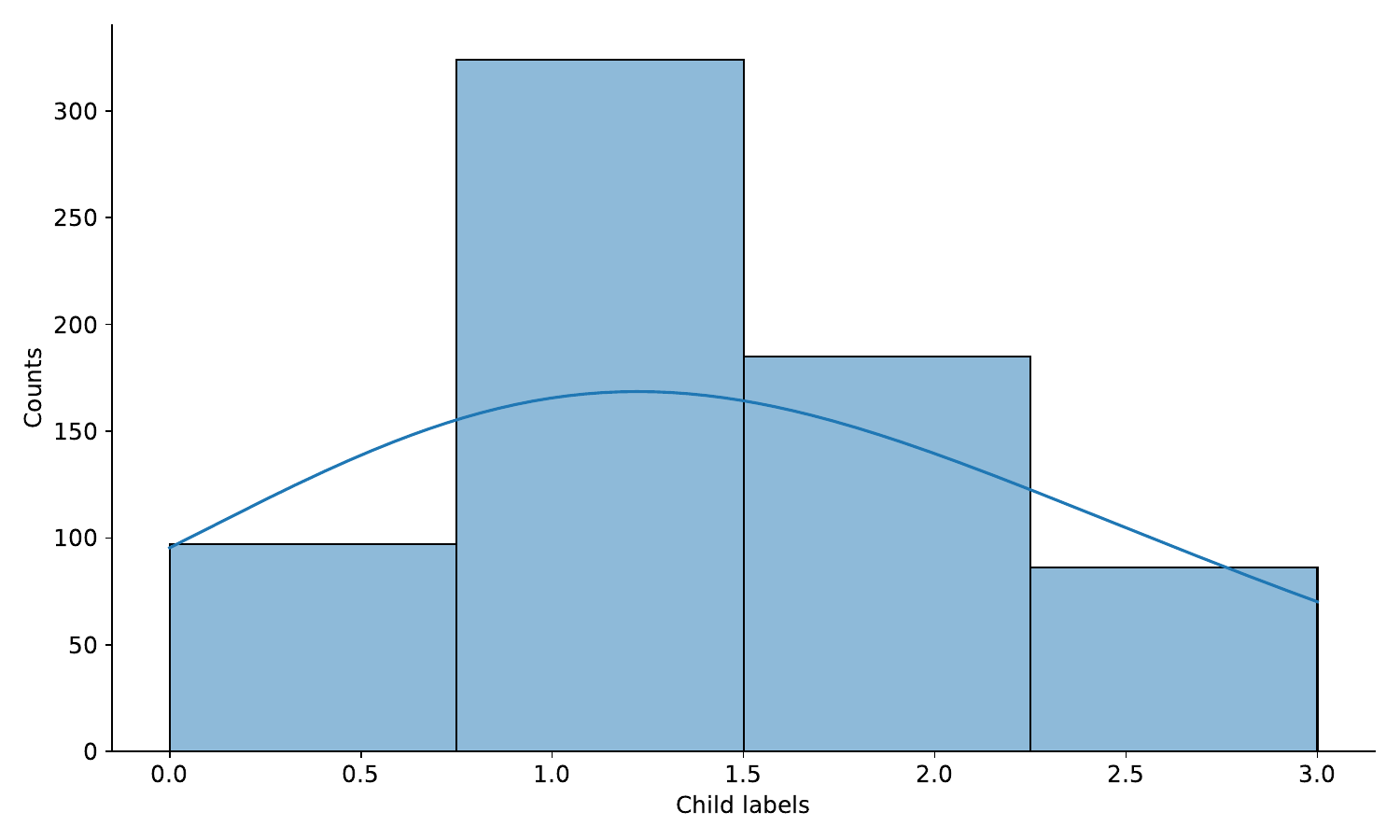} 
        \caption{Journalist J: Histogram of posterior predictive outcome}
    \end{subfigure}
    \hfill
    \begin{subfigure}[b]{0.45\textwidth}
        \centering
        \includegraphics[width=\textwidth]{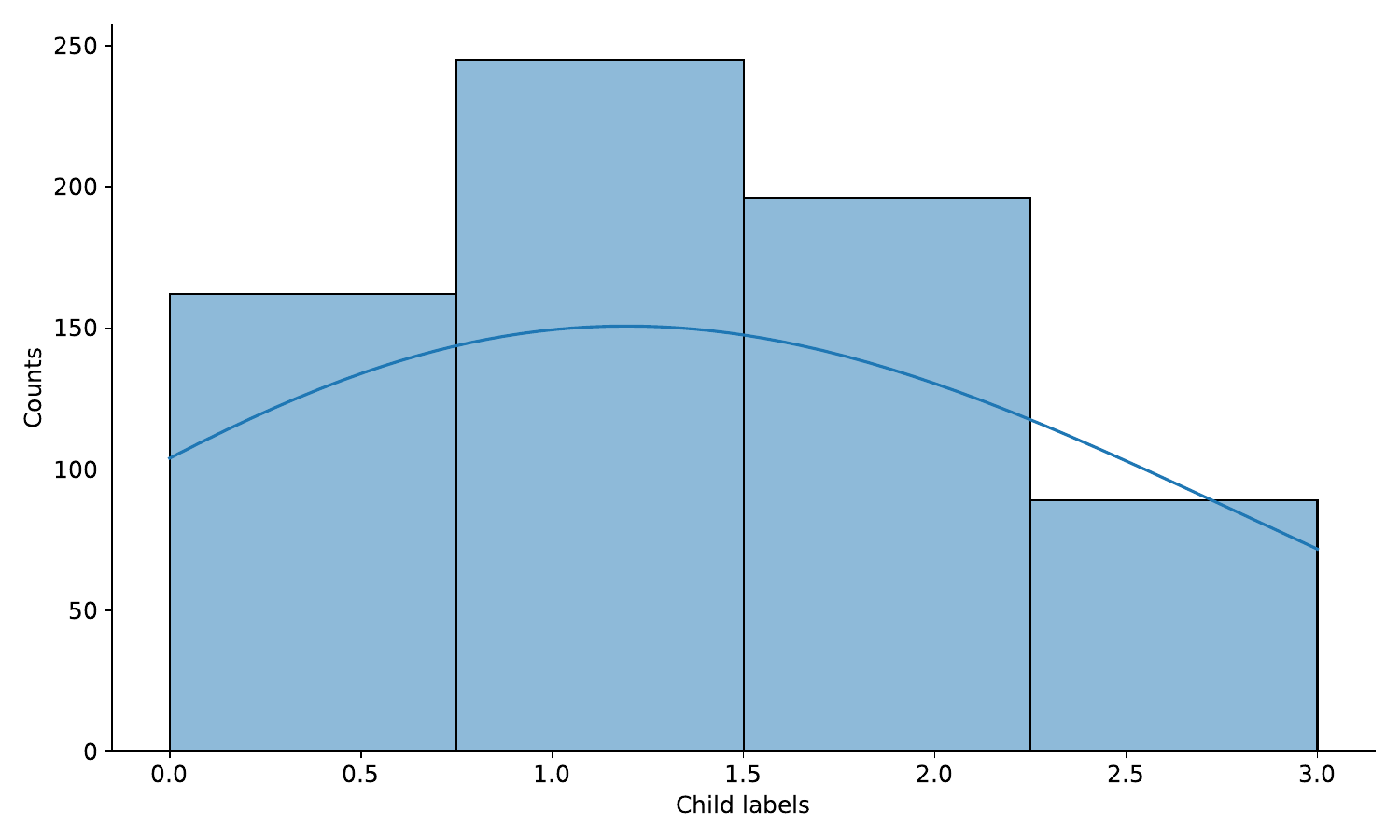} 
        \caption{Journalist J: Histogram of ground truth outcome distribution}
    \end{subfigure}
    \caption{Journalist J: Posterior predictive outcome distribution vs. ground truth outcome distribution}
\end{figure}

\begin{figure}[htbp]
    \begin{subfigure}[b]{\textwidth}
        \centering
        \includegraphics[width=\textwidth]{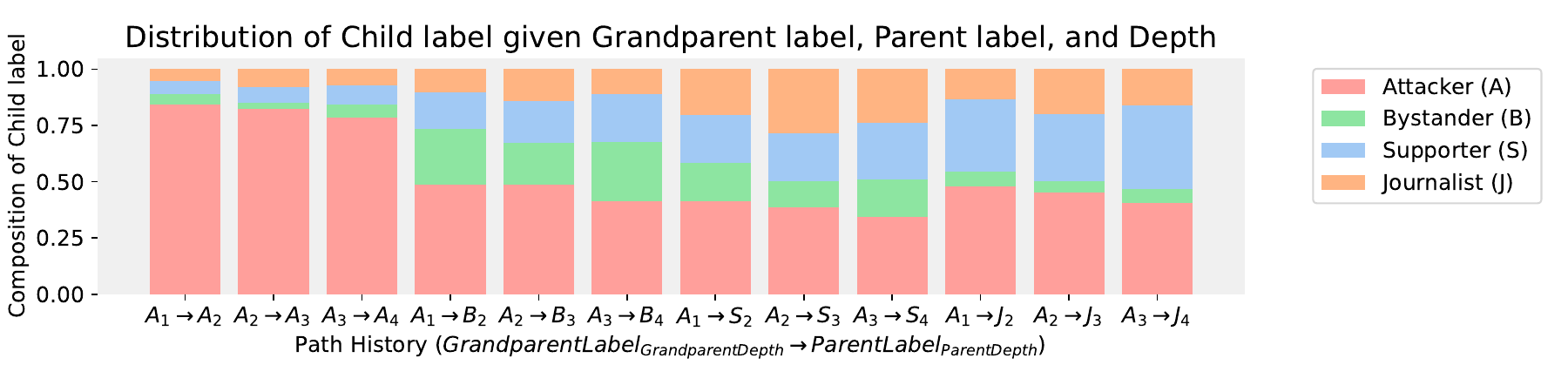} 
        \caption{Journalist K: attacker grandparent.}
    \end{subfigure}
    \begin{subfigure}[b]{\textwidth}
        \centering
        \includegraphics[width=\textwidth]{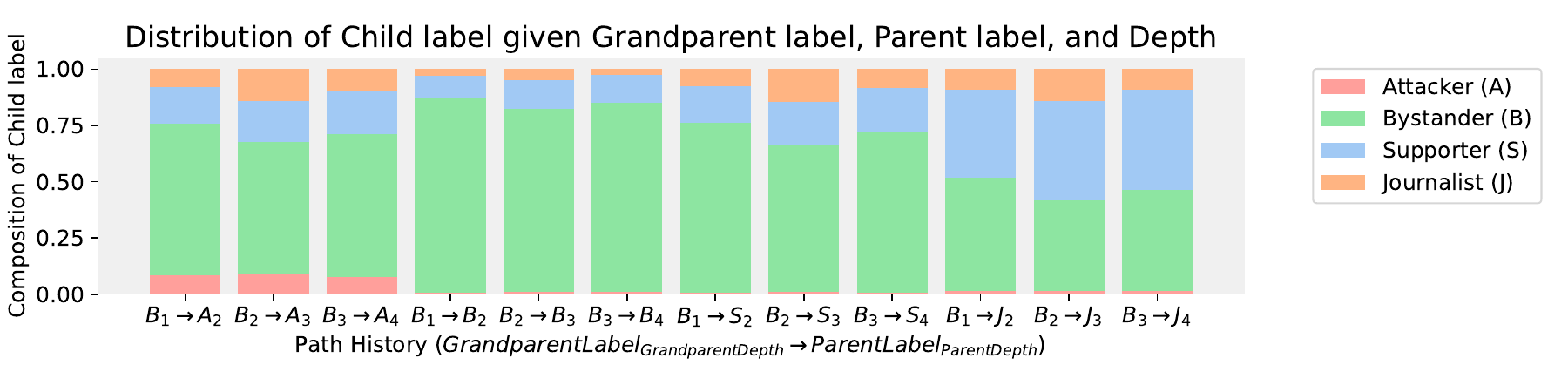} 
        \caption{Journalist K: bystander grandparent.}
    \end{subfigure}
    \begin{subfigure}[b]{\textwidth}
        \centering
        \includegraphics[width=\textwidth]{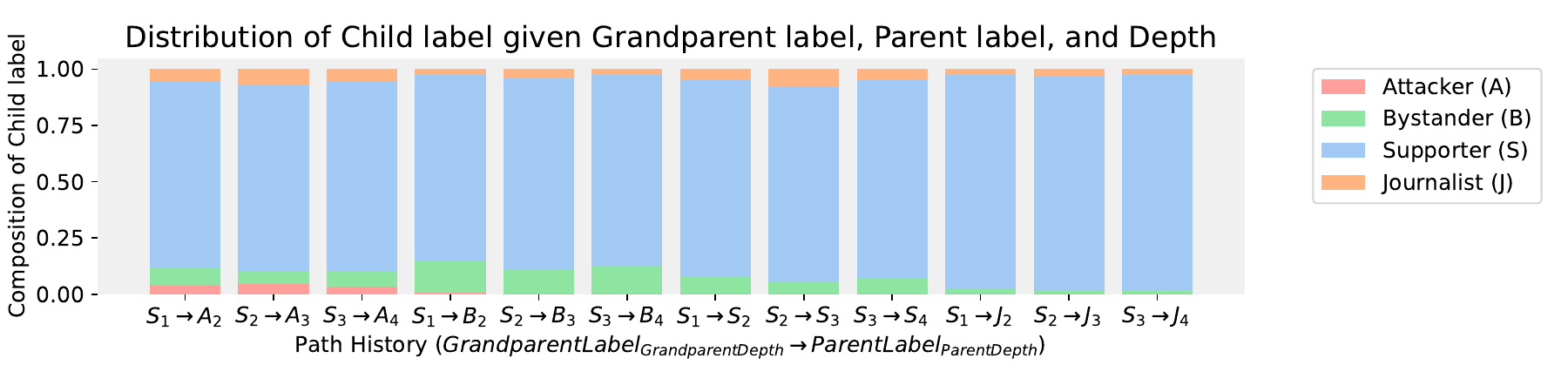} 
        \caption{Journalist K: supporter grandparent.}
    \end{subfigure}
    \begin{subfigure}[b]{\textwidth}
        \centering
        \includegraphics[width=\textwidth]{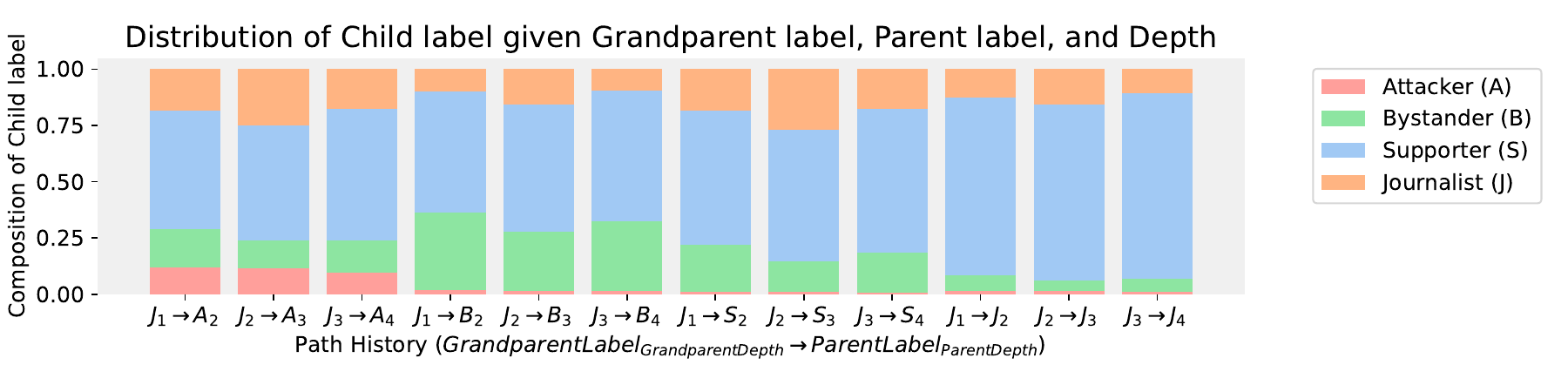} 
        \caption{Journalist K: journalist grandparent.}
    \end{subfigure}
    \caption{Journalist K: Effect of grandparent and parent label on child label distribution}
\end{figure}

\begin{figure}[ht]
    \centering
    \begin{subfigure}[b]{0.45\textwidth}
        \centering
        \includegraphics[width=\textwidth]{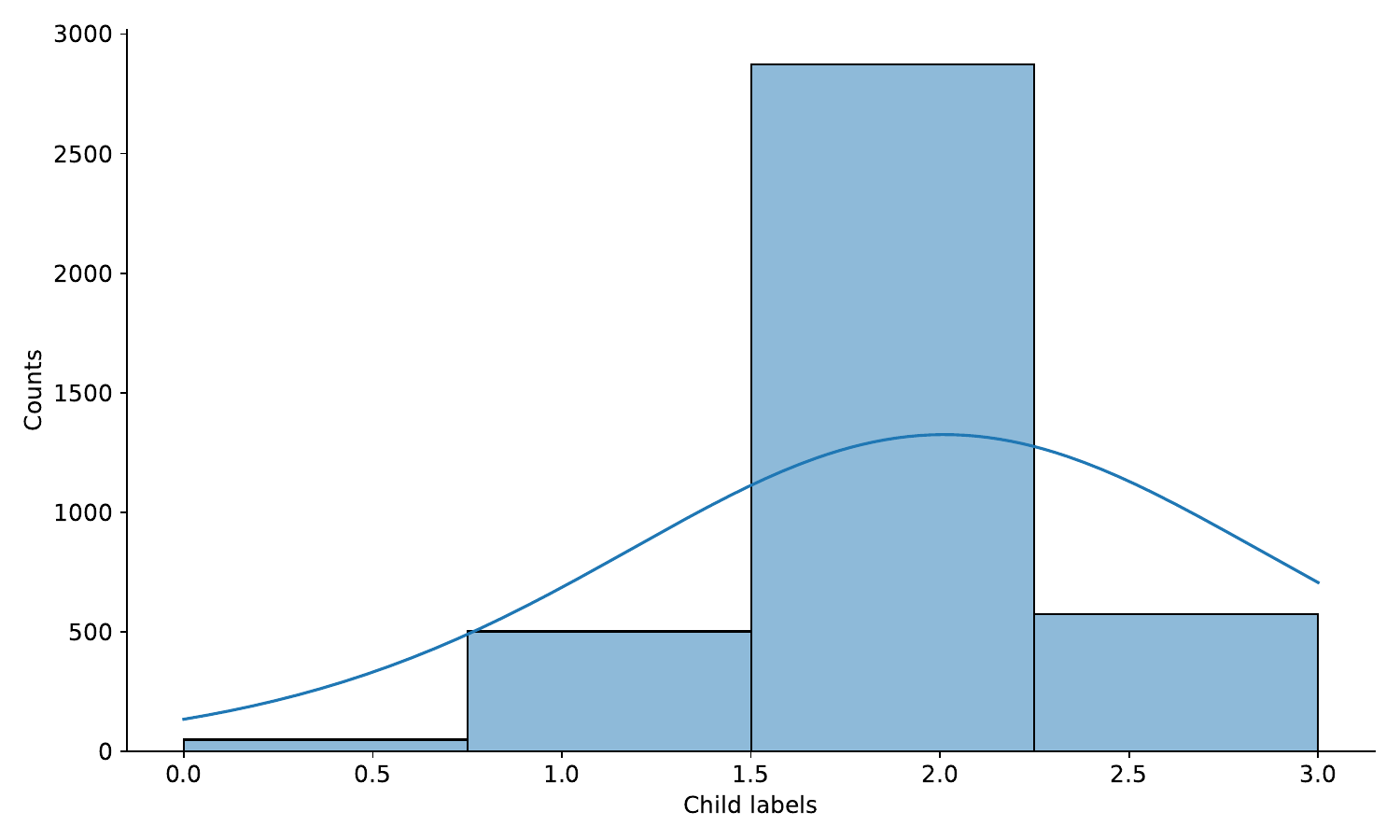} 
        \caption{Journalist K: Histogram of posterior predictive outcome}
    \end{subfigure}
    \hfill
    \begin{subfigure}[b]{0.45\textwidth}
        \centering
        \includegraphics[width=\textwidth]{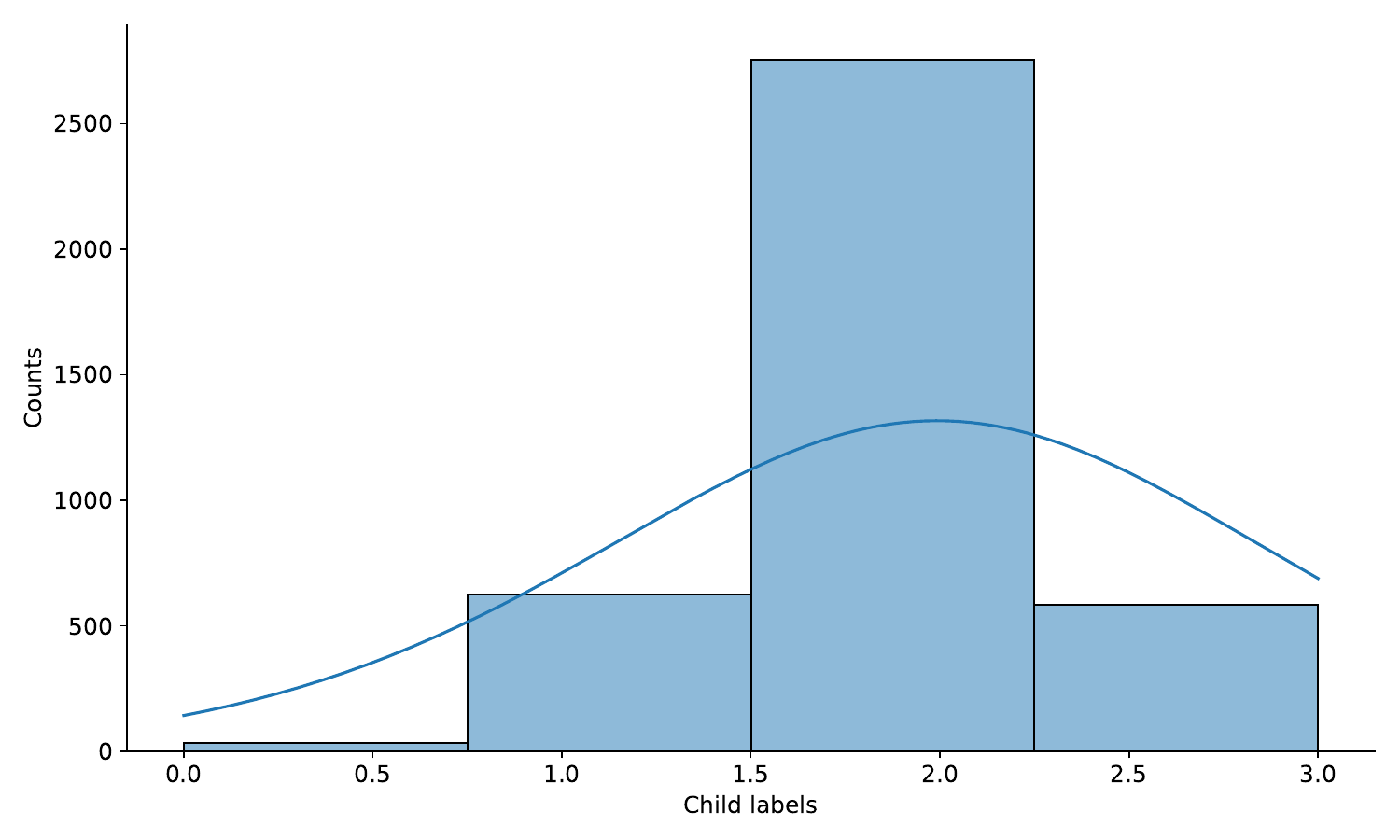} 
        \caption{Journalist K: Histogram of ground truth outcome distribution}
    \end{subfigure}
    \caption{Journalist K: Posterior predictive outcome distribution vs. ground truth outcome distribution}
\end{figure}

\begin{figure}[htbp]
    \begin{subfigure}[b]{\textwidth}
        \centering
        \includegraphics[width=\textwidth]{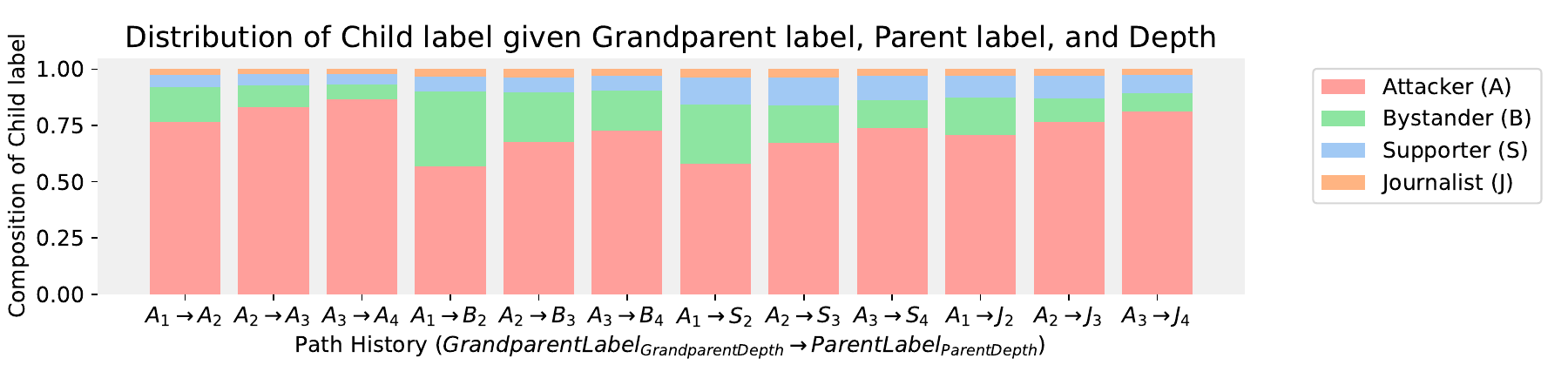} 
        \caption{Journalist L: attacker grandparent.}
    \end{subfigure}
    \begin{subfigure}[b]{\textwidth}
        \centering
        \includegraphics[width=\textwidth]{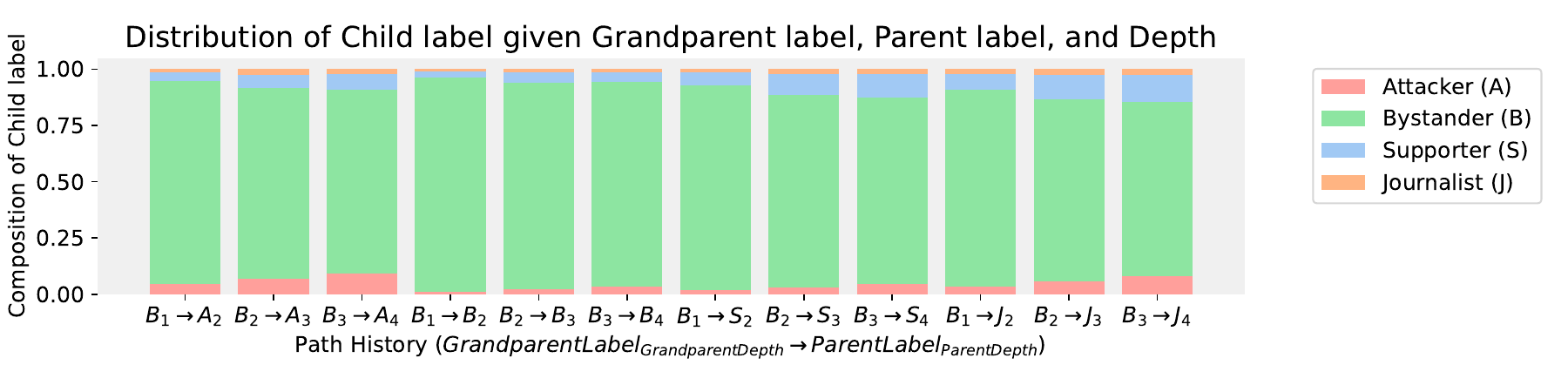} 
        \caption{Journalist L: bystander grandparent.}
    \end{subfigure}
    \begin{subfigure}[b]{\textwidth}
        \centering
        \includegraphics[width=\textwidth]{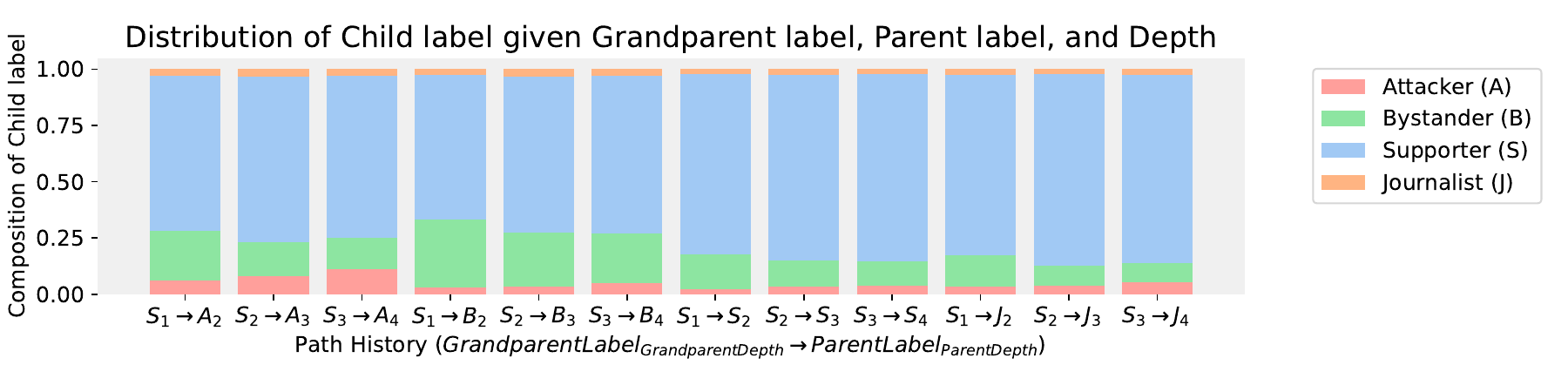} 
        \caption{Journalist L: supporter grandparent.}
    \end{subfigure}
    \begin{subfigure}[b]{\textwidth}
        \centering
        \includegraphics[width=\textwidth]{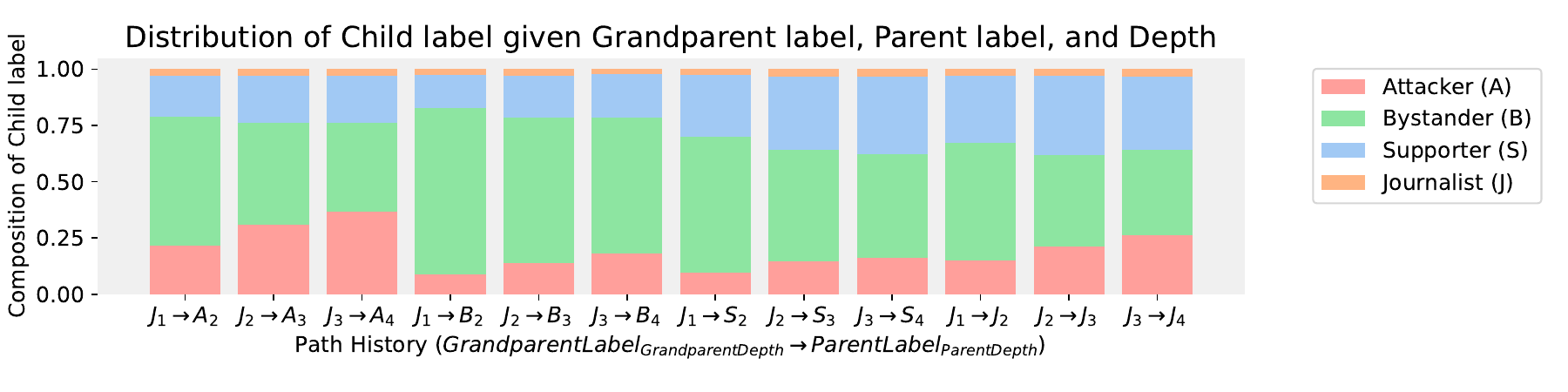} 
        \caption{Journalist L: journalist grandparent.}
    \end{subfigure}
    \caption{Journalist L: Effect of grandparent and parent label on child label distribution}
\end{figure}

\begin{figure}[ht]
    \centering
    \begin{subfigure}[b]{0.45\textwidth}
        \centering
        \includegraphics[width=\textwidth]{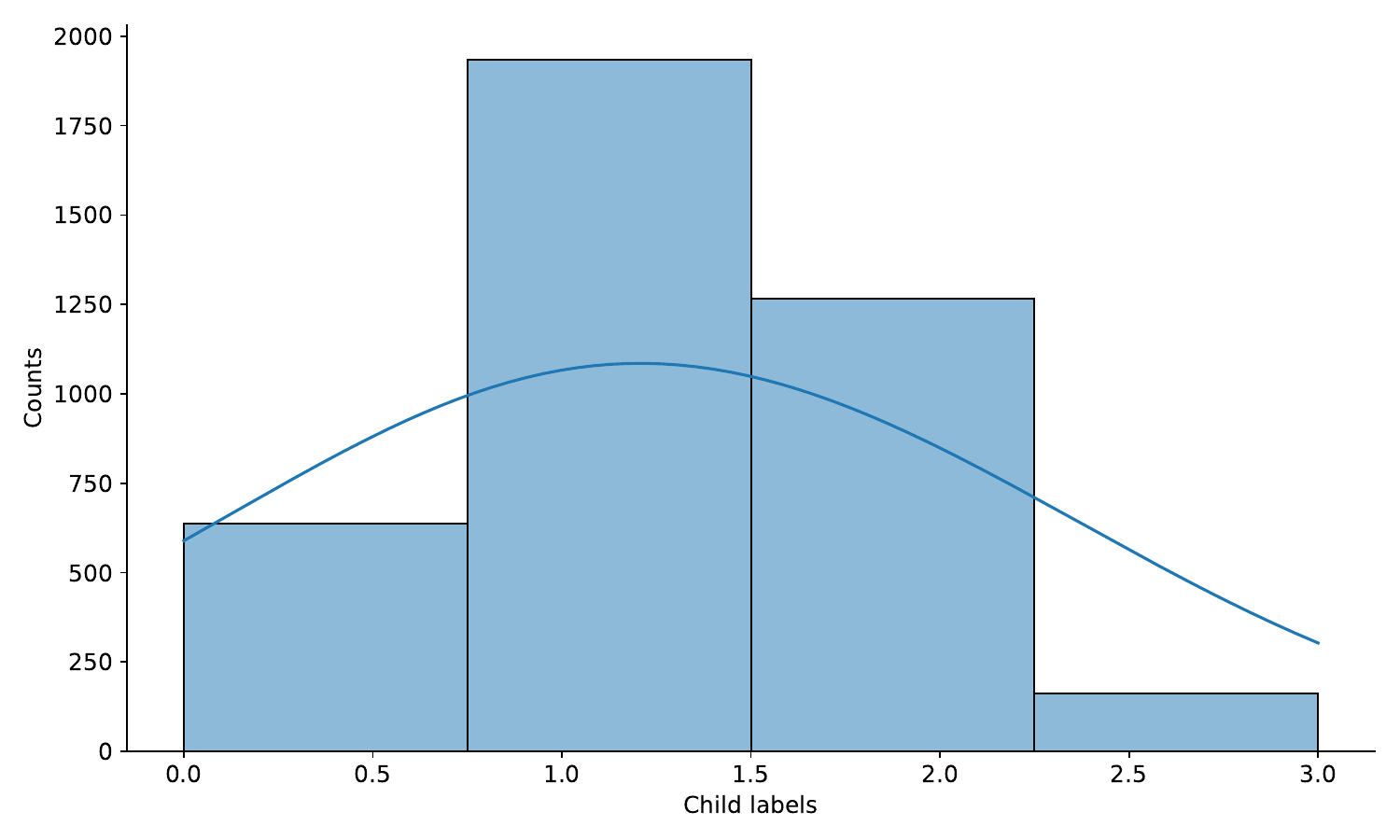} 
        \caption{Journalist L: Histogram of posterior predictive outcome}
    \end{subfigure}
    \hfill
    \begin{subfigure}[b]{0.45\textwidth}
        \centering
        \includegraphics[width=\textwidth]{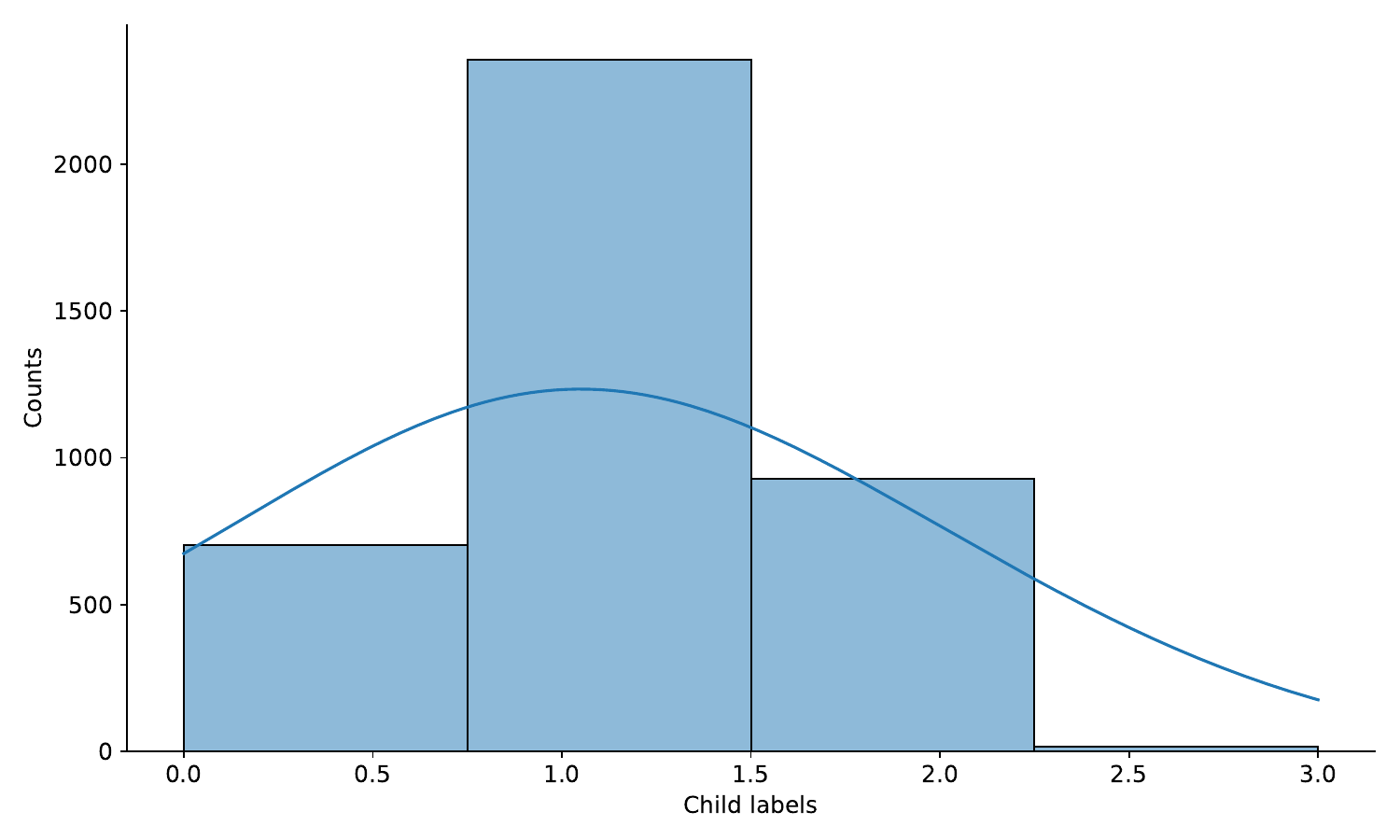} 
        \caption{Journalist L: Histogram of ground truth outcome distribution}
    \end{subfigure}
    \caption{Journalist L: Posterior predictive outcome distribution vs. ground truth outcome distribution}
\end{figure}

\begin{figure}[htbp]
    \begin{subfigure}[b]{\textwidth}
        \centering
        \includegraphics[width=\textwidth]{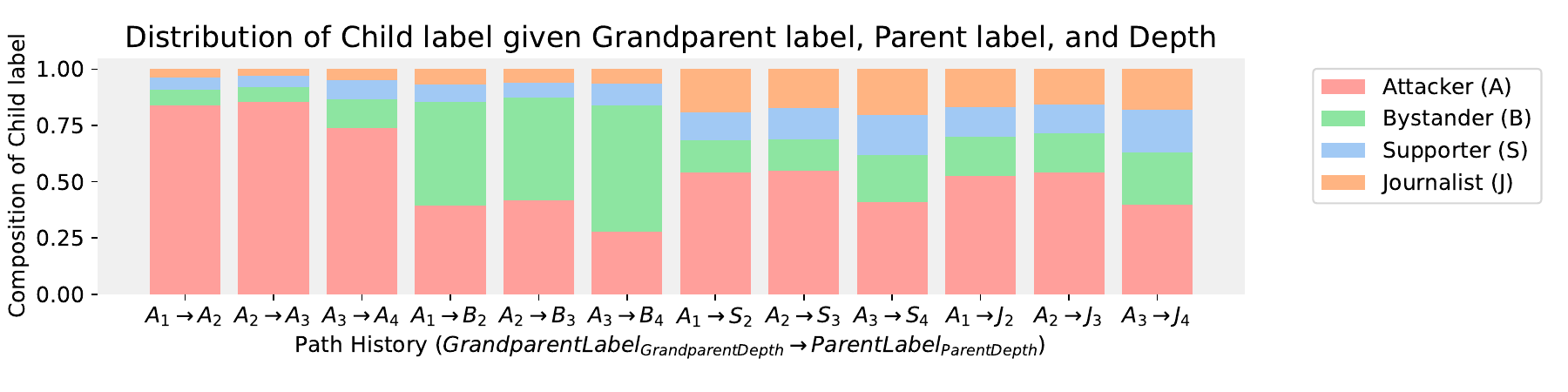} 
        \caption{Journalist M: attacker grandparent.}
    \end{subfigure}
    \begin{subfigure}[b]{\textwidth}
        \centering
        \includegraphics[width=\textwidth]{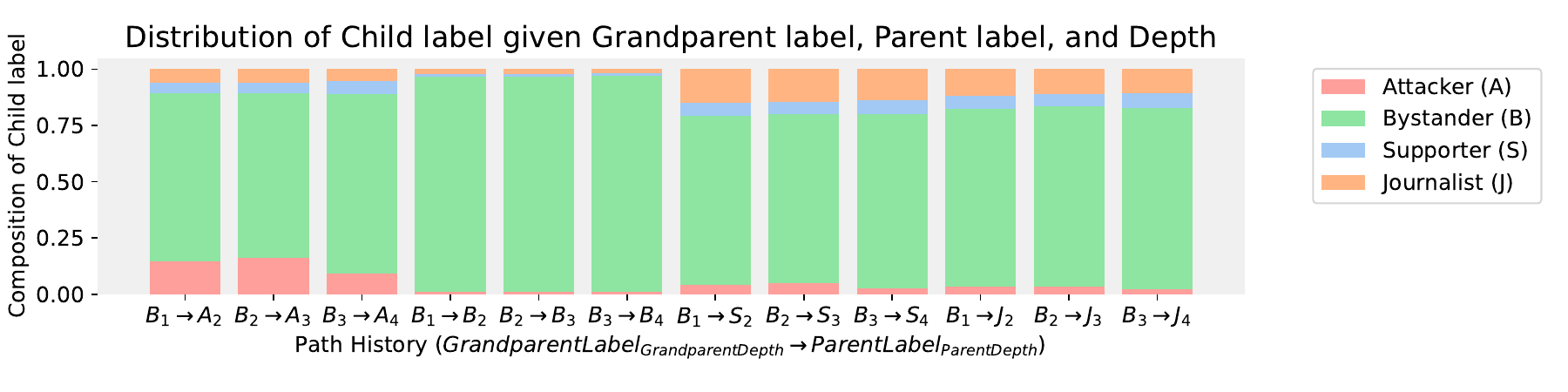} 
        \caption{Journalist M: bystander grandparent.}
    \end{subfigure}
    \begin{subfigure}[b]{\textwidth}
        \centering
        \includegraphics[width=\textwidth]{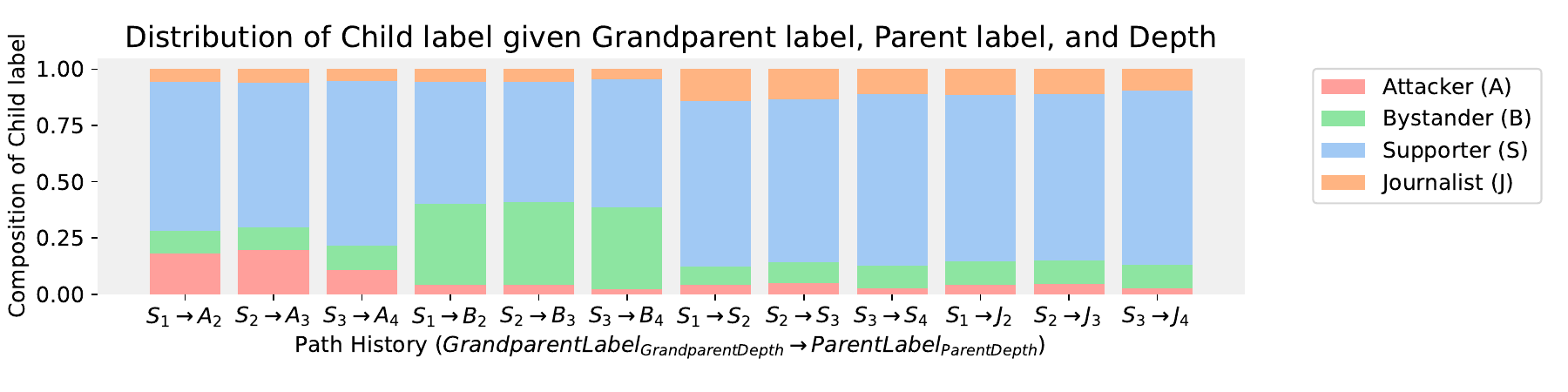} 
        \caption{Journalist M: supporter grandparent.}
    \end{subfigure}
    \begin{subfigure}[b]{\textwidth}
        \centering
        \includegraphics[width=\textwidth]{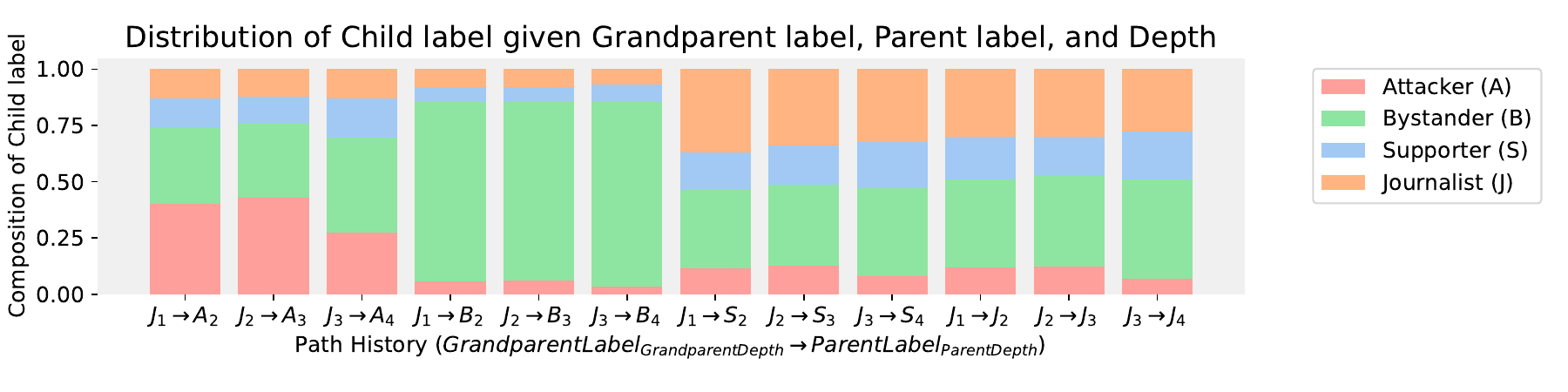} 
        \caption{Journalist M: journalist grandparent.}
    \end{subfigure}
    \caption{Journalist M: Effect of grandparent and parent label on child label distribution}
\end{figure}

\begin{figure}[ht]
    \centering
    \begin{subfigure}[b]{0.45\textwidth}
        \centering
        \includegraphics[width=\textwidth]{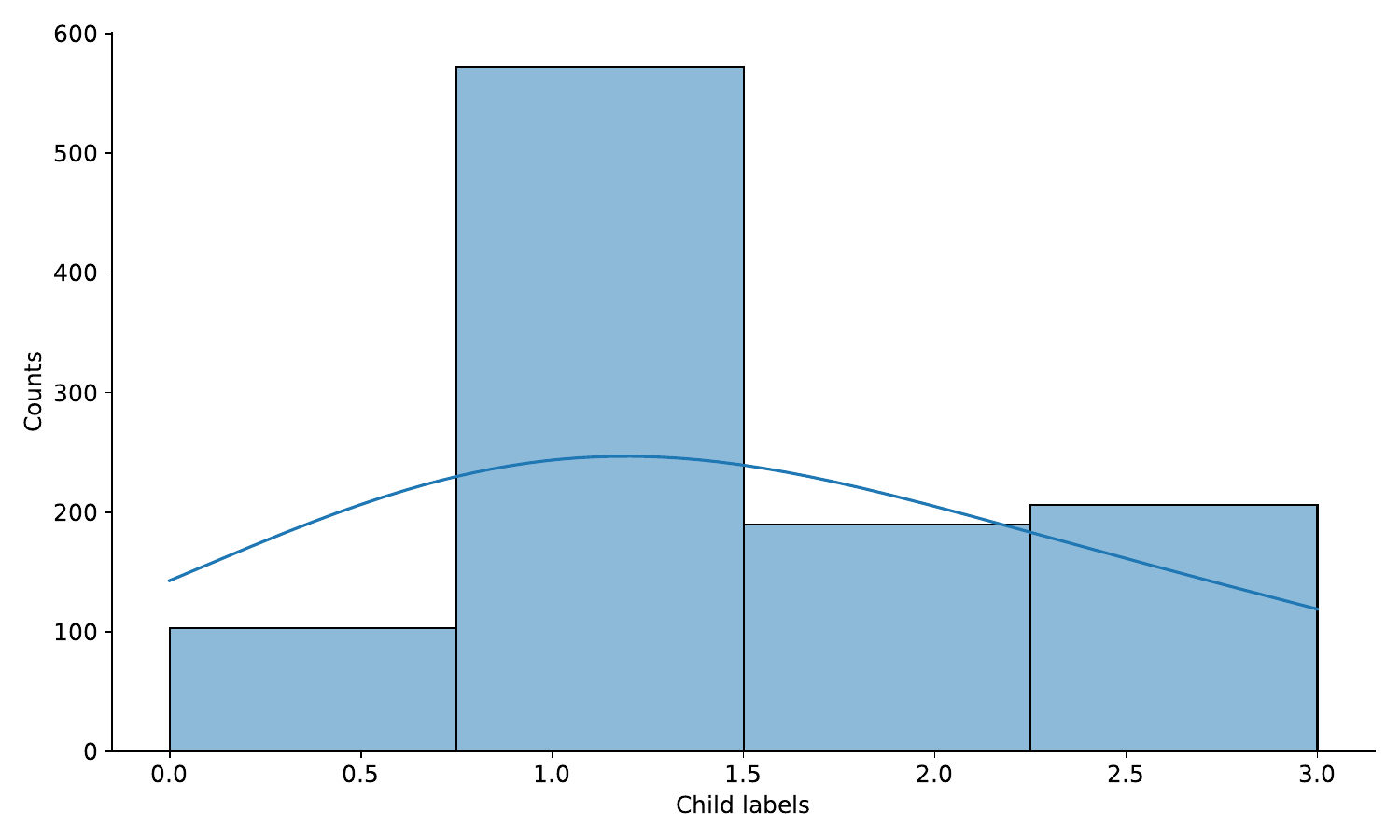} 
        \caption{Journalist M: Histogram of posterior predictive outcome}
    \end{subfigure}
    \hfill
    \begin{subfigure}[b]{0.45\textwidth}
        \centering
        \includegraphics[width=\textwidth]{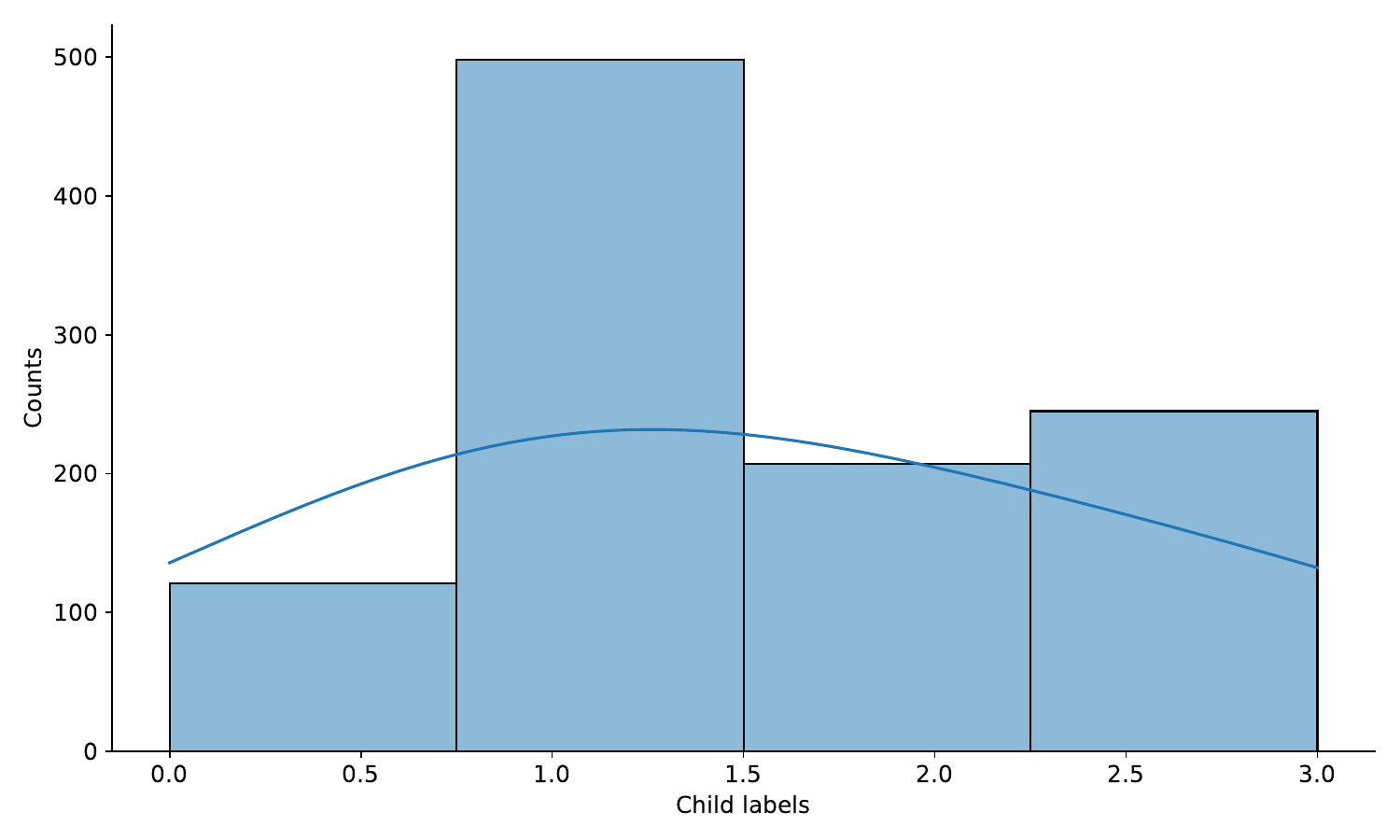} 
        \caption{Journalist M: Histogram of ground truth outcome distribution}
    \end{subfigure}
    \caption{Journalist M: Posterior predictive outcome distribution vs. ground truth outcome distribution}
\end{figure}

\end{document}